%% file: PhDThesis.tex
\documentclass[11pt,twoside,openleft]{book}

\usepackage[letterpaper,lmargin=2.9cm,rmargin=1.9cm,bmargin=2.8cm,tmargin=2.8cm,includeheadfoot]{geometry} 

\usepackage{graphicx}  
\usepackage{tikz, tikz-3dplot}
\usepackage{flafter}  
\usepackage[list=off,font=small,skip=3pt]{caption}
\usepackage{subcaption}

\usepackage{lipsum}

\usepackage{cite}

\usepackage{amsthm}
\usepackage{amsmath,amssymb}  
\usepackage{textcomp} 
\usepackage{bm}  

\usepackage[utf8]{inputenc} 
\usepackage[T1]{fontenc}
\usepackage[spanish,slovene]{babel}

\usepackage{booktabs}
\usepackage{enumitem}
\setlist{itemsep=0.5pt}

\usepackage[plain]{fancyref}
\usepackage{hyperref}  

\usepackage{fancyhdr}
\usepackage{aas_macros} 
\usepackage[]{tocbibind} 

\makeatletter 
\DeclareRobustCommand*{\bfseries}{%
  \not@math@alphabet\bfseries\mathbf
  \fontseries\bfdefault\selectfont
  \boldmath
}
\makeatother

\begin{document}
	
	\selectlanguage{spanish}

    \renewcommand{\tablename}{Tabla}

    \renewcommand{\thefootnote}{\Roman{footnote}}

    \lefthyphenmin=2
    \righthyphenmin=2

    \frontmatter

    \input{./titlepage}
   \input{./dedicatoria}
	\input{./agradecimientos}

    \pagestyle{empty}

\input{./abstract}
    \tableofcontents


    \mainmatter

    \pagestyle{fancy}
    \pagestyle{plain}
    \input{./introduccion}
    \input{./cap1}

    \input{./cap2}

	\input{./cap2_5}
    \input{./cap3}

    \input{./cap4}

    \input{./cap5}
    \input{./conclusiones}

    \appendix
    \addcontentsline{toc}{chapter}{Apéndices}
     \input{./appA}
     \input{./appC}

     \input{./appD}

      \input{./appH}

  \backmatter

  \bibliographystyle{ieeetr}  
  \bibliography{bibG} 	

\end{document}

%% file: titlepage.tex
\pagestyle{empty}

\newcommand{\HRule}{\rule{\linewidth}{1pt}} 

\begin{titlepage}
	
				
\begin{center}

\vfill

\Large

Universidad de La Habana

\medskip

Facultad de Física

\vspace{10mm}


\vfill

{\Large TESIS  \\[5pt] \emph{presentada en opción al grado de}  \\[5pt] \textbf{Doctor en Ciencias Físicas}} \\[30pt]


{\LARGE \bfseries \MakeUppercase{Efectos del campo magn\'etico en un gas de bosones vectoriales neutros: aplicaciones astrof\'isicas} \\[5pt] \MakeUppercase{}}




\vfill

\begin{tabular}{rl}

\textbf{Autora:} & M.C. Gretel Quintero Angulo \\
\noalign{\vspace{10pt}}

\textbf{Tutora:} & Dra. Aurora Pérez Martínez, ICIMAF \\[5pt]
			      
\noalign{\vspace{2mm}}
\end{tabular}

\vfill

\Large
La Habana, 2019

\end{center}


\end{titlepage}
\cleardoublepage

%% file: dedicatoria.tex
\indent

\vfill

\begin{quote}
\begin{flushright}
	\textit{"Que así viviendo de respirar, no piensas que respiras, y solo ves la fruta en la bandeja,\\ y caminas bajo las estrellas como si fuera una costumbre y no un prodigio."}
	
\vspace{12pt}
\textit{Dulce María Loynaz,\\}
\textit{Tr\'iptico de San Mart\'in}
\end{flushright}

\end{quote}

\vfill
\cleardoublepage

\pagestyle{plain}

%% file: agradecimientos.tex
\begin{center}
	\section*{Agradecimientos}
\end{center}


\begin{quote}
	
Uno de los momentos más felices durante la escritura de la tesis lo tuve al darme cuenta de que los agradecimientos sobrepasarían las dos páginas. En él fui consciente de la gran cantidad de personas que me han ayudado y acompañado en estos últimos años. 

A todas expreso ahora el deseo de que mi paso por sus vidas esté siendo para ellas tan beneficioso como lo ha sido para mí.

\begin{flushright}
	
	Hecho esto, comenzaré agradeciendo a mi familia, en especial a mis abuelas, dos mujeres excepcionales a quienes está dedicada la tesis; 
	
	a mi mamá, por su cariño y porque sin su apoyo doméstico yo no podría investigar;
	
	a mi papá, que además del apoyo doméstico, contribuyó muchísimo a la calidad final de la tesis escrita, la cual tuvo la paciencia de leer más de una vez a la caza de errores tipográficos y de redacción;
	
	a mi hermana, que siempre supo que yo lo lograría;
	
	a Dariel, que casi es otro hermano;
	
	y a mis tías Celia, Aleida, Herminia y Carmín, a mi tío Coqui y a mis primas Adrianee, Ayleem e Ingrid, por su sustento tanto espiritual como material, muchas veces desde la distancia.
	
	\vspace{12pt}
	
	Igualmente agradezco a mis amigos, dondequiera que estén, y especialmente:
	
	a Martica y Akemi, siempre;
	
	a Etién, básicamente por ser Etién;
	
	a Marlon, que compartió mis momentos más oscuros y me acompañó por el largo camino hasta la luz;
	
	a Ibis, Mirna y Cecilia, siempre preocupadas por mi familia y por mí;
	
	a Iris y Erick, por esas buenas y largas conversaciones sobre lo mundano y lo esotérico;
	
	a Diana, por las infinitas horas de trabajo compartidas y por ser de esas personas para quienes decir y hacer vienen siendo lo mismo;
	
	a Sandra, porque hacía rato que ninguna de las dos tenía una amiga nueva, y ahora podemos odiar y amar juntas, y chismear principalmente;
	
	a Lismary, que tuvo el valor de ponerse en mis manos para que la dirigiera en su maestría, algo que yo nunca hubiera hecho pero que va saliendo muy bien;
	
	a Joeluis, por una amistad en la que sobran las palabras;
	
	a David, quien a pesar de sus perretas, o precisamente por ellas, me recuerda mucho a mí cuando tenía su edad;
	
	a mis compañeros de los locales de adiestrados, quienes aguantaron con más paciencia de la que merezco mi estrés de la predefensa, en particular a Elena y Lídice, las chicas del café, a Yanela, la chica de la cerveza, y a los chicos del 32 (el local más \textit{cool} de la Facultad de Física): Antonio, Adrián, Frank, Vicente, Andy, Joe, Laciel, Ernesto, Michael, Alfredito y Gustavo; 
	
	y a toda la gente que ha experimentado el teatro a través del Monopolo Magnético: Sandra, Isabel, Joeluis, Ale Ramos, David, Lismary, Lídice, Adrián, Landy, Gustavo, Antonio, Vicente, Víctor y Gabriel, por seguirme en esta aventura gracias a la cual ahora nos conocemos mejor. 
	
	\vspace{12pt}		
	
	Quiero además agradecer a los trabajadores y profesores de la Facultad Física, en particular a dos personas cuyas magnificas clases recuerdo siempre, Eddy Jiménez y Medel Pérez, y a aquellos con los que no solo aprendí en el aula, sino trabajando juntos en las más disimiles tareas: Maite, Marín, Maruchy y Carlos Rodríguez;
	
	y también doy ahora gracias a una persona sin la cual nunca hubiera logrado defender la tesis, Mercy, el ángel que nos salva de morir aplastados por la burocracia.
	
	\vspace{12pt}
	
	Para finalizar mencionaré al Departamento de Física Teórica del ICIMAF, pero no al lugar, se entiende, sino a las personas que lo componen, porque allí encontré el entorno ideal para la investigación científica, un ambiente en el que prima la solidaridad sobre la competencia y en el que cada quien pone su granito de arena para que los cálculos, los artículos, los seminarios, las presentaciones en eventos, los viajes, los conciertos de rock, las obras de teatro, las fiestas y hasta las salidas con los novios salgan siempre lo mejor posible. 
	
	Por ello quiero dar las gracias a Nilda, que escaneó todos los cálculos de la tesis y que, sobre todo en las etapas finales de su escritura, más de una vez esperó por Aurora y por mí hasta pasadas las ocho de la noche para cerrar el edificio;
	
	a Yamila y su sexto sentido para detectar cuando estamos necesitados de café;
	
	a Eru, que siempre nos recibe con una sonrisa;
	
	al Gabi, que todos los días llegaba al instituto dispuesto a terminar su programa, pero acaba sumándose a lo que fuera que Aurora, Diana, Daryel y yo estuviéramos discutiendo;
	
	a Duvier, que para toda situación tensa tiene un chiste;
	
	a Daryel, porque todo pasa;
	
	a Frank, por ayudarme a armar la historia que conté en la presentación final;
	
	a Lídice, Diana, Lismary y Samantha, las estrellitas de Aurora; 
	
	a Elizabeth, por la dulzura con que en los seminarios hace siempre las preguntas más difíciles;
	
	a los profes Cabo y Augusto, y a Vicky por sus buenas sugerencias y críticas;
	
	a nuestros colaboradores Ernesto Contreras y Pedro Bargueño, con quienes aprendí mucho de Relatividad General, y comprobé que investigar vía WhatsApp es posible;
	
	al profesor Omair Zubairi, quien muy amablemente accedió a contestar por esta misma vía todas mis dudas acerca de sus trabajos en la modelación de objetos compactos no esféricos;

	a Hugo, porque en un momento en el que mi amor por la Fisica había muerto, su pasión por el conocimiento me hizo recuperarlo de golpe en una de nuestras primeras conversaciones sobre esos vacíos cuánticos poblados de partículas; 
	
	y muy especialmente a Aurora, cuyas energías inagotables son las que inspiran el trabajo de todos los que estamos a su alrededor; cuya voluntad de excelencia es la responsable de que mi tesis haya estado lista en dos años y dos meses; y cuya bondad y sentido de lo humano la hacen tener en cada colega o estudiante un amigo.

\end{flushright}

\end{quote}

\vfill
\cleardoublepage

\pagestyle{plain}

%% file: abstract.tex
\vfill

\begin{center}
\section*{Resumen}
\addcontentsline{toc}{chapter}{Resumen}
\end{center}
\medskip

\begin{quote}
	
Esta tesis est\'a dedicada al estudio de los efectos del campo magn\'etico en la materia que forma las Estrellas de Neutrones suponiendo que los neutrones en su interior pueden estar parcial o totalmente apareados formando bosones vectoriales neutros. 

Las propiedades termodinámicas de un gas ideal de bosones vectoriales neutros magnetizado se investigan a partir del espectro obtenido de la teor\'ia de Proca para una y tres dimensiones espaciales. Los principales efectos del campo magn\'etico sobre el gas son, primero, favorecer la transici\'on de fase al condensado de Bose-Eintein, y segundo, separar la presi\'on en dos componentes, una paralela y otra perpendicular al eje magn\'etico, dando lugar a ecuaciones de estado anisotr\'opicas. A bajas densidades o altos campos magnéticos la menor de las presiones puede ser negativa, y el sistema de bosones deviene susceptible de sufrir un colapso magn\'etico. Por otra parte, cuando la temperatura es lo suficientemente baja, el gas puede automagnetizarse.

Las ecuaciones de estado del gas magnetizado de bosones vectoriales neutros se aplican, en primer lugar, al estudio del colapso magn\'etico de un gas ideal de neutrones, electrones y protones (gas \textit{npe}) en el que los nucleones est\'an parcialmente bosonizados. A partir de los resultados obtenidos, se propone un modelo para la expulsi\'on de materia de la estrella y la formaci\'on de \textit{jets} astrof\'isicos. 

Posteriomente, las ecuaciones de estado del gas vectorial magnetizado se utilizan para el estudio de la influencia del campo magn\'etico en la estructura (masa, radio y deformaci\'on) de Estrellas de condensado de Bose-Einsten, es decir, de objetos compactos compuestos en su totalidad por un gas de neutrones completamente bosonizados. A fin de tener en cuenta de manera apropiada la anisotrop\'ia en las ecuaciones de estado de este objeto, en la tesis se construye un sistema de ecuaciones para describir el equilibrio hidrodin\'amico -la estructura- de objetos compactos esferoidales a partir de una m\'etrica axisim\'etrica. Los resultados obtenidos validan a la automagnetizaci\'on de los bosones como un mecanismo plausible para la generaci\'on del campo magnético en los Objetos Compactos.

\end{quote}

\vfill
\cleardoublepage

\begin{center}
\section*{Abstract}
\addcontentsline{toc}{chapter}{Abstract}
\end{center}
\medskip

\begin{quote}

In this thesis, the effects of the magnetic field on neutron stars matter are studied, assuming that the neutrons inside these stars are partially or totally paired forming neutral vector bosons.

The thermodynamic properties of an ideal gas of magnetized neutral vector bosons are investigated in the frame of Proca theory, for one and three spatial dimensions. The main effects of the magnetic field on the gas are, first, to favor the phase transition to the Bose-Eintein condensation, and second, to split the pressure in two components, one parallel and the other perpendicular to the magnetic axis, giving rise to anisotropic equations of state. At low densities or high magnetic fields, the lowest pressure can be negative, and the boson system becomes susceptible to suffer a transverse magnetic collapse. On the other hand, when the temperature is low enough, the gas can spontaneously magnetize.

The equations of state of the magnetized gas of neutral vector bosons are applied, firstly, to study the magnetic collpase of an ideal gas of neutrons, electrons and protons (\textit{npe} gas) in which the nucleons are partially bosonized. From the results obtained, a model for the ejection of matter out of the star and the formation of astrophysical \textit {jets} is proposed.

Subsequently, the equations of state of the magnetized vector gas are used to study the magnetic field influence on the structure (mass, radius and deformation) of Bose-Einsten condensate stars, that is, of compact objects composed entirely by a neutron gas fully bosonized. In order to appropriately take into account the magnetic anisotropy in the equations of state of this object, we develope a system of equations to describe the hydrodynamic equilibrium -the structure- of spheroidal compact objects starting from an axisymmetric metric. The obtained results validate the bosons self-magnetization as a feasible source for the magnetic field of compact objects.

\end{quote}

\vfill
\cleardoublepage

%% file: introduccion.tex
\chapter*{Introducción} \label{intro}
\addcontentsline{toc}{chapter}{Introducción}

Con masas en el orden de la masa del Sol\footnote{ La masa del Sol es igual a $M_{\odot} = 1.989 \times 10^{30}$~kg, en tanto su radio es $R_{\odot} = 6.959 \times 10^{5}$~km. } confinadas en una pocas decenas de kil\'ometros, las Estrellas de Neutrones (ENs) son los \'unicos objetos conocidos en cuyo interior se alcanzan densidades mayores que la densidad nuclear $\rho_{nuc} = 2.7 \times 10^{14}$g$/$cm$^3$ \cite{Camezind,Lattimerprognosis}. Como a tan altas densidades y masas ni los efectos de las interacciones electromagn\'etica, d\'ebil y fuerte, ni los de la relatividad general pueden ser despreciados, ellas han devenido laboratorios naturales para la comprobaci\'on de las leyes m\'as b\'asicas de la F\'isica en unas  condiciones que a\'un est\'an lejos del alcance de la experimientaci\'on. Recientemente, las ENs han sido las protagonistas de una de las comprobaciones m\'as espectaculares de la Teor\'ia General de la Relatividad: la detecci\'on de manera directa, con los detectores gemelos del Observatorio de Ondas Gravitacionales de Interferometr\'ia Láser (LIGO) \cite{GW}, de ondas gravitacionales provenientes de la fusión de dos de estas estrellas. De hecho, la detecci\'on de ondas gravitacionales ha generado una serie de posibilidades observacionales que se espera devengan en los pr\'oximos a\~nos en una nueva forma de explorar las ENs y el Universo \cite{GW}. Por otra parte, a  partir de observatorios como el NICER \footnote{ El \textit{Neutron star Interior Composition Explorer} (NICER) es un observatorio astron\'omico espacial lanzado por la NASA (National Aeronautics and Space Administration of the United States) en junio de 2017. https://www.nasa.gov/nicer.}, construidos espec\'ificamente para recolectar informaci\'on acerca de la composici\'on de las ENs, se espera encontrar evidencias de la existencia en su interior de nuevas fases y tipos de materia  \cite{Lattimerprognosis}. Todo ello hace de las ENs unos de los escenarios astrof\'isicos m\'as fascinantes, a cuyo estudio se dedican numerosos esfuerzos y recursos tanto te\'oricos como observacionales.

Las ENs pueden encontrarse en los remanentes de las supernovas, solas o en sistemas binarios, e incluso se conoce una que tiene planetas \cite{Camezind}. Ellas forman parte de los llamados Objetos Compactos (OCs), al igual que las Enanas Blancas (EBs)  y los Agujeros Negros (ANs). Los OCs (considerando solo las EBs y las ENs) se diferencian de las estrellas t\'ipicas, tales como el Sol o las que podemos observar de noche a simple vista, en cuatro aspectos que son fundamentales para comprender la f\'isica que determina sus propiedades \cite{Shapiro,Camezind,103}:

\begin{itemize}

\item En general, las estrellas mantienen su forma aproximadamente esf\'erica gracias al equilibrio que se establece entre la  fuerza de gravedad, que empuja la materia hacia  el centro  de la estrella, y la presión que la materia ejerce hacia afuera. En las estrellas t\'ipicas,  la gravedad es contrarrestada por la presi\'on t\'ermica que proviene de las reacciones termonucleares que tienen lugar en su interior. 
En un OC la gravedad es contrarrestada por presiones de origen cu\'antico \cite{Shapiro,Camezind}. 
    
    \item Los OCs no son visibles a ojos desnudos. En el caso de las EBs ello se debe a su baja luminosidad, mientras que las ENs emiten mayormente en las regiones del espectro electromagn\'etico correspondientes a los rayos X, los rayos gamma, y las radiofrecuencias \cite{Shapiro,Camezind}.

\item Los OCs son mucho m\'as pequeños que las estrellas normales y en consecuencia tienen campos gravitatorios superficiales mucho más fuertes. A modo de comparación, téngase en cuenta que si bien tanto las EBs como las ENs tienen masas del orden de la masa del Sol, el orden de los radios típicos para las EBs es de $\sim 10^{-2}R_{\odot}$, mientras que para ENs es de $\sim 10^{-5}R_{\odot}$ \cite{Shapiro}.

\item Los OCs son fr\'ios. Esto puede sonar contradictorio, pues la temperatura de un OC justo despu\'es de su nacimiento est\'a en el orden de los $10^{11}$~K$\sim 10$~MeV, y posteriormiente disminuye hasta estabilizarse en el orden de los keV.  Sin embargo, estas temperaturas son peque\~nas si se comparan con el potencial qu\'imico de los bariones que forman la estrella: $T \leq 10$~MeV $\ll \mu$, pues $\mu \geq m_n \sim 939$~ MeV, y $m_n$ es la masa del neutr\'on. Por ello tomar $T=0$ cuando se modela un OC es, la mayor\'ia de las veces, una buena aproximaci\'on \cite{103}.

\item Con frecuencia los objetos compactos tienen campos magn\'eticos mucho m\'as intensos que los de las estrellas t\'ipicas. El campo magn\'etico superficial del Sol puede llegar hasta los $10^{3}$~G \cite{106}. En cambio, los campos magn\'eticos superficiales observados para EBs oscilan entre los $10^6$-$10^9$G, en tanto que para ENs pueden alacanzar valores de hasta $10^{15}$G \cite{Malheiro:2013loa}.

Comp\'arense estos campos magn\'eticos adem\'as con el campo magn\'etico de la Tierra, $B\sim 0.6$~G, el de un im\'an com\'un, $B \sim 100$~G, o los m\'as intensos campos magn\'eticos estacionarios obtenidos en los laboratorios terrestres, $B \sim 10^5 $~G \cite{103}.

\end{itemize}

La vida de los OCs comienza con la muerte de una estrella t\'ipica. A medida que la estrella envejece, el combustible nuclear en su interior disminuye y esta se enfr\'ia. Cuando el combustible nuclear se agota, la estrella colapsa -muere- ante la imposibilidad de contrarrestar el empuje de la gravedad. Durante el colapso, sus capas externas son expulsadas hacia afuera, mientras que el n\'ucleo, compuesto por uno o varios elementos met\'alicos en estado s\'olido, se comprime hasta  formar un objeto compacto \cite{Shapiro,Camezind}. Que el resultado del colapso sea una EB, una EN o un AN depender\'a de la masa de la estrella progenitora. Estrellas progenitoras con masas entre $M_{\odot}$ y $4 M_{\odot}$ aproximadamente dan lugar a EBs, mientras que si la masa est\'a entre $10 M_{\odot}$ y $25 M_{\odot}$ el resultado ser\'a una EN; estrellas con masas mayores se convierten, al morir, en agujeros negros \cite{Shapiro}.

La predicci\'on te\'orica de la existencia de las ENs precedi\'o en mas de treinta a\~nos a su descubrimiento observacional en 1967 \cite{1968Natur217709H}. Hacia los incios de la d\'ecada del $30$ del siglo pasado, ya las propiedades de las EBs hab\'{\i}an sido exitosamente descritas en los trabajos de Fowler \cite{1926MNRASF} y Chandrasekar \cite{1931ApJ....74...81C}, a partir de suponer que la estrella se sustenta por la presi\'on del gas degenerado de electrones en su interior, o presi\'on de Pauli. La clave para comprender c\'omo es posible que la presi\'on de un gas degenerado balancee a la gravedad en un objeto tan masivo radica en las altas densidades que se alcanzan durante la contracci\'on del n\'ucleo estelar que da lugar a la EB.

A medida que el n\'ucleo de la estrella progenitora se comprime la densidad electr\'onica aumenta, aumentando tambi\'en la energ\'ia de Fermi del sistema y, en consecuencia, la energ\'ia cin\'etica media de las part\'iculas y la presi\'on, hasta que esta \'ultima se hace lo suficientemente alta como para frenar la contracci\'on. Uno de los resultados fundamentales de los trabajos arriba mencionados fue el establecimiento te\'orico de una masa m\'axima para las EBs, ya que la presi\'on de Pauli ejercida por un gas de electrones solo puede sostener masas menores que $\sim 1.4  M_{\odot}$ \footnote{El valor exacto del l\'imite de Chandrasekar depende de la composici\'on de la EB. Por lo general, estas estrellas se suponen compuestas de carbono y ox\'igeno, pero otros elementos m\'as pesados como el hierro podr\'ian estar presentes \cite{Camezind}.}, un l\'imite que es hoy conocido como masa de Chandrasekar. 

En n\'ucleos estelares colapsantes con masas mayores que la de Chandrasekar, como el gas de electrones no puede frenar la contracci\'on, esta contin\'ua y con ella el aumento de la densidad y la energ\'ia de Fermi de dichas part\'iculas. La producci\'on del neutr\'on en los laboratorios en 1932 \cite{Chadwick:1932ma},  hizo comprender a varios cient\'ificos que, sin nada que se oponga a la contracci\'on del n\'ucleo estelar, eventualmente los electrones ganan energ\'ia suficiente como para que se produzca el decaimiento beta inverso, $e^-+p \rightarrow n + \nu_e$, dando lugar a una neutronizaci\'on de la materia que colapsa \cite{1934PNAS20254B,2013PhyU56289Y}. La densidad necesaria para que la neutronizaci\'on empiece es de aproximadamente  $1.4 \times 10^9$g/cm$^3$. En este punto, debido a que los electrones comienzan a combinarse con los protones de los n\'ucleos at\'omicos que conforman el n\'ucleo de la estrella, la presi\'on dismunuye y la contracci\'on del n\'ucleo estelar contin\'ua. Cuando la densidad en el interior alcanza los $4 \times 10^{11}$g/cm$^3$, los n\'ucleos at\'omicos comienzan a desintegrarse a trav\'es de un proceso conocido como goteo de neutrones; pasados los $10^{14}$g/cm$^3$ los n\'ucleos at\'omicos desaparecen y la materia que colapsa es en s\'i misma como un gran n\'ucleo formado mayormente por neutrones. A medida que la contracci\'on contin\'ua, la presi\'on del gas degenerado de neutrones aumenta hasta hacerse  comparable a la gravedad y el colpaso se detiene: se ha formado la EN.

Los primeros c\'alculos de la estructura, masas y radios, de una estrella compuesta totalmente por un gas degenerado de neutrones se deben a Oppenheimer, Volkoff y Tolman \cite{PhysRev55374,PhysRev55364}.
No obstante, si bien originalmente la denominaci\'on Estrellas de Neutrones hac\'ia referencia a un objeto estricta, o al menos mayoritariamente formado por neutrones degenerados, su composici\'on actual dista bastante de la idea primigenia. De un lado el equilibrio $\beta$ exige la presencia en las ENs de cierta fracci\'on de electrones y protones. De otro, y siguiendo el mismo argumento utilizado al introducir la neutronizaci\'on, las altas densidades de las ENs hacen que, al menos en teor\'ia, sea posible la existencia en ellas de toda una serie de otras part\'iculas como muones, mesones $\pi$ y $\rho$ o hyperones, adem\'as de la aparici\'on de fases m\'as o menos ex\'oticas, como el superfluido de nucleones o el plasma de quarks deconfinados \cite{Camezind,Lattimerprognosis,Page:2011yz}. En relaci\'on con esto, diversos modelos ex\'oticos, como las Estrellas de Quarks (EQs) \cite{Weber:2004kj}, las Estrellas de Hyperones o las Estrellas de condensado de Bose-Einstein (EBE) \cite{Chavanis2012,latifah2014bosons,113}  han sido desarrollados  como alternativas para la descripci\'on de las partes m\'as densas de las ENs.

Una Estrella de condensado de Bose-Eisntein es un objeto compacto formado completamente por bosones que interact\'uan entre s\'i \cite{Chavanis2012,latifah2014bosons,113}. Este tipo de modelo se basa a su vez en otros que consideran el interior de la EN compuesto por un superfluido de nucleones \cite{Migdal,Sauls:1989,PageSC,Page:2011yz,Shternin:2010qi}, de forma tal que los bosones que constituyen una EBE son el resultado del apareamiento de dos neutrones \cite{Chavanis2012}. Ya que en el escenario m\'as simple posible, el interior de una EN consiste, como m\'inimo, en una mezcla de neutrones, protones, electrones (gas \textit{npe}), y nucleones apareados \cite{Chavanis2012,113}, las Estrellas de condesando de Bose-Einntein son un caso l\'imite en el cual el gas de neutrones estar\'ia totalmente bosonizado \cite{Chavanis2012}. En ellas la fracci\'on de electrones y protones es despreciable debido a que la bosonizaci\'on de los neutrones inhibe el decaimiento $\beta$ \cite{Page:2011yz,PageSC}.

Aunque los bosones resultantes del apareamiento de dos neutrones son el\'ectricamente neutros, cuando la densidad es lo sufcientemente alta, los neutrones se aparean con los spines paralelos dando lugar a una part\'icula de spin uno \cite{Sauls:1989,PageSC}. Estos bosones vectoriales compuestos reaccionar\'an necesariamente a la presencia de los intensos campos magn\'eticos que existen en el interior de las ENs, al igual que el resto de las part\'iculas all\'i presentes \cite{Lattimerprognosis}.

La presencia del campo magn\'etico en un OC plantea, por lo general, tres problem\'aticas:

\begin{itemize}
	\item La ecuaciones de estado de los gases cu\'anticos magnetizados son anisotr\'opicas.
	
	Los gases cu\'anticos sometidos a un campo magn\'etico sufren un desdoblamiento de la presi\'on en las direcciones paralela y perpendicular al eje magn\'etico \cite{Ferrer}. En particular, si se trata de los gases que constituyen un OC, la anisotrop\'ia en las presiones se reflejar\'a en su forma, que ya no ser\'a esf\'erica como sucede en el caso no magnetizado.
	La descripci\'on en el marco de la Teor\'ia General de la Relatividad del equilibrio hidrodin\'amico de un OC con simetr\'ia axial, como es el caso de una estrella deformada por la presencia del campo magn\'etico, es un problema que a\'un no ha sido resuelto totalmente (v\'ease por ejemplo \cite{Paret2014,Chatterjee,Zubairi2017_88}).
	
	\item Bajo ciertas condiciones, los gases cu\'anticos magnetizados pueden sufrir un colapso magn\'etico.
	
	En dependencia del campo magn\'etico, la temperatura y la densidad de part\'iculas, la menor de las presiones del gas puede hacerse negativa dando lugar a la inestabilidad  conocida como colapso magn\'etico \cite{Chaichian1999gd,Aurora2003EPJC,Felipe:2002wt,Quintero2017PRC}. Aunque desde el punto de vista de la estabilidad gravitacional de un OC el colpaso magn\'etico constituye un problema, este fen\'omeno podr\'ia ser crucial para la explicaci\'on de la expulsi\'on de materia hacia el exterior de la estrella y de la formaci\'on de los llamados \textit{jets} astrof\'isicos: chorros de materia fr\'ia y densa que luego de ser expulsados de la estrella permanecen colimados mientras se alejan de ella \cite{108,deGouveiaDalPino:2005xn}.
	
	\item El origen de los intensos campos magn\'eticos estelares es uno de los grandes misterios de la astrof\'isica.
	
A pesar de que varios mecanismos  han sido propuestos, hasta el momento ninguno alcanza a explicar de manera satisfactoria la generaci\'on del campo magn\'etico en entornos astrof\'isicos \cite{106,107}. En este sentido, la presencia de bosones de spin uno en el OC ser\'ia conveniente, pues es conocido que los gases vectoriales pueden automagnetizarse, o sea, generar su propio campo magn\'etico, que en depedencia de la densidad del gas alcanza valores en el orden de los estimados para los OCs \cite{Yamada,Elizabeth,Quintero2017IJMP,Quintero2017PRC,Lismary}. Este fen\'omeno pudiera estar conectado con la generaci\'on de los altos campos magn\'eticos de las ENs.

\end{itemize}

De acuerdo con todo lo anterior, y teniendo en cuenta que la construcci\'on de modelos te\'oricos consistentes es esencial para la interpretaci\'on de las observaciones astrof\'isicas, el objetivo principal de esta tesis es \textbf{estudiar las propiedades termodin\'amicas de un gas ideal magnetizado de bosones vectoriales neutros y aplicarlas al estudio de los efectos del campo magn\'etico en la materia que forma las ENs, considerando como casos particulares los \emph{jets} astrof\'isicos y las Estrellas de condensado de Bose-Einstein.}

En particular, investigaremos los efectos del campo magn\'etico en la estabilidad de un gas \textit{npe} parcialmente bosonizado, as\'i como en las ecuaciones de estado (EdE) y en la estructura de las Estrellas de condensado de Bose-Einstein, tomadas estas como una descripci\'on alternativa de las regiones m\'as internas de las ENs. Al mismo tiempo, se pretende explorar la posibilidad de que la automagnetizaci\'on de los bosones sea una de las causas de los altos campos magn\'eticos presentes en estos objetos. Para ello nos planteamos los siguientes objetivos espec\'{\i}ficos:

\begin{itemize}

\item Estudiar las propiedades termodin\'amicas de un gas de bosones vectoriales neutros en presencia de un campo magn\'etico.

\item Obtener las EdE de un gas \textit{npe} parcialmente bosonizado y estudiar los efectos del campo magn\'etico en ellas, en particular en lo que concierne a su estabilidad, a fin de evaluar si el colapso magn\'etico y la automagnetizaci\'on consituyen mecanismos v\'alidos para la generaci\'on y mantenimiento de los \textit{jets} astrof\'isicos.

\item Construir las ecuaciones de estado de una Estrella de condensado de Bose-Einstein magnetizada y estudiar los efectos del campo magn\'etico en ellas.

\item Obtener un sistema de ecuaciones de estructura anisotr\'opicas que permita la descripci\'on de objetos compactos magnetizados.

\item Estudiar los efectos del campo magn\'etico en la masa y la forma de las Estrellas de condensado de Bose-Einstein a partir de la relaci\'on masa-radio obtenida con el uso de las ecuaciones de estructura derivadas en el punto anterior.

\item A partir de suponer que el campo magn\'etico en la EBE es generado por los bosones que la componen, obtener los perfiles del campo magn\'etico como funci\'on del radio interno de la estrella y compararlos con los valores observacionales para el campo magn\'etico en la superficie.

\end{itemize}

La tesis est\'a organizada de la forma siguiente: El primer cap\'itulo es introductorio y se dedica al estudio de algunas cuestiones acerca de la estructura interna, la descripci\'on te\'orica y los campos magn\'eticos de las ENs. En el Cap\'itulo \ref{cap2} se investigan las propiedades termodin\'amicas del gas de bosones vectoriales neutros en presencia de un campo magn\'etico. En el Cap\'itulo \ref{cap2_5} se propone un mecanismo para la formaci\'on y mantenimiento de los \textit{jets} astrof\'isicos a partir del estudio termodin\'amico del gas \textit{npe} parcialmente bosonizado en presencia de campo magn\'etico. El Cap\'itulo \ref{cap3}, por su parte, est\'a dedicado a la obtenci\'on de las EdE de una EBE magnetizada. Para describir la estructura de estas estrellas, en el Cap\'itulo \ref{cap4} se estudia el equilibrio hidrodin\'amico de objetos esferoidades a partir de combinar las ecuaciones de Einstein con una m\'etrica axisim\'etrica. Las ecuaciones de estructra all\'i obtenidas se aplican en el Cap\'itulo \ref{cap5} al c\'alculo de las curvas masa-radio de las EBE magnetizadas. En este cap\'itulo y debido a su importancia, se dedica una secci\'on al estudio por separado de las Estrellas de czondensado de Bose-Einstein automagnetizadas as\'i como a los perfiles de campo magn\'etico interno que se obtienen durante la integraci\'on de las ecuaciones de estructura.

Al final de la tesis, las conclusiones y recomendaciones presentan un resumen de los principales resultados as\'i como de los caminos por los cuales creemos debe continuarse la investigaci\'on. Se adjuntan adem\'as cuatro ap\'endices. El primero contiene las constantes y unidades f\'isicas utilizadas. Los otros tres  recogen detalles de los c\'alculos que han sido eliminados del texto principal a fin de no recargarlo y facilitar su lectura.

En todos los c\'aculos te\'oricos y num\'ericos de la tesis se ha utilizado el sistema de unidades naturales (UN), $\hbar=c=k_B = 1$. Todas las ecuaciones est\'an escritas en este sistema de unidades y todas magnitudes que se encuentran en ellas est\'an expresadas en potencias de MeV (ver Ap\'endice \ref{appA}), excepto la Ec.~(\ref{Ec1}) por motivos pr\'acticos, y la Ec.~(\ref{PI}) por motivos hist\'oricos.  En los gr\'aficos y tablas, a fin de hacer nuestros resultados fácilmente comparables con los reportados en la literatura especializada, las magnitudes f\'isicas fueron reportadas en las unidades más usuales en cada caso. El campo magnético y las magnetizaciones se muestran en Gauss (G), las densidades de part\'iculas están dadas en cm$^{-3}$ y las de masa en g/cm$^3$, las densidades de presi\'on y energía en MeV/fm$^{3}$, las temperaturas en Kelvin (K) y las masas de las micropartículas en MeV. Las masas de los objetos astronómicos tratados se expresan en t\'erminos de la masa del sol $M_{\odot}$ y sus radios en km.

%% file: cap1.tex
\chapter{Estrellas de neutrones: fenomenolog\'ia y descripci\'on te\'orica}
\label{cap1}

Este cap\'itulo tiene car\'acter introductorio. En \'el se presenta un conjunto de aspectos relacionados con la fenomenolog\'ia y la modelaci\'on de las Estrellas de Neutrones y los \textit{jets} astrof\'isicos, todos ellos necesarios para la comprensi\'on de la tesis as\'i como de sus motivaciones.

\section{El interior de una Estrella de Neutrones}

Una Estrella de Neutrones es el Objeto Compacto resultante de la muerte de una estrella para la cual la masa del n\'ucleo colapsante supera la masa de Chandrasekar, pero sin llegar a ser tan masivo como para conducir a la formaci\'on de un agujero negro \cite{Shapiro,Camezind}. Sus carater\'isticas generales en lo que respecta a masa ($M$), radio ($R$), densidad de masa ($\rho$), temperatura ($T$) y campo magn\'etico superficial ($B_s$) e interno ($B_i$) han sido resumidas en la Tabla \ref{tablaneutrones}.
\\
\begin{table}[ht]
	\centering
	\begin{tabular}{|c|c|c|c|c|r|}
		\hline
		$M$($M_{\odot}$) &
		$R$(km)&
		$\rho (\text{g}/\text{cm}^3)$ &
		$T$(K) &
		$B_s\text{(G)}$&
		$B_i\text{(G)}$ \\
		\hline
		$\sim 1.5$ &
		$\sim 10$&
		$\sim 10^7$-$10^{15}$ &
		$\sim 10^5$-$10^{11}$ &
	    $10^{9}$-$10^{15}$ &
		$\lesssim 10^{18}$ \cr
		\hline
	\end{tabular}
\vspace{10pt}
	\caption{Valores t\'ipicos de la masa, el radio, la densidad, la temperatura y el campo magn\'etico de las Estrellas de Neutrones \cite{Camezind,Lattimerprognosis,Page:2011yz,PageSC,Malheiro:2013loa}.}
	\label{tablaneutrones}
\end{table}

\begin{figure}[h]
	\centering
	\includegraphics[width=1\linewidth]{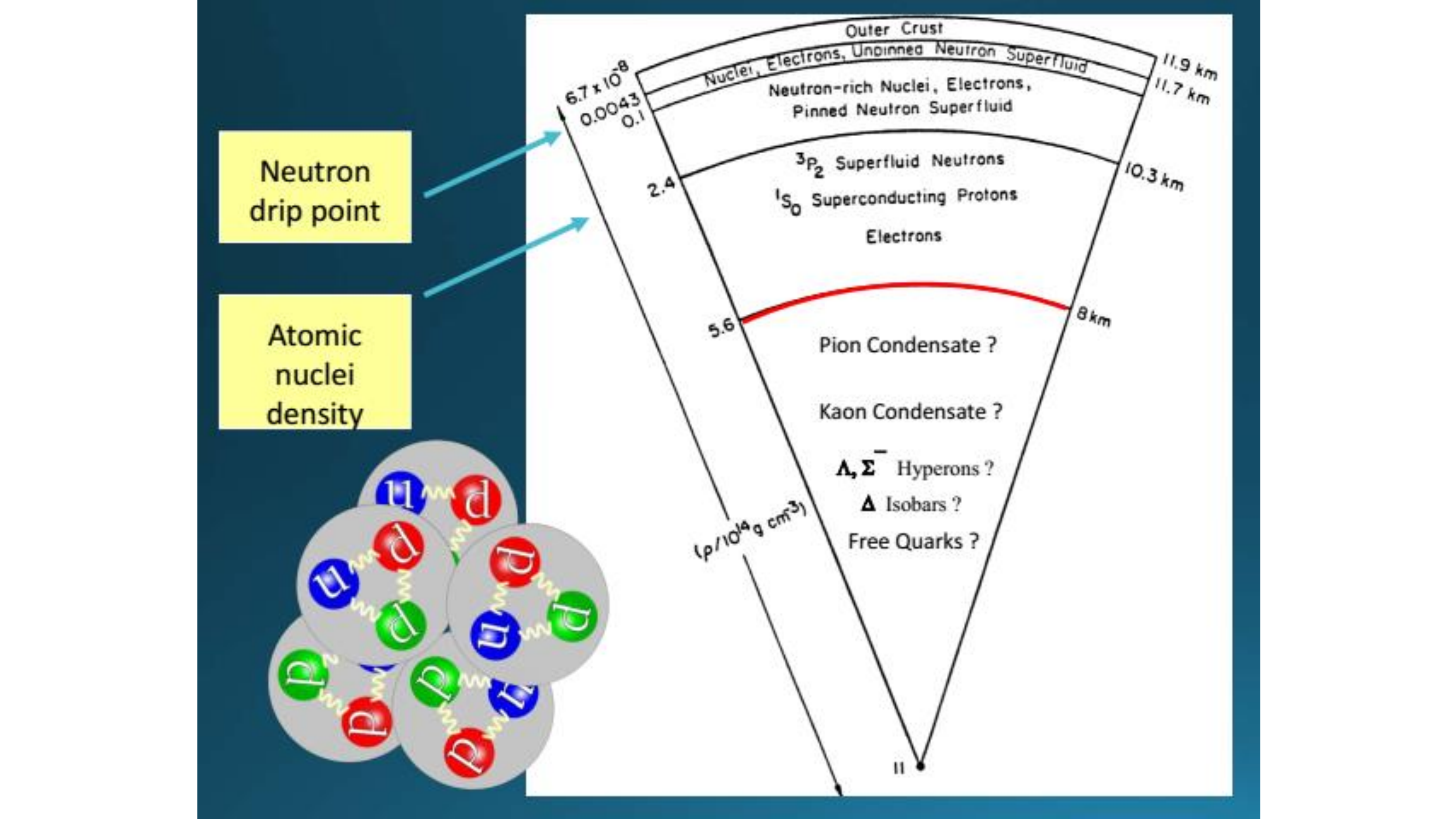}
	\caption{\label{fig01} Representaci\'on diagram\'atica aproximada del interior de una Estrella de Neutrones. Cortes\'ia del Prof. W. Becker.}
\end{figure}

Si comenzando desde la capa más externa de una EN, nos movemos hacia su centro, transitaremos sucesivamente por las siguientes regiones \cite{Camezind,Page:2011yz} (ver Fig.~\ref{fig01}):

\begin{itemize}
	\item La atm\'osfera de unos centímetros de espesor.
	
	\item La envoltura, que es una capa s\'olida formada por una red cristalina de hierro de alrededor de unos cientos de metros.
	
	\item  La corteza, con un grosor de alrededor de $1$km y que de acuerdo a su densidad se divide en exterior e interior.
	
	En la corteza exterior se alcanzan densidades de hasta $4 \times 10^{11}$g/cm$^3$ y est\'a formada por una red de n\'ucleos embebidos en un mar de Fermi de electrones. En la corteza interior, el mar de Fermi en que se encuentran los n\'ucleos contiene tanto electrones como neutrones y las densidades pueden llegar hasta alrededor de $1.2 \times 10^{14}$g/cm$^3$.
	
	\item El n\'ucleo, que ocupa el $90 \%$ del volumen de la EN y contiene la mayor\'ia de su masa.
	
Para densidades aproximadamente entre $1.2 \times 10^{14}$g/cm$^3$ y $7.5 \times 10^{14}$g/cm$^3$ el n\'ucleo se supone compuesto principalmente de una mezcla de neutrones con una peque\~na proporci\'on de protones y electrones -la justa para mantener el equilibrio $\beta$-, y muones, todos degenerados. Para densidades mayores que $7.5 \times 10^{14}$g/cm$^3$, se ha conjeturado que gran variedad de materia ex\'otica como fermiones apareados, mesones, hyperones, o quarks u, d y s deconfinados pudieran estar presentes en su interior, con una probabilidad que aumenta a medida que nos aproximamos al centro de la estrella.
	
\end{itemize}

La materia que forma la corteza exterior de la EN, es decir, lo n\'ucleos cristalinos embebidos en un mar de Fermi de electrones, es similar a la que compone a las Enanas Blancas y puede describirse como ellas \cite{Camezind}. Por eso, la mayor\'ia de los esfuerzos te\'oricos en lo que concierne a las ENs se dedican a la descripci\'on de la materia que forma su corteza interior y, m\'as frecuentemente, su n\'ucleo, ya que este contiene la mayor parte de la masa de la estrella. De manera que, cuando en la literatura especializada se habla de uno u otro modelo de ENs, a lo que se hace referencia es a distintas combinaciones de part\'iculas, interacciones y fases que podr\'ian coexistir en los n\'ucleos de estos objetos, de unos $10$~km de radio.

Toda descripci\'on te\'orica de las ENs parte de un gas de neutrones con una peque\~na fracci\'on de protones y electrones, el gas \textit{npe} \cite{Camezind,103}. A partir de este gas, diversos modelos para la composici\'on interna de la estrella se construyen a\~nadiendo interacciones o nuevas part\'iculas, dando lugar cada combinaci\'on a EdE distintas \cite{Camezind}. Otros modelos para el interior de las ENs se obtienen, en cambio, de suponer que sus condiciones internas son tales que favorecen la aparici\'on de nuevas fases de la materia. Tal es el caso de las Estrellas de Quarks o Estrellas Extra\~nas, compuestas por quarks $u$, $d$ y $s$ deconfinados y que pueden contener fases supercoductoras de color \cite{102,103}, y de las Estrellas de condensado de Bose-Einstein formadas por un condensado de neutrones apareados \cite{Chavanis2012}. Igualmente, existen modelos llamados de  Estrellas H\'ibridas, en los que al n\'ucleo de la EN se le supone una estructura a capas. Por ejemplo, el n\'ucleo podr\'ia estar compuesto a su vez por un n\'ucleo de quarks deconfinados envuelto en una corteza de materia nuclear \cite{101,103}, pero podr\'ia tambien contener capas intermedias con mezcla de las dos fases \cite{102}.

Para saber si un modelo de ENs es o no gravitacionalmente estable, lo usual es combinar sus ecuaciones de estado con las llamadas ecuaciones de estructura. Las EdE est\'an relacionadas con la f\'isica micr\'oscopica de la estrella y contienen toda la informaci\'on de la materia y los campos que la componen, as\'i como de las interacciones entre ellos. Las ecuaciones de estructura, en cambio, est\'an relacionadas con la f\'isica macrosc\'opica del objeto compacto. Ellas son consecuencia de las ecuaciones de Einstein y expresan el equilibrio hidrodin\'amico que se establece en el interior de la estrella entre la fuerza de gravedad y la presi\'on que ejerce la materia \cite{Shapiro,Camezind}. La relaci\'on entre las EdE y las ecuaciones de estructura viene dada a trav\'es de la presi\'on y densidad de energ\'ia  de la materia que compone la estrella.

Los modelos de ENs que hasta el momento se consideran m\'as realistas est\'an basados en la f\'isica nuclear experimental y en c\'alculos relativamente precisos de muchos cuerpos para modelos de materia nuclear densa \cite{Camezind,Lattimerprognosis,Nandi}. No obstante, todos ellos est\'a limitados por la imposibilidad actual de alcanzar densidades superiores a la densidad nuclear en el laboratorio \cite{Lattimerprognosis}. Esto significa que la descripci\'on de la materia y sus interacciones en reg\'imenes en los cuales $\rho>\rho_{nuc}$ se hace siempre de manera aproximada, a partir de extrapolar los comportamientos conocidos para densidades menores \cite{103}. Estas limitaciones reafirman la importancia de las observaciones y los modelos astrof\'isicos, pues ambos son necesarios para una mejor comprensi\'on de la f\'isica de la materia superdensa.

A pesar de la larga lista de procesos de generaci\'on de especies y de transiciones de fase que se ha conjeturado pueden darse en el interior de una ENs (ver por ejemplo \cite{Page:2011yz,101}), hasta el momento resulta casi imposible saber cu\'ales llegan a producirse y cu\'ales no \cite{Lattimerprognosis}. Ello se debe, de un lado a las limitaciones observacionales, y del otro al hecho de que la mayoría de las EdE propuestas cumplen con las de restricciones te\'oricas generales que pueden imponerse sobre las masas y radios de los OCs a fin de acotar sus  valores posibles.  Dichas rectricciones son \cite{Lattimerprognosis,103,Weinberg}:

\begin{itemize}
	
	\item Requerimiento de estabilidad gravitacional: Para que un OC sea estable su radio $R$ debe ser mayor que el radio de Schwarzschild $R_s = 2 G M$, donde $G = 6.711 \times 10^{-45}$~MeV$^{-2}$ es la constante de gravitaci\'on y $M$ la masa de la estrella. Para $R<R_s$ la estrella colapsa formando un AN \cite{103}.
	
	 \item Requerimiento de presi\'on finita: Para que la presi\'on permanezca finita, y suponiendo que la estrella es un objeto esf\'erico, las Ecuaciones de Einstein requieren que $R >9 /4 G M$ \cite{Weinberg}.
	
	\item Requerimiento de causalidad: Exigencia a las EdE de que la velocidad del sonido en la estrella no supere a la velocidad de la luz, lo cual implica que $R > 2.9 G M$ \cite{Lattimerprognosis}.
	
	\item Requirimiento de estabilidad rotacional: Para que un OC sea rotacionalmente estable es necesario que
	\begin{equation}\label{Ec1}
	R < 10.4 \left( \frac{100 \text{Hz}}{\nu} \right)^{3/2} \left(\frac{M}{M_{\odot}}\right)^{1/2}\text{km},
	\end{equation}
	\noindent donde $\nu$ es la frecuencia de rotaci\'on de la estrella \cite{Lattimerprognosis}.

\end{itemize}

Adem\'as de las restricciones te\'oricas ya mencionadas, las observaciones astron\'omicas aportan constantemente nuevas cotas para las masas y los radios de las ENs. En estos momentos, las mediciones m\'as robustas y confiables han resultado en valores observacionales para las masas de las ENs por encima de las $2M{\odot}$. En ese grupo se encuentran las masas de los p\'ulsares\footnote{Un púlsar (\textbf{puls}ating st\textbf{ar}) es una EN en rotaci\'on y altamente magnetizada que emite radiaci\'on electromagn\'etica de forma peri\'odica.} PSR J1614-2230 y J03487+0432 \cite{Nandi}, y las estimadas para las ENs cuya fusi\'on dio lugar a la observaci\'on de la onda gravitacional GW170817.
Para la EN PSR J1614-2230, usando el llamado \emph{Shapiro delay}, se obtiene una  masa máxima posible de $(1.928 \pm 0.0017) M_{\odot}$ \cite{105}, mientras que para el p\'ulsar J03487+0432 las mediciones arrojan una masa de $(2.01 \pm 0.04) M_{\odot}$ \cite{Antoniadis:2013pzd}. Por otro lado, a las ENs que dieron lugar a la onda gravitacional GW17082017 se les asocia una masa m\'axima de $\sim2.26 M_{\odot}$, mientras que la masa de la EN resultante de la fusi\'on se estima en alrededor de $\sim2.67 M_{\odot}$  \cite{GW,2016PhRvL.116x1102A,2017arXiv171005836T}. 
Estas observaciones han hecho que dentro de la comunidad astrofísica haya una efervescencia en la b\'usqueda de  modelos de ENs cuyas EdE produzcan estrellas con masas m\'aximas  $M \geqslant 2 M_{\odot}$ \cite{Nandi}. No obstante,  las EdE que producen estrellas de masas menores siguen siendo importantes para la descripci\'on de algunas de las capas internas de la estrella (como en una Estrella H\'ibrida) o para explicar observaciones que no respondan a las caracter\'isticas de las ENs can\'onicas.  

Por otro lado, si bien las masas han sido medidas con bastante exactitud, los radios contin\'uan resisti\'endose a mediciones robustas. Se espera que la informaci\'on que aporte el NICER sobre ellos ayude a complementar las cotas actuales sobre las EdE.

\begin{figure}[h]
	\centering
	\includegraphics[width=1\linewidth]{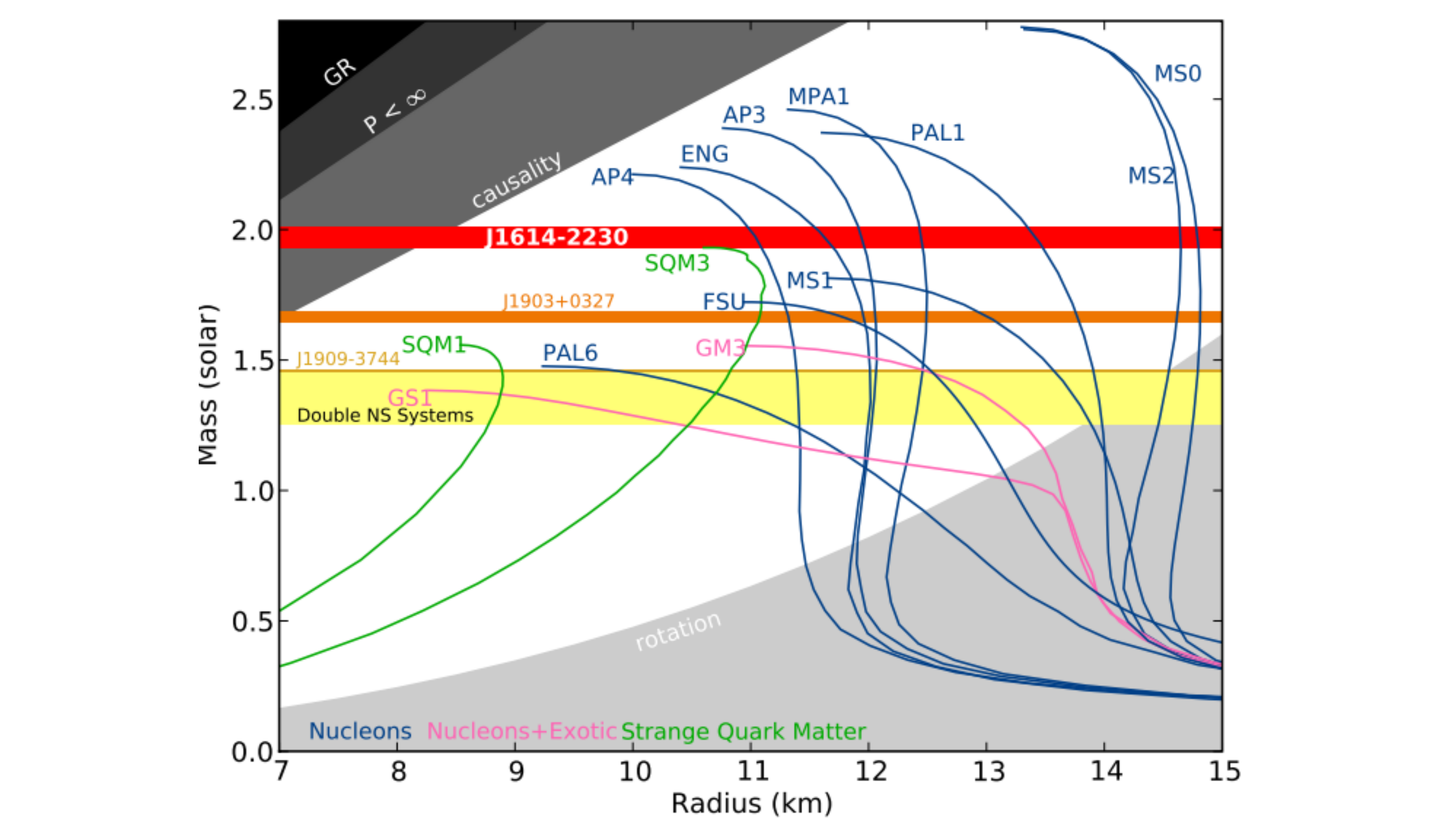}
	\caption{\label{fig02} Diagrama masa-radio de las Estrellas de Neutrones \cite{105} para varias EdE t\'ipicas (v\'ease la Ref.~25 de \cite{105}). En \'el se muestran adem\'as las zonas prohibidas por las restricciones te\'oricas mencionadas en el texto: Requerimiento de estabilidad gravitacional (GR); Requerimiento de presi\'on finita ($P<\infty$); Requerimiento de causalidad (causality); Requerimiento de estabilidad rotacional (rotation). La curva que determina la estabilidad rotacional est\'a dibujada teniendo en cuenta la frecuencia de rotaci\'on, $\nu = 716$~Hz, del p\'ulsar PSR J1748-2446ad que es el m\'as r\'apido de los conocidos \cite{Lattimerprognosis}. Las bandas horizontales muestran las restricciones observacionales derivadas de la medici\'on de las masas de los p\'ulsares J1615-2230, J1903+0327 y J1909-3744. }
\end{figure}

Aunque la plausibilidad de las EdE propuestas para las ENs puede evaluarse en funci\'on de las restricciones te\'oricas y observacionales mencionadas arriba, as\'i como de de argumentos provenientes de la f\'isica de altas energ\'ias experimental \cite{Nandi}, dada la similitud de las curvas masa-radio generadas por diversos modelos es todav\'ia extremadamente dif\'icil establecer cu\'ales son los mejores \cite{101}. N\'otese en la  Fig.~\ref{fig02} que las curvas masa-radio de las doce EdE representadas poseen al menos una parte en la regi\'on no restringida (la regi\'on blanca). En esta misma figura puede verse c\'omo la mayor\'ia de las curvas masa-radio intersecta las bandas horizontales definidas por las masas medidas a los los p\'ulsares J1615-2230, J1903+0327 y J1909-3744. En consecuencia, la composici\'on interna de esas estrellas podr\'ia, en principio, estar decrita por cualquiera de las EdE que pasan por sus franjas de masa. Esto significa que, por el momento, es imposible discriminar unas de otras a partir de argumentos te\'oricos ni tampoco a trav\'es de las cotas observacionales.

\section{Estrellas de Neutrones con interior superfluido: Estrellas de bosones}

Una de las teor\'ias m\'as interesantes acerca del interior de las ENs considera que cuando la estrella est\'a lo suficientemente fr\'ia, los nucleones en la corteza y el n\'ucleo sufren una transici\'on hacia una fase superfluida en el caso de los neutrones, o superconductora en la caso de los protones \cite{Sauls:1989,112}. Si bien el surgimiento de esta idea data de m\'as de cincuenta a\~nos atr\'as, siendo incluso previo al descubrimiento de la primera estrella de neutrones \cite{Migdal,GK}, hasta el momento el argumento observacional m\'as fuerte que tiene a su favor proviene del ajuste de la curva de enfriamiento de la EN en el centro de Casiopea A. Este fue hecho en el a\~no 2011 de manera independiente por dos grupos de investigaci\'on, \cite{Page:2011yz,PageSC} y \cite{Shternin:2010qi}. Para la obtenci\'on te\'orica de la curva de enfriamiento, ambos grupos partieron de suponer que en el interior de la EN los nucleones se encontraban apareados formando un superfluido, y obtuvieron excelentes resultados al comparar sus curvas t\'eoricas con la curva de enfriamiento observacional. Aunque la suposici\'on de un interior superfluido no es la \'unica explicaci\'on posible para esta curva de enfriamiento\footnote{Para la correcta discriminaci\'on entre ambos modelos ser\'a necesaria al menos una d\'ecada m\'as de de observaci\'on \cite{Blaschke:2011gc}.} \cite{Blaschke:2011gc}, ella de conjunto con los avances hechos en la primera d\'ecada de este siglo en el enfriamiento de sistemas at\'omicos, ha dado lugar al resurgimiento de otra idea que tambi\'en data de m\'as de cincuenta a\~nos, la de la existencia de estrellas formadas total o mayoritariamente por gases de bosones (fermiones apareados en este caso) \cite{Chavanis2012,latifah2014bosons,113}.
\phantom{Casiopea A es un remanente de supernova en la constelaci\'on de Casiopea y la fuente de radio m\'as brillante fuera del sistema solar. }

\begin{figure}[h] 
	\centering
	\includegraphics[width=0.8\linewidth]{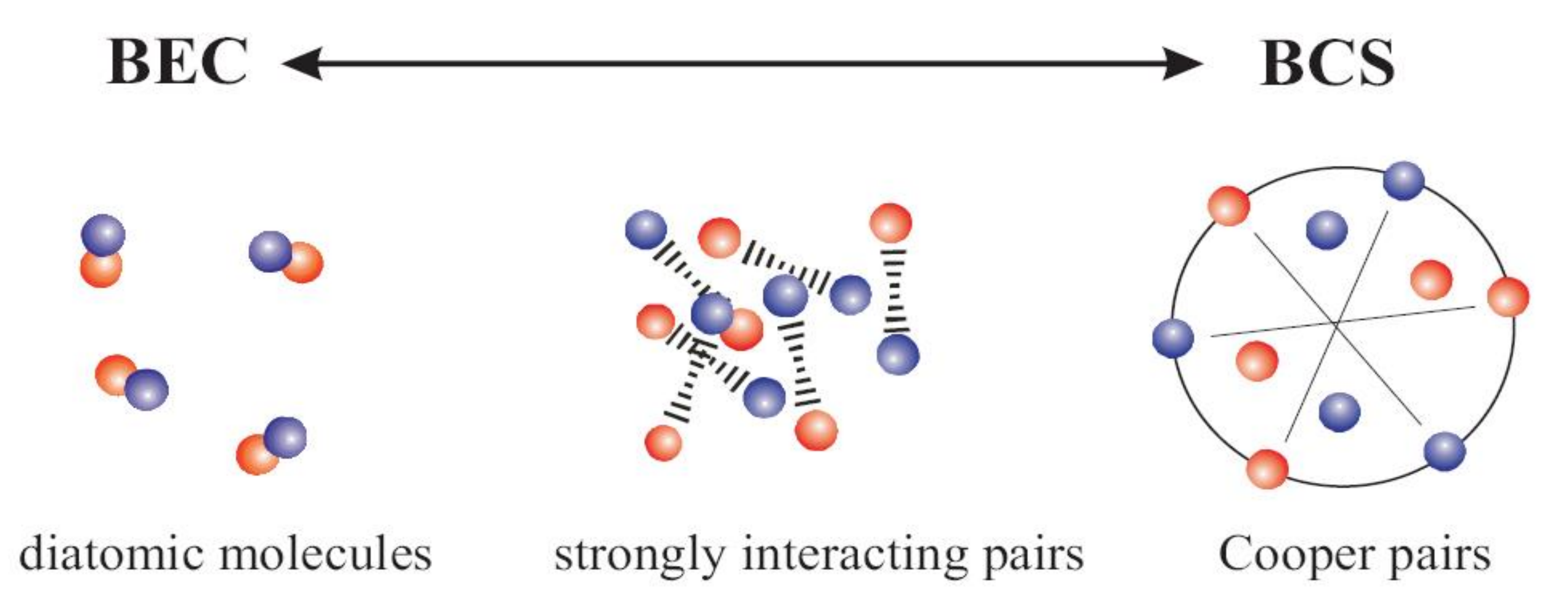}
	\caption{\label{fig025} Representaci\'on diagram\'atica del apareameinto de fermiones y sus estados l\'imite: la condensaci\'on de Bose o CBE, y la superfluidez/superconductividad o l\'imite BCS. \cite{Regal_Thesis}}
\end{figure}

La formaci\'on de pares es un fen\'omeno gen\'erico de los fermiones con interacciones atractivas entre ellos \cite{Camezind,Migdal,103,116,117,118,119,120,121,122,123,124,125} y que se manifiesta, por ejemplo, para los electrones de un superconductor \cite{BCS1,BCS2}, los nucleones en los n\'ucleos at\'omicos \cite{Bohr,Migdal}, para \'atomos con spin semientero como el $^3$He \cite{103} y para quarks de colores diferentes \cite{103}. A partir de una serie de experimentos llevados a cabo alrededor del a\~no $2000$ en el enfriamiento de sistemas at\'omicos, ha sido posible probar experimentalmente que la superfluidez/superconductividad y la condensaci\'on de Bose-Einstein (CBE) no son m\'as que los estados extremos de este f\'enomeno (Fig.~\ref{fig025}) \cite{Randeria,Leggett,Parish,116,118,123}.

Para comprender la naturaleza f\'isica del apareamiento de fermiones consideremos un gas ideal de estas part\'iculas en un volumen finito. Debido al Principio de exclusi\'on de Pauli, al a\~nadir part\'iculas a este sistema cada nuevo fermi\'on debe ponerse en un nivel energ\'etico superior. A temperatura cero, o muy baja, este proceso implica que todos los niveles por debajo de la energ\'ia de Fermi est\'an llenos, ya que para un gas ideal de fermiones este es el estado de menor energ\'ia posible. Sin embargo, si entre los fermiones hay una interacci\'on atractiva -como la mediada por  la red entre los electrones de un superconductor-, existe un estado menos energ\'etico: aquel en el cual los fermiones est\'an apareados y por tanto tienen un comportamiento bos\'onico, o sea, dejan de cumplir el Principio de exclusi\'on de Pauli y pueden condensarse.

En el l\'imite de superconductividad/superfluidez, o l\'imite BCS, los fermiones se aparean formando pares de Cooper que est\'an d\'ebilmente ligados y cuyo tama\~no es del orden de la distancia entre part\'iculas. Por el contrario, en el l\'imite CBE los fermiones est\'an fuertemente ligados formando part\'iculas que se comportan como bosones de manera efectiva. La temperatura cr\'itica para el apareamiento coincide con la temperatura cr\'itica para la aparici\'on de la superfluidez/superconductividad en el limite BCS, pero es superior a la temperatura de condensaci\'on en el l\'imite CBE. Esto significa que el gas de fermiones puede estar bosonizado aunque no condensado.

En las ENs, y a partir de los conocimientos que se han podido extraer de los experimentos, se supone que  los neutrones de la corteza se aparean en un estado no magn\'etico con los spines antiparalelos, mientras que en el n\'ucleo se aparean con los spines paralelos dando lugar a un bos\'on/par de Cooper magn\'etico \cite{Sauls:1989,PageSC,103,114,115,119,121}. Los protones de la corteza no est\'an apareados, mientras que los del n\'ucleo se aparean con los spines antiparalelos \cite{Sauls:1989,PageSC,103}. Las temperaturas cr\'iticas de los distintos procesos de apareamiento que tienen lugar en el interior de una EN se estiman el orden de los $10^8-10^{10}$K \cite{PageSC,Shternin:2010qi,115}; estas temperaturas se alcanzan en el interior de la estrella algunos cientos de a\~nos despu\'es de su nacimiento. Como la transici\'on a la fase apareada es de segundo orden, a menos que la temperatura de la estrella sea mucho menor que la temperatura de transici\'on, caso en el cual los nucleones estar\'ian totalmente bosonizados, los procesos de formaci\'on y separaci\'on t\'ermica de los pares estar\'an balanceados, y una mezcla de nucleones libres y apareados coexistir\'a en su interior \cite{PageSC}.

Por lo tanto, una estrella de neutrones puede pensarse como compuesta por una mezcla de electrones, neutrones, protones, neutrones apareados y protones apareados (gas \textit{npe} parcialmente bosonizado) \cite{Chavanis2012,113}. La fracci\'on de nucleones apareados depende, en principio, de la densidad, la temperatura y la fortaleza de la interacci\'on entre los fermiones. No obstante, la forma expl\'icita de esta dependencia es dif\'icil de predecir debido a las condiciones tan distintas que tienen los escenarios astrof\'isicos con respecto a aquellas en la cuales se llevan a cabo los experimentos.
Por otra parte, al estar  las ENs, como todo objeto del Universo, en constante evoluci\'on y, m\'as espec\'ificamente, en constante enfriamiento, a priori no puede descartarse que en alg\'un momento de su evoluci\'on estelar ellas est\'en compuestas mayormente por neutrones bosonizados \cite{Chavanis2012}.

\section{Modelaci\'on de estrellas de bosones: Estrellas de condensado de Bose-Einstein}

Aunque los OCs han sido extensamente estudiados considerando que la gravedad es balanceada por la presi\'on de un gas degenerado de fermiones, la existencia de sistemas autogravitantes formados por gases fr\'ios de bosones ha sido explorada desde la decada de los $60$s del siglo pasado \cite{PhysRev.187.1767,PhysRevD.38.2376}.

En una estrella de fermiones, el empuje de la gravedad es contrarrestrado en \'ultima instancia por la presi\'on del gas degenerado o presi\'on de Pauli. Dadas las altas densidades que existen en el interior del los OCs, incluso en el caso de que el gas de fermiones se suponga ideal y a temperatura cero, la presi\'on de Pauli es lo suficientemente alta para balancear la gravedad.

Por el contrario, un gas ideal de bosones a temperatura cero no ejerce presi\'on porque todas las part\'iculas est\'an en el condensado de Bose-Einstein \cite{PhysRev.187.1767,Landau}. Pero a\'un en este caso el colapso gravitacional puede evitarse gracias a la llamada presi\'on de Heisenberg, es decir, al hecho de que un gas bos\'onico condensado no puede ser comprimido infinitamente porque, debido al Principio de Incertidumbre:

\begin{equation}\label{PI}
\Delta p \Delta x \thickapprox \hbar, 
\end{equation}

\noindent a medida que la incertidumbre en el volumen decrece, la del momentum aumenta, aumentando con ella la presi\'on. Como a temperatura cero la presi\'on del gas de bosones es nula, el esquema de combinar las EdE con ecuaciones de estructura no puede ser usado para encontrar las masas y radios de las configuraciones estables de estrellas bos\'onicas. No obstante, describir la estructura de dichos objetos es posible a partir de resolver directamente las ecuaciones de Einstein para los coeficientes m\'etricos utilizando como fuente de materia el promedio mecanocu\'antico (y no el estad\'istco como se hace en el caso de las estrellas de fermiones) del tensor energ\'ia-momento de los bosones en el estado b\'asico \cite{PhysRev.187.1767}.

La masa m\'axima de las estrellas de bosones as\'i construidas puede estimarse si se compara la energ\'ia cin\'etica media del gas de bosones con su energ\'ia gravitacional \cite{Takasugi1984}. Dicho c\'alculo arroja que $M_{max} \sim m_p/m$, donde $m$ es la masa del bos\'on y $m_p = 2.1 \times 10^{-8}$kg$=1.22\times 10^{22}$MeV es la masa de Planck. Para estrellas de bosones formadas por part\'iculas con masa en el orden de la masa del neutr\'on, la masa de la estrella es de alrededor de $10^{-20} M_{\odot}$, un resultado descorazonador cuando lo que se quiere es describir un objeto compacto, a pesar de que la masa de estas estrellas puede aumentarse a\~nadiendo a los modelos interacciones repulsivas entre las part\'iculas \cite{Grandclement}.

De hecho, no fue hasta despu\'es de la obtenci\'on experimental del condensado para gases de Bose no ideales \cite{Anderson1995} (enfriamiento en trampas magn\'eticas de \'atomos de rubidio), y del \'exito en el ajuste de la curva de enfriamiento de Cassiopea A, que la presi\'on resultante de la interacci\'on repulsiva vino a jugar el papel principal en la construcci\'on de modelos de estrellas de bosones \cite{Chavanis2012,latifah2014bosons}. En estos modelos relativamente recientes, conocidos en la literatura como Estrellas de condensado de Bose-Einstein, la gravedad es balanceada por la presi\'on que resulta de las fuerzas repulsivas, y el esquema de combinar EdE con ecuaciones de estructura es retomado para la b\'usqueda de las configuraciones estelares estables.

Tanto en \cite{Chavanis2012} como en  \cite{latifah2014bosons}, las Estrellas de condensado de Bose-Einstein se proponen como una alternativa para la descripci\'on del n\'ucleo de una EN en la cual los bosones son formados por el apareamiento de dos neutrones. Los par\'ametros que definen el tama\~no y la masa de la estrella resultante son la masa de los bosones y la fortaleza de la repulsi\'on entre ellos. Y precisamente parte del \'exito  de estos modelos se debe a que, a pesar de ser relativamente sencillos -solo incluye un tipo de part\'iculas y dos par\'ametros-, es posible, a trav\'es de una selecci\'on apropiada de la masa y la fortaleza de la interacci\'on, lograr con \'el estrellas con masas del orden de dos masas solares \cite{Chavanis2012}.

Sin embargo, hasta donde conocemos, ning\'un modelo de Estrellas de condensado de Bose-Einstein ha considerado la influencia del campo magn\'etico en sus EdE o su estructura, a pesar de que en el n\'ucleo de una EN los neutrones se aparean con spin paralelo, dando lugar a un bos\'on vectorial, y de que los campos magn\'eticos de dichas estrellas suelen ser muy intensos. Por ello, la modelaci\'on de EBE magnetizadas fue abordada en la tesis. Los resultados obtenidos se muestran en los Cap\'itulos \ref{cap3} y \ref{cap5}.

\section{Campo magn\'etico en el interior de las estrellas de neutrones}

\begin{figure}[h]
	\centering
	\includegraphics[width=1\linewidth]{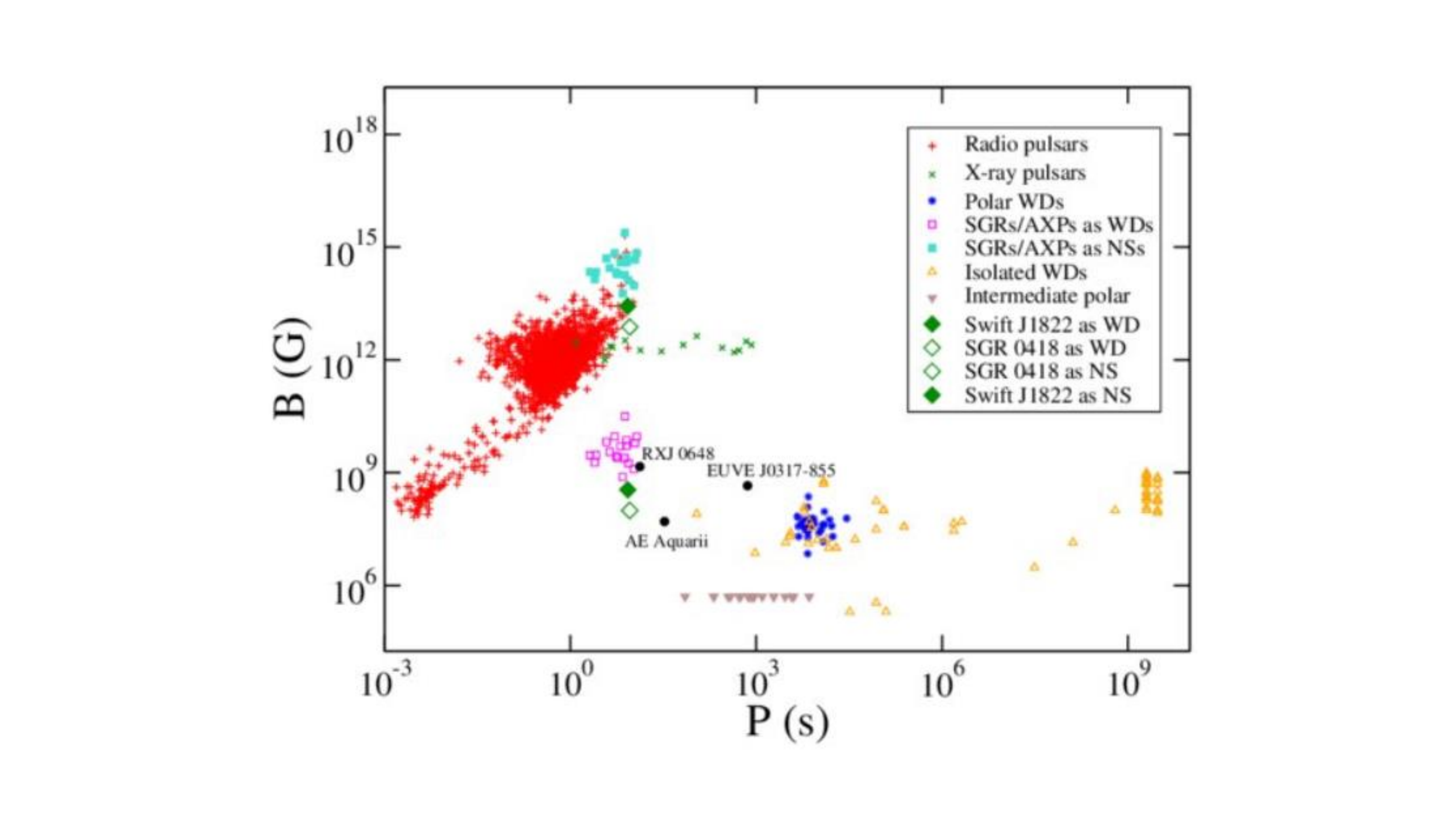}
	\caption{\label{fig03} Campo magn\'etico superficial de EBs y ENs como funci\'on del per\'iodo de rotaci\'on \cite{Malheiro:2013loa}.}
\end{figure}

Como ya fue mencionado anteriomente, la mayoría de las estrellas de neutrones conocidas tienen campos magnéticos. Los valores del campo magn\'etico superficial de las ENs estimados a partir de las observaciones astron\'omicas, Fig.~\ref{fig03}, se encuentran entre los $10^{9}-10^{13}$~G para los p\'ulsares, y en el orden de los $10^{15}$~G para las magnetars\footnote{Una magnetar (\textbf{magnet}ic st\textbf{ar}) es una EN  cuyo campo magn\'etico es extremadamente fuerte.} \cite{Malheiro:2013loa}.

Para el campo magn\'etico interno puede establecerse una cota m\'axima a partir de comparar la energ\'ia magn\'etica de la estrella con su energ\'ia gravitacional $(4\pi R^3/3)(B_{\text{max}}^2/8\pi) \sim GM^2/R$ \cite{Lattimerprognosis}. De esta forma un estimado del campo magn\'etico m\'aximo que puede sostener un OC se obtiene como funci\'on de la masa y el radio de la estrella:

\begin{equation}\nonumber
B_{\text{max}} \sim 2 \times  10^{8} \left( \frac{M}{M_{\odot}} \right) \left( \frac{R}{R_{\odot}} \right)^{-2} \text{G}.
\end{equation}

\noindent Para una EN con una masa de $M = 1.44 M_\odot$  y el radio en el orden de los kil\'ometros $R = 10^{-4}R_\odot$ el campo magn\'etico m\'aximo as\'i obtenido es de alrededor de $B_{\text{max}} \sim 10^{18}$ G.

El origen de los campos magn\'eticos presentes en los entornos astrof\'isicos a\'un no ha podido ser completamente dilucidado. Este es, no obstante, un problema de gran importancia, ya que los campos magn\'eticos son ubicuos en el Universo y han demostrado ser claves en la explicaci\'on de numerosos fen\'omenos \cite{106,107}.

En el caso de las ENs, campos magn\'eticos en el orden de hasta $10^{12}$~G pueden ser explicados por la conservaci\'on del flujo magn\'etico de la estrella progenitora durante el colapso que da lugar al OC~\cite{Shapiro,2005ASPC..334..281F,2015SSRv..191...77F}. Dado que durante la compresi\'on el radio de la estrella progenitora puede dismuir hasta en cinco \'ordenes, campos magn\'eticos en el orden de los $100$~G se ver\'ian, en el objeto compacto resultante, amplificados hasta los $10^{12}$ G. Con este efecto pueden explicarse, por ejemplo, el campo magn\'etico observado para algunos radio-p\'ulsares, como el p\'ulsar del Cangrejo, y los p\'ulsares de rayos X 
 \cite{Shapiro}. Sin embargo la explicaci\'on de campos magn\'eticos superficiales en el orden de  $10^{15}$ G, como los presentes en las magnetars,  precisan de modelos más elaborados que aún est\'an en discusi\'on \cite{1996AIPC..366..111D,2015SSRv..191...77F}.

Una de las explicaciones m\'as aceptadas para los campos magn\'eticos m\'as intensos observados en ENs se basa en el efecto dinamo \cite{1993ApJ...408..194T,2015SSRv..191...77F}. Este efecto consiste en la generaci\'on de un campo magn\'etico en un fluido conductor el\'ectricamente neutro a partir de corrientes que circulan en \'el. En el caso espec\'ifico de un OC las corrientes el\'ectricas circular\'ian debido a las inhomogeneidades de la temperatura y al campo magn\'etico en su interior \cite{106}. No obstante, este efecto es incapaz de explicar todas las observaciones por lo que el problema contin\'ua abierto \cite{1993ApJ...408..194T,107,2015SSRv..191...77F}.

Por la importancia que tiene el campo magn\'etico en las propiedades microsc\'opicas de la materia as\'i como en los obervables macrosc\'opicos, sus efectos han sido considerados en numerosos estudios de OCs sin tener en cuenta c\'omo se origina. En nuestro trabajo, para la modelaci\'on de los OCs magnetizados consideraremos dos situaciones. En la primera, el campo magn\'etico se supondr\'a constante en el interior del objeto y nos referiremos a \'el como un campo externo en el sentido de que su intensidad no depende expl\'icitamente de las EdE de la estrella. En la segunda, tomaremos el campo magn\'etico como generado por la automagnetizaci\'on de los  bosones, en este caso su intensidad depender\'a de la densidad de part\'iculas y diremos de \'el que es un campo magn\'etico interno. Esta segunda situaci\'on permitar\'a en el Cap\'itulo \ref{cap5} evaluar la plausibilidad de la automagnetizaci\'on como mecanismo para la generaci\'on de los campos magn\'eticos estelares. 

\section{Efectos del campo magn\'etico en las ecuaciones de estado de gases cu\'anticos: objetos compactos deformados y \textit{jets} astrof\'isicos}

En el interior de un OC magnetizado, todas las part\'iculas que lo forman est\'an sujetas a los efectos del campo magn\'etico sobre ellas. En consecuencia, la presencia del campo magn\'etico puede modificar la composici\'on, la estructura y la estabilidad de la estrella \cite{Lattimerprognosis,Felipe2008,Paret2014,Paret2015,Chatterjee,Dexheimer2017}. La presencia del campo magn\'etico rompe la simetría esférica de la materia que compone la estrella y da lugar a un desdoblamiento de las presiones en las direcciones paralelas y perpendicular a \'el \cite{Ferrer,Chatterjee}. Esta anisotrop\'ia tiene un impacto directo en la forma del OC, que no podr\'a ser considerada esf\'erica, como es usual en la descripci\'on de la estructura de objetos compactos no magnetizados.

\begin{figure}[h]
	\centering
	\includegraphics[width=0.8\linewidth]{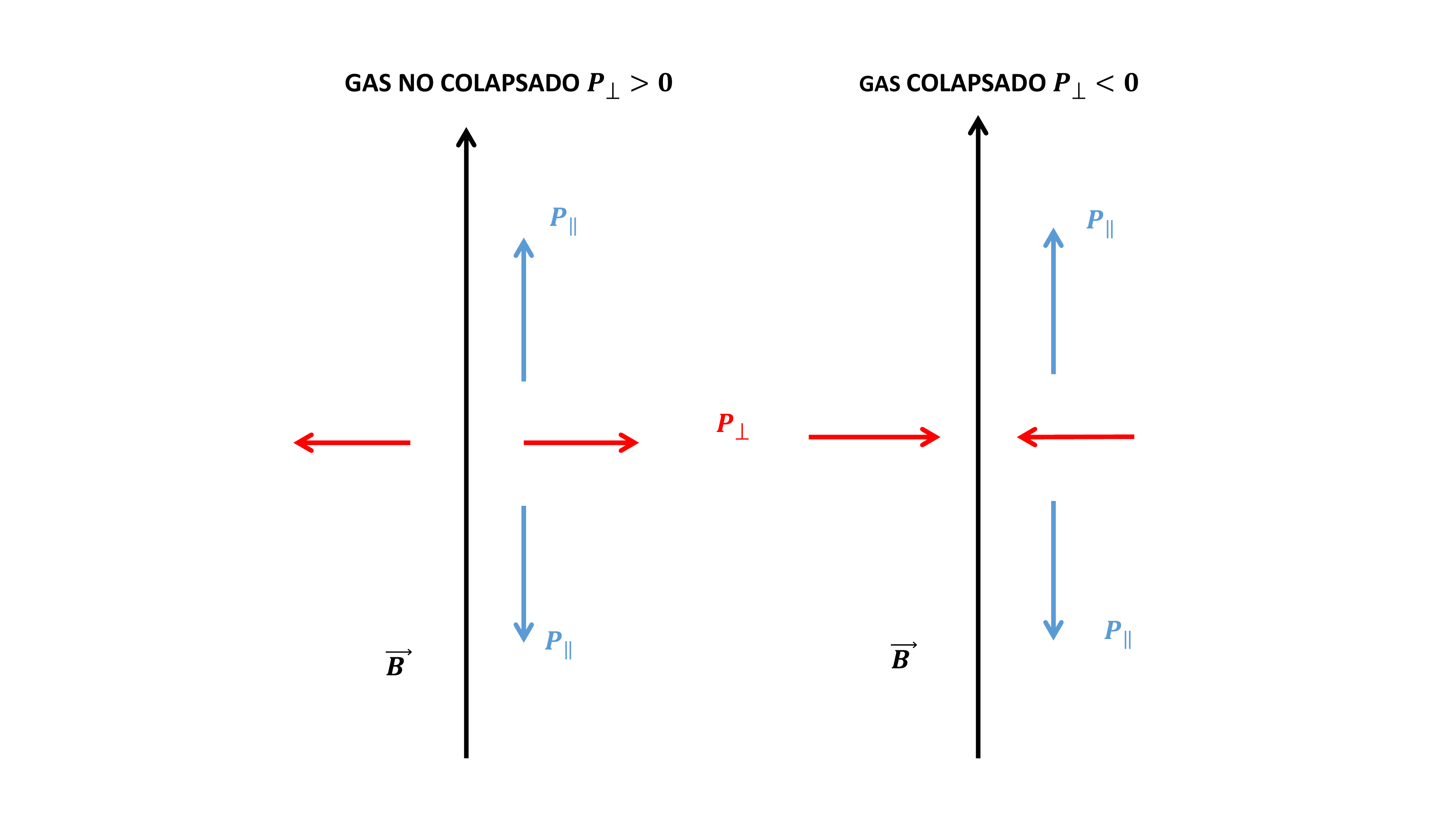}
	\caption{\label{fig04} Representaci\'on diagram\'atica de la anisotrop\'ia en las presiones de un gas cu\'antico magnetizado y del colapso magn\'etico transversal. Las flechas negras indican la direcci\'on del campo magn\'etico, mientras que las azules y rojas el sentido de las presiones paralela y perpendicular respectivamente.}
\end{figure}

Por otra parte, para varios gases cu\'anticos magnetizados ha sido posible comprobar que, bajo ciertas condiciones, la menor de las presiones puede hacerse cero e incluso negativa, dando lugar a una inestabilidad que es conocida como colapso magn\'etico \cite{Chaichian1999gd,Aurora2003EPJC,Felipe:2002wt,Quintero2017PRC}. En el caso de que sea la presi\'on perpendicular al eje magn\'etico la que devenga inestable, las part\'iculas del gas se ver\'ian forzadas a moverse hacia \'este, dando lugar a una estructura axisim\'etrica y alargada (Fig.~\ref{fig04}). Aunque el colpaso magn\'etico constituye un problema desde el punto de vista de la estabilidad de los objetos compactos, podr\'ia ser crucial en la explicaci\'on de otros fen\'omenos, como los \textit{jets} astrof\'isicos.

\subsection{\textit{Jets} astrof\'isicos}

\begin{figure}[h]
	\centering
	\includegraphics[width=1\linewidth]{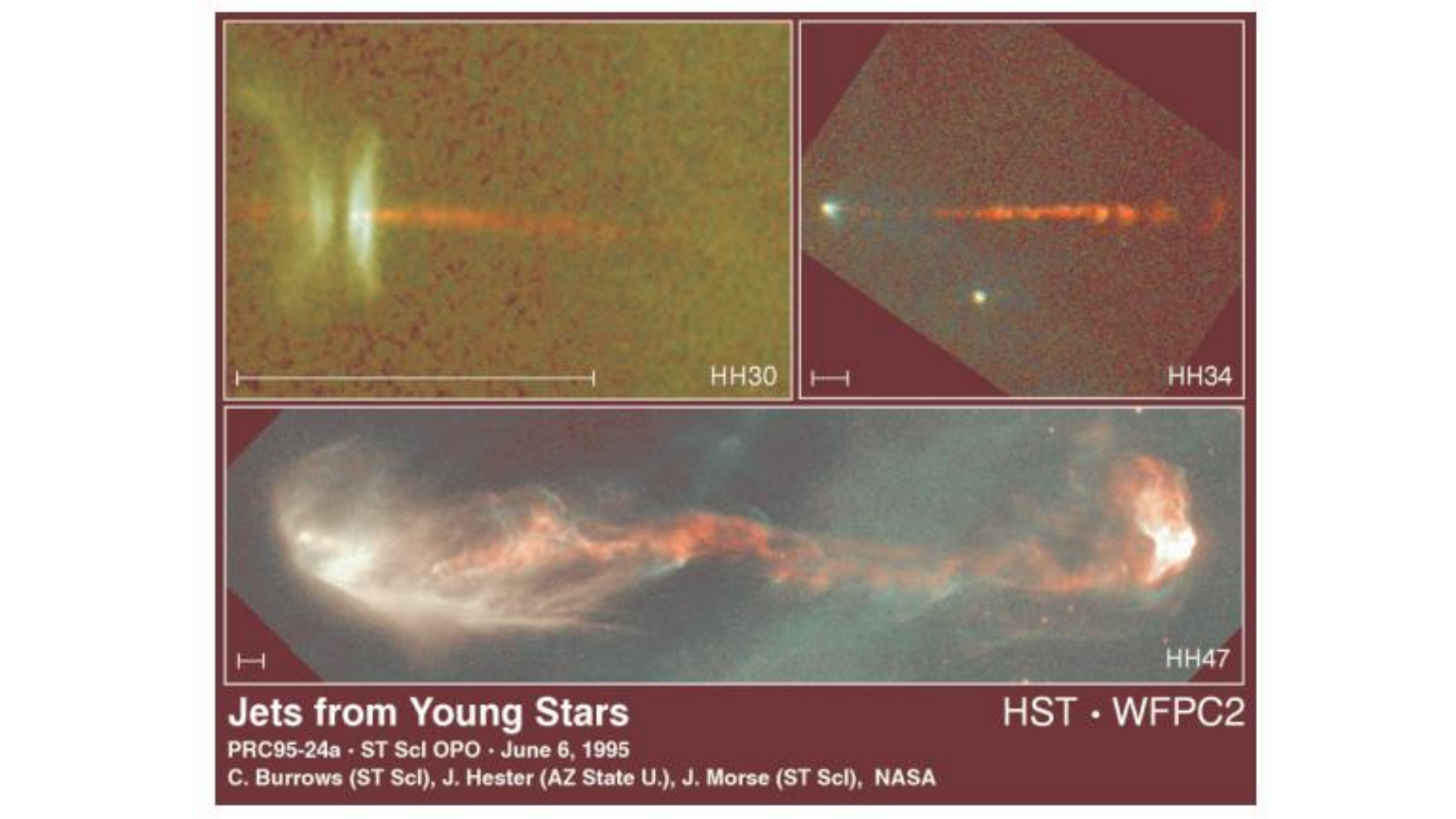}
	\caption{\label{fig045} Im\'agenes del Telescopio Espacial Hubble de tres \textit{jets} astrof\'isicos emitidos por estrellas j\'ovenes \cite{deGouveiaDalPino:2005xn}.}
\end{figure}

\begin{figure}[h]
	\centering
	\includegraphics[width=1\linewidth]{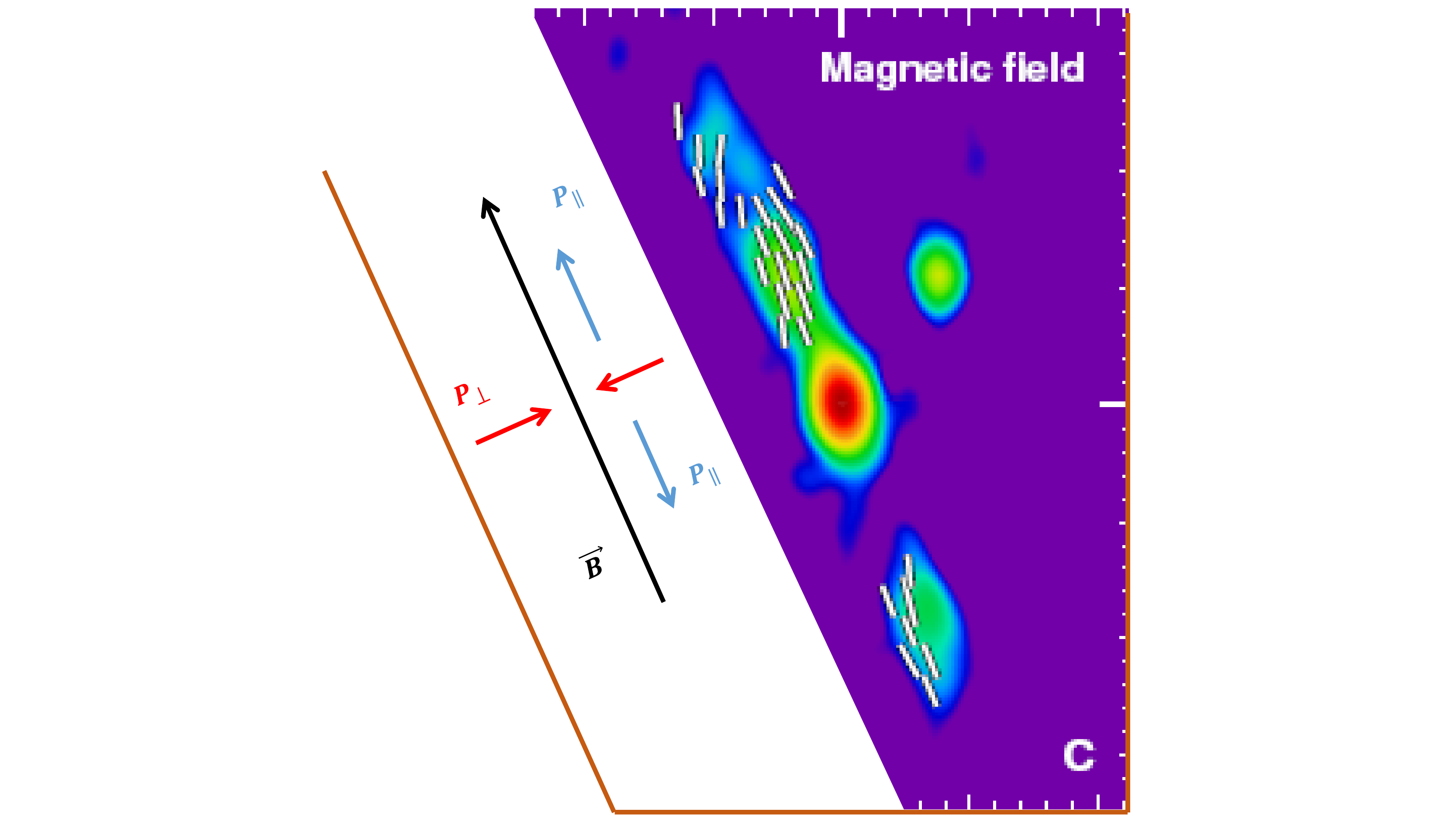}
	\caption{\label{fig05} Imagen del \textit{jet} HH 80-81 a 6 cm de longitud de onda tomada de \cite{Carrasco2010_20}. Las barras blancas indican la dirección del campo magnético aparente. El diagrama del colpaso magn\'etico transversal se ha a\~nadido a la imagen para resaltar la similitud de la geometr\'ia del \textit{jet} con la de este fen\'omeno.}
\end{figure}

Un \textit{jet} astrof\'isico es un chorro de materia densa, fría y colimada que luego de ser expulsado de un objeto astrof\'isico (estrellas, protoestrellas, nebulosas protoplanetarias, OCs, galaxias, etc.), se aleja de su fuente sin dispersarse \cite{deGouveiaDalPino: 2005xn,Carrasco2010_20,Beall_21,Elizabeth}. Este fen\'omeno puede ser visto en escalas que cubren m\'as de siete \'ordenes de magnitud \cite{108,deGouveiaDalPino: 2005xn}. Por ejemplo, la longitud de los \textit{jets} emitidos por N\'ucleos Gal\'acticos Activos est\'a en el orden de los $10^6$ pc\footnote{$1 \text{pc} = 3.086 \times 10^{16}$~m}, y sus velocidades son cercanas a la velocidad de la luz; los jets emitidos por Objetos Estelares J\'ovenes, en cambio, tienen longitudes en el orden de los pc y velocidades tres \'ordenes por debajo de la velocidad de la luz \cite{108,deGouveiaDalPino: 2005xn}. Con independencia de la fuente que los produce, los \textit{jets} astrof\'isicos comparten una serie de propiedades \cite{108,deGouveiaDalPino: 2005xn,Carrasco2010_20}:

\begin{itemize}
	
	\item Son estructuras altamente colimadas con foma aproximadamente cil\'indrica o c\'onica (Fig.~\ref{fig045}).
	
	\item Muestran una cadena de nudos que se mueven a altas velocidades alej\'andose de la fuente (Fig.~\ref{fig045}).
	
	\item Est\'an asociados con campo magn\'eticos cuya direcci\'on puede inferirse a partir de mediciones de polarizaci\'on (Fig.~\ref{fig05}). Las observaciones sugieren que el campo magn\'etico de un \textit{jet} est\'a alineado con su eje \cite{Carrasco2010_20}.

\end{itemize}

\noindent Por ello, a pesar de su origen diverso y de sus diferencias de escala se cree  que todos los \textit{jets} astrof\'isicos son producidos por mecanismos similares \cite{108,deGouveiaDalPino:2005xn}. Aunque dichos mecanismos  est\'a a\'un en debate, s\'i hay consenso acerca de que el campo magnético debe jugar un papel importante en ellos \cite{deGouveiaDalPino:2005xn,108,Carrasco2010_20}.

La característica más intrigante de los \emph{jets} astrof\'isicos es su forma. La gravitación es la fuerza de largo alcance dominante en el Universo. Su simetría central combinada con la conservación del momento angular determina la forma aproximadamente esf\'erica  de planetas y estrellas, y la forma de disco de las galaxias. Conviviendo con estas estructuras encontramos los jets, que rompen con estas simetr\'ias al estar totalmente alejados de la forma esf\'erica o de disco (Fig.~\ref{fig045}). En consideraci\'on a su forma at\'ipica, cuya simetr\'ia es similar a la simetr\'ia axial de un gas cu\'antico que ha sufrido un colpaso magn\'etico transversal, se ha sugerido que tanto la producci\'on como la estructura de los \textit{jets} podr\'ia estar determinada por la f\'isica que subyace detr\'as de este tipo de colapso \cite{Elizabeth}.

Los estudios teóricos de los \textit{jets} astrof\'isicos se pueden dividir en dos tipos. Aquellos, que se dedican a estudiar la estabilidad hidrodin\'amica y gravitacional del chorro \cite{Tucker:2016wvt,Chicone:2010aa,Bini:2017uax}, y los dedicados a dilucidar sus mecanismos de producci\'on \cite{deGouveiaDalPino:2005xn,108}. Incluso existen experimentos orientados a la producci\'on de chorros estables de materia colimada en el laboratorio \cite {Hartigan}. En el Cap\'itulo \ref{cap2_5} de la tesis se sugiere un mecanismo para producir y mantener estructuras colimadas largas basado en las propiedades de los sistemas cu\'anticos fuertemente magnetizados, pero sin detenernos a estudiar el problema de la estabilidad gravitacional de la estructura que se forma.

\subsection{Modelaci\'on de la estructura de objetos compactos magnetizados}

Si bien en lo que concierne a las EdE la anisotrop\'ia magn\'etica puede ser apreciada de una manera relativamente directa, el estudio de sus efectos en la masa y la forma de los objetos astron\'omicos requiere el uso de ecuaciones de estructura que vayan m\'as all\'a de las usuales ecuaciones de Tolman–Oppenheimer–Volkoff (TOV) en simetr\'ia esf\'erica.


El problema de resolver las Ecuaciones de Einstein (acopladas a las de Maxwell en el caso de estrellas magnetizadas) para modelos espec\'ificos de materia, como los dados por unas EdE, ha sido atacado de manera n\'umerica, en particular en relaci\'on con el estudio de estrellas magnetizadas y/o en rotaci\'on, dos casos que responden al mismo tipo de simetr\'ia axial \cite{Bocquet,Konno,Grandclement,Chatterjee}. Igualmente, diversos esfuerzos se han dedicado a la b\'usqueda de ecuaciones de estructura axisim\'etricas que puedan ser tratadas de manera (semi-)anal\'itica \cite{Paret2014,Paret2015,ZubairiRomero_2015_89,Zubairi_2015_90,Zubairi2017_88,Zubairi_2017_91,Zubairi_2017_92}. Sin embargo, debido al car\'acter m\'as o menos aproximado y/o restringido de las soluciones encontradas hasta ahora, la b\'usqueda, ya sea anal\'itica o num\'erica, de soluciones de las ecuaciones de Einstein en espacio-tiempos no esf\'ericos contin\'ua siendo de gran inter\'es para la f\'isica.

En particular, en el grupo de investigaci\'on al que se adscribe esta tesis, se han realizado varios intentos dirigidos a lograr una descripci\'on del OC magnetizado en simetr\'ia cil\'indrica \cite{Paret2014}. No obstante, las ecuaciones obtenidas no permiten calcular la masa total del objeto compacto, lo cual constituye una limitaci\'on al ser este uno de los observables m\'as importantes. El problema de encontrar ecuaciones de estructura para objetos esferoidales a partir de una m\'etrica axisim\'etrica en coordenadas esf\'ericas ser\'a tratado en el quinto cap\'itulo de la tesis.

%% file: cap2.tex
\chapter{Propiedades termodin\'amicas de un gas magnetizado de bosones vectoriales neutros en el l\'imite de baja temperatura}
\label{cap2}

En este capítulo se estudian las propiedades termodin\'amicas de un gas magnetizado de bosones vectoriales neutros en el límite de baja temperatura. La descripci\'on de los bosones se hace en el marco de la teor\'ia de Proca para las part\'iculas de spin 1, y el espectro energ\'etico es obtenido a partir del hamiltoniano de Sakata-Taketani \cite{PhysRev.131.2326,PhysRevD.89.121701}. Posteriormente, se muestra el cálculo del potencial termodinámico en tres y una dimensiones, y se discuten la condensación de Bose-Einstein, las propiedades magnéticas y las ecuaciones de estado de este gas. La obtenci\'on del espectro para bosones vectoriales neutros así como los cálculos relativos a sus propiedades termodinámicas son resultados originales de la autora de la tesis \cite{Quintero2017IJMP,Quintero2017PRC,Quintero2017AN}.

\section{Gas de bosones vectoriales en presencia de campo magn\'etico}

A pesar de su potencial importancia en la modelaci\'on de OCs y, en particular, para la generaci\'on de los campos magn\'eticos estelares, las propiedades termodin\'amicas del gas magnetizado de bosones vectoriales neutros han sido relativamente poco estudiadas, a diferencia de las de su an\'alogo cargado que han recibido mucha m\'as antenci\'on
\cite{ROJAS1996148,ROJAS1997,PEREZROJAS2000,Khalilov:1999xd,Khalilov1997,Bargueno,Jian,Chernodub:2010qx,Chernodub:2012fi,Satarov:2017jtu}.

El estudio de la condensaci\'on de Bose-Einstein y de las propiedades magn\'eticas de gases bos\'onicos cargados, escalares o vectoriales, puede verse en \cite{ROJAS1996148,PEREZROJAS2000,Khalilov:1999xd,Khalilov1997,Bargueno,Jian}.
Para bajas temperaturas, el gas de bosones vectoriales cargados es paramagn\'etico, puede automagnetizarse y experimenta una transici\'on de fases difusa al condensado de Bose-Einstein \cite{ROJAS1996148,PEREZROJAS2000}.

En una transici\'on
de fase difusa no existe una temperatura cr\'itica, sino un intervalo de temperaturas a lo largo del cual la transici\'on ocurre gradualmente \cite{ROJAS1997}. En particular, la transici\'on de fase difusa al condensado de
Bose-Einstein se caracteriza por la presencia de una fracci\'on finita de la densidad total de part\'iculas en una vecindad del estado b\'asico a cierta temperatura $T>0$. En este sentido,
el criterio para definir un condensado de Bose-Einstein difuso es m\'as d\'ebil que el usado para definir el condensado de Bose-Einstein usual, pues este \'ultimo requiere la existencia de una temperatura cr\'itica por debajo de la
cual exista una cantidad macrosc\'opica de part\'iculas en el estado b\'asico. 

Que la transici\'on al condensado sea usual o difusa depende de las dimensiones del sistema \cite{ROJAS1997}. En el caso de
un gas de bosones cargados en presencia de un campo magn\'etico, la transici\'on difusa aparece debido a la reducci\'on dimensional que sufre el gas de tres a una dimensi\'on en el r\'egimen de campo fuerte. En este r\'egimen los bosones son forzados por el campo magn\'etico a concentrarse en el primer nivel de Landau. Como en este nivel la componente del momento perpendicular al eje magn\'etico se anula, los bosones solo pueden moverse en la direcci\'on paralela
a \'el \cite{ROJAS1996148,Khalilov:1999xd}. 

La aparici\'on de niveles de Landau es un efecto directo del acoplamiento entre la carga el\'ectrica de las part\'iculas y el campo magn\'etico \cite{Pathria}. Por tanto, es de esperar que en un gas de bosones neutros ellos
no aparezcan y  en consecuencia, que algunas de sus propiedades termodin\'amicas sean esencialmente diferentes de la de un gas de bosones cargado. Por ello se impone estudiarlas si se pretende utilizar este tipo de gas en la modelaci\'on de objetos compactos. Adem\'as de que si bien en lo que respecta a los OCs, la anisotrop\'ia en las presiones del gas magnetizado ha sido hasta el momento lo m\'as relevante, otros fen\'omenos como la condensaci\'on de Bose-Einstein y el comportamiento ferromagn\'etico que se ha observado a bajas temperaturas en los gases vectoriales, son interesantes en s\'i mismos. 

El ferromagnetismo de Bose-Einstein para bosones vectoriales neutros ha sido estudiado previamente en \cite{Yamada} considerando a los bosones como no relativistas. En \cite{Chernodub:2010qx} y \cite{Chernodub:2012fi}
la superconductividad y la superfluidez inducida por el campo magn\'etico en un gas formado por la mezcla de mesones vectoriales cargados y neutros fue estudiada pero sin tener en cuenta el acoplamiento entre los mesones neutros y
el campo magn\'etico. M\'as recientemente, la condensaci\'on de Bose-Einstein para un gas de bosones vectoriales interactuantes a campo magn\'etico cero fue estudiada en \cite{Satarov:2017jtu}.

En este cap\'itulo, nos proponemos estudiar en detalle las propiedades termodin\'amicas de un gas de bosones vectoriales neutros en presencia de un campo magn\'etico. Adem\'as obtendremos las ecuaciones del estado de dicho gas, partiendo de las cuales los efectos del campo magn\'etico en la materia que compone las ENs, y en las EdE y estructura de las estrellas de condensado de Bose-Einstein, ser\'an investigados.

\section{Lagrangiano, hamiltoniano y espectro energ\'etico del gas de bosones vectoriales neutros en interacci\'on con un campo magnético}

Los bosones neutros con spin 1 en presencia de un campo magnético se describen a través de una extensión del lagrangiano de Proca que incluya la interacci\'on entre las partículas ($\rho_{\mu}$) y el campo magn\'etico \cite{PhysRev.131.2326,PhysRevD.89.121701}:

\begin{eqnarray}\label{Lagrangian}
  L = -\frac{1}{4}F_{\mu\nu}F^{\mu\nu}-\frac{1}{2} \rho^{\mu\nu}\rho_{\mu\nu}
       + m^2 \rho^{\mu}\rho_{\mu}
      +i m \kappa(\rho^{\mu} \rho_{\nu}-\rho^{\nu}\rho_{\mu}) F_{\mu\nu}.
\end{eqnarray}

En la Ec.~(\ref{Lagrangian}) los índices $\mu$ y $\nu$ van de $1$ a $4$, $m$ es la masa de la part\'icula, $\kappa$ su momento magn\'etico, $F^{\mu\nu}$ es el tensor del campo electromagnético, y  $\rho_{\mu\nu}$ y $\rho_{\mu}$ cumplen \cite{PhysRev.131.2326}:

\begin{equation}\label{fieldeqns}
  \partial_{\mu} \rho_{\mu\nu}-m^2 \rho_{\nu}+ 2i \kappa m \rho_{\mu} F_{\mu\nu}=0,\quad\quad
  \rho_{\mu\nu} = \partial_{\mu} \rho_{\nu} - \partial_{\nu} \rho_{\mu}.
\end{equation}

De considerar la variación del langragiano con respecto a  $\rho_{\mu}$ se obtienen las ecuaciones de movimiento, que en el espacio de los momentos $p_{\mu}$ pueden escribirse como:

\begin{equation}
\left((p_{\mu}^2  + m^2)\delta_{\mu\nu} -p_{\mu} p_{\nu}  - 2  i \kappa m F_{\mu \nu}\right)\rho_{\mu} = 0.
 \end{equation}

Por tanto, el propagador de los bosones es:

\begin{equation}\label{propagator}
D_{\mu\nu}^{-1}=((p_{\mu}^2  + m^2)\delta_{\mu\nu}-p_{\mu} p_{\nu}  - 2  i \kappa m F_{\mu \nu}).
\end{equation}

En lo que sigue del cap\'itulo el campo magnético se considerará uniforme, constante y en la dirección $p_3$: $\textbf{B}=B\textbf{e}_3$. $p_j$, con $j = 1, 2, 3$ son las componentes del momento de la part\'icula. Con ello, a partir de la Ec.~(\ref{fieldeqns}) se obtiene el hamiltoniano generalizado de Sakata-Taketani para la función de onda de seis componentes que describe el sistema de bosones vectoriales \cite{PhysRev.131.2326, PhysRevD.89.121701} siguiendo el procedimiento descrito en la Ref. \cite{PhysRev.131.2326}. El hamiltoniao es:

\begin{equation}\label{hamiltonian}
\hat{H} = \sigma_3 m + (\sigma_3 + i \sigma_2) \frac{\textbf{p}^2}{2 m} -
    i \sigma_2 \frac{(\textbf{p}\cdot\textbf{S})^2}{m}
    -(\sigma_3 - i \sigma_2) \kappa \textbf{S} \cdot \textbf{B},
\end{equation}

\noindent con $\textbf{p}=(p_{\perp},p_3)$ y $p_{\perp}=p_1^2 + p_{2}^2$.  $\sigma_{i}$ son las matrices de Pauli\footnote{
$\begin{array}{ccc}\sigma_1=
 \left(
\begin{array}{cc}0 & 1  \\1 & 0 \end{array}
\right),
& i\sigma_2=
\left(
\begin{array}{cc} 0 & 1  \\ \text{-}1 & 0\end{array}
\right),
&\sigma_3=
\left(
\begin{array}{cc}1&0\\0&\text{-}1\end{array}
\right)
\end{array}$ \\[1pt]}, $S_{i}$ son las matrices de $3\times3$ de spin uno en una representación en la que $S_3$ es diagonal y $\textbf{S} = \{S_1,S_2,S_3\}$\footnote{
$\begin{array}{ccc} S_1=\frac{1}{\sqrt{2}}
\left( \begin{array}{ccc}
0 & 1& 0\\
1 & 0 & 1\\
0 & 1 & 0
\end{array}\right),
& S_2=\frac{i}{\sqrt{2}}
\left( \begin{array}{ccc}
0 & \text{-}1& 0\\
1 & 0 & \text{-}1\\
0 & 1 & 0
\end{array} \right),
& S_3=
\left( \begin{array}{ccc}
1 & 0& 0\\
0 & 0 & 0\\
0 & 0 & \text{-}1\end{array}\right)\end{array}$}.

Las ecuaciones de movimiento para el momento $\textbf{p}$ y la posición $\textbf{r}$ son:

\begin{subequations}
	\begin{align}
\frac{\partial \textbf{p}}{\partial t} &= i [\hat{H},\textbf{p}], \\
\frac{\partial \textbf{r}}{\partial t} &= i [\hat{H},\textbf{r}].
\end{align}
\end{subequations}

\noindent Con el uso de la Ec.~(\ref{hamiltonian}) ellas quedan:

\begin{subequations} \label{motion}
 \begin{align}
 \frac{\partial \textbf{p}}{\partial t} &=\vec{0},\label{motionp}
\\
 \label{motionr}
m  \frac{\partial \textbf{r}}{\partial t} &= (\sigma_3 - i \sigma_2) \textbf{p} + i \sigma_2 [\textbf{S}, \textbf{p}],
 \end{align}
 \end{subequations}

\noindent con $[a,b] = ab-ba$ el conmutador de $a$ y $b$.

De la Ec.~(\ref{motionp}) se sigue que los bosones vectoriales neutros se mueven libremente tanto en la dirección paralela al campo magn\'etico como en la perpendicular. Esto es una diferencia con respecto al caso de los bosones vectoriales cargados, en el cual la componente del momento perpendicular al campo está cuantizada \cite{Elizabeth}.

El espectro energ\'etico del gas de bosones vectoriales neutros es:

\begin{equation}
\varepsilon(p_3,p_{\perp}, B,s)=\sqrt{m^2+p_3^2+p_{\perp}^2-2\kappa s B\sqrt{p_{\perp}^2+m^2}},\label{spectrum}
\end{equation}

\noindent donde $s=0, \pm 1$ son los autovalores del spin. N\'otese que la intensidad del campo magnético $B$ entra en la energía acoplada a la componente del momento de la part\'icula que es perpendicular a \'el (último término en la Ec.~(\ref{spectrum})). Este acoplamiento refleja la simetría axial impuesta al sistema por el campo magnético.

La energía del estado básico ($s=1$ y $p_3=p_{\perp}=0$) para los bosones vectoriales de spin 1 es:
\begin{equation}
\varepsilon(0, B)=\sqrt{m^2-2\kappa B m}=m\sqrt{1-b}.\label{massrest}
\end{equation}
\noindent con $b=B/B_c$ y $B_c=m/2\kappa$.

 De la Ec.~(\ref{massrest}) se sigue que la energía del estado básico del sistema decrece con el campo magnético y se hace cero para $B=B_c=m/2 \kappa$. Este tipo de comportamiento ha sido obtenido para otros gases cuánticos en interacción con un campo magnético (v\'ease por ejemplo \cite{ROJAS1996148}). Para los valores de $m$ y $\kappa$ aquí considerados, $m \cong 2 m_n$ y $\kappa\cong 2 \mu_n$, con $m_n$ la masa del neutr\'on y $\mu_n$ su momento magn\'etico, $B_c \cong 1.98 \times 10^{20}~G$. 

\section{Propiedades termodinámicas del gas de bosones vectoriales neutros}
\label{thermo}

Para calcular el potencial termodinámico por unidad de volumen partiremos de su definici\'on:

 \begin{equation}\label{Grand-Potential-4}
\Omega(B,\mu,T)= \sum_{s=-1,0,1}\frac{1}{\beta}\left[\sum_{p_4}\int \frac{p_{\perp}dp_{\perp}dp_3}{(2\pi)^2} \ln \det D^{-1}(\overline{p}^*)\right],
\end{equation}
 donde en este caso  $D^{-1}(\overline{p}^*)$ es el propagador de los bosones neutros vectoriales dado por (\ref{propagator}).  $\beta = 1/T$  denota al inverso de la temperatura, $\mu$ es el potencial químico de los bosones y ${\overline{p}}^*=(ip_{4}-\mu,0,p_{\perp},p_{3})$.
Luego de hacer la suma de Matsubara, la Ec.(\ref{Grand-Potential-4}) puede escribirse como \cite{fradkin67}:

\begin{equation}\label{potencial}
\Omega(B,\mu,T)= \Omega_{st}+\Omega_{vac},
\end{equation}

\noindent donde $\Omega_{st}$  es la contribución estadística de los bosones/antibosones y depende de $B, T$ y $\mu$:

\begin{equation}\label{potencialvst}
\Omega_{st}(B,\mu,T)=  \sum_{s=-1,0,1} \frac{1}{\beta}\left(\int\frac{p_{\perp}dp_{\perp}dp_3}{(2\pi)^2} \ln \left((1-e^{-(\varepsilon(p_3,p_{\perp}, B,s)- \mu)\beta})(1-e^{-(\varepsilon(p_3,p_{\perp}, B,s)+ \mu)\beta})\right)  \right ),
\end{equation}

\noindent mientras que $\Omega_{vac}$ solo depende de $B$ y se le conoce como t\'ermino o contribuci\'on del vac\'io.

\begin{equation}\label{potencialvac}
\Omega_{vac}=\sum_{s=-1,0,1}\int\limits\frac{p_{\perp}dp_{\perp}dp_3}{(2\pi)^2}\varepsilon(p_3,p_{\perp}, B,s).
\end{equation}

A fin de facilitar el cálculo, $\Omega_{st}$ puede ser reescrito como:

\begin{equation}
\Omega_{st}(B,\mu,T)=  \sum_{s=-1,0,1} \Omega_{st}(s),
\end{equation}

\noindent siendo

\begin{equation}\label{Grand-Potential-sst}
\Omega_{st}(s)=\frac{1}{\beta}\left(\int\frac{p_{\perp}dp_{\perp}dp_3}{(2\pi)^2} \ln \left((1-e^{-(\varepsilon(p_3,p_{\perp}, B,s)- \mu)\beta})(1-e^{-(\varepsilon(p_3,p_{\perp}, B,s)+ \mu)\beta})\right)  \right )
\end{equation}

\noindent la contribución de cada estado del spin al término estadístico del potencial temodinámico.

Utilizando la expansión de Taylor del logaritmo, la Ec.~(\ref{Grand-Potential-sst}) puede escribirse como:

\begin{equation}\label{Grand-Potential-sst1}
\Omega_{st}(s)= - \frac{1}{4 \pi^2 \beta}  \sum_{n=1}^{\infty} \frac{e^{n \mu \beta}+e^{- n \mu \beta }}{n} \int\limits_{0}^{\infty} p_{\perp}  dp_{\perp} \int\limits_{-\infty}^{\infty} dp_3 e^{-n \beta \varepsilon(p_3,p_{\perp}, B,s)}.
\end{equation}

En la Ec.~(\ref{Grand-Potential-sst1}) el término $e^{n \mu \beta}$ corresponde a las partículas mientras que el término $e^{- n \mu \beta}$ a las antipartículas.
En ella la integración sobre $p_3$ puede hacerse, mientras que la expresión solo es integrable analíticamente en $p_{\perp}$ de manera parcial. Con ello se obtiene:

\begin{eqnarray}\label{Grand-Potential-sst2}
\Omega_{st}(s)= - \frac{z_0^2}{2 \pi^2 \beta^2}  \sum_{n=1}^{\infty} \frac{e^{n \mu \beta}+e^{- n \mu \beta }}{n^2} K_2 (y z_0)
- \frac{\alpha}{2 \pi^2 \beta}  \sum_{n=1}^{\infty} \frac{e^{n \mu \beta}+e^{- n \mu \beta }}{n} \int\limits_{z_0}^{\infty} dz
\frac{z^2}{\sqrt{z^2+\alpha^2}} K_1 (y z),
\end{eqnarray}

\noindent donde $K_n(x)$ es la función de McDonald de orden $n$, $ y = n \beta$, $z_0= m \sqrt{1-s b}$ y $\alpha=s m b/2$. Para llegar a la ecuación Ec.~(\ref{Grand-Potential-sst2}) se hizo además el cambio de variables $z^2 = (m^2 + p_{\perp}^2 + \alpha^2)^2 - \alpha^2$.

\noindent Para calcular la integral en el segundo término de la Ec.~(\ref{Grand-Potential-sst2}):

\begin{eqnarray}
I = \int\limits_{z_0}^{\infty} dz
\frac{z^2}{\sqrt{z^2+\alpha^2}} K_1 (y z),
\end{eqnarray}

\noindent se sigue el procedimiento presentado en el Apéndice A. Finalmente, $\Omega_{st} (s)$ queda:

\begin{eqnarray}\label{Grand-Potential-sst3}
\Omega_{st}(s)= - \frac{z_0^2}{2 \pi^2 \beta^2} \left(1+\frac{\alpha}{\sqrt{z_0^2 + \alpha^2}}\right) \sum_{n=1}^{\infty} \frac{e^{n \mu \beta}+e^{- n \mu \beta }}{n^2} K_2 (y z_0)
+ \frac{\alpha z_0^2}{ \pi^2 \beta^2 \sqrt{z_{0}^2 + \alpha^2}} \\ \times \sum_{n=1}^{\infty} \frac{e^{n \mu \beta}+e^{- n \mu \beta }}{n^2} \sum_{w=1}^{\infty} \frac{(-1)^w(2 w -1)!!}{(z_{0}^2 + \alpha^2)^w} \left(\frac{z_0}{y}\right)^w K_{-(w+2)}(y z_0).\nonumber
\end{eqnarray}

Tomar en la Ec.~(\ref{Grand-Potential-sst3}) el límite de baja temperatura $T \ll m$ (que para $m=2 m_n\cong10^3 MeV$ significa $T \ll 10^{13} K$ y por tanto en el caso de OCs est\'a bien justificado), es equivalente a hacer el límite $y = n \beta \rightarrow \infty$. En este límite, todos los términos de $\Omega_{st}(s)$ tienden a cero salvo el primero, por tanto:

\begin{eqnarray}
\Omega_{st}(s) \cong - \frac{z_0^2}{2 \pi^2 \beta^2} \left(1+\frac{\alpha}{\sqrt{z_0^2 + \alpha^2}}\right) \sum_{n=1}^{\infty} \frac{e^{n \mu \beta}+e^{- n \mu \beta }}{n^2} K_2 (y z_0).\nonumber
\end{eqnarray}

\noindent para $T \ll m$. En este límite además $K_{2}(y z_0)$ puede sustituirse por su fórmula asintótica \cite{Marin}:

\begin{equation}
K_2 (y z_0) \cong \frac{\sqrt{\pi} e^{-y z_0}}{\sqrt{2 y z_0}} = \frac{\sqrt{\pi} e^{-n \beta z_0}}{\sqrt{2 n \beta z_0}},\nonumber
\end{equation}

\noindent con lo cual $\Omega_{st}(s)$ puede escribirse como:

\begin{eqnarray}
\Omega_{st}(s) \cong - \frac{z_0^{3/2}}{2^{3/2} \pi^{3/2} \beta^{5/2}} \left(1+\frac{\alpha}{\sqrt{z_0^2 + \alpha^2}}\right) \sum_{n=1}^{\infty} \frac{e^{n \beta (\mu - z_0)}+e^{- n \beta (\mu + z_0) }}{n^{5/2}} .\nonumber
\end{eqnarray}

Para $\beta \gg 1$, el término de las antipartículas $e^{- n \beta (\mu + z_0)}$ tiende a cero para todos los valores del spin $s = 0, \pm 1$, porque $\mu + z_0$ ($z_0= m \sqrt{1-s b}$) es siempre positivo. De esta forma vemos que en el límite de baja temperatura la contribución de las antipartículas al potencial termodinámico puede ser despreciada.

Por el contrario, el término de las partículas $e^{ n \beta (\mu - z_0)}$ solo tiende a cero para $s = 0, -1$. En estos casos, como $\mu \leq \varepsilon(0,B) = z_0(s=1)= m \sqrt{1-b}$ y $z_0(s=1)$ es menor que $z_0(0) = m$ y $z_0(-1)= m \sqrt{1+b} $, $\mu - z_0$ es negativo. Sin embargo, para $s=1$, cuando $T \rightarrow 0$, el término de las partículas $e^{ n \beta (\mu - z_0)}$ tiende a $1$, porque $\mu \rightarrow \varepsilon(0,B)$, y en consecuencia $\mu - z_0(1) = \mu - \varepsilon(0,B)$ tiende a cero. Por tanto, las contribuciones a $\Omega_{st}(s)$ que vienen de las partículas con $s=0,-1$ son también despreciables, y $\Omega_{st} \cong \Omega_{st}(1)$. De modo que finalmente, para las bajas temperaturas asumidas y luego de evaluar $z_0(1)$ y $\alpha(1)$, la parte estadística del potencial termodinámico es:

\begin{eqnarray}\label{Grand-Potential-stfinal}
\Omega_{st}(B,\mu,T) = -\frac{\varepsilon(0,B)^{3/2}}{2^{1/2} \pi^{3/2} \beta^{5/2} (2-b)} Li_{5/2}(e^{\beta \mu^{\prime}}),
\end{eqnarray}

\noindent donde $Li_{n}(x)$ es la función polilogarítmica de orden $n$ y $ \mu^{\prime} = \mu - \varepsilon(0,B)$. La cantidad $\mu^{\prime}$ es función de la temperatura y del campo magnético. De acuerdo a como ha sido definida $\mu^{\prime}$, la existencia de una temperatura distinta de cero $T_{cond}$ para la cual $\mu^{\prime} = 0$ es el requirimiento para la condensación de Bose-Einstein usual.

La contribución del vacío al potencial termodinámico después de ser regularizada es (ver c\'alculos en el Apéndice B):

\begin{align}\label{Grand-Potential-vac}
\Omega_{vac} &= -\frac{m^4}{288 \pi}\left( b^2(66-5 b^2)-3(6-2b-b^2)(1-b)^2 \log(1-b)\right.
\\\nonumber
& \left. -3(6+2b-b^2)(1+b)^2\log(1+b) \right).
\end{align}

Sumando las Ecs.~(\ref{Grand-Potential-Tcero}) y (\ref{Grand-Potential-vac}) se obtiene que el potencial termodinámico total para el gas de bosones vectoriales neutros en el límite de baja temperatura es:

\begin{align}\label{Grand-Potential-Tcero}
\Omega(B,\mu,T) & =-\frac{\varepsilon(0,B)^{3/2}}{2^{1/2} \pi^{3/2} \beta^{5/2} (2-b)} Li_{5/2}(e^{\beta \mu^{\prime}})-\frac{m^4}{288 \pi}\left( b^2(66-5 b^2) \right. \\\nonumber
& \left. -3(6-2b-b^2)(1-b)^2 \log(1-b)-3(6+2b-b^2)(1+b)^2\log(1+b) \right).
\end{align}

\subsection{Densidad de partículas y condensación de Bose-Einstein}

Para obtener la densidad de partículas $N$ calcularemos la derivada de la Ec.(\ref{Grand-Potential-Tcero}) con respecto al potencial químico $\mu$:

\begin{equation}\label{Particle-Density-Tcero}
N = -\frac{\partial \Omega(B,\mu,T)}{\partial \mu}= \frac{\varepsilon(0,B)^{3/2}}{2^{1/2} \pi^{3/2} \beta^{3/2} (2-b)} Li_{3/2}(e^{\mu^\prime  \beta}).
\end{equation}

La expresión anterior permite a sustitución $\mu^{\prime} = 0$ (porque $Li_{3/2}(1)=\zeta(3/2)$, donde $\zeta(x)$ es la función zeta de Riemann). En consecuencia, una temperatura crítica para la condensación puede ser definida (Ec.~\ref{Tcond}) y el gas de bosones vectoriales neutros muestra una condensación de Bose-Eisntein usual.

\begin{equation}\label{Tcond}
T_{cond} = \frac{1}{\varepsilon(0,B)} \left ( \frac{2^{1/2} \pi^{3/2} (2-b) N}{\zeta(3/2)}\right)^{2/3}.
\end{equation}

Aunque este comportamiento es similar al obtenido para bosones a campo magnético cero -la relación funcional entre $T_{cond}$ y $N$ es la misma-, cuando el campo magnético está presente la temperatura crítica depende de él a través de $\varepsilon(0,B)$ y $b$, y lo que es más interesante, $T_{cond}$ diverge cuando $b \rightarrow 1$ ($B \rightarrow B_c$). La dependencia de la temperatura crítica con el campo pone en evidencia que la condensación puede ser alcanzada no solo por el decrecimiento de la temperatura o el aumento de la densidad de part\'iculas, sino tambi\'en por el incremento del campo magnético. Esto puede verse f\'acilmente si se calcula la densidad de partículas fuera del condensado $N_{oc}$ (para $T<T_{cond}$):

\begin{equation}\label{Particle-Density-oc}
N_{oc} = \frac{\varepsilon(0,B)^{3/2}  T^{3/2}}{2^{1/2} \pi^{3/2} (2-b)} Li_{3/2}(e^{\mu^\prime  \beta}) = N \left(\frac{T}{T_{cond}}\right)^{3/2},
\end{equation}

\noindent porque de la Ec.~(\ref{Particle-Density-oc}) se sigue que $N_{oc} \rightarrow 0$ cuando $T \rightarrow 0$ pero también cuando $b \rightarrow 1$
($ \varepsilon(0,B)\rightarrow 0$).

\begin{figure}[h]
\centering
\includegraphics[width=0.49\linewidth]{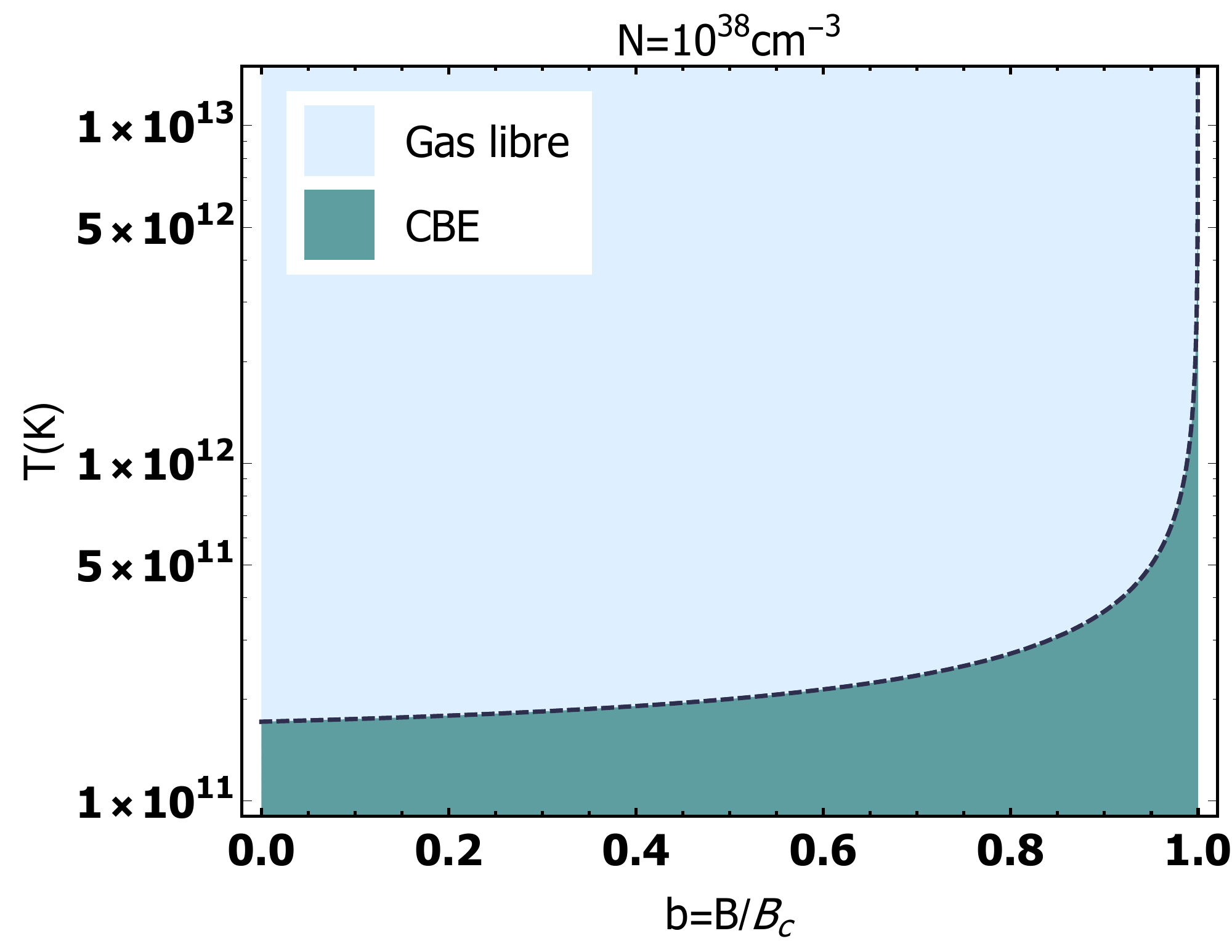}
\includegraphics[width=0.49\linewidth]{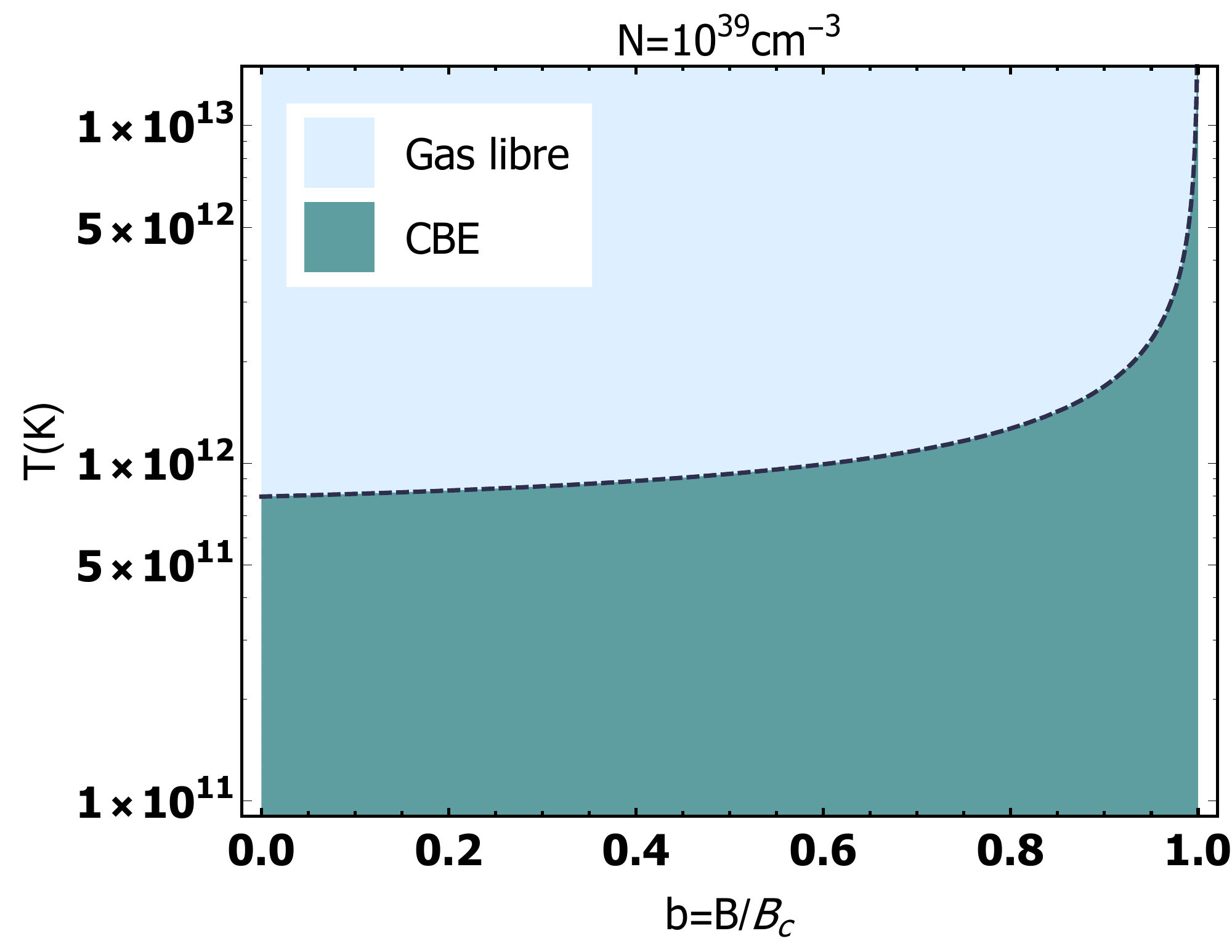}	
\caption{\label{fig5} Diagramas de fases para el condensado de Bose-Einsten en el plano $T-b$ para dos valores distintos de la densidad de part\'iculas. La l\'inea punteada corresponde a la curva $T_{cond}(b)$ y separa la regi\'on en la que el gas est\'a condensado (azul oscuro) de aquella en la que no lo est\'a (azul claro). N\'otese que la temperatura cr\'itica tambi\'en depende de la densidad de part\'iculas.}
\end{figure}

Dado que la temperatura crítica está bien definida para cada valor del campo magnético (como mismo hay un campo crítico para cada temperatura), es posible dibujar un diagrama de fases $T$ vs $b$. Esto fue hecho en la Fig.~\ref{fig5} para dos valores fijos de la densidad de part\'iculas: $N=10^{38} cm^{-3}$ y $N=10^{39} cm^{-3}$. Estas altas densidades de bosones pueden ser asumidas para objetos compactos, en particular para el interior de ENs, pues a la densidad de masa nuclear $\rho_{nuc} = 2.4 \times 10^{14}$g$/$cm$^3$ corresponde una densidad de part\'iculas de aproximadamente $10^{38}$cm$^{-3}$.  Los valores de la temperatura crítica están en el orden de los $T=10^{11}-10^{12}$~K. Estas temperaturas son ligeramente mayores que las temperaturas típicas de los objetos compactos. Por lo tanto es de esperar que una fracci\'on de los bosones vectoriales presentes en este tipo de objetos se encuentre en el condensado.

La transici\'on al condensado también puede ser examinada a través del comportamiento del calor específico. En particular, consideraremos el calor específico a volumen constante, definido como:

\begin{equation}\label{specificheat}
C_v = \frac{\partial E}{\partial T},
\end{equation}

\noindent donde $E = T S+\Omega+\mu N$ es la densidad de energía interna del sistema.
La entropía del gas es:

\begin{equation}\label{entropy}
S=-\frac{\partial \Omega}{\partial T}= -\beta \left(\mu^{\prime} N +\frac{5}{2} \Omega_{st} + \beta \frac{\partial \mu^{\prime}}{\partial \beta} N\right ),
\end{equation}

\noindent con

\begin{equation}
\mu^{\prime} \cong -\frac{\zeta(3/2)T}{4 \pi} \left ( 1-\left(\frac{T_{cond}}{T}\right)^{3/2} \right )\Theta(T-T_{cond})
\end{equation}

\noindent en el límite de baja temperatura. Aquí $\Theta(x)$ es la función paso unitario de Heaviside.

Con el uso de la Ec.~(\ref{entropy}) la energía interna puede escribirse como:

\begin{equation}\label{energy}
E = \varepsilon(0,B) N + \Omega_{vac}- \frac{3}{2} \Omega_{st} -\beta \frac{\partial \mu^{\prime}}{\partial \beta} N.
\end{equation}

La Ec.~(\ref{energy}) permite obtener para el calor específico la expresión siguiente:

\begin{equation}\label{specificheat-1D}
C_v = -\beta \left(\frac{15}{4} \Omega_{st} + \frac{3}{2} \mu^{\prime} N +\frac{1}{2} \beta \frac{\partial \mu^{\prime}}{\partial \beta} N- \beta^2 \frac{\partial^2 \mu^{\prime}}{\partial \beta^2}N\right ).
\end{equation}

\begin{figure}[h]
\centering
\includegraphics[width=0.5\linewidth]{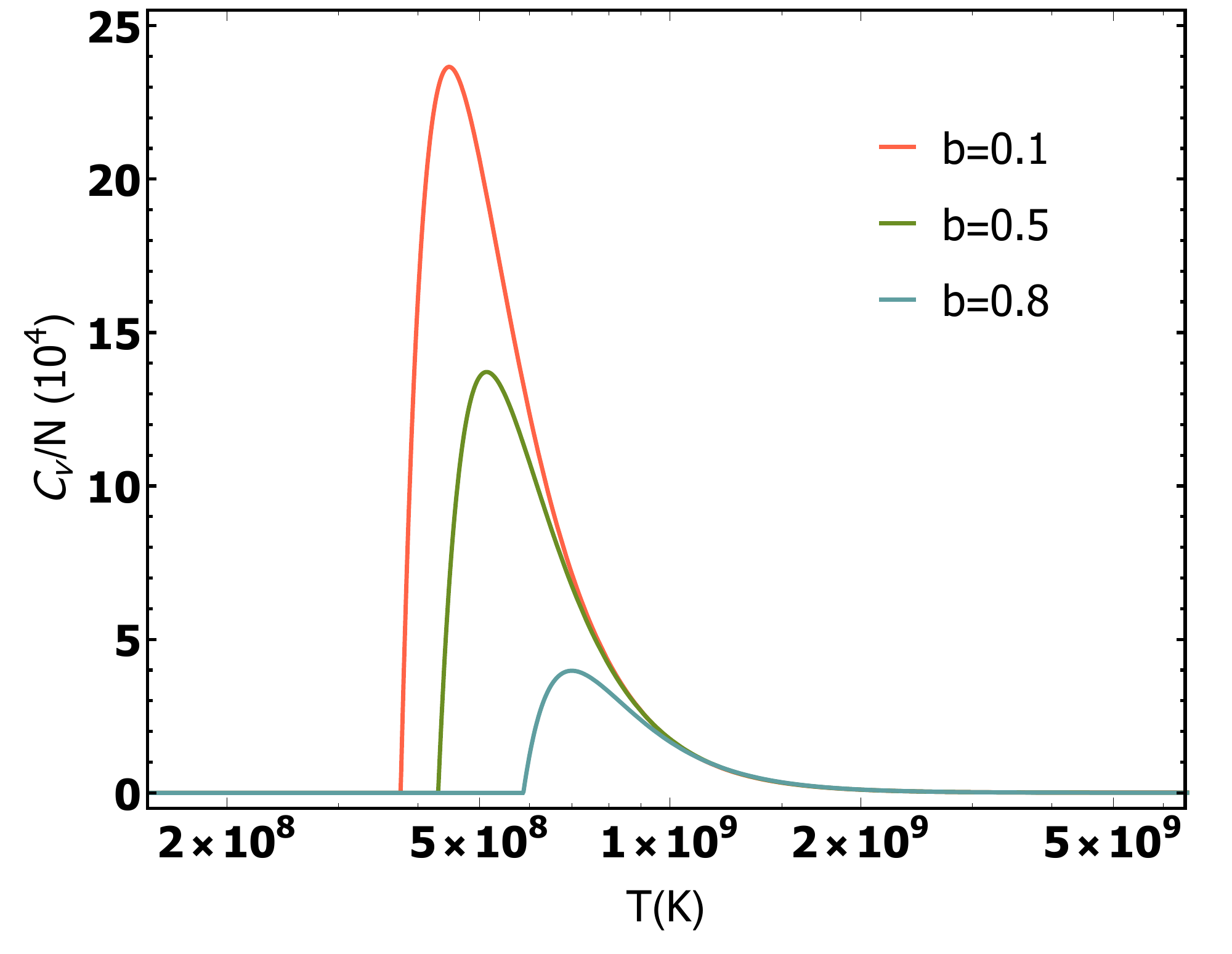}
\caption{\label{fig2} El calor específico como función de la temperatura para $N=10^{34} cm^{-3}$ y varios valores del campo magnético.}
\end{figure}

En la Fig.~\ref{fig2} se muestra el calor espec\'ifico como función de la temperatura para un valor fijo de la densidad de partículas $N=10^{34}$ cm$^{-3}$ y tres valores del campo magnético. Como puede observarse, $C_v$ tiene un máximo; este es una se\~nal de la transición al condensado. El máximo decrece a medida que $B \rightarrow B_c$ ($b \rightarrow 1$), porque cuando $B=B_c$ el gas está condensado para toda temperatura y densidad de part\'iculas.

\subsubsection{Condensación de Bose-Einstein para un gas magnetizado de bosones vectoriales neutros en una dimensión}

Consideremos ahora un gas de bosones vectoriales neutros con $p_{\perp}=0$. A pesar de que en nuestro caso la reducci\'on a una dimensi\'on no se da de manera natural, como ocurre para un gas cargado en el régimen de campo magnético fuerte, el estudio del gas en una dimensi\'on no deja de tener interés pues podría ser útil en los primeros pasos para la modelación de estructuras lineales tales como los \textit{jets} astrofísicos.

El el caso unidimensional, la contribuci\'on estad\'istica al potencial termodinámico tiene la forma:

\begin{equation}\label{Grand-Potential-2Dst}
\Omega^{1D}_{st}(B,\mu,T)= \sum_{s=-1,0,1} \frac{1}{2 \pi \beta} \int\limits_{-\infty}^{\infty} dp_3 \ln \left((1-e^{-(\varepsilon(p_3, B,s)- \mu)\beta})(1-e^{-(\varepsilon(p_3,B,s)+ \mu)\beta})\right).
\end{equation}

Usando el desarrollo en serie de Taylor del logaritmo podemos escribir la Ec.~(\ref{Grand-Potential-2Dst}) como:

\begin{equation}\label{Grand-Potential-2Dsst1}
\Omega^{1D}_{st}(B,\mu,T)= -\sum_{s=-1,0,1} \frac{1}{2 \pi \beta}  \sum_{n=1}^{\infty} \frac{e^{n \mu \beta}+e^{- n \mu \beta }}{n} \int\limits_{0}^{\infty} dp_3 e^{-n \beta \varepsilon(s)},
\end{equation}

\noindent donde los términos $e^{n \mu \beta}$ y $e^{- n \mu \beta}$ corresponden a las partículas y las antipartículas respectivamente. La integración por $p_3$ puede hacerse, y con ello se obtiene para $\Omega^{1D}_{st}(B,\mu,T)$ la expresión:

\begin{equation}\label{Grand-Potential-2Dsst3}
\Omega^{1D}_{st}(B,\mu,T)= -\sum_{s=-1,0,1} \frac{m \sqrt{1- s b}}{ \pi \beta}  \sum_{n=1}^{\infty} \frac{e^{n \mu \beta}+e^{- n \mu \beta }}{n} K_1(n \beta m \sqrt{1-s b}),
\end{equation}

\noindent $K_1(x)$ es la función de McDonald de primer orden. 

Sumando sobre el spin se tiene que:

\begin{eqnarray}\label{Grand-Potential-2Dstexact}
\Omega^{1D}_{st}(B,\mu,T)= - \frac{m \sqrt{1+ b}}{ \pi \beta}  \sum_{n=1}^{\infty} \frac{e^{n \mu \beta}+e^{- n \mu \beta }}{n} K_1(n \beta m \sqrt{1+b})\\ \nonumber
-\frac{m}{ \pi \beta}  \sum_{n=0}^{\infty} \frac{e^{n \mu \beta}+e^{- n \mu \beta }}{n} K_1(n \beta m)\\ \nonumber
-\frac{m \sqrt{1-b}}{ \pi \beta}  \sum_{n=0}^{\infty} \frac{e^{n \mu \beta}+e^{- n \mu \beta }}{n} K_1(n \beta m \sqrt{1-b}). \nonumber
\end{eqnarray}

\noindent Como en el caso tridimensional, en el límite de baja temperatura, el término dominante en la Ec.~(\ref{Grand-Potential-2Dstexact}) viene de las partículas con $s=1$. Por ello $\Omega^{1D}_{st}$ puede reescribirse como:

\begin{equation}\label{Grand-Potential-2DstTcero}
\Omega^{1D}_{st}(B,\mu,T)= - \frac{\sqrt{\varepsilon(0,B)}}{\sqrt{2 \pi} \beta^{3/2}} Li_{3/2}(e^{\mu^{\prime 1D} \beta}),
\end{equation}

\noindent con  $\mu^{\prime 1D} =\mu-\varepsilon(0,B)$. (Aunque $\mu^{\prime 1D}$ y $\mu^{\prime}$ responden a la misma definición, la distinci\'on es necesaria para evitar confusiones en lo que sigue.)

La contribución del vacío al potencial termodinámico en una dimensión es, luego de ser regularizada (Apéndice A):

\begin{equation}\label{Grand-Potential-2Dvacreg}
\Omega^{1D}_{vac}(B)= -\frac{m^2}{2 \pi} ((1-b) \log (1-b)+ (1+b)\log (1+b)).
\end{equation}

Finalmente, para el potencial termodinámico total en el límite de baja temperatura tenemos:

\begin{equation}\label{Grand-Potential-2DTcero}
\Omega^{1D}(B,\mu,T) =  - \frac{\sqrt{\varepsilon(0,B)}}{\sqrt{2 \pi} \beta^{3/2}} Li_{3/2}(e^{\mu^{\prime 1D} \beta}) -\frac{m^2}{2 \pi} \left ( (1-b) \log (1-b)+ (1+b)\log (1+b) \right ).
\end{equation}

La densidad de partículas obtenida de la Ec.~(\ref{Grand-Potential-2DTcero}) es:

\begin{equation}\label{Particle-Density-2DTcero}
N^{1D} = -\frac{\partial \Omega^{1D}(B,\mu,T)}{\partial \mu}= \frac{\sqrt{\varepsilon(0,B)}}{\sqrt{2 \pi \beta^{1/2}}} Li_{1/2}(e^{\mu^{\prime 1D}  \beta}).
\end{equation}

Esta expresión es similar a la obtenida para un gas tridimensional de bosones vectoriales cargados en presencia de un campo magnético constante (ver Ec.~(23) de  \cite{Khalilov1997}). Ella no admite la sustitución $\mu^{\prime 1D} = 0$ (porque $Li_{1/2}(1)\rightarrow\infty$). En consecuencia, una temperatura crítica no puede ser definida y el gas no condensa en el sentido usual. Sin embargo, en el límite de muy baja temperatura, en el cual $\mu\sim \varepsilon(0,B)\gg T$ pero $T \gg \mu^{\prime 1D}$ la expresión (\ref{Particle-Density-2DTcero}) puede aproximarse como:

\begin{equation}\label{Particle-Density-2Dcondensate}
N^{1D} \simeq \frac{1}{2 \beta} \sqrt{\frac{2 \varepsilon(0,B)}{-\mu^{\prime 1D}}}.
\end{equation}

La Ec.~(\ref{Particle-Density-2Dcondensate}) diverge cuando $\mu'\rightarrow 0$ a pesar de que la densidad de part\'iculas debe permanecer finita. Esto indica que la interpetaci\'on correcta de esta expresión no es como una manera de calcular $N^{1D}$, sino como una manera de hallar $\mu^{\prime 1D}$ en función de $N$, $B$ y $T$:

\begin{equation}\label{miuprima-1D}
\mu^{\prime 1D} = -\varepsilon(0,B)\frac{T^2}{ N^2}= -m \sqrt{1-b}\frac{T^2}{ N^2}.
\end{equation}

De la Ec.~(\ref{miuprima-1D}) se sigue que $\mu^\prime$ es una función decreciente de $B$ que va a cero cuando $B \rightarrow B_c$ a\'un en el caso de que la temperatura sea finita. Como  $\mu^{\prime 1D} = 0$ es la condición para la ocurrencia de la condensación de Bose-Eintein, esto significa que a pesar de ser imposible definir una temperatura cr\'itica, el condensado a\'un puede alcanzarse a temperatura finita.

Para analizar esto en m\'as detalle, tengamos en cuenta que con la ayuda de la Ec.~(\ref{miuprima-1D}) la distribución de Bose-Einsten $n^{+}(p_3)$ puede aproximarse en una vecindad de $p_3 =0$ como:

\begin{eqnarray}
n^+(p_3) = \frac{1}{e^{\beta(\varepsilon(p_3,B)-\mu)-1}} \simeq \frac{2 \varepsilon(0,B) T}{p^2 +\varepsilon(0,B)^2 - \mu^2} \simeq \frac{2 \varepsilon(0,B) T}{p^2 -2 \varepsilon(0,B) \mu^{\prime 1D}}, \\\nonumber
\end{eqnarray}

\begin{eqnarray}\label{nmas}
n^+(p_3) \simeq 2 N \frac{\gamma}{p_{3}^2 + \gamma^2},
\end{eqnarray}

\noindent donde $\gamma = \sqrt{- 2 \varepsilon(0,B) \mu^{\prime 1D}}$. La Ec.(\ref{nmas}) significa que para $p_3 \approx 0$, $n^+(p_3)$ tiende a una distribución de Cauchy centrada en $p_3=0$. Ahora el equivalente a los límites $T \rightarrow 0$ o $B\rightarrow B_c$  es el límite $\gamma \rightarrow 0$ y:

\begin{equation}\label{nmas1}
\lim_{\gamma\rightarrow 0} n^+(p_3) = 2 \pi N \delta(p_3).
\end{equation}

Usando la Ec.(\ref{nmas1}) tenemos la siguiente expresión para la densidad de partículas $N$ en la vencindad de $p_3=0$:

\begin{equation}\label{densityTcero}
N \simeq \frac{1}{2 \pi} \lim_{\gamma\rightarrow 0}\int_{-\infty}^{\infty} 2 \pi N \frac{\gamma}{p_{3}^2 + \gamma^2} dp_3 = N \int_{-\infty}^{\infty} \delta(p_3)dp_3.
\end{equation}

El comportamiento tipo $\delta$ de Dirac de la distribución de Bose-Einstein $n^+(p_3)$ para bajas temperaturas o campos magn\'eticos altos se muestra en la Fig.~3.
En su panel izquierdo $n^+(p_3)$ ha sido representada como función de $p_3$ para un valor fijo de la temperatura $T=10^7 K$ y varios valores del campo magnético. El máximo de las curvas aumenta y se mueve hacia la derecha cuando el campo magnético crece. En el  panel derecho, $n^+(p_3)$ se presenta como función de $p_3$ para un valor fijo del campo magnético $B=0.1 B_c$ y varios valores de la temperatura. El máximo de la curva aumenta y se mueve a la derecha con la disminución de la temperatura.
La Fig.~3 ilustra c\'omo, aunque no se puede definir una temperatura crítica, a medida que $T \rightarrow 0$ o $B \rightarrow B_c$ los bosones se concentran en el estado básico y sus estados vecinos. Por tanto, el sistema muestra una condensación difusa.
	
\begin{figure}[h!]
\centering
\includegraphics[width=0.45\linewidth]{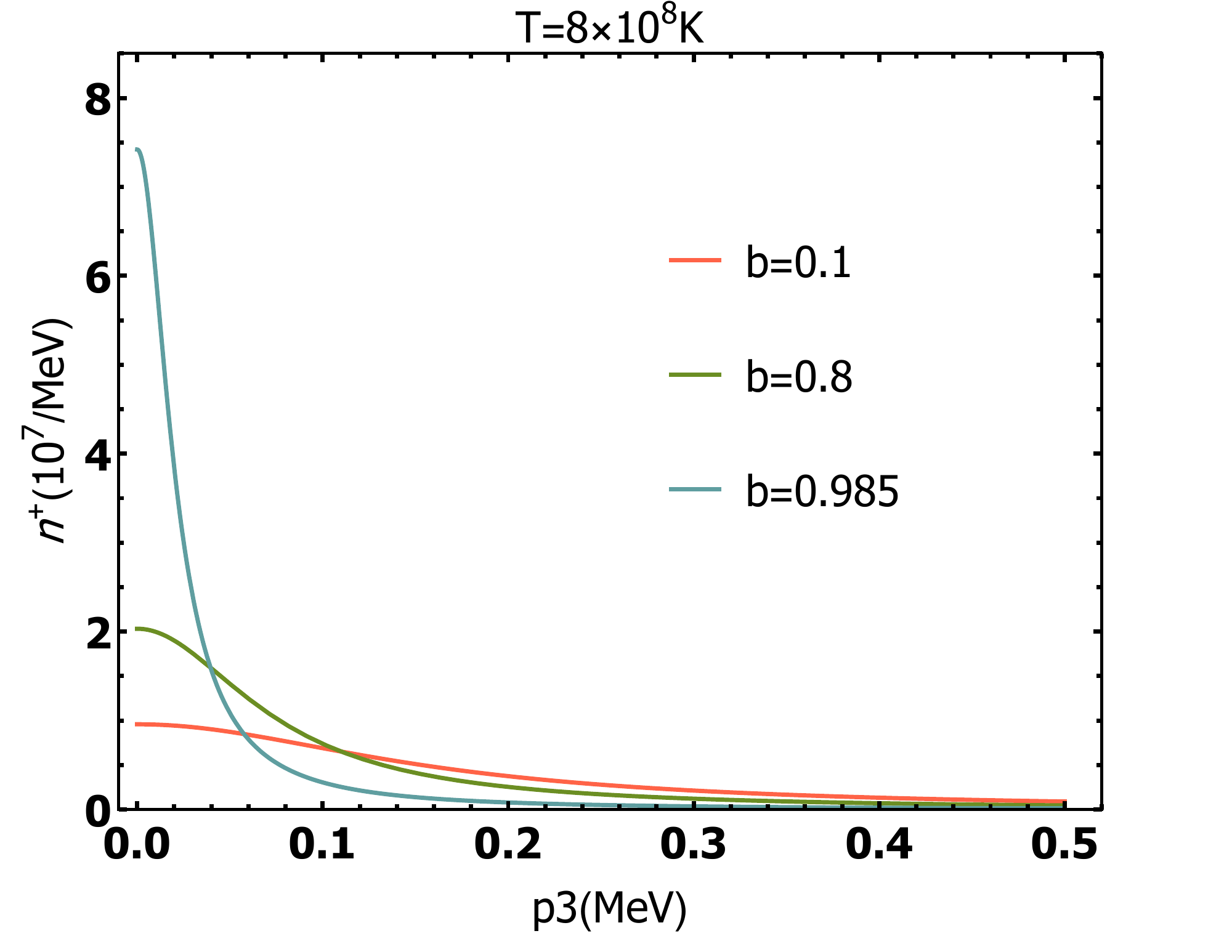}
\includegraphics[width=0.45\linewidth]{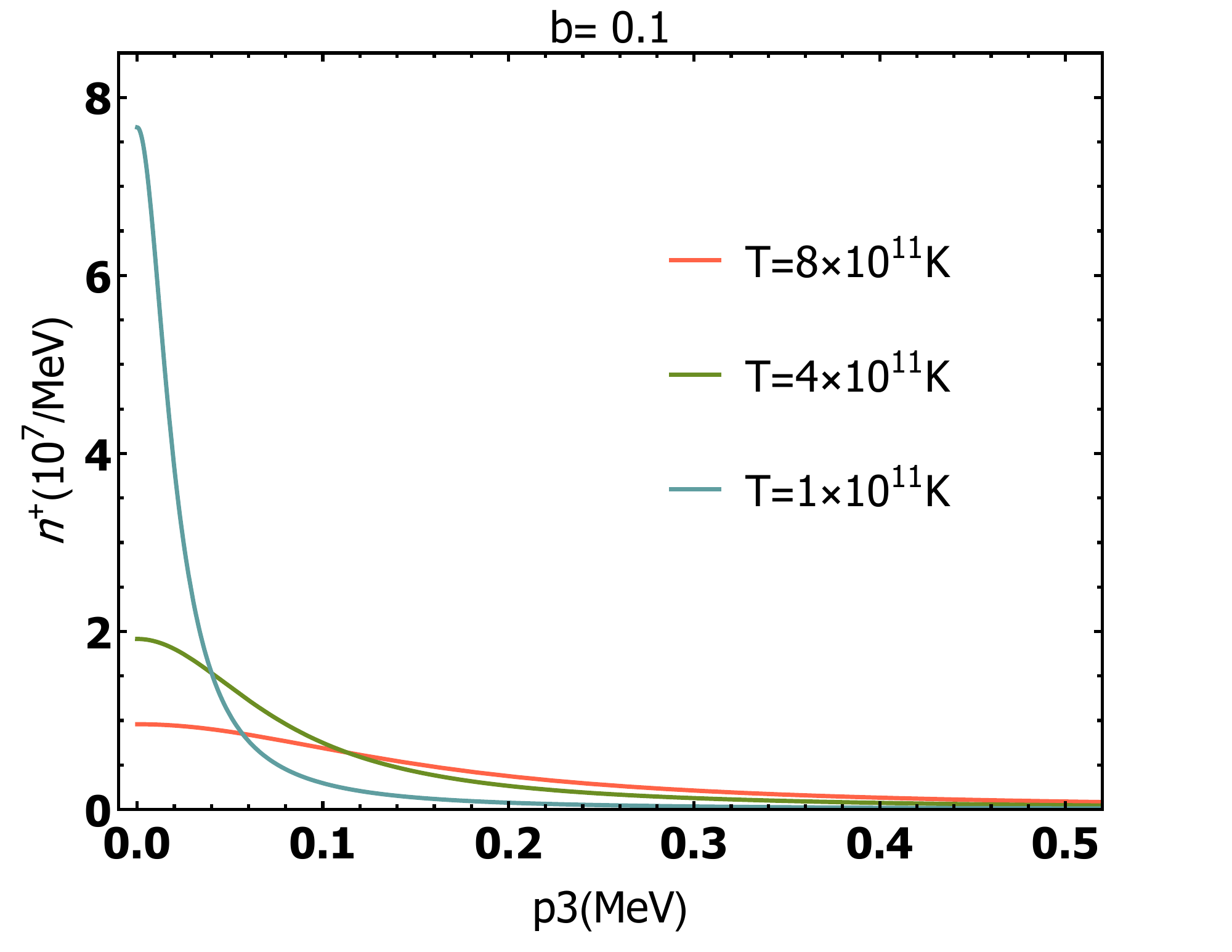}	
\caption{\label{fig1} Panel izquierdo: distribución de Bose-Einstein como función de $p_3$ para $T$ fija y varios valores del campo magnético. Panel derecho: distribución de Bose-Einstein como función de $p_3$ para campo magnético fijo y varios valores de la temperatura. En ambos gráficos $N=10^{38} cm^{-3}$.}
\end{figure}

El orden de las temperaturas alrededor de las cuales la transición de fase difusa ocurre puede ser estimado a partir del calor específico. Para calcularlo hacen falta la energía y la entropía del gas unidimensional. Ellas son:

\begin{equation}\label{entropy-1D}
S^{1D}=-\frac{\partial \Omega^{1D}}{\partial T}= -\frac{1}{T}\left(\mu^{\prime} N +\frac{3}{2} \Omega^{1D}_{st}+ \frac{2 \varepsilon(0,B)T^2}{N}\right),
\end{equation}

\begin{equation}\label{energy-1D}
E^{1D} = T S^{1D}+\Omega^{1D}+\mu N = \varepsilon(0,B) \left ( N - \frac{2 T^2}{N}\right ) + \Omega^{1D}_{vac}- \frac{1}{2} \Omega^{1D}_{st}.
\end{equation}

De la Ec.(\ref{energy-1D}) el calor específico es:

\begin{equation}\label{specificheat-1D}
C^{1D}_v= \frac{\partial E^{1D}}{\partial T} = -\frac{1}{2 T} \left ( \mu^{\prime 1D} N +\frac{3}{2} \Omega_{st} + 10 \frac{\varepsilon(0,B) T^2}{N}\right ).
\end{equation}

\begin{figure}[h]
\centerline{
\includegraphics[width=0.45\linewidth]{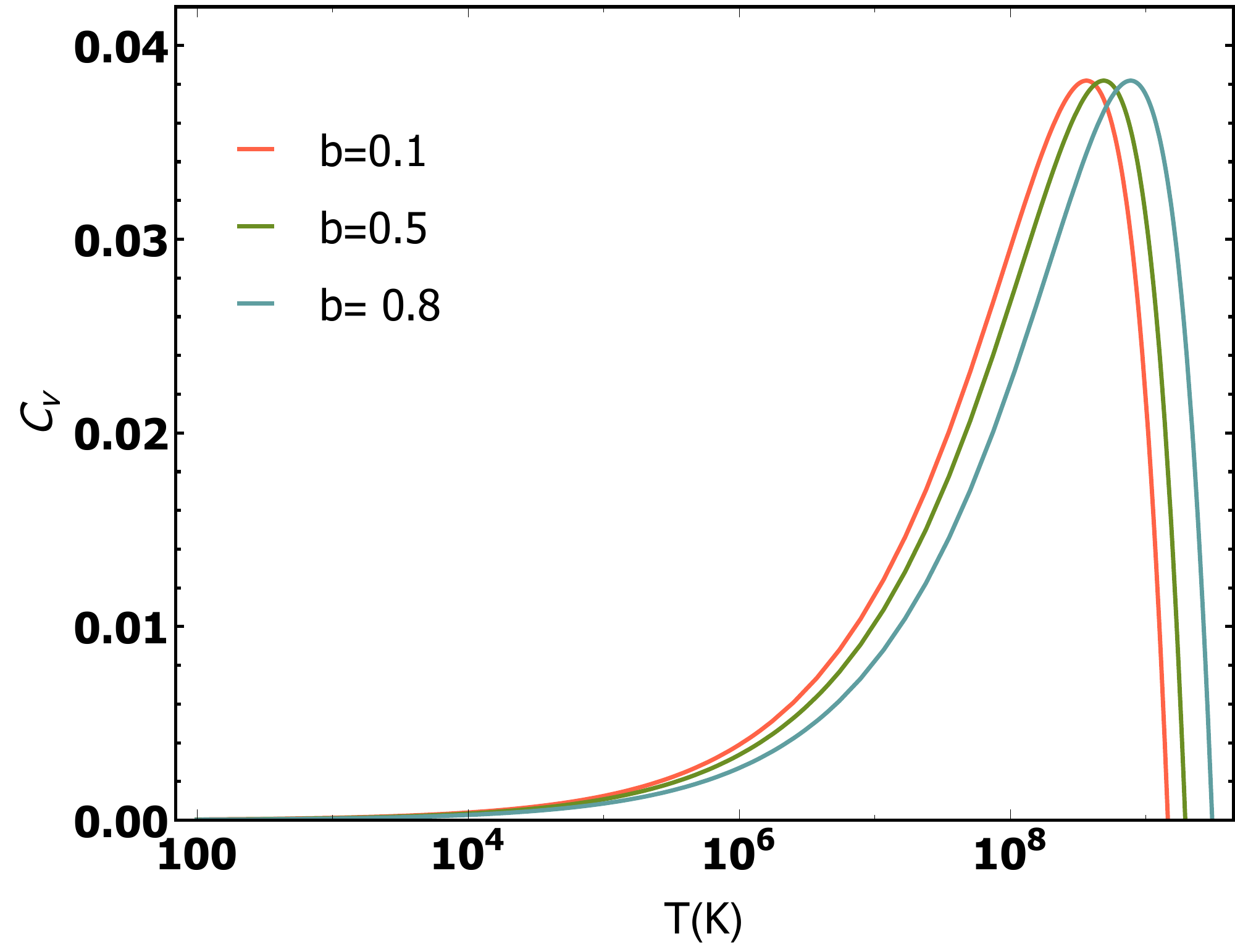}
\includegraphics[width=0.45\linewidth]{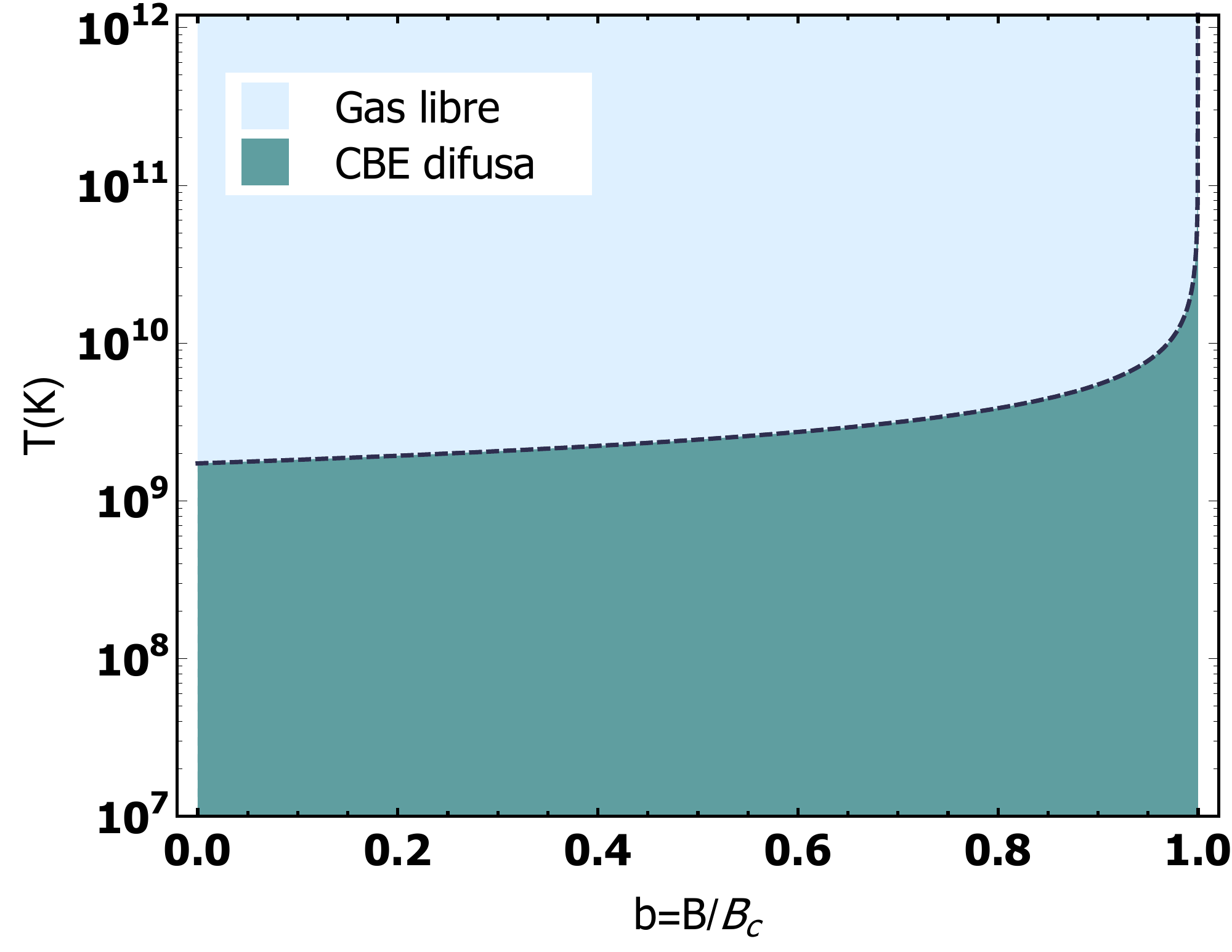}}	
\vspace*{8pt}
\caption{\label{fig4} Panel izquierdo: El calor específico como función de la temperatura para el caso unidimensional. Panel derecho: $T_{max}$ vs. $b$. En ambos gráficos $N= 10^{34}$cm$^{-3}$.}
\end{figure}

$C^{1D}_v$ ha sido representado como función de la temperatura en el panel derecho de la Fig.~\ref{fig4}. De esta figura puede verse que el calor específico tiene un máximo. Su posición en el eje de las abscisas puede calcularse como función de $b$:

\begin{equation}\label{Tc-1D}
T_{max} = \frac{(\zeta (\frac{3}{2}) N)^2}{144 \varepsilon(0,B)}.
\end{equation}

\noindent Al igual que $T_{cond}$, $T_{max}$ crece con la densidad y diverge con el campo magnético. En el panel izquierdo de la Fig.~\ref{fig4} se muestra $T_{max}$ como función de $b$ para $N = 10^{34}cm^{-3}$. Aunque las regiones condensada y no condensada fueron sombreadas, es conveniente no olvidar que $T_{max}$ no define una temperatura crítica. Por lo tanto, el panel izquierdo de la Fig.~\ref{fig4} es solo un diagrama de fases aproximado que indica el rango de temperaturas alrededor de las cuales el n\'umero de part\'iculas en una vecindad del estado básico comienza a crecer.

\subsection{Magnetizaci\'on}

La magnetización de los gases en una y tres dimensiones se obtiene a partir de las ecuaciones para sus potenciales termodinámicos, Ec.~(\ref{Grand-Potential-Tcero}) y Ec.~(\ref{Grand-Potential-2DstTcero}), si se deriva en ellas con respecto al campo magnético:

\begin{equation}\label{magnetization}
\mathcal M = -\frac{\partial \Omega_{st}}{\partial B}-\frac{\partial \Omega_{vac}}{\partial B},
\end{equation}

\begin{equation}\label{magnetization-1D}
\mathcal M^{1D}=-\frac{\partial \Omega^{1D}_{st}}{\partial B}-\frac{\partial \Omega^{1D}_{vac}}{\partial B}.
\end{equation}

Las contribuciones estadísticas a la magnetización en una y tres dimensiones son:

\begin{eqnarray}\label{magnetization-st2}
\mathcal M_{st}= -\frac{\partial \Omega_{st}}{\partial B}= \frac{\kappa m}{\varepsilon(0,B)} N - \frac{2 \kappa m T^{5/2}}{(4 \pi)^{5/2} (2-b)^2 \varepsilon(0,B)^{1/2}} Li_{5/2}(e^{\beta \mu^{\prime}}),
\end{eqnarray} y

\begin{equation}\label{magnetization-1Dst0}
\mathcal M^{1D}_{st}=-\frac{\partial \Omega^{1D}_{st}}{\partial B}= \frac{\kappa m}{\varepsilon(0,B)} N - \frac{\kappa m T^{3/2} }{2^{3/2} \pi^{1/2} \varepsilon(0,B)^{3/2}} Li_{3/2}(e^{\beta \mu^{\prime}}).
\end{equation}

En las Ecs.~(\ref{magnetization-st2}) y (\ref{magnetization-1Dst0}) la cantidad:

\begin{equation}
\frac{\kappa m}{\varepsilon(0,B)} = \frac{\kappa}{\sqrt{1-b}}
\end{equation}

\noindent juega el papel de un momento magnético efectivo que aumenta con la intensidad del campo magn\'etico. 

Las contribuciones de vacío son:

\begin{equation}\label{magnetization-vac}
\mathcal M_{vac}= -\frac{\kappa m^3}{72 \pi} \left( 7 b(b^2-6) + 3(2 b^2+2 b-7)(1-b)\log(1-b)-3(2b^2-2b-7)(1+b)\log(1+b)\right ),
\end{equation}

\begin{equation}\label{magnetization-1Dvac}
\mathcal M^{1D}_{vac}= \frac{\kappa m}{\pi} \log \left(\frac{1+b}{1-b}\right ).
\end{equation}

Dadas las altas densidades de part\'iculas y bajas temperaturas con las que estamos tratando, los segundos términos en las Ecs.~(\ref{magnetization-st2}) y (\ref{magnetization-1Dst0}) son despreciables. Asimismo, se puede demostrar que las magnetizaciones de vacío (Ecs. (\ref{magnetization-vac}) y (\ref{magnetization-1Dvac})) solo son relevantes para densidades de partículas bajas a campos muy altos, por lo que también pueden ser despreciadas. De manera que para una y tres dimensiones la magnetizaci\'on queda igual a: 

\begin{equation}\label{magnetizationtotal}
\mathcal M = \frac{\kappa m}{\varepsilon(0,B)} N = \frac{\kappa}{\sqrt{1-b}} N.
\end{equation}

\begin{figure}[h]
\centering
\includegraphics[width=0.49\linewidth]{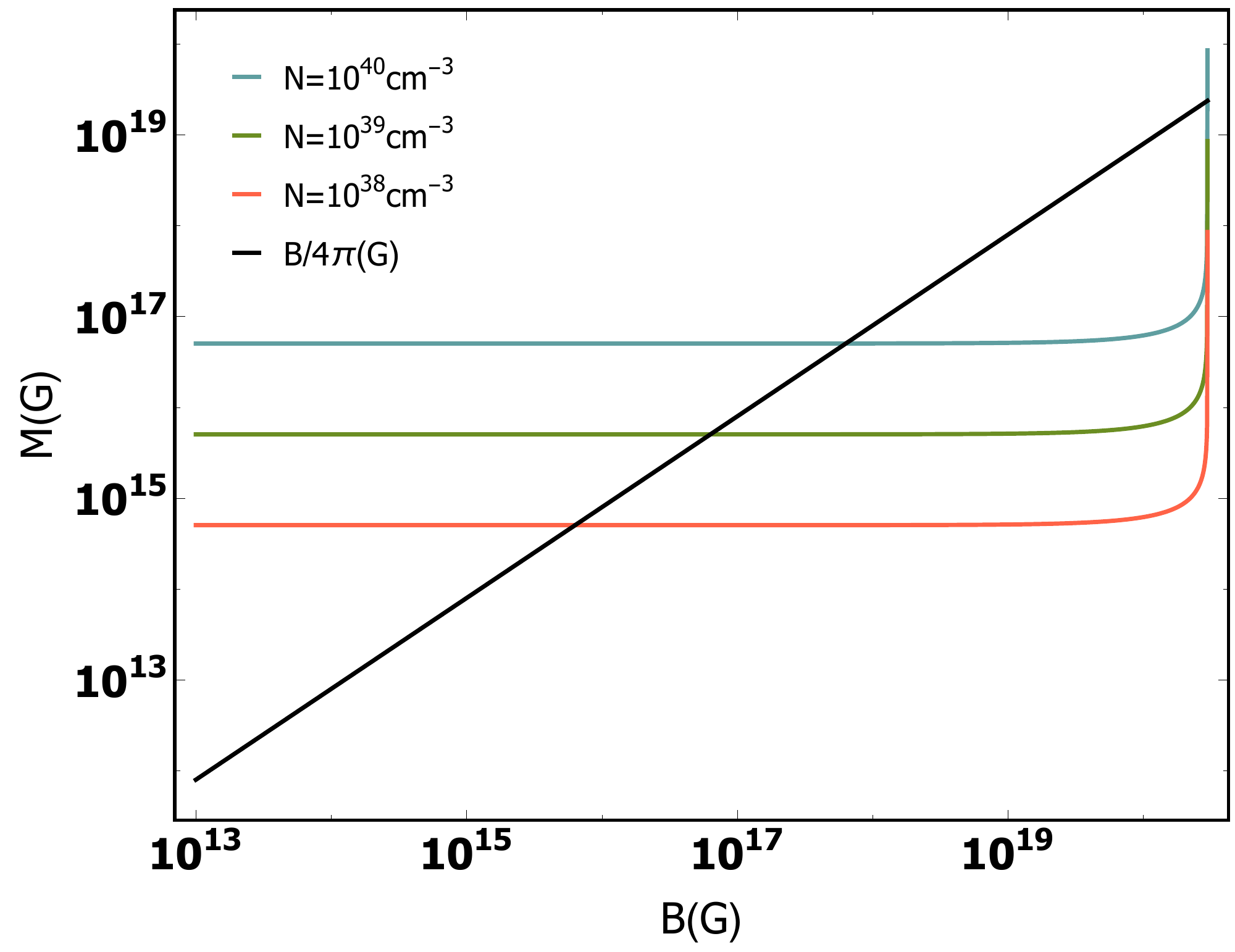}
\includegraphics[width=0.49\linewidth]{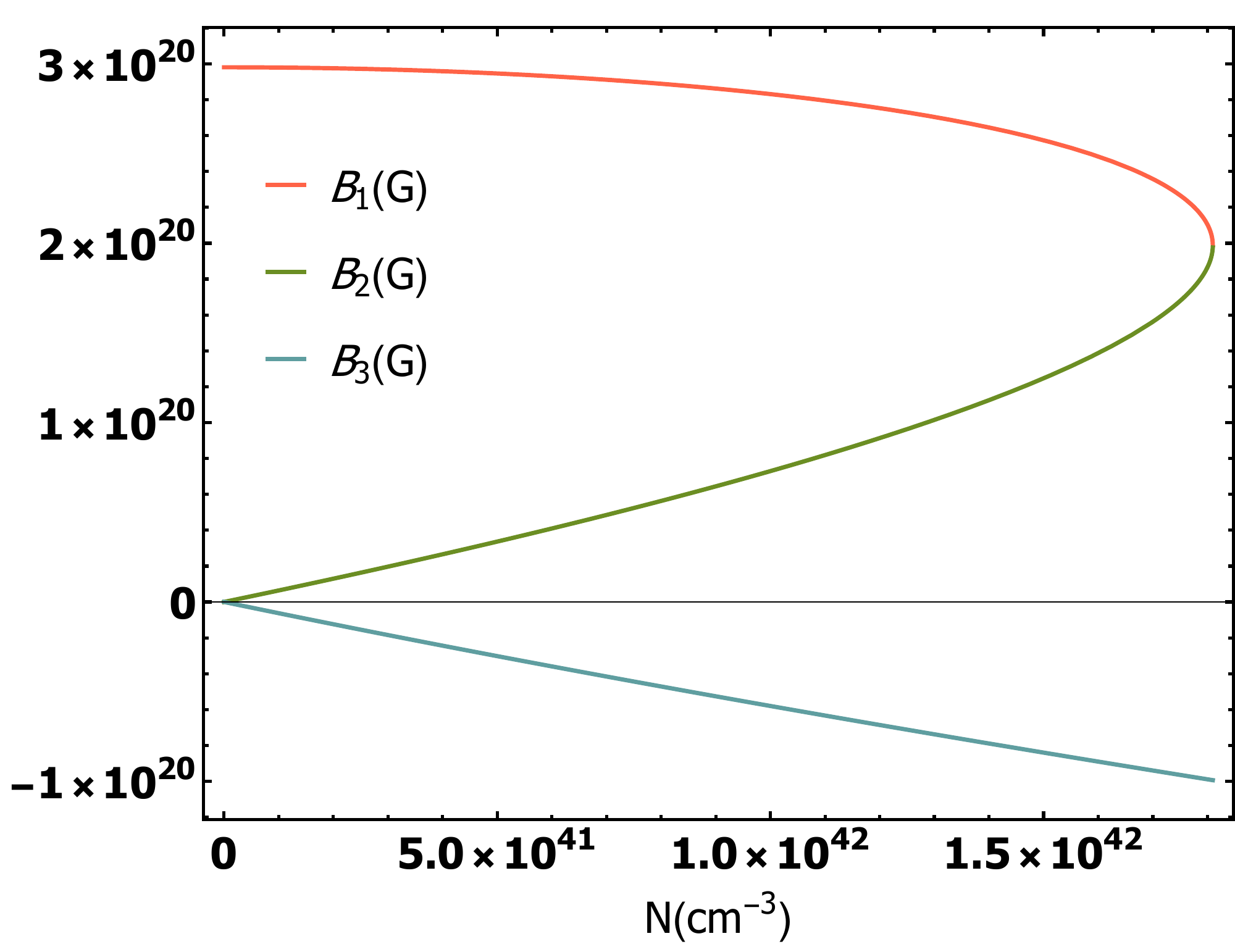}
\caption{\label{fig3} Panel izquierdo: La magnetizaci\'on como funci\'on del campo magn\'etico para varios valores de la densidad de part\'iculas. En este gr\'afico tambi\'en se ha trazado la curva $B/4\pi$. Panel derecho: Las tres soluciones de la ecuaci\'on de automagnetizaci\'on como funci\'on de la densidad de part\'iculas.}
\end{figure}

La expresi\'on anterior no es más que el producto del momento magnético efectivo por la densidad de partículas y refleja el hecho de que a baja temperatura todas las partículas están en el estado de spin $s=1$.  No obstante, un incremento en el campo magnético aún aumenta la magnetización porque el momento magnético efectivo $\kappa/\sqrt{1-b}$ crece con $B$ y diverge cuando $B\rightarrow B_c$ ($b \rightarrow 1$). N\'otese adem\'as que la magnetización es siempre positiva y diferente de cero incluso cuando $B=0$ ($\mathcal M(B=0)=\kappa N$). Esto es una evidencia del comportamiento ferromagnético del gas de bosones vectoriales neutros a baja temperatura. Este fen\'omeno, conocido como ferromagnetismo de Bose-Einstein \cite{Yamada}, no es el resultado de una interacci\'on entre los spines de las part\'iculas, sino una consecuencia de la aparici\'on de la condensaci\'on de Bose-Einstein, un estado en el cual todos los bosones tienen $s=1$. El comportamiento descrito para la magnetizaci\'on se muestra en el panel izquierdo de la Fig.~\ref{fig3}.

En la b\'usqueda de fuentes para los campos magnéticos astrof\'isicos, estamos ahora interesados en explorar si el sistema alcanza la condición de automagnetización, es decir, si la curva sólida en el panel izquierdo de la Fig.~\ref{fig3} intersecta las curvas de la magnetización. Para ello consideraremos la ecuaci\'on $H = B-4 \pi \mathcal M$ sin campo magnético externo $H=0$, y resolvemos la ecuación de automagnetizaci\'on:

\begin{equation}\label{selfmag1}
B = 4 \pi {\mathcal M},
\end{equation}

\noindent que en este caso es una ecuación cúbica en $B$:

\begin{eqnarray}\label{selfmag}
B =  \frac{ 4 \pi \kappa}{\sqrt{1-B/B_c}} N,
\end{eqnarray}

En el panel derecho de la Fig.~\ref{fig5} se muestran las tres soluciones de Ec.~(\ref{selfmag}), pero solo una de ellas tiene sentido físico. Para una de las raíces, el campo magnético es negativo (línea azul), mientras que para la otra el campo decrece con la densidad de partículas, alcanzando el valor $B_c$ cuando $N$ va a cero (línea roja). Esas soluciones implican ambas que la magnetización decrece con $N$, por tanto, ellas se contradicen con la Ec.~(\ref{magnetizationtotal}) y deben ser desechadas. Por tanto, la única solución admisible para la ecuación de automagnetización es la dada por la línea verde. Los puntos de esta curva son los valores del campo magnético automantenido por el gas. No obstante, esta solución deviene compleja para densidades mayores que $N_c = 1.81 \times 10^{42}cm^{-3}$. En consecuencia, $N_c$ acota los valores de la densidad de partículas para las cuales la automagnetización es posible. El campo máximo que puede ser sostenido por el gas corresponde a esta densidad crítica y es igual a $2/3\times B_c \sim 10^{20}$~G. Este l\'imite es similar al obtenido para el gas de bosones vectoriales cargados \cite{Elizabeth}. Por otra parte, los valores de $B \sim 10^{12}-10^{18}$~G y $N \sim 10^{38}$~cm$^{-3}$ t\'ipicos de las ENs est\'an en los rangos para los cuales la automagnetizaci\'on del gas es posible.  

\section{Presiones anisotr\'opicas}

Para calcular las presiones del gas partiremos del tensor de energ\'ia-momento de nuestro sistema. El tensor de energ\'ia-momento $T^i_j$ es un tensor diagonal cuya parte espacial contiene a las presiones y la componente temporal a la densidad de energ\'ia interna $E$. A partir del potencial termodin\'amico $T^i_j$ puede calcularse como: 

\begin{equation}\label{emtensor}
T^i_j=\frac{\partial\Omega}{\partial a_{i,\lambda}}a_{j,\lambda}-\Omega\delta_{j}^i,\quad\quad T_4^4=-E,
\end{equation}

\noindent donde $a_{i}$ denota los campos presentes (fermiones, bosones, electromagn\'etico, etc.) \cite{PerezRojas:2006dq}.
Para un potencial termodin\'amico que depende de un campo externo, la Ec.(\ref{emtensor}) conduce a t\'erminos de presi\'on con la forma:

\begin{equation}
\textit{T}^i_j=-\Omega-F_k^i\left( \frac{\partial\Omega}{\partial F_k^j}\right),\quad  i=j.
\end{equation}

Si teniendo en cuenta que el campo magn\'etico est\'a dirigido en la direcci\'on $\textbf{e}_3$, se calcula la presi\'on a lo largo de cada una de las direcciones, la anisotrop\'ia se hace expl\'icita:

\begin{subequations}\label{pressures}
	\begin{align}
P_{\parallel}&=\textit{T}^3_3=-\Omega = -\Omega_{st} -\Omega_{vac},\\
P_{\perp}&=\textit{T}_1^1=\textit{T}_2^2=-\Omega-B \mathcal M = P_{\parallel}-B \mathcal M. 
\end{align}
\end{subequations}

\begin{figure}[h!]
	\centering
	\includegraphics[width=0.45\linewidth]{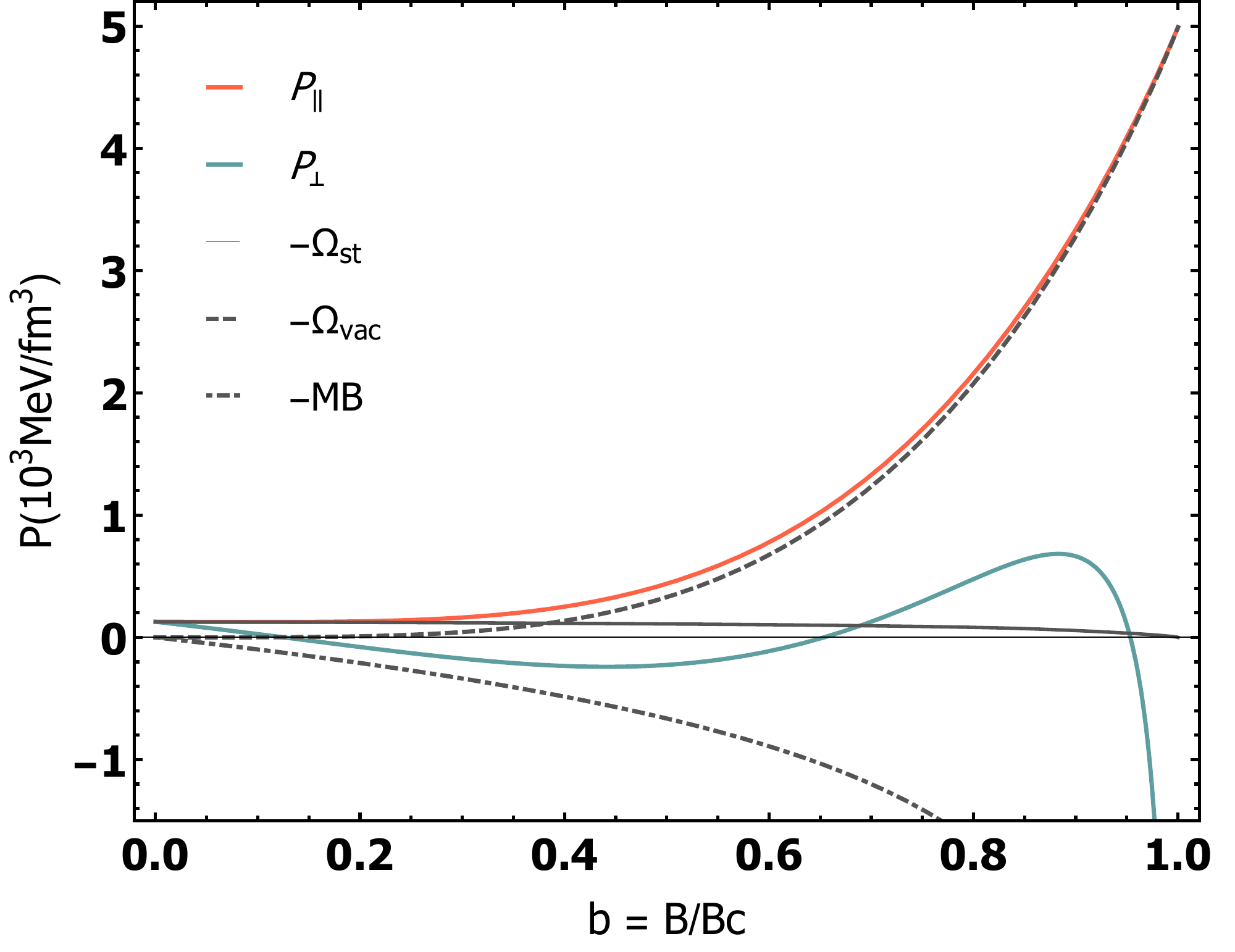}
	\includegraphics[width=0.45\linewidth]{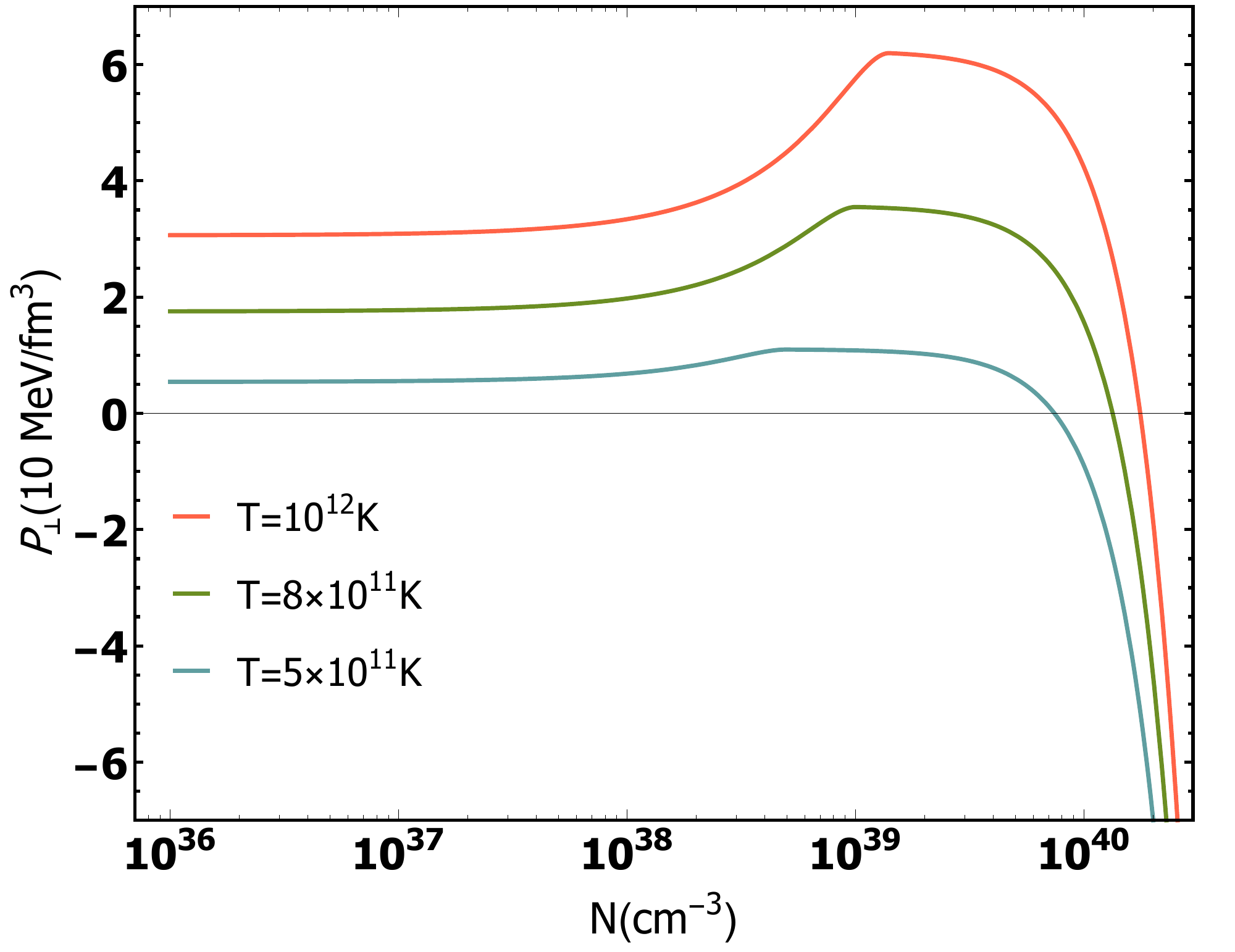}	
	\caption{\label{fig8} Panel derecho: Las presiones paralela (l\'inea roja) y perpendicular (l\'inea azul) como funci\'on del campo magn\'etico para $N = 10^{39}$cm$^{-3}$ y $T = 1.5 \times 10^{12}$K; las contribuciones estad\'istica (l\'inea gris continua), de vac\'io (l\'inea gris a rayas) y magn\'etica (l\'inea gris a puntos y rayas) a las presiones han sido tambi\'en representadas. Panel derecho: La presi\'on perpendicular del gas automagnetizado como funci\'on de la densidad de part\'iculas para varios valores de la temperatura.}
\end{figure}

En el panel derecho de la Fig.~\ref{fig8} se muestran las presiones paralela y perpendicular como funci\'on del campo magn\'etico (Ecs.(\ref{pressures})) para $T=1.5 \times 10^9 K $ y $N=10^{39} cm^{-3}$. Mostramos adem\'as las contribuciones estad\'istica y de vac\'io a la presi\'on paralela. Los valores de la presi\'on paralela y su parte estad\'istica ($-\Omega_{st}$) coinciden para $B=0$, pero su comportamiento es diferente a medida que el campo magn\'etico crece. Ambas son siempre positivas, pero la presi\'on paralela total aumenta con el campo y tiende a la contribuci\'on del vac\'io $-\Omega_{vac}$, mientras que la parte estad\'istica decrece y se va a cero para $B=B_c$ -cuando todas las part\'iculas est\'an en el condensado el gas no ejerce presi\'on. Es importante resaltar que la presi\'on paralela se mantiene diferente de cero gracias a la contribuci\'on de vac\'io.

Por el contrario, la presi\'on perpendicular (l\'inea azul en el panel izquierdo de la Fig.~6) en dependencia del campo magn\'etico puede alcanzar valores negativos. Esto sucede porque la contribuci\'on principal a $P_{\perp}$ viene del t\'ermino $-\mathcal M B$ que es siempre negativo y diverge en el campo cr\'itico. Tener una presi\'on negativa significa que el sistema se hace inestable. Como el efecto de una presi\'on perpendicular negativa es empujar a las partículas hacia el eje del campo magn\'etico, estar\'iamos en presencia de un colapso magn\'etico transversal. Este tipo de colapso ha sido descrito previamente para gases fermi\'onicos \cite{Chaichian1999gd,Ferrer}. 

Que la presi\'on perpendicular se haga negativa o no depende del campo magn\'etico pero tambi\'en de la temperatura y la densidad de part\'iculas. Esto puede verse si examinamos en m\'as detalle la presi\'on para el gas automagnetizado. Para hacerlo sustituimos en $P_{\perp}$  la soluci\'on de la condici\'on de automagnetizaci\'on $B = 4 \pi \mathcal M$, y graficamos la presi\'on perpendicular como funci\'on de la densidad de part\'iculas para varios valores de $T$ en el panel derecho de la Fig.~\ref{fig8}. Si nos movemos de la zona de bajas densidades hacia la de densidades mayores, es posible apreciar que en un inicio aumentar $N$ incrementa la presi\'on paralela, pero esto también aumenta el campo autogenerado y $T_{cond}$. Una vez que $T_{cond}$ supera la temperatura del gas, la presión disminuye porque una fracción de partículas pasa al condensado. Además, a medida que el campo magn\'etico aumenta, la contribución a $P_{\perp}$ del término magn\'etico $- \mathcal M B$ es cada vez m\'as relevante, de forma tal que seguir aumentando $N$ hace al sistema inestable. Un decremento en la temperatura disminuye el valor de la densidad de partículas para el cual la inestabilidad aparece.

Cuando el gas no está automagnetizado, sino sometido a un campo mágnetico externo, aumentar la densidad de partículas continuamente también lleva al sistema a una inestabilidad. En este caso un aumento de $N$ no aumenta el campo magn\'etico, pero s\'i la magnetización y la temperatura cr\'itica. Por tanto, el gas de bosones vectoriales neutro será inestable en dependencia de los valores de la temperatura, la densidad y el campo magnético, sea este autogenerado o no.

\section{Conclusiones del capítulo}

En este cap\'itulo se obtuvo el espectro energético de un gas de bosones vectoriales neutros en interacción con un campo magnético uniforme y constante en el marco del formalismo de Proca \cite{PhysRev.131.2326,PhysRevD.89.121701}. En dicho espectro se encontró un acoplamiento entre el campo magnético y la componente del momento de las partículas que es perpendicular a la dirección de este. Este acoplamiento es similar al que aparece para bosones cargados, aunque en el caso que aquí nos ocupa la ausencia de la carga implica la no aparición de niveles de Landau. La energía del estado básico del sistema Ec.~(\ref{massrest}) resultó ser una función decreciente del campo magnético que se anula cuando este alcanza el valor crítico $B_c=m/2 k$. 

Una vez conocido el espectro calculamos el potencial termodinámico del gas de bosones vectoriales neutros y sus propiedades termodin\'amicas fueron estudiadas en una y tres dimensiones. Los resultados obtenidos durante este estudio arrojaron las siguientes conclusiones:

\begin{itemize}

\item Cuando la temperatura es lo suficientemente baja, el gas de bosones vectoriales neutros sufre una transición de fase al condensado de Bose-Einstein. En dependencia de si el gas es uni o tridimensional, esta transición es difusa o usual. En ambos casos la transición al condensado está determinada no solo por el decrecimiento de la temperatura o el aumento de la densidad de part\'iculas, sino también por el aumento del campo magnético. 

\item La magnetización del gas es una cantidad positiva que crece con el campo magnético y diverge cuando $B=B_c$ para ambos casos.

\item Para densidades de partículas por debajo de cierto valor crítico, $N_c \cong 1.81 \times 10^{42}cm^{-3}$, la condición de automagnetización se satisface y el gas puede generar su propio campo magnético. El campo magn\'etico máximo que puede ser alcanzado por automagnetización es $2/3 \times B_c \sim 10^{20} G$. Ambos valores cr\'iticos, $N_c$ y $B_c$, est\'an por encima de los valores m\'aximos esperados para la densidad de part\'iculas y el campo magn\'etico en las ENs. Esto significa que la condici\'on de automagnetizaci\'on es alcanzable para los neutrones apareados en el interior de un EN.

\item El cambio de simetr\'ia esférica a simetría axial provocado en el sistema por la presencia del campo magnético se manifiesta, adem\'as de en el espectro, en la separaci\'on de las presiones en una componente paralela y otra perpendicular al eje magnético, tal como ocurre para gases magnetizados fermi\'onicos \cite{Chaichian1999gd}.

\item Cuando el campo magnético es débil, la presión ejercida por las partículas tiene el mayor peso en ambas componentes de la presi\'on, que son iguales. En cambio, a medida que el campo magnético se hace más fuerte, la presión paralela aumenta dominada por la presión del vacío, en tanto la presión perpendicular disminuye guiada por la presi\'on magn\'etica. 

\item Bajo ciertas condiciones de campo magnético, temperatura y densidad de partículas, la presión perpendicular se hace negativa y el sistema deviene inestable. Esto indica que el gas de bosones vectoriales neutros es susceptible de sufrir un colapso magnético transversal.

\end{itemize}

Todos estos fen\'omenos se dan para campos magn\'eticos, densidades de partículas y temperaturas en el orden de las típicas para las Estrellas de Neutrones, por lo que podrían ser relevantes en la descripción de su interior. Independientemente de ello, el estudio aqu\'i presentado tiene un car\'acter general que va m\'as all\'a de las aplicaciones astrof\'isicas, pues las conclusiones a las que hemos llegado son v\'alidas para cualquier gas magnetizado de bosones vectoriales neutros y masivos.

%% file: cap2_5.tex
\chapter{Colapso magn\'etico transversal en un gas \textit{npe} parcialmente bosonizado. Implicaciones astrof\'isicas}
\label{cap2_5}

En este cap\'itulo estudiaremos los efectos del campo magn\'etico en la materia que forma el interior de las Estrellas de Neutrones, considerada esta como un gas \textit{npe} parcialmente bosonizado. En particular, centraremos la atenci\'on en la posibilidad de que el colapso magn\'etico y la automagnetizaci\'on del gas se produzcan para las condiciones t\'ipicas del interior de estos objetos compactos. A partir de los resultados obtenidos en este estudio, se propone un modelo fenomenol\'ogico para la formaci\'on y mantenimiento de los \textit{jets} astrof\'isicos originados en el interior de una ENs. Este cap\'itulo comprende los resultados originales de la autora recogidos en \cite{Quintero2019jets}.

\section{Gas \textit{npe} parcialmente bosonizado}

El interior de una Estrella de Neutrones est\'a compuesto, al menos, por una mezcla de electrones ($ e $), neutrones ($ n $), protones ($ p $) -el gas \textit{npe} \cite{Camezind, 103}-, y neutrones y protones apareados ($ nn $) y ($ pp $) respectivamente \cite{PageSC,Chavanis2012,113}. Si adem\'as estas part\'iculas est\'an bajo la acci\'on de un campo magn\'etico, este contribuye a la bosonizaci\'on porque para $B\neq0$ la energía del estado b\'asico de los bosones compuestos es menor que la de los nucleones no apareados, es decir $\Delta\varepsilon = 2\varepsilon_F - \varepsilon_B > 0$, donde las energ\'ias del estado b\'asico de los fermiones $\varepsilon_F$ y los bosones $\varepsilon_B$ vienen dadas por:

\begin{equation}\label{fermionsGS}
\varepsilon_F = \Bigg \{
\begin{array}{lr}
\sqrt{m_F ^ 2 + 2 q B}, &\hbox{Fermiones cargados} \\
m_F - \kappa B,  &\hbox{Fermiones neutros}  \\
\end{array},
\end{equation}

\begin{equation} \label {bosonsGS}
\varepsilon_b = \Bigg \{
\begin {array} {lr}
  \sqrt{m_B ^ 2 - q B}, & \hbox{Bosones escalares cargados}\\
  \sqrt{m_B ^ 2 - 2 m_B\kappa B}, & \hbox{Bosones vectoriales neutros}
  \\
\end{array}.
\end{equation}

\begin{figure}[h]
	\centering
	\includegraphics[width=0.49\linewidth]{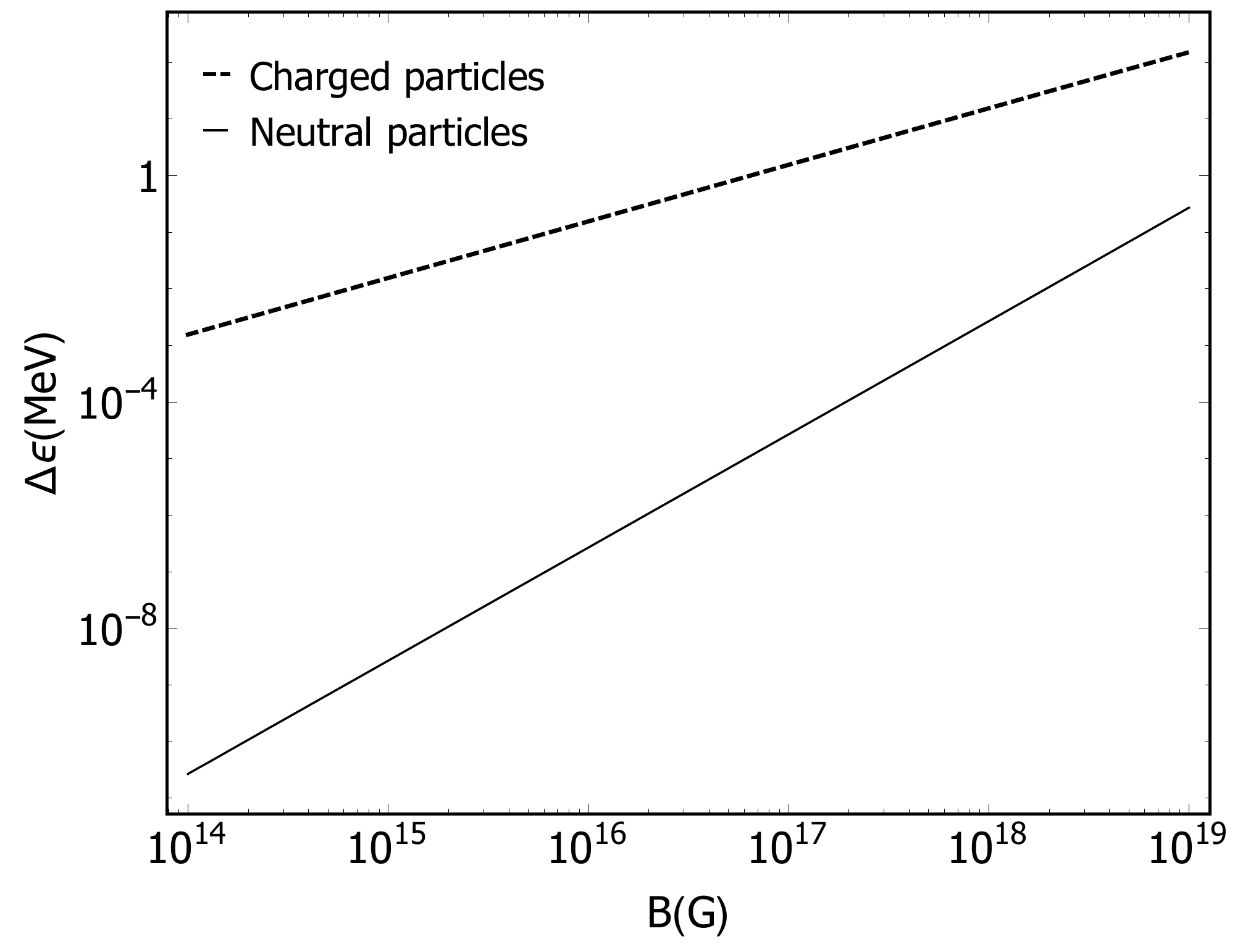}		
	\caption{\label{groundstate} Diferencia entre la energ\'ia del estado b\'asico de dos fermiones libres con respecto a la de dos fermiones apareados como funci\'on del campo magn\'etico, para part\'iculas cargadas y neutras.}
\end{figure}

En las Ecs.(\ref{fermionsGS}) y (\ref{bosonsGS}), $m_B$ y $m_F$ denotan las masas del bos\'on y el fermi\'on; $q$ es la carga eléctrica. La Fig.~\ref{groundstate}, muestra a $\Delta\varepsilon$ como función del campo magnético para neutrones (fermiones/bosones neutros) y protones (fermiones/bosones cargados). $\Delta\varepsilon$ es siempre mayor que cero y aumenta con $B$.

La presencia del campo magn\'etico en la estrella podr\'ia actuar como un desencadenante de la expulsión de la materia a través del colapso magnético transversal \cite{Elizabeth}. La condici\'on de colapso magn\'etico transversal viene dada por la condici\'on 
 
\begin{equation} \label{colapsop}
P_{\perp}=-\Omega-{\mathcal M} B\leq 0,
\end{equation}
 
\noindent para uno o varios de los gases que componen la estrella. 

Por otra parte, el campo magn\'etico generado por la automagnetizaci\'on de la materia expulsada podr\'ia colimarla evitando su dispersi\'on y favoreciendo la formaci\'on del \textit{jet}. En lo que sigue exploraremos ambas posibilidades para el gas \textit{npe}, suponiendo este como una mezcla de gases ideales donde las part\'iculas interact\'uan con el campo magn\'etico pero no entre ellas.

La descripción termodinámica de cada uno de los gases de la mezcla se har\'a a trav\'es de las EdE que se obtienen de calcular sus potenciales termodin\'amicos Ec.~(\ref{potencial}) con los espectros correspondientes. En el caso de los electrones, los protones y los neutrones, las EdE de los gases magnetizados fueron tomadas de trabajos previos del grupo de investigaci\'on  \cite{Felipe:2002wt,Aurora2003EPJC}. Los neutrones apareados ser\'an descritos con los resultados del Cap\'itulo \ref{cap2} \cite{Quintero2017IJMP,Quintero2017PRC,Quintero2017AN}, mientras que las EdE del gas degenerado de bosones escalares cargados fueron calculadas espec\'ificamente por la autora para esta parte de la tesis (Ap\'endice \ref{appH}).

Para un gas de fermiones cargado (electrones y protones), las EdE en el límite de baja temperatura son \cite{Felipe:2002wt}:

\begin{subequations}\label{EoSCF}
\begin{align}
E^{e,p} &= \frac{m_{e,p}^2}{4\pi^2}\frac{B}{B^{e,p}_c} \sum_{l=0}^{l_{max}} g_{l}\! \left( \mu^{e,p}\,p^{e,p}_F +{\mathcal {E}_{l}}^2\ln\frac{ {\mu^{e,p}}+ {p^{e,p}_F}}{\mathcal {E}_{l}}\right),\\
\label{EoSCF2}
P^{e,p}_{\parallel}&= \frac{m_{e,p}^2}{4\pi^2}\frac{B}{B^{e,p}_c}\sum_{l=0}^{l_{max}}g_{l}\!\left[ \mu^{e,p} \,p^{e,p}_F -{\mathcal {E}_{l}}^2\ln\!\left(\frac{\mu^{e,p}+ p^{e,p}_F}{\mathcal {E}_{l}}\right)\right]\!,\\
P^{e,p}_{\perp} &= \frac{m_{e,p}^4}{2\pi^2}\left(\frac{B}{B^{e,p}_c}\right)^{\!\!2}\,\sum_{l=0}^{l_{max}}g_l l\ln\left (\frac{\mu^{e,p}+p^{e,p}_F}{\mathcal{E}_l}\right),\\
\mathcal M^{e,p} &= \frac{m_{e,p}^2}{4\pi^2 B_c^{e,p}} \sum_{l=0}^{l_{max}}g_{l}\!\left[ \mu^{e,p} \mathcal {E}_{l} - \mathcal {E}_{l} \ln (\frac{\mu^{e,p} + \mathcal {E}_{l}}{\mathcal {E}_{l}}) \right]\!  ,\\
 N^{e,p} &= \frac{m_{e,p}^2}{4\pi^2} \frac{B}{B^{e,p}_c} \sum_{l=0}^{l_{max}} 2 g_{l} p^{e,p}_F.
 \end{align}
\end{subequations}

En las Ecs.~(\ref{EoSCF}) los \'indices $e$ y $p$ hacen referencia a los electrones y los protones respectivamente; $l$ denota los niveles de Landau y $g_{l}=2-\delta_{0,l}$ su degeneración; $l_{max}= I[\frac{\mu^2-m^2}{2qB}]$ e $I[z]$ es la parte entera de $z$. El momento de Fermi es ${p^{e,p}_F}=\sqrt{({\mu^{e,p}})^2-\mathcal{E}_{l}^2}$, el espectro viene dado por $\mathcal{E}_{l}=\sqrt{2|q^{e,p} B|l+m_{e,p}^2}$ y $B^{e,p}_c = m_{e,p}^2/q^{e,p}$.

Las EdE del gas magnetizado de neutrones (fermiones neutros con momento magn\'etico) en cambio tienen la forma \cite{Aurora2003EPJC}:

\begin{subequations}\label{EoSNF1}
\begin{align}
E^{n} &= - P^{n}_{\parallel} + \mu^{n} N^{n},\\
\label{EoSNF2}
P^{n}_{\parallel} &=  \frac{m_{n}^4}{2\pi^2} \sum_{\eta=1,-1} \left\{ \frac{\mu^{n} f^3}{12 m_n} + \frac{(1+\eta b^n)(5 \eta b^{n} -3)\mu^n f}{24 m_{n}} + \frac{(1+\eta b^{n})^3 L}{24} -\frac{\eta b^{n} (\mu^{n})^3 s}{6 m_{n}^3} \right\} ,\\
\label{EoSNF3}
P^{n}_{\perp} &=  P^{n}_{\parallel} - \mathcal M^{n} B,\\
\label{EoSNF4}
\mathcal M^{n} &=   \frac{m_{n}^3 \kappa_n}{2\pi^2} \sum_{\eta=1,-1} \eta \left\{ \frac{(1-2 \eta b^{n})}{6} f + \frac{(1 + \eta b^{n})^2(1-\eta b^{n}/2)}{3} L - \frac{(\mu^{n})^3}{6 m_{n}^2} s \right\},\\
\label{EoSNF5}
N^{n} &=  \frac{m_{n}^3}{2\pi^2} \sum_{\eta=1,-1} \left\{ \frac{f^3}{3} + \frac{\eta b^{n}(1+\eta b^{n})}{2}f - \frac{\eta b^{n} (\mu^{n})^2}{2 m_{n}^2}s  \right\} ,
\end{align}
\end{subequations}

\noindent donde $n$ denota las cantidades pertenecientes a los neutrones y $b^{n}=B/B^{n}_c$ con $B^{n}_c=m_{n}/\kappa_{n}$, siendo $m_n$ y $\kappa_n$, la masa y el momento magn\'etico del neutr\'on. Las funciones $f$, $L$ y $s$ se definen como:

\begin{eqnarray}\label{EoSNF}
f &=& \frac{(\mu^n)^2-(\varepsilon^n(\eta))^2}{m_n},\\
L &=& \frac{1}{1+\eta b^n} \ln \left(\frac{\mu^n + \sqrt{(\mu^n)^2 - \varepsilon^n(\eta)^2}}{m_n}\right),\\
s &=& \frac{\pi}{2}-\frac{m_n}{\mu^n} \arcsin(1+\eta b^n),
\end{eqnarray}

\noindent con $\varepsilon^n(\eta) = m_n + \eta \kappa_n B$.

Para obtener las EdE del gas de bosones escalares cargados (protones apareados $pp$) se sigui\'o el mismo procedimiento mostrado en el Cap\'itulo \ref{cap2} para los bosones vectoriales neutros. Los detalles del c\'alculo del potencial termodin\'amico correspondiente (Ec.~(\ref{omegasc})) se muestran en el Ap\'endice \ref{appH}. Las EdE obtenidas en los reg\'imenes de campo d\'ebil ({\bf CD}) $T > m_{pp} b^{pp}$  y campo fuerte ({\bf CF}) $T < m_{pp} b^{pp}$ son:

\begin{subequations}\label{EoSCSB1}
\begin{align}
E^{pp} &= \left \{
\begin{array}{ll}
\varepsilon^{pp} N^{pp} - \frac{3}{2} \Omega^{pp}, & \, {\bf CD} \\[10pt]
\varepsilon^{pp} N^{pp} - \frac{1}{2} \Omega^{pp}, & \, {\bf CF}
\end{array}
\right.   ,\\[10pt]
\label{EoSCSB2}
P^{pp}_{\parallel} &= -\Omega^{pp} ,\\[10pt]
\label{EoSCSB3}
P^{pp}_{\perp} &= P^{pp}_{\parallel} - \mathcal M^{pp} B ,\\[10pt]
\label{EoSCSB4}
\mathcal  M^{pp} &=  - \frac{q^{pp}}{2 \varepsilon^{pp}} N^{pp},\\[10pt]
\label{EoSCSB5}
N^{pp} &= \left \{
\begin{array}{ll}
\frac{3 (\varepsilon^{pp})^{3/2} Li_{3/2}(e^{\beta(\mu^{pp}-\varepsilon^{pp})})}{(2 \pi \beta)^{3/2}}, & \, {\bf CD} \\[10pt]
\frac{3 m_{pp}^2 b^{pp} (\varepsilon^{pp})^{1/2} Li_{1/2}(e^{\beta(\mu^{pp}-\varepsilon^{pp})})}{(2 \pi)^{3/2} \beta^{1/2}}, & \, {\bf CF}
\end{array}
\right. 
\end{align}
\end{subequations}

En las Ecs.~(\ref{EoSCSB1}), $\varepsilon^{pp} = m_{pp} \sqrt{1+b^{pp}}$, $b^{pp}= B/B^{pp}_c$, $B^{pp}_c = m_{pp}^2/q$, $q$ es la carga el\'ectrica de los bosones y:

\begin{equation}\label{omegasc}
\centering
\Omega^{pp} = \left \{
\begin{array}{ll}
-\frac{3 (\varepsilon^{pp})^{3/2} Li_{5/2}(e^{\beta(\mu^{pp}-\varepsilon^{pp})})}{(2 \pi)^{3/2} \beta^{5/2}} , & \, {\bf CD} \\[10pt]
-\frac{3 m_{pp}^2 b (\varepsilon^{pp})^{1/2} Li_{3/2}(e^{\beta(\mu^{pp}-\varepsilon^{pp})})}{(2 \pi)^{3/2} \beta^{3/2}}, & \, {\bf CF}
\end{array}
\right.,
\end{equation}

\noindent con:

\begin{eqnarray}
\mu^{pp} &=& \left \{
\begin{array}{ll}
-\frac{\zeta(3/2) T}{4 \pi} \left[ 1-(\frac{T^{pp}_{cond}}{T})^{3/2}\right] \Theta(T - T^{pp}_{cond}), & \, {\bf CD} \\[10pt]
\varepsilon^{pp} \left( 1- \frac{m_{pp}^4 (b^{pp})^2 T^2}{8 \pi^2 (N^{pp})^2} \right ), & \, {\bf CF}
\end{array}
\right.,
\end{eqnarray}
\\

\begin{eqnarray}
T^{pp}_{cond} &=& \frac{2 \pi}{\varepsilon^{pp}} \left ( \frac{N^{pp}}{3 \zeta(3/2)}\right )^{2/3}.
\end{eqnarray}

$T^{pp}_{cond}$ es la temperatura de condensaci\'on de los protones apareados, $Li_k(x)$ la funci\'on polilogar\'itmica de orden $k$ y $\zeta(x)$ es la funci\'on zeta de Reimann. N\'otese que la Ec.~(\ref{EoSCSB5}) implica que la magnetizaci\'on de los bosones cargados escalares es siempre negativa \cite{PEREZROJAS2000}.  Por tanto la presi\'on transversal de este gas nunca ser\'a cero y los pares de protones nunca sufrir\'an un colapso magn\'etico transversal. 

Como los neutrones apareados se describen en funci\'on de los resultados del cap\'itulo previo, sus EdE vienen dadas por las Ecs.~(\ref{energy}), (\ref{magnetizationtotal}) y (\ref{pressures}). En el caso de los fermiones se ha tomado el l\'imite de temperatura cero pero para los bosones trabajaremos a temperatura finita ya que ellos son m\'as sensibles a los cambios de esta magnitud.

Las cantidades termodinámicas totales de la mezcla de gases, en particular las presiones totales, $P^T_{\parallel}$ y $P_{\perp}^T$, la magnetización total $\mathcal M^T$ y la densidad de energía total $E^T$, se calculan como la suma de las contribuciones de cada gas:

\begin{subequations}\label{edenpe}
	\begin{align}
P^T_{\parallel} &= \sum_{i = e, p, n, pp, nn} P^i_{\parallel},\\
P^T_{\perp} &= \sum_{i = e, p, n, pp, nn} P^i_{\perp},\\
\mathcal M^T &= \sum_{i = e, p, n, pp, nn} \mathcal M^i,\\
E^T &= \sum_{i = e, p, n, pp, nn} E^i.
\end{align}
\end{subequations}

Nuestro modelo requiere imponer la cantidad de fermiones que est\'an bosonizados. Llamaremos $x_n$ y $x_p $ a la fracción de neutrones y protones bosonizados respectivamente. Esta fracci\'on depende, en principio, de la densidad, la temperatura y la fortaleza de la interacci\'on entre los fermiones. No obstante, la forma expl\'icita de esta dependencia es dif\'icil de predecir debido a las condiciones tan distintas que tienen los escenarios astrof\'isicos con respecto a aquellas en la cuales se llevan a cabo los experimentos terrestres. Por ello, en este trabajo las fracciones  $x_n$ y $x_p$ ser\'an tomadas como  par\'ametros libres e iguales entre s\'i,  $x_n =x_p =0.5$, a menos que se indique lo contrario.

\begin{figure}[h]
	\centering
	\includegraphics[width=0.49\linewidth]{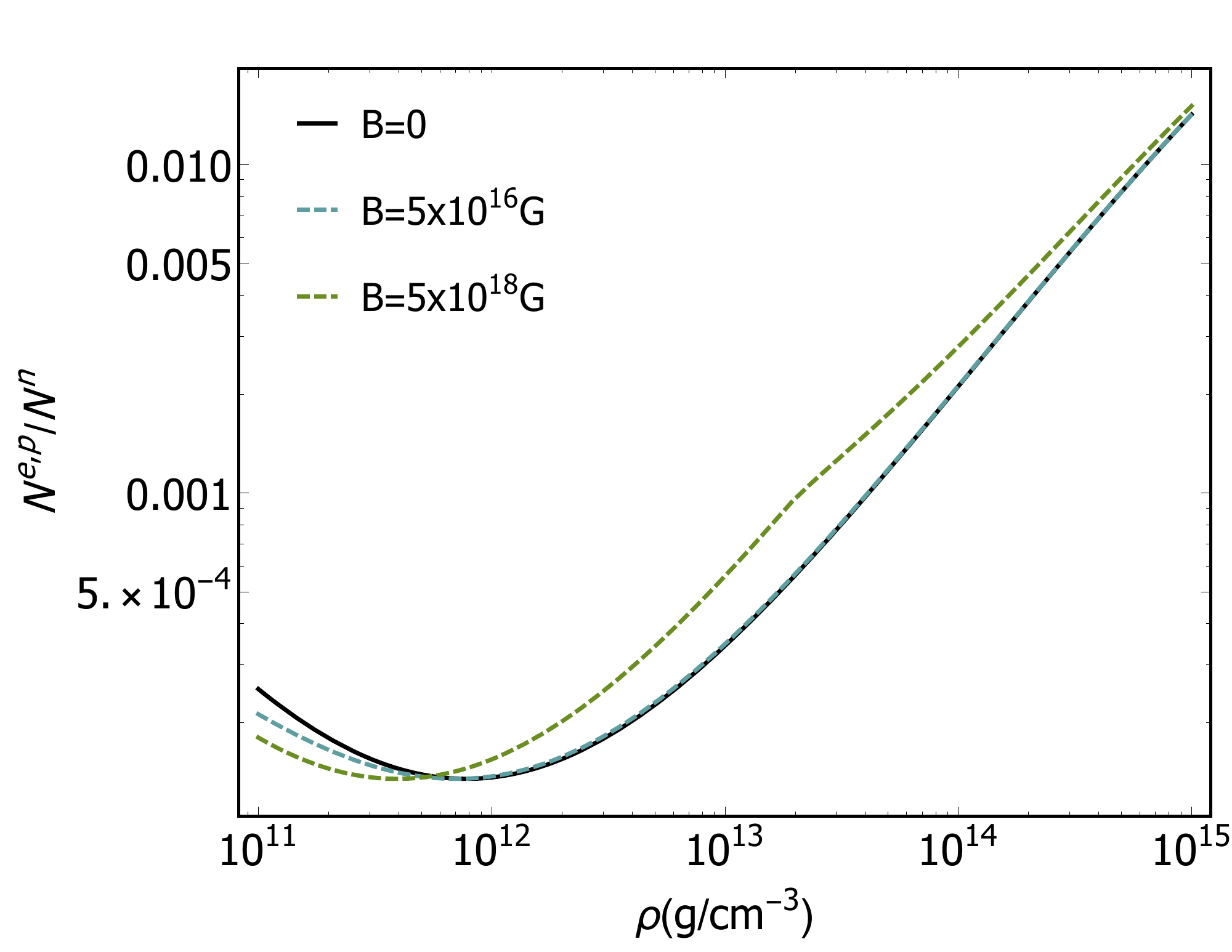}
	\caption{\label{protonelectronfraction} Fracci\'on de protones y electrones como  funci\'on de la densidad de masa de los neutrones para  varios valores del campo magn\'etico.}
\end{figure}

Consideraremos adem\'as que el gas \textit{npe} parcialmente bosonizado se encuentra en condiciones de equilibrio estelar, es decir, que el n\'umero bar\'ionico $N^n + N^p$ se conserva (equilibrio $\beta$) y que el gas es neutro $N^p = N^e$. En dichas condiciones, el equilibrio $\beta$ conduce adem\'as a que, dada una densidad de neutrones $N^n$, la fracci\'on de electrones y protones es \cite{Weinberg, Shapiro}:

\begin{equation}\label{protonfraction}
\frac{N^{e,p}}{N^n} = \frac{1}{8} \left( \frac{1 + 4 Q/(m_n y_n^2) + 4 (Q^2 - m_e^2)/(m_n^2 y_n^2)}{1+1/y_n^2} \right)^{3/2}
\end{equation}

\noindent donde $m_n$, $m_p$ y $m_e$ son las masas del neutr\'on, el prot\'on y el electr\'on respectivamente, $Q=m_n-m_p$, $y_n =p^n_F /m_n$ y $p^n_F = m_n (f(\eta=1)+f(\eta=-1))/2$ es el momento de Fermi de los neutrones.  La fracción de protones y electrones (\ref{protonfraction}) depende de la intensidad del campo magn\'etico a trav\'es de $p^n_F$. Esta dependencia puede apreciarse en la Fig.~\ref{protonelectronfraction}, donde se ha graficado la Ec.~(\ref{protonfraction}) como funci\'on de la densidad de masa de los neutrones $\rho = m_n N^n$ para varios valores de $B$. Aumentar el campo mueve el mínimo de la fracción protón/electrón hacia la izquierda y, en consecuencia, en la regi\'on de altas densidades, mientras mayor sea el campo, mayor ser\'a dicha fracción.

\section{Colapso magnético transversal: expulsi\'on de materia hacia el exterior de la estrella}

Como ya hemos visto, el colapso magn\'etico de un gas consiste en que alguna de sus presiones se haga negativa. En el caso de que la presi\'on que deviene negativa es la perpendicular al eje magn\'etico, las part\'iculas del gas son empujadas hacia \'el dando lugar a la formaci\'on de una estructura alargada y axisim\'etrica con forma de cigarro. De ah\'i nuestra propuesta de que la raz\'on  de la expulsión de masa de la estrella se debe al colapso magnético transversal de al menos uno de los gases que la componen.

\begin{figure}[h!]
	\centering
	\includegraphics[width=0.45\linewidth]{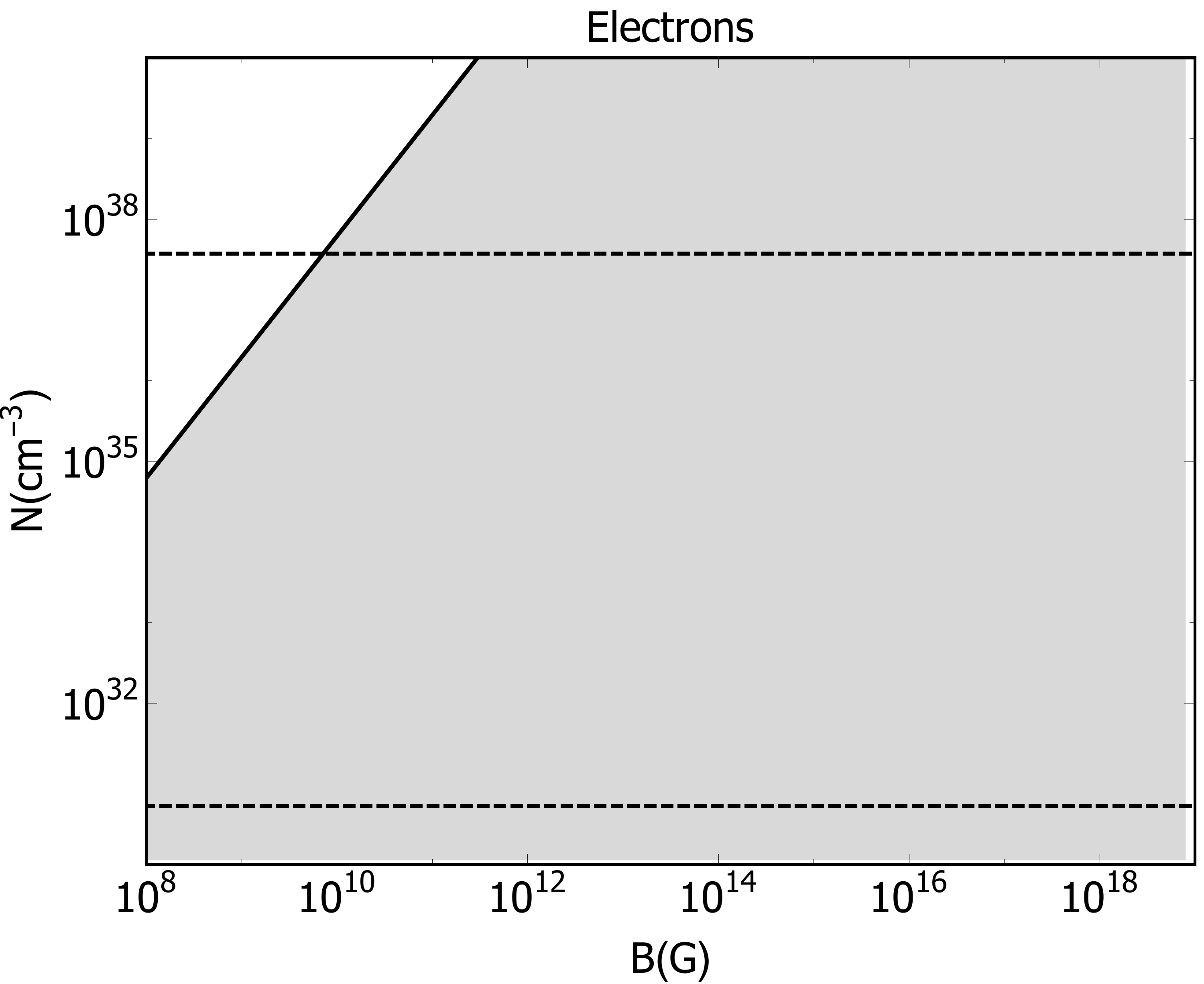}
	\includegraphics[width=0.45\linewidth]{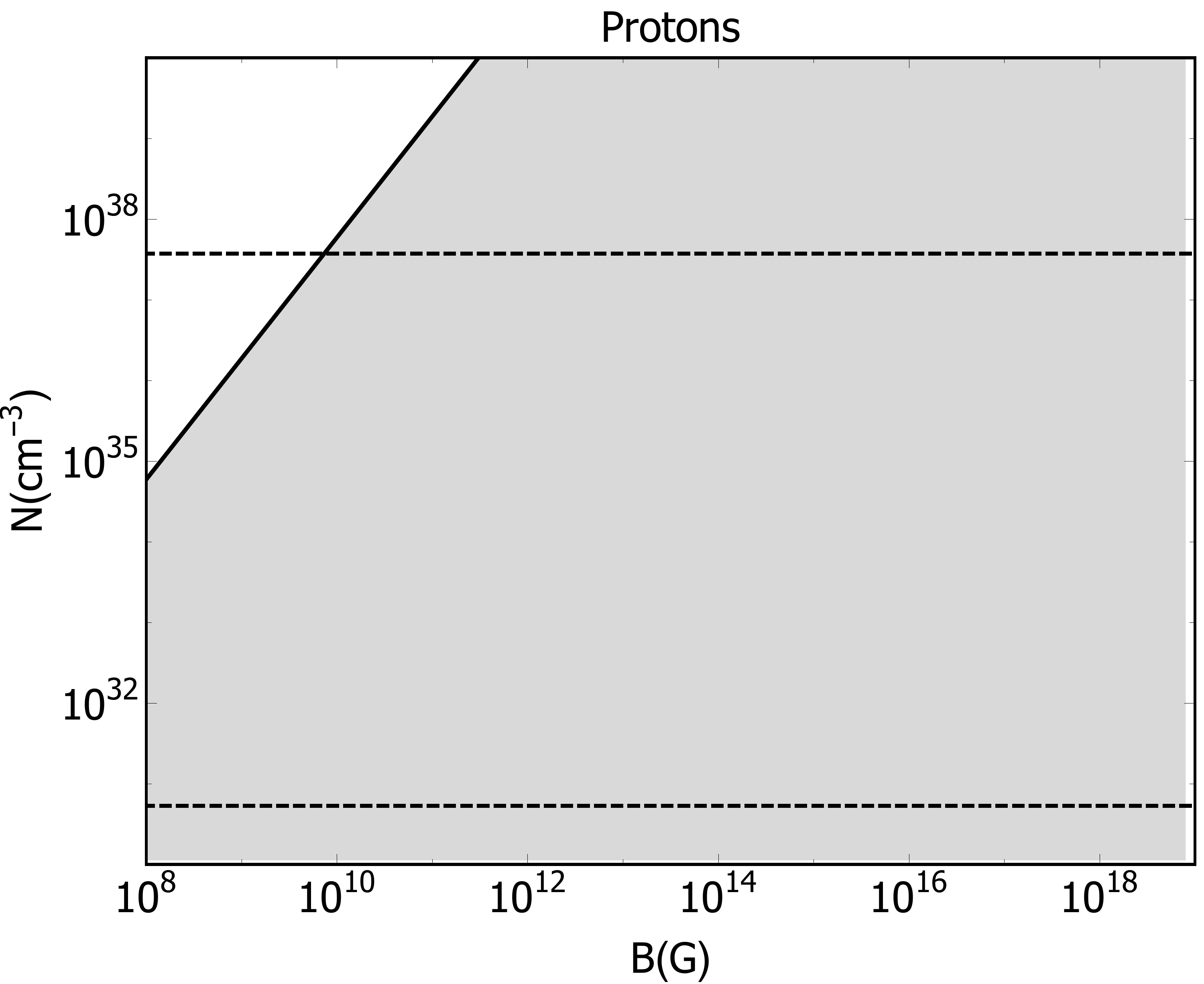}\\
	\includegraphics[width=0.45\linewidth]{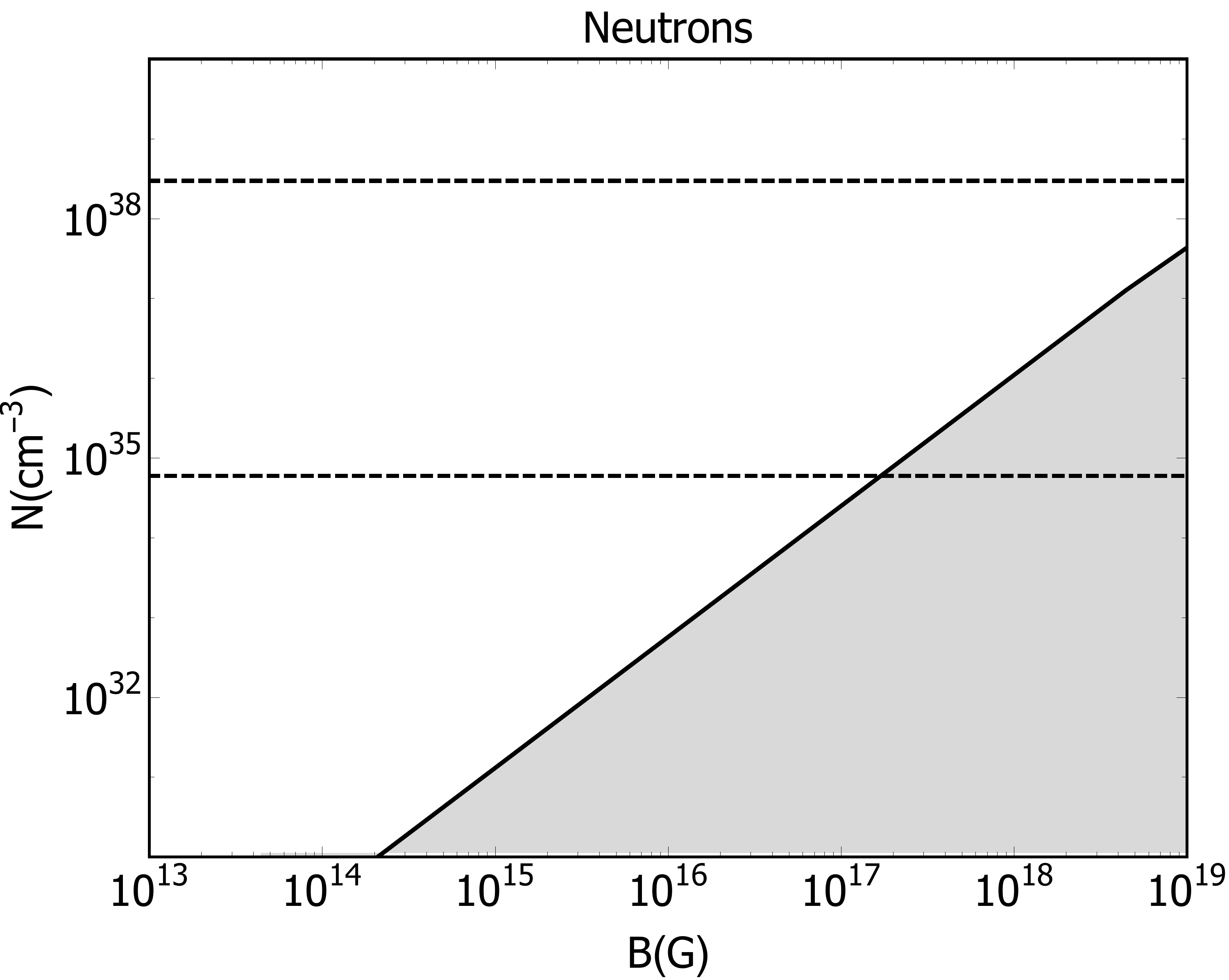}
	\includegraphics[width=0.45\linewidth]{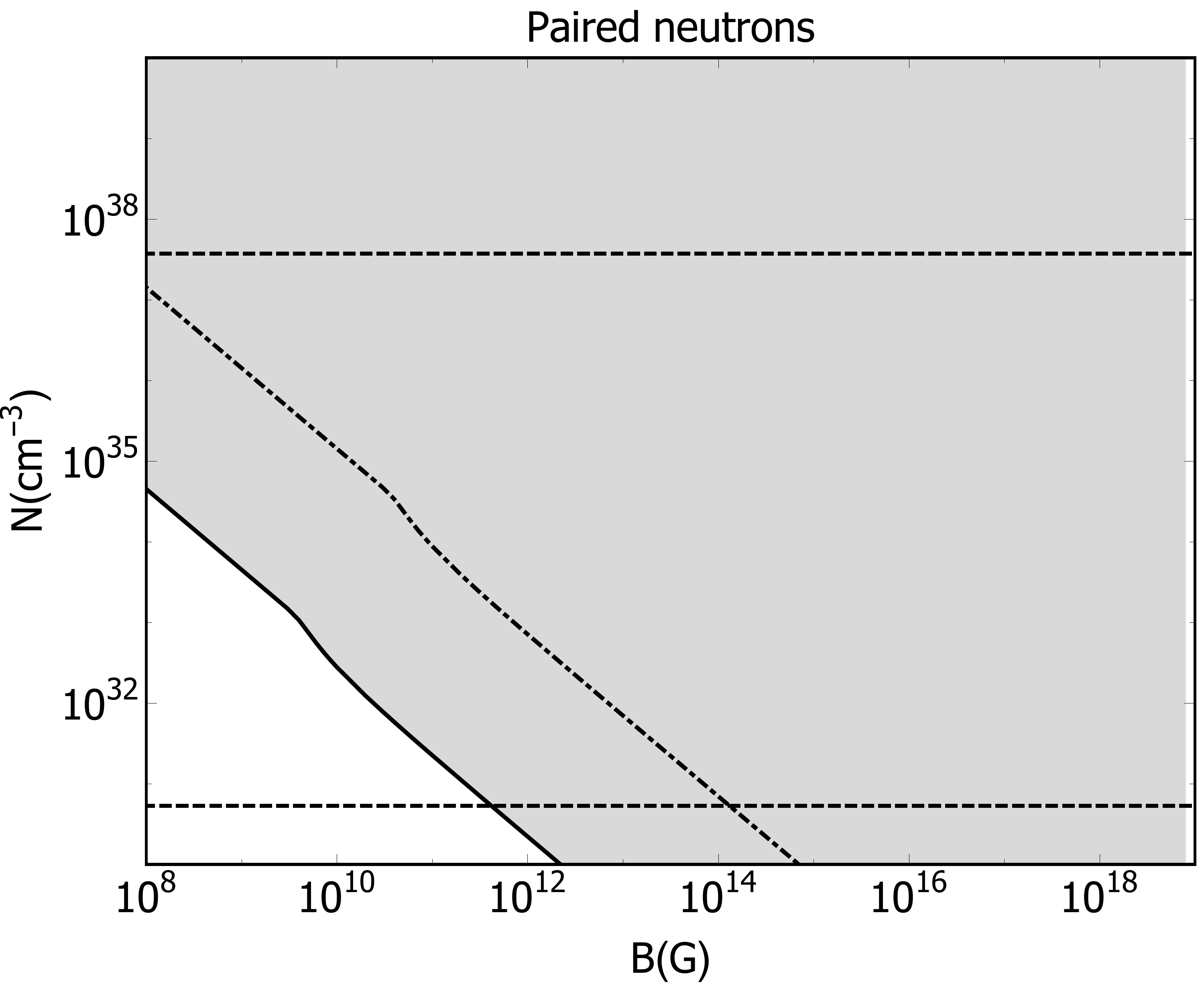}
	\caption{\label{colapso} Diagramas de fase para el colapso magn\'etico transversal de electrones, protones, neutrones y neutrones apareados en el interior de una EN. En la regi\'on sombreada $P_{\perp} (B,N)<0$ y el gas es inestable. Las l\'ineas horizontales delimitan el rango en el que se encuentra la densidad de cada part\'icula en el interior de la EN.}
\end{figure}

Para evaluar la plausibilidad de esta propuesta el primer paso es comprobar que, efectivamente, el colapso magn\'etico transversal puede producirse en el interior de la EN. Dado que la condici\'on para que dicho colapso se produzca es que la presi\'on perpendicular de los gases sea nula, en la figura Fig.~\ref{colapso} se han dibujado las curvas sobre las cuales $P^i_ {\perp}(B, N)=0$ para $i = n,p,e,nn$. La zona sombreada corresponde a los valores de densidad y campo para los cuales $P^i_{\perp} < 0$ y el gas es inestable. Las l\'ineas horizontales delimitan el rango en el que se mueve de la densidad de cada part\'icula en el interior de la EN.

Para electrones, protones y neutrones, dado un campo magnético, hay una densidad de part\'iculas crítica por debajo de la cual el gas es inestable. Esto es consecuencia de que aumentar la densidad de un sistema fermi\'onico aumenta su presi\'on estad\'istica, lo que ayuda a equilibrar la presi\'on magn\'etica ($- \mathcal M B$) y estabiliza el gas. Las densidades críticas de electrones y protones son del orden de las esperadas para estos fermiones en estrellas de neutrones (las líneas verticales discontinuas) y, en particular, para campos magnéticos de más de $10^{10}$~G son siempre más grandes. Para los neutrones, en cambio, la densidad cr\'itica es  varios órdenes menor que los estimados para la corteza y el n\'ucleo de las EN, salvo cuando el campo magn\'etico es muy alto ($B \geq 5 \times 10^{17} $~G). De esta manera, la posibilidad de que ocurra el colapso magn\'etico transversal queda confirmada para los gases de fermiones.

En el caso de los neutrones apareados, el diagrama de fases se invierte, el gas es estable por debajo de la densidad de part\'iculas  cr\'itica. Esto se debe a que aumentar la densidad un gas de bosones vectoriales incrementa su magnetizaci\'on a la vez que favorece la aparici\'on del condensado de Bose-Einstein, lo que disminuye la presi\'on estad\'istica y desestabiliza el gas. Por otra parte, la densidad de part\'iculas cr\'itica de los neutrones apareados es bastante sensible a la temperatura, como muestra el panel inferior derecho de la Fig. \ref{colapso}. En este gr\'afico la curva  $P^{nn}_ {\perp}(B, N)=0$ se muestra para dos temperaturas, $10^8$~K (l\'inea s\'olida) y $10^9$~K (l\'inea punteada). Las densidades críticas de los neutrones apareados tambi\'en est\'an en el orden de las esperadas en el interior de una EN, en particular para $B \geq 10^{12}$~G. Por tanto, la posibilidad del colapso del gas de neutrones apareados tambi\'en queda confirmada. De hecho, n\'otese que el r\'egimen de colapso es el mas probable para ellos en el interior de la estrella.

El diagrama de fases del colapso de los protones apareados no se muestra porque el gas de bosones escalares que ellos forman es diamagn\'etico,  $\mathcal M <0$, como puede verse de la Ec. ~(\ref{EoSCSB5}) y ya discutimos con anterioridad. En consecuencia, la presi\'on magn\'etica en lugar de disminuir, aumenta la presi\'on perpendicular. Por tanto para este gas el colapso magn\'etico transversal no puede ocurrir.

De modo que todos los componentes del gas \textit{npe} parcialmente bosonizado, excepto los protones  apareados, son susceptibles de colapsar en las condiciones internas de una EN. En un gas \textit{npe} la presi\'on que sostiene  la estrella es la presi\'on del gas degenerado de neutrones \cite{Camezind}, por tanto, cuando los neutrones est\'an en r\'egimen de colapso el gas \textit{npe} se vuelve inestable. Como estamos interesados en la expulsi\'on de materia hacia el exterior de la EN en lo que sigue trabajaremos para campos magn\'eticos tales que el gas de neutrones sea estable y supondremos que las condiciones para el colapso del resto de los gases son locales en el interior de la estrella.

\subsection{¿Pueden los gases colapsados superar la gravedad?}

En la secci\'on anterior encontramos los candidatos a producir la expulsi\'on de materia hacia el exterior de la EN. Ellos  son los gases de electrones, protones y neutrones bosonizados. Sin embargo, para que la expulsi\'on de materia se produzca es necesario que los gases colapsados ejerzan a lo largo de la dirección del campo magnético una presi\'on mayor que la presi\'on gravitacional en la estrella. Para evaluar si esto puede suceder o no, compararemos las presiones paralelas de los gases colapsados con la presi\'on gravitacional media $P_{GAV}$ para una estrella con masa t\'ipica de las ENs, $M = 1.5 M_{\odot}$, en el l\'imite de compacidad relacionado con la tercera cota te\'orica presentada en el cap\'itulo introductorio.

El l\'imite de compacidad de un objeto compacto es una de las pocas soluciones anal\'iticas de las ecuaciones TOV y se obtiene al resolverlas suponiendo que la densidad de energ\'ia ($E$) de la estrella es constante \cite{Weinberg}. En este l\'imite el radio de la estrella $R$ y la presi\'on en su interior pueden obtenerse de manera anal\'itica como funci\'on de la masa total de la estrella y de su densidad de energ\'ia  \cite{Weinberg}:

\begin{subequations}\label{compacidad}
\begin{align}
R &= \frac{1}{3 P_c + E} \sqrt{\frac{3 P_c (P_c + 2 E)}{G E}}\\
P(r) &= E \frac{\sqrt{1-2 G M/R}-\sqrt{1-2 G M r^2/R^3}}{\sqrt{1-2 G M r^2/R^3}-3 \sqrt{1-2 G M/R}},
\end{align}
\end{subequations}

\noindent donde $G$ es la constante de gravitaci\'on universal y $r$ el radio interno de la estrella. La densidad (constante) de energ\'ia dentro de la estrella $E$ y la la presi\'on central de la estrella  $P_c$ son tomadas de las EdE del gas \textit{npe} parcialmente bosonizado Eqs.~(\ref{edenpe}). Dadas las Ecs.~(\ref{compacidad}) la presi\'on media en el interior de la estrella es:

\begin{equation}\label{Pmed}
P_{GAV} = \frac{1}{R} \int_{0}^{R} P(r) dr.
\end{equation}

\begin{figure}[h]
\centering
\includegraphics[width=0.45\linewidth]{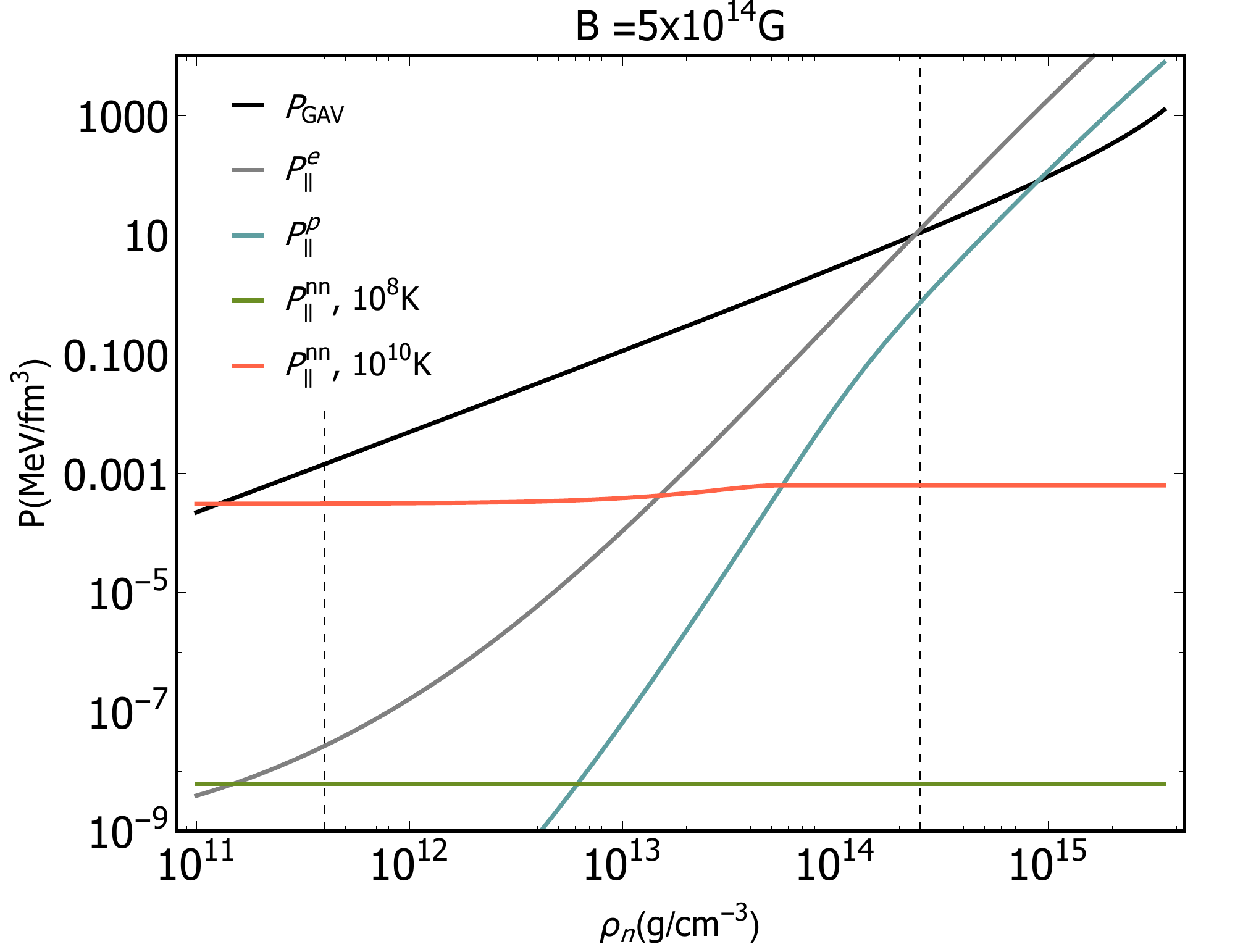}
\includegraphics[width=0.45\linewidth]{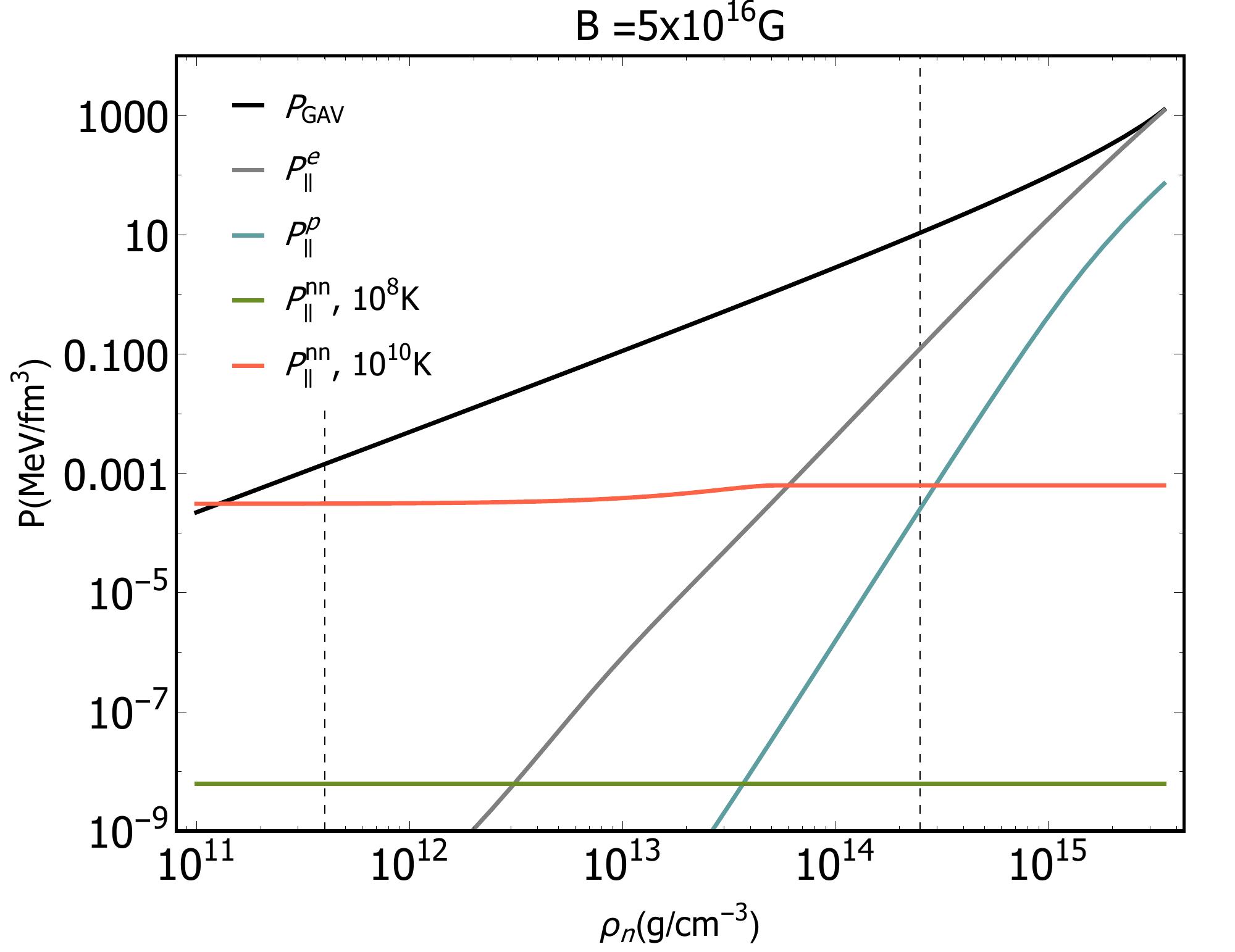}
\caption{\label{collapse3} La presi\'on perpendicular de los gases de electrones, protones y neutrones bosonizados en funci\'on de la densidad de masa neutr\'onica en el interior de la estrella. Las l\'ineas verticales se\~nalan las densidades en las que comienzan la corteza y el n\'ucleo respectivamente.}
\end{figure}

En la Fig.~\ref{collapse3} se muestra la comparaci\'on de $P_{GAV}$  con las presiones paralelas de los gases de electrones, protones y neutrones apareados. Al hacerlo se ha garantizado que estos gases se encuentren en r\'egimen de colapso. En la figura se aprecia que para campos magn\'eticos relativamente bajos $P_{GAV}$ puede ser superada por la presi\'on paralela de los electrones y protones que se encuentran en el n\'ucleo de la estrella ($\rho>\rho_{nuc}$). No obstante, a medida que el campo magn\'etico aumenta las presiones de los gases fermi\'onicos disminuyen llegando a ser menores que $P_{GAV}$.

El gas de neutrones apareados solo puede superar a $P_{GAV}$ para temperaturas relativamente altas ($10^{10}$K) y densidades de masa del orden de $10^{11}$g/cm$^3$, que se corresponden con las regiones exteriores de la estrella, en particular, con la corteza. Pero como se vio en el Cap\'itulo \ref{cap1}, el apareamiento con spin paralelo de los neutrones solo se da en el n\'ucleo de la estrella, por lo cual no es probable encontrar estos bosones vectoriales en sus capas m\'as externas.

Por tanto, en las condiciones adecuadas de campo magn\'etico y densidad de part\'iculas, solo los gases colapsados de electrones y protones que se encuentras en el n\'ucleo de la EN podrían superar la gravedad y desencadenar la expulsión de la materia. Esto es consistente con ciertas observaciones que indican que los \textit{jets} cuya fuente es un OC se originan cerca de su centro y no en sus capas externas \cite{108,deGouveiaDalPino:2005xn}.


Para los c\'alculos que siguen  supondremos que la composici\'on de la materia expulsada es similar a la de la EN, es decir, que tambi\'en est\'a formada por un gas \textit{npe} parcialmente bosonizado, pero en el cual hay al menos un gas colapsado.

\section{Forma del chorro y estabilidad: generación del campo magnético}

Una vez que la materia deja a la estrella y forma el \emph{jet}, surge una pregunta: ¿por qué permanece colimada en lugar de dispersarse? En nuetsro modelo, la forma alargada del \textit{jet} podr\'ia explicarse si uno o varios de los gases que lo forman continuara en r\'egimen de colapso, disminuyendo as\'i la presi\'on perpendicular con respecto a la paralela y ayudando a la gravedad a comprimir la materia hacia el eje del chorro. Pero esto solo es posible si el campo magn\'etico en el \textit{jet} es tan intenso como el de la EN. Desde el punto de vista observacional, ha sido demostrado que el campo magnético del chorro está en el orden del campo magnético de su fuente y alineado con el eje del \textit{jet} \cite{Carrasco2010_20,deGouveiaDalPino:2005xn}. Por tanto, ahora la pregunta es si la mezcla de gases que estamos proponiendo como modelo para la composici\'on del chorro puede permanecer magnetizada una vez que abandona la estrella.

\begin{figure}[h!]
	\centering
	\includegraphics[width=0.45\linewidth]{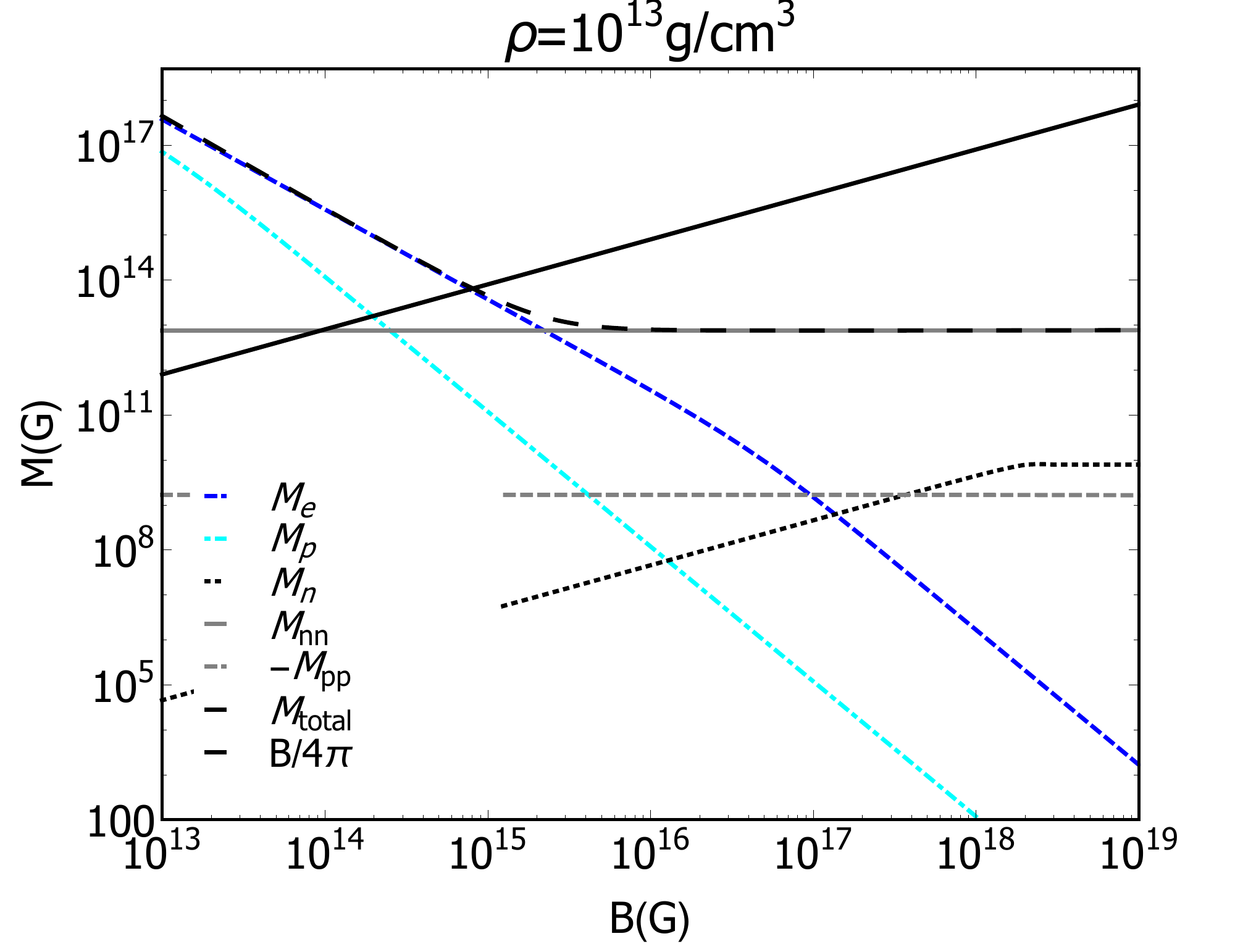}
	\includegraphics[width=0.45\linewidth]{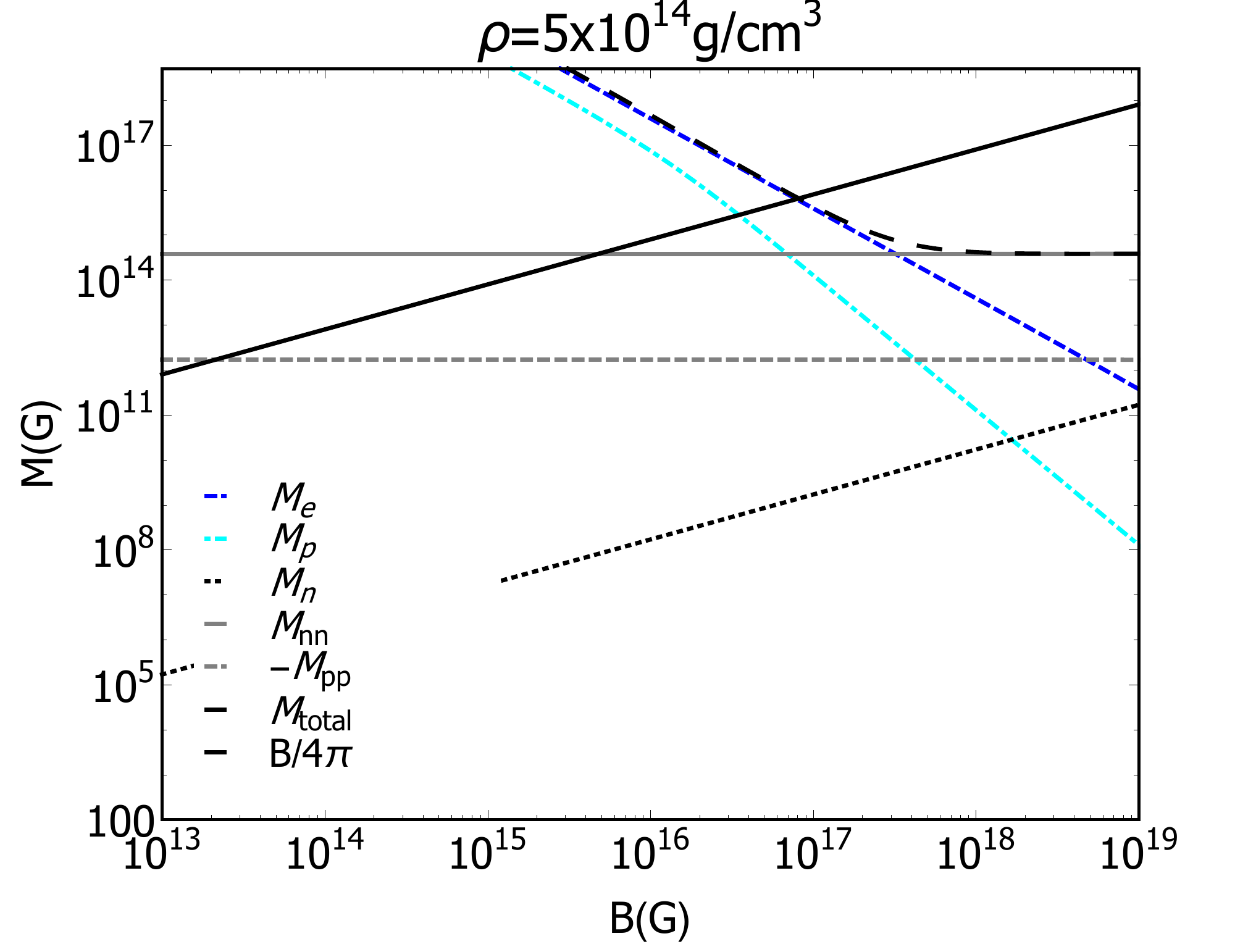}
	\caption{\label{magnetizationnpe} La magnetizaci\'on del gas \textit{npe} parcialmente bosonizado, as\'i como el aporte que a esta hacen cada uno de sus componentes, como funci\'on del campo magn\'etico, para dos densidades de masa neutr\'onicas distintas.}
\end{figure}

La Fig.~\ref{magnetizationnpe} muestra la magnetizaci\'on del gas \textit{npe} parcialmente bosonizado como funci\'on del campo magn\'etico para dos densidades de masa neutr\'onica distintas. Además, mostramos la línea $B/4\pi$, debido a que la condición para la automagnetización es $ \mathcal M =B /4\pi$. La magnetización total de la mezcla de gases está dominada por la magnetización de los electrones para los campos magn\'eticos m\'as bajos, y por la de los neutrones apareados para los m\'as altos. La magnetización de los otros gases est\'a varios órdenes por debajo de estas dos. En cuanto a la automagnetizaci\'on del gas, esta es posible, pues la l\'inea $B/4\pi$ intersecta a la curva de la magnetizaci\'on. 

\begin{figure}[h!]
	\centering
	\includegraphics[width=0.5\linewidth]{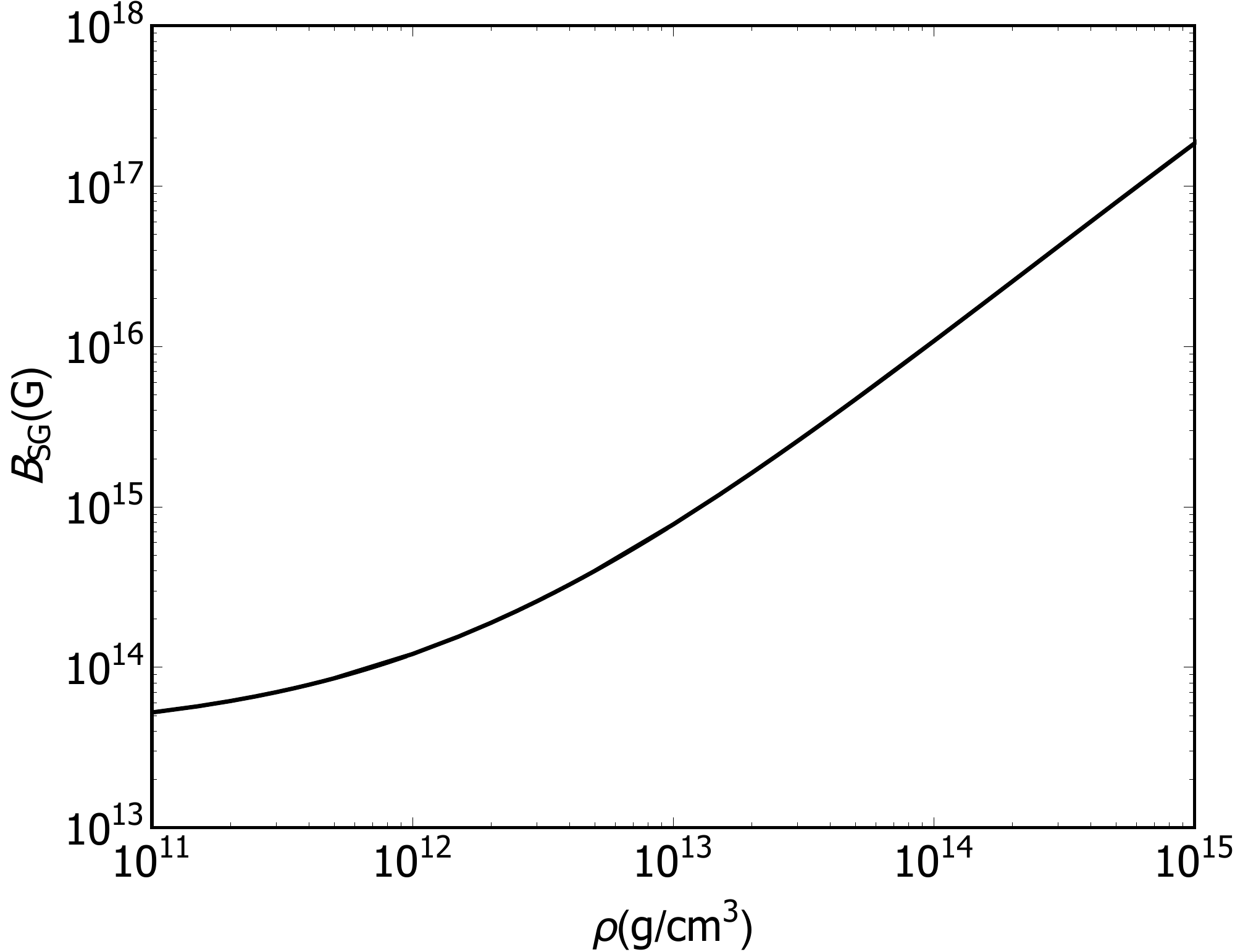}
	\caption{\label{magnetizacionnpe1} El campo magn\'etico autogenerado por el gas \textit{npe} parcialmente bosonizado como funci\'on de la densidad de masa neutr\'onica.}
\end{figure}

El campo autogenerado por el gas \textit{npe} parcialmente bosonizado fue calculado y se muestra en la Fig.~\ref{magnetizacionnpe1}. Los valores de $B_{SG}$ son lo suficientemente altos ($\geq 10^{13}$~G) como para mantener a los gases que forman el chorro en r\'egimen de colapso. Por tanto, que el campo magn\'etico del \textit{jet} contin\'ue siendo tan alto como el de su fuente al alejarse de ella, puede ser justificado por la automagnetizaci\'on del gas que lo compone.

\section{Ecuaciones de estado de la materia colimada}

Una vez comprobado que el colapso magn\'etico transversal es un mecanismo v\'alido para la expulsi\'on de materia hacia el exterior de la estrella de neutrones, y que la materia expulsada puede producir un campo magn\'etico tal que la mantenga colimada, es posible obtener ecuaciones de estado para el \textit{jet} y estudiar c\'omo dependen las presiones con la densidad de masa y el campo magn\'etico. Aunque estamos suponiendo que la composici\'on del \textit{jet} es la misma que la de la EN donde se origina, cabr\'ia esperar que las proporciones de cada uno de los gases de la mezcla fuera distinta dadas las diferencias geom\'etricas del \textit{jet} con la estrella.

\begin{figure}[!h]
	\centering
	\includegraphics[width=0.42\linewidth]{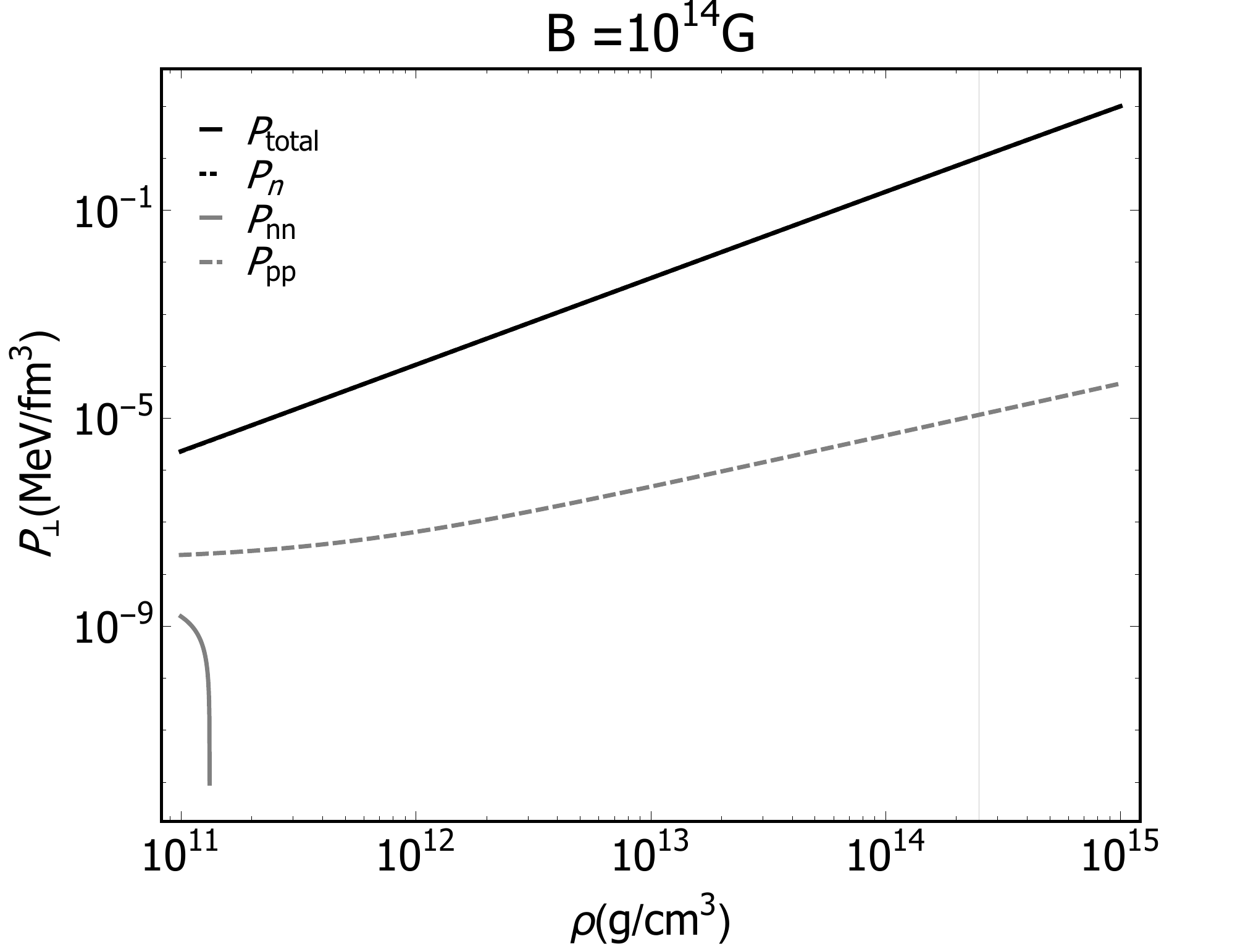}
	\includegraphics[width=0.42\linewidth]{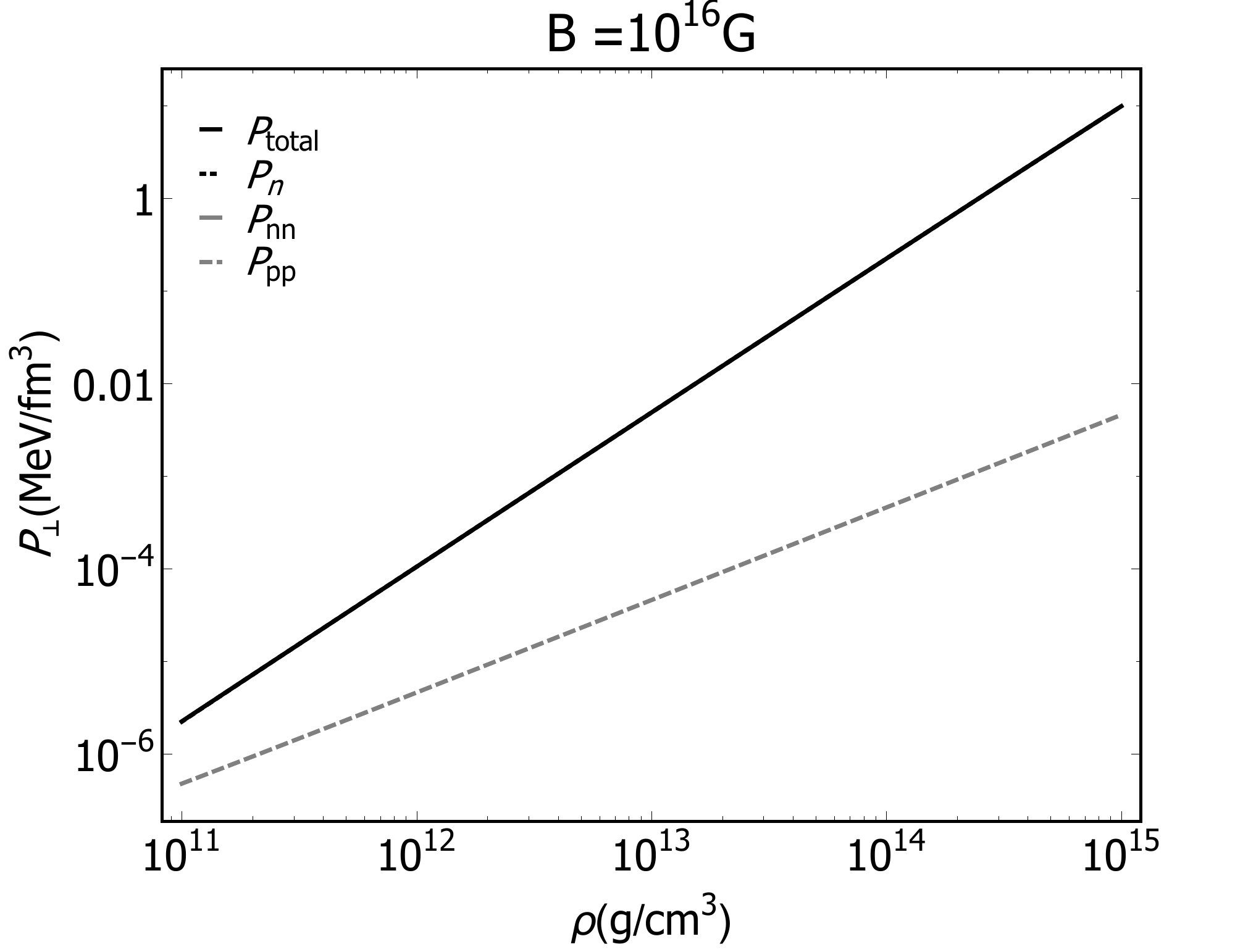}
	\includegraphics[width=0.42\linewidth]{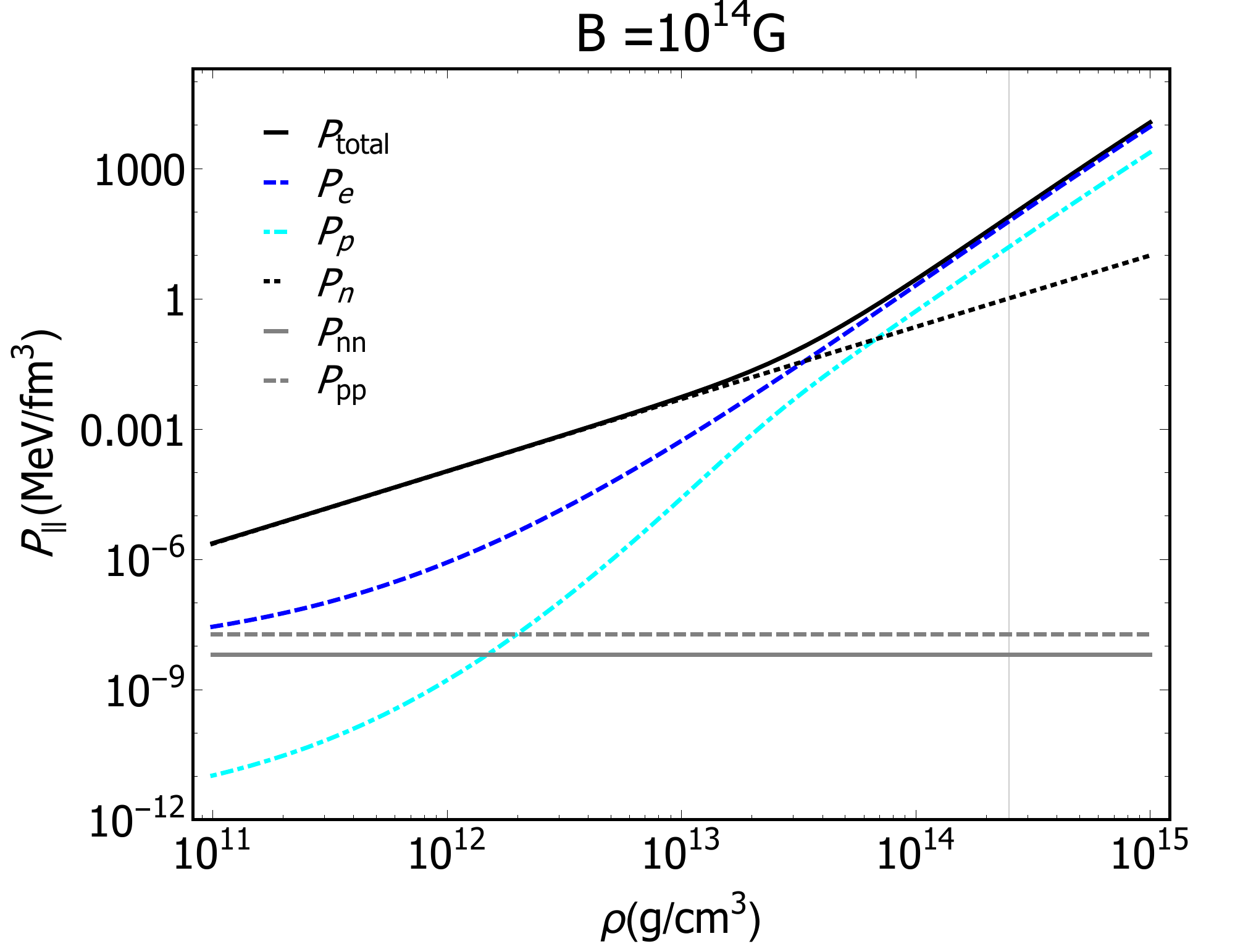}
	\includegraphics[width=0.42\linewidth]{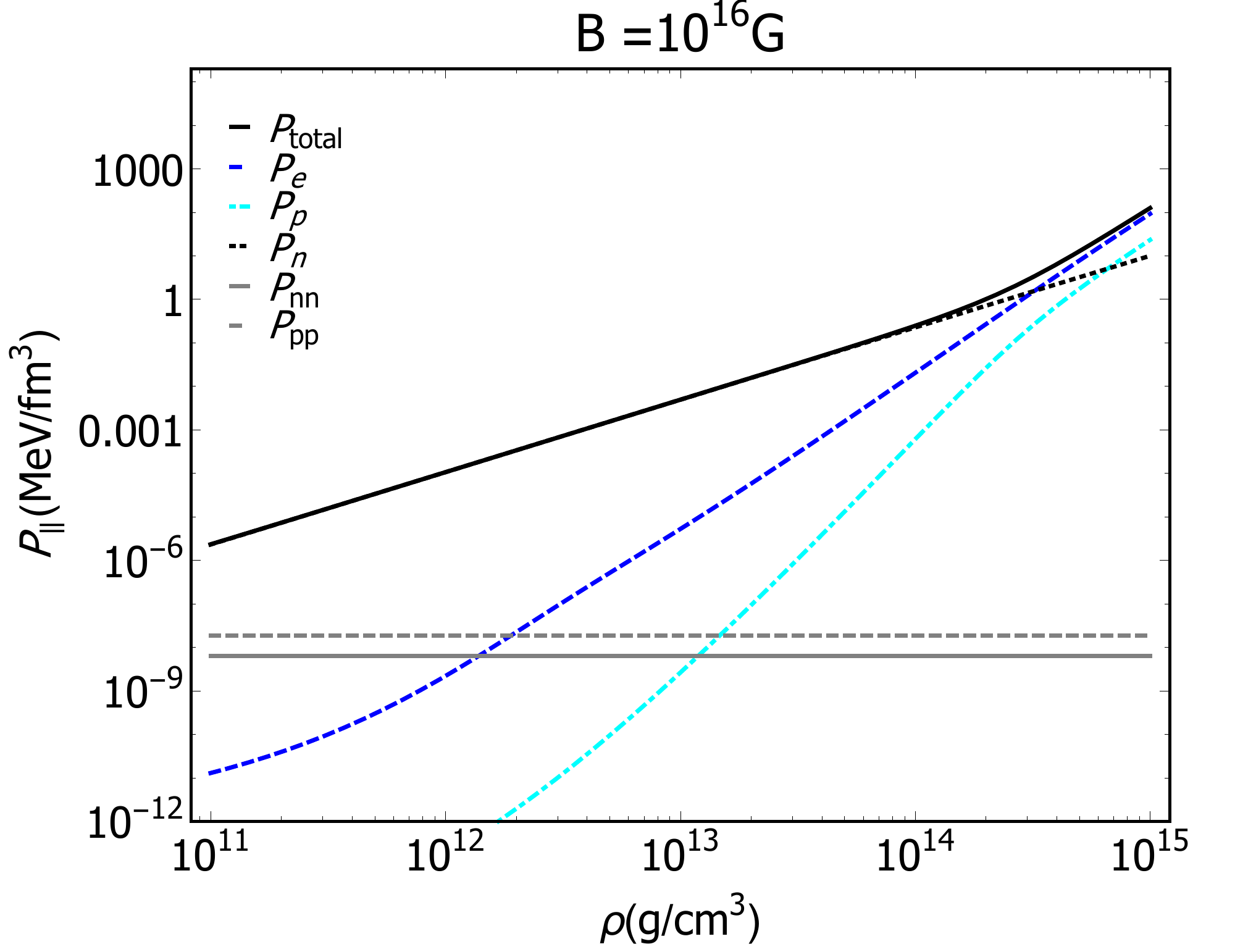}
	\includegraphics[width=0.42\linewidth]{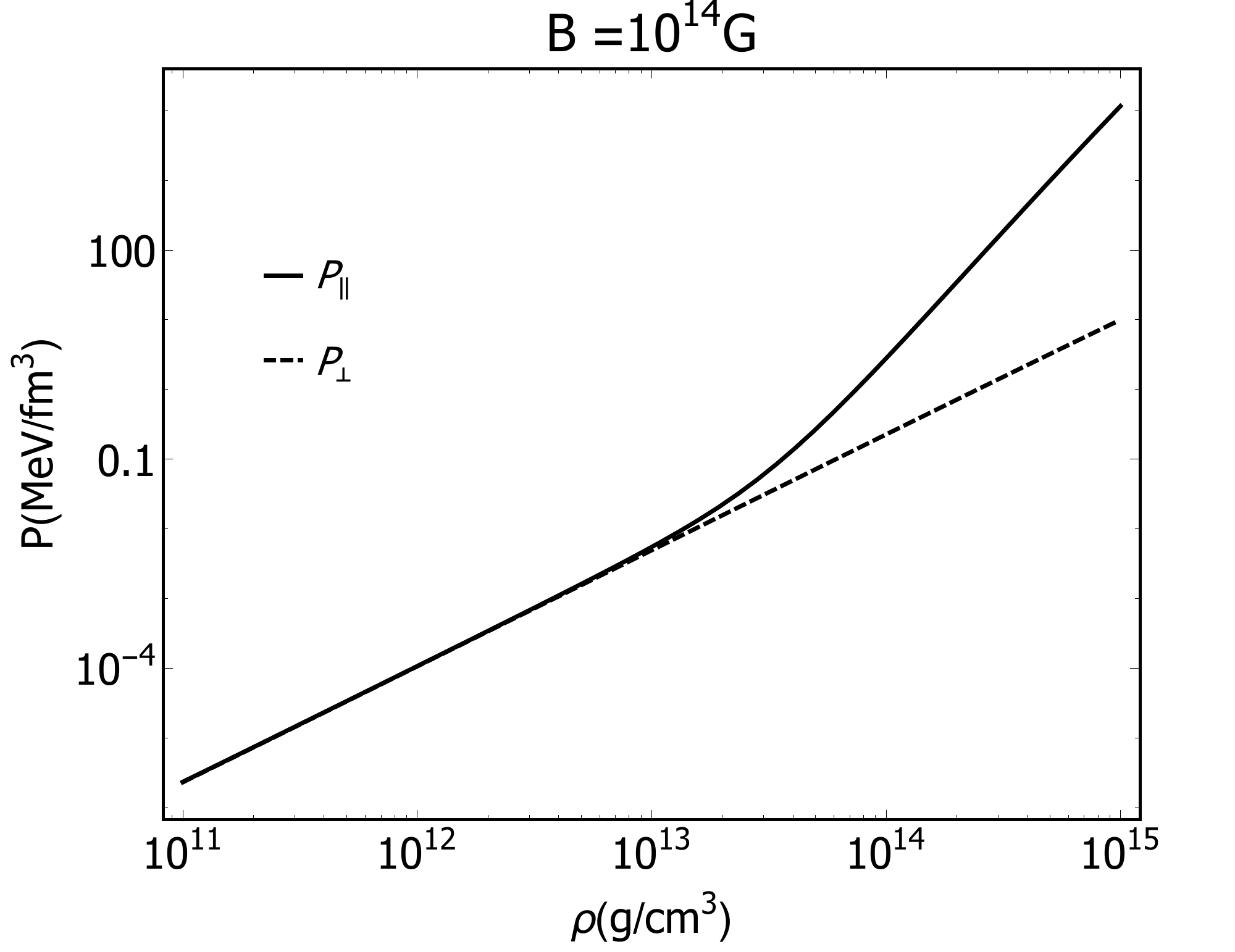}
	\includegraphics[width=0.42\linewidth]{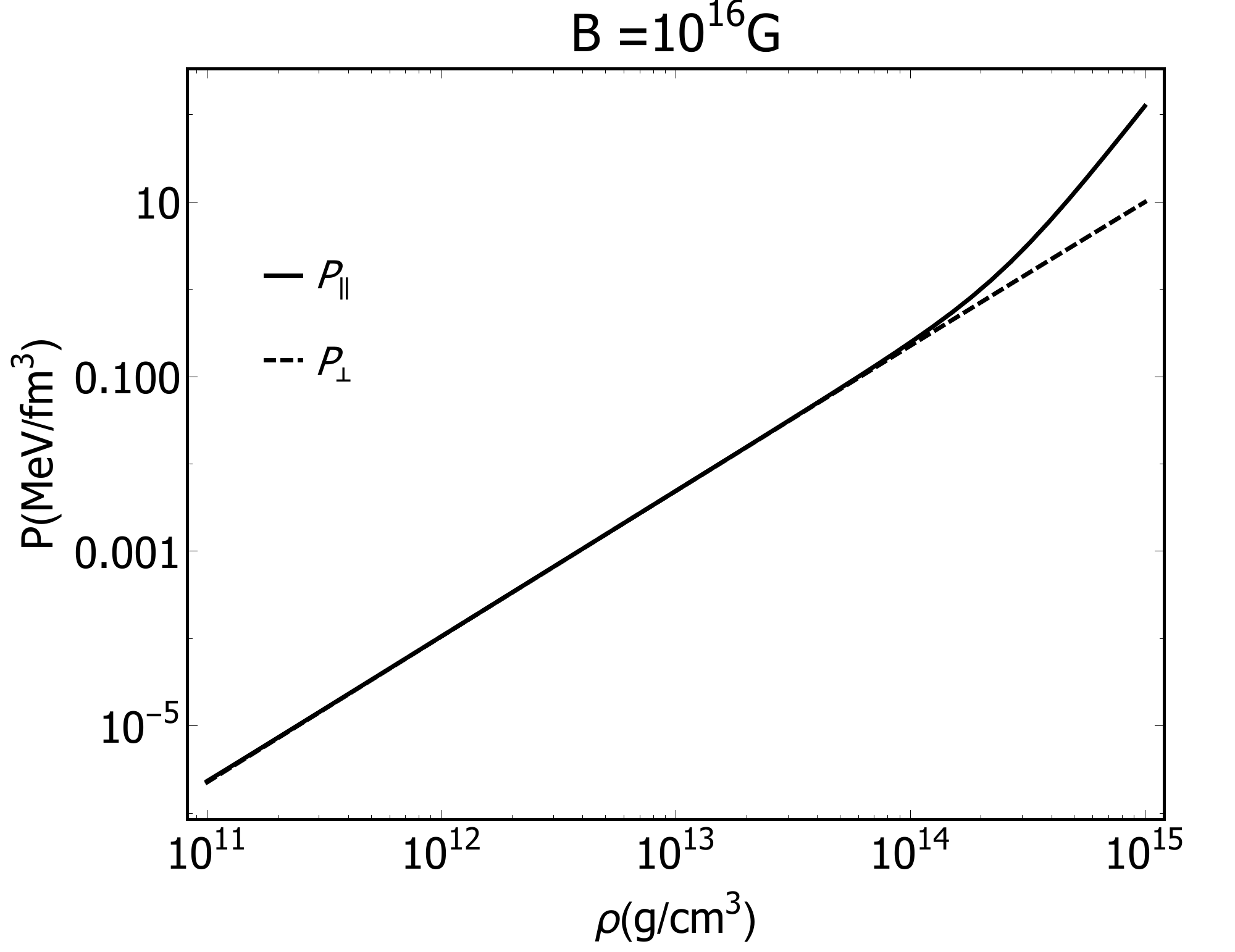}
	\caption{\label{presionesjet} Paneles superiores: Las presiones paralela y perpendicular total del \textit{jet} y sus coponentes, como funci\'on de la densidad de masa de neutrones para varios valores de $B$. Panel inferior: Las dos presiones totales en un mismo gr\'afico a fin de apreciar mejor las diferencias entre ellas.}
\end{figure}

\begin{figure}[!h]
	\centering
	\includegraphics[width=0.42\linewidth]{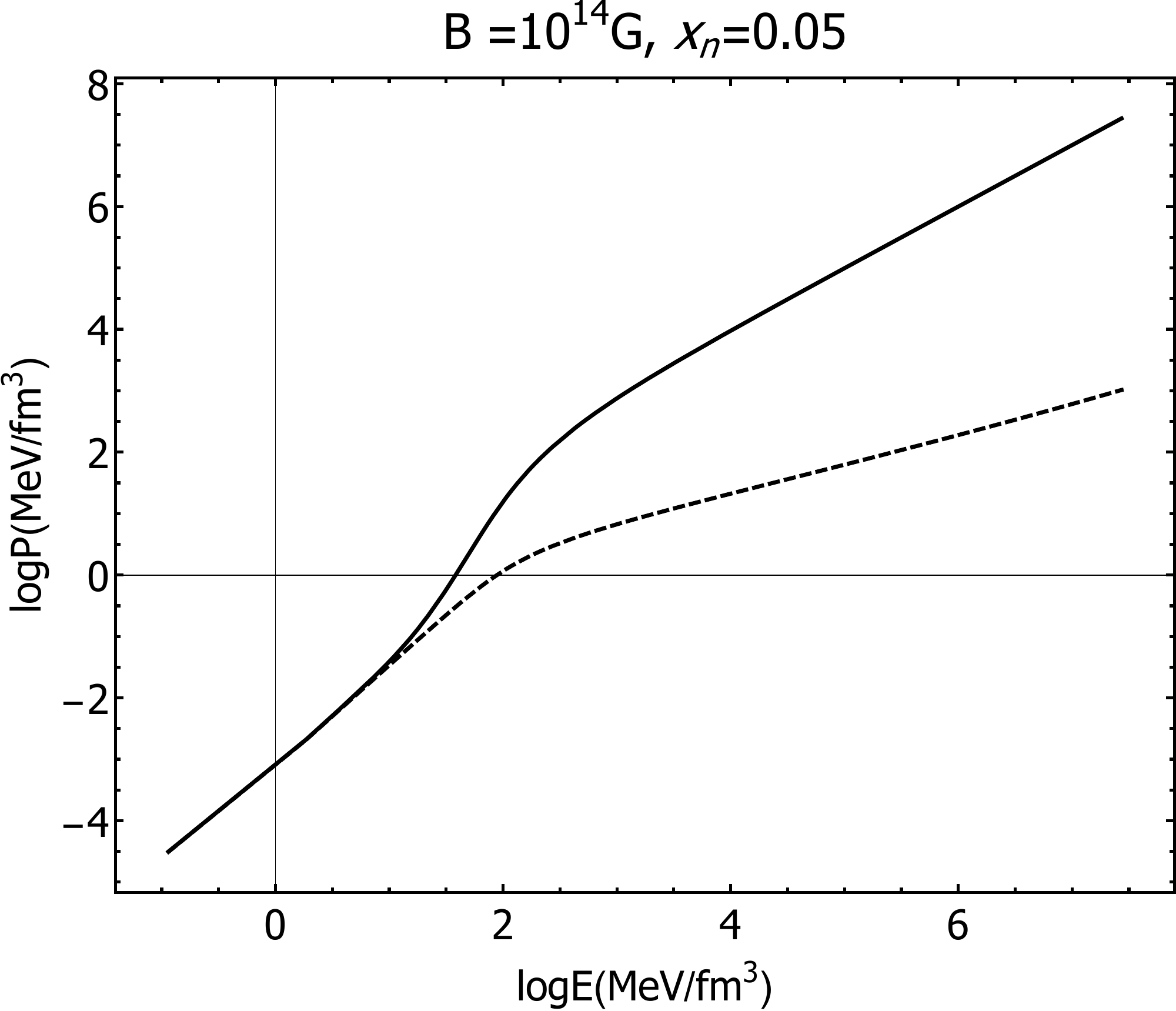}
	\includegraphics[width=0.42\linewidth]{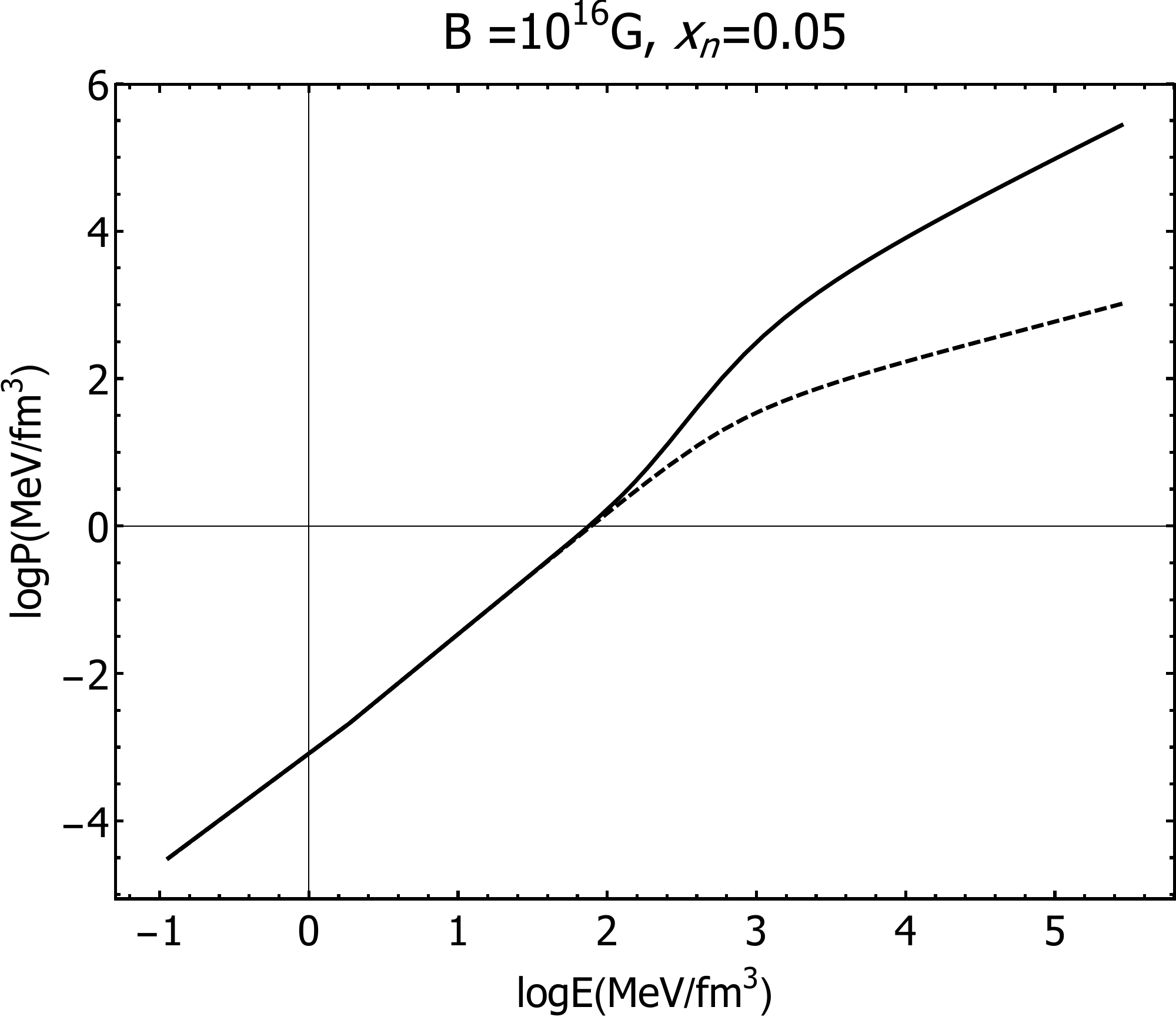}
	\includegraphics[width=0.42\linewidth]{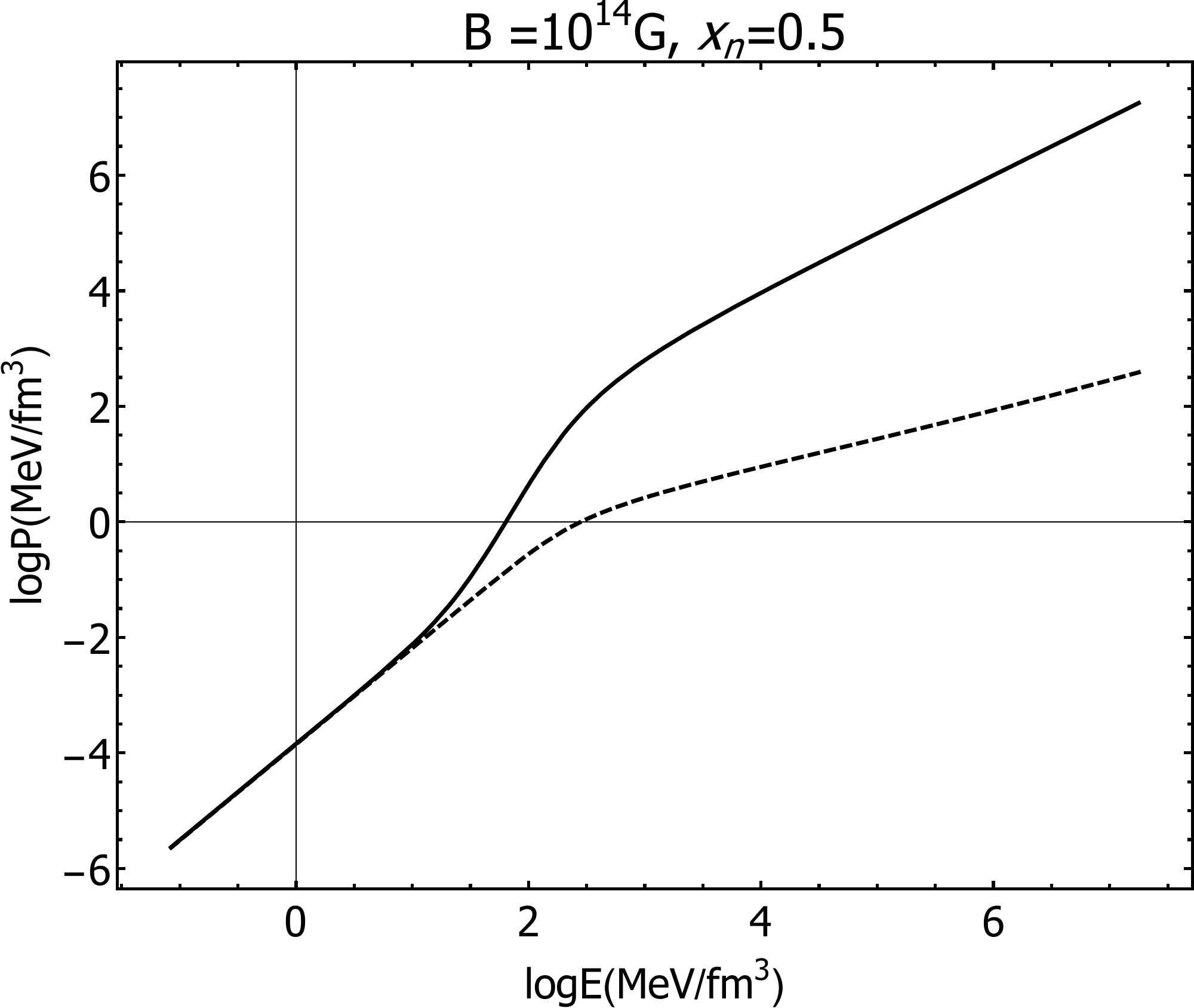}
	\includegraphics[width=0.42\linewidth]{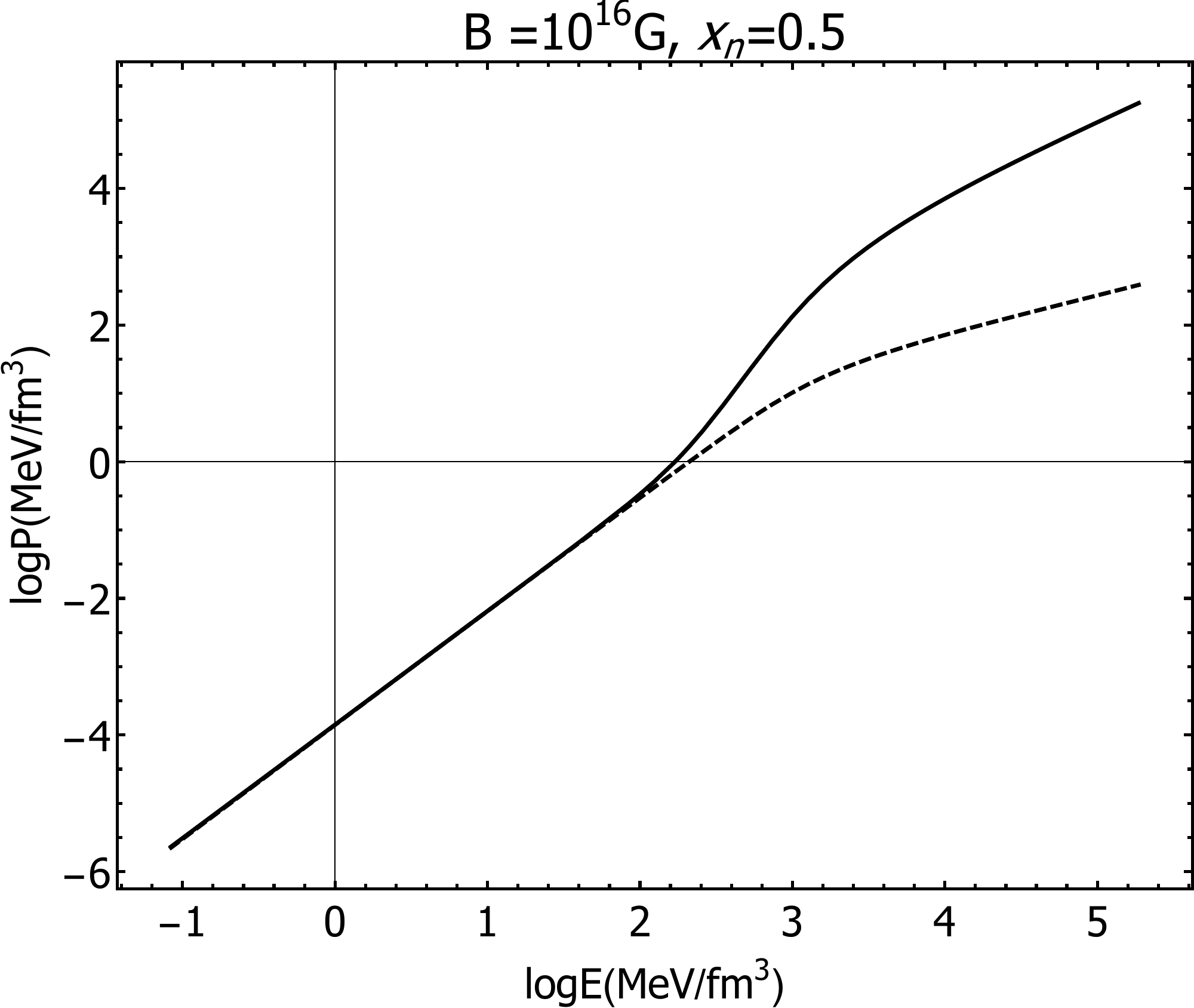}
	\includegraphics[width=0.42\linewidth]{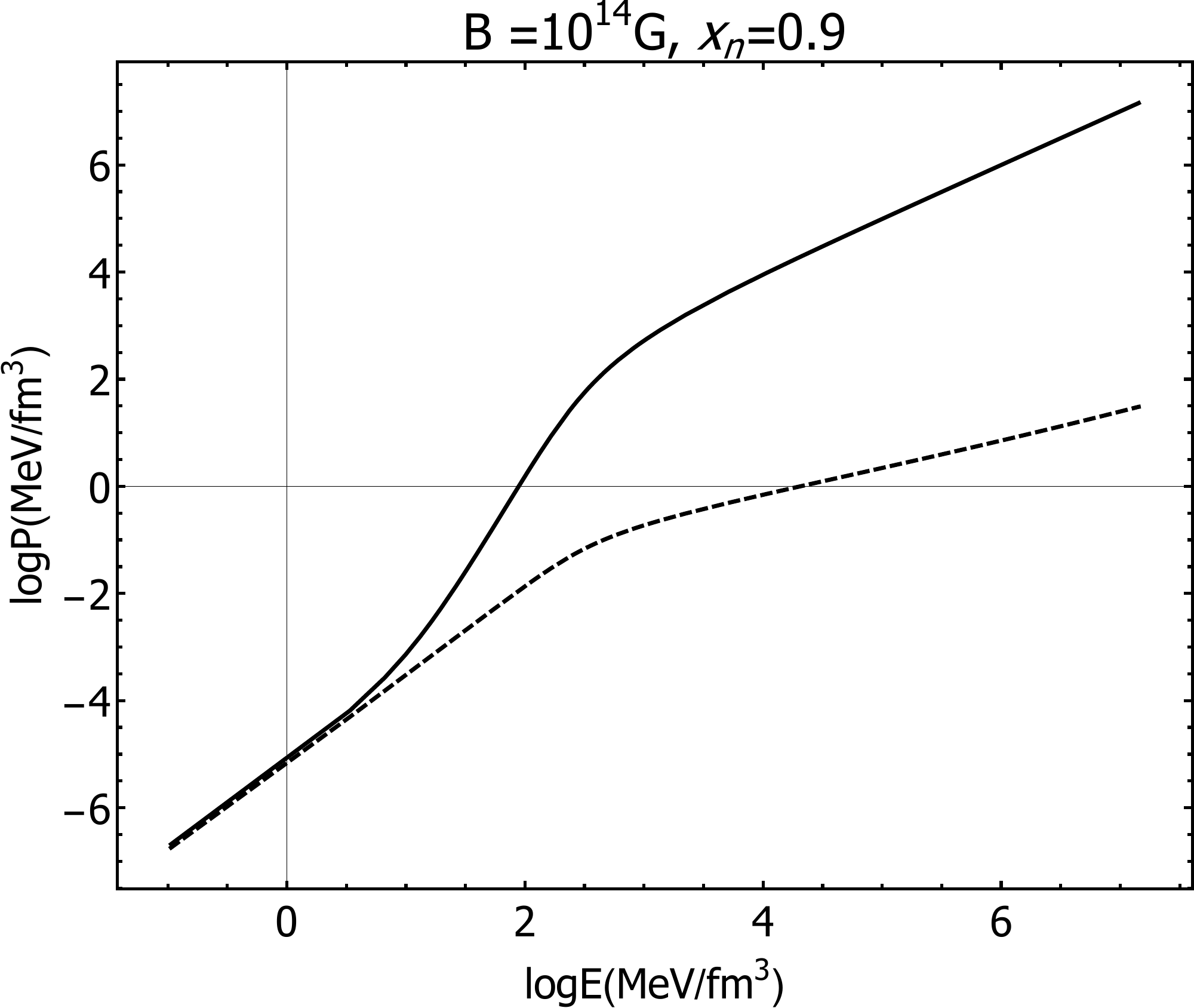}
	\includegraphics[width=0.42\linewidth]{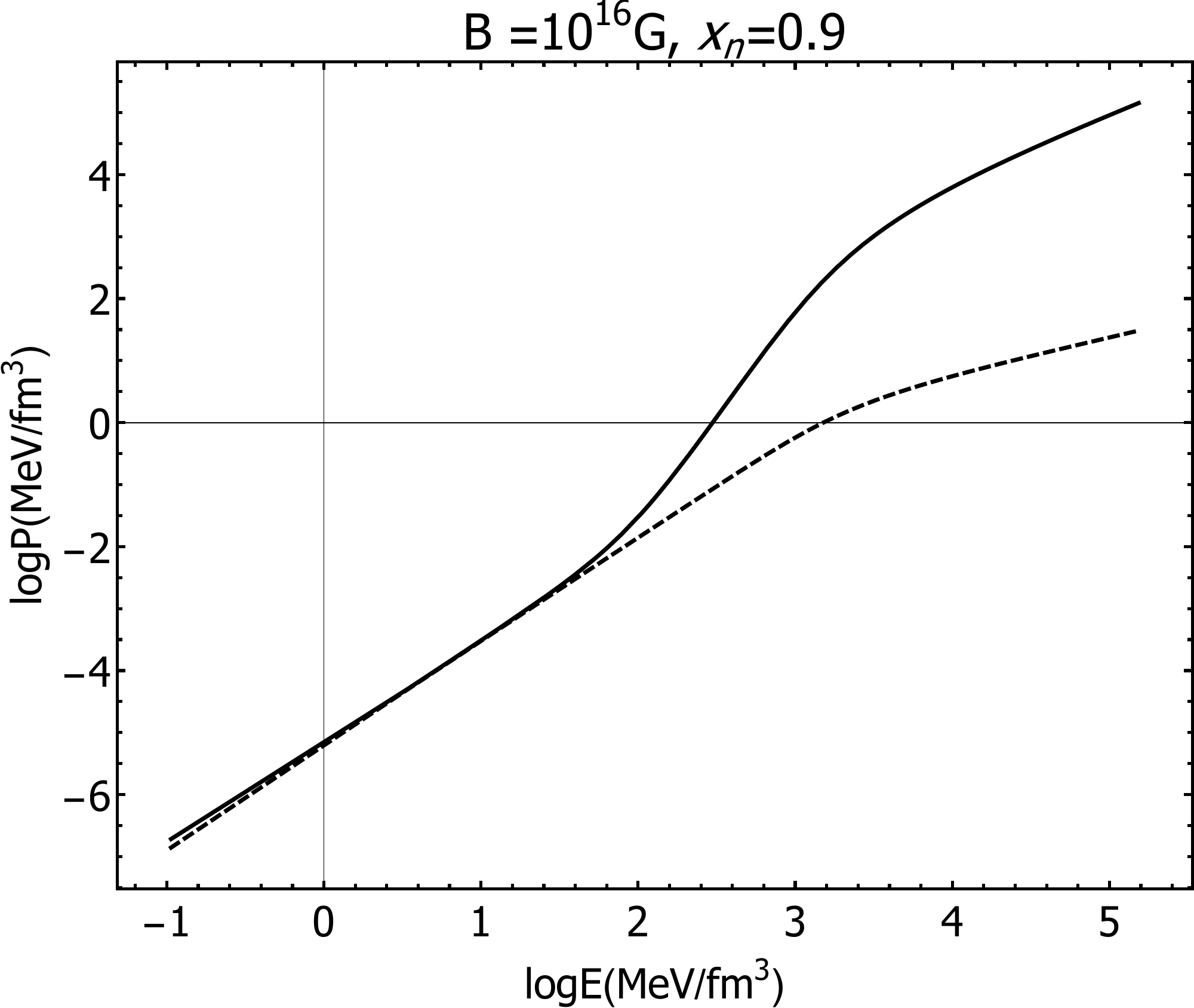}
	\caption{\label{EoS2} Las presiones paralela y perpendicular total del \textit{jet} y sus coponentes, como funci\'on de la densidad de masa de neutrones para varios valores de $B$ y de la fracci\'on de nucleones bosonizada.}
\end{figure}

La Fig.~\ref{presionesjet} recoge estas dependencias para las presiones del \textit{jet}. El panel superior muestra la presi\'on perpendicular del gas \textit{npe} parcialmente bosonizado en r\'egimen de colapso, as\'i como la aportaci\'on a ella de cada uno de los gases. Las presiones de los gases colapsados no aparecen porque son negativas y el gr\'afico es logar\'itmico. Como puede apreciarse $P^{T}_{\perp}$ est\'a dominada completamente por la presi\'on del gas de neutrones. Lo mismo sucede, en la regi\'on de bajas densidades, con $P_{\parallel}^T$ (paneles intermedios). Por el contrario, para las densidades m\'as altas y los campos magn\'eticos m\'as bajos, $P_{\parallel}^T$ est\'a determianda por las presiones paralelas de los gases colapsados de electrones y protones, que son considerablemente mayores que la presi\'on del gas de neutrones, llegando a ser la diferencia entre las dos presiones de hasta dos \'ordenes de magnitud.

Para que las ecuaciones de estado que estamos proponiendo puedan describir realmente al \textit{jet}, las densidades de masa neutr\'onica en su interior deben estar en la zona en que las presiones son anisotr\'opicas, de manera que el objeto resultante se aleje de la forma (cuasi-)esf\'erica de una estrella.
La densidades para las que aparece la anisotrop\'ia as\'i como la diferencia entre las presiones dependen tambi\'en de la fracci\'on de nucleones bosonizados, como muestra la Fig.~\ref{EoS2}. N\'otese que aumentar la fracci\'on de nucleones bosonizados disminuye la presi\'on perpendicular. Esto quiere decir que quiz\'as la clave para construir EdE altamente anisotrópicas partiendo del gas \textit{npe} radique en suponerlo muy bosonizado.

\begin{figure}[h!]
	\centering
	\includegraphics[width=0.49\linewidth]{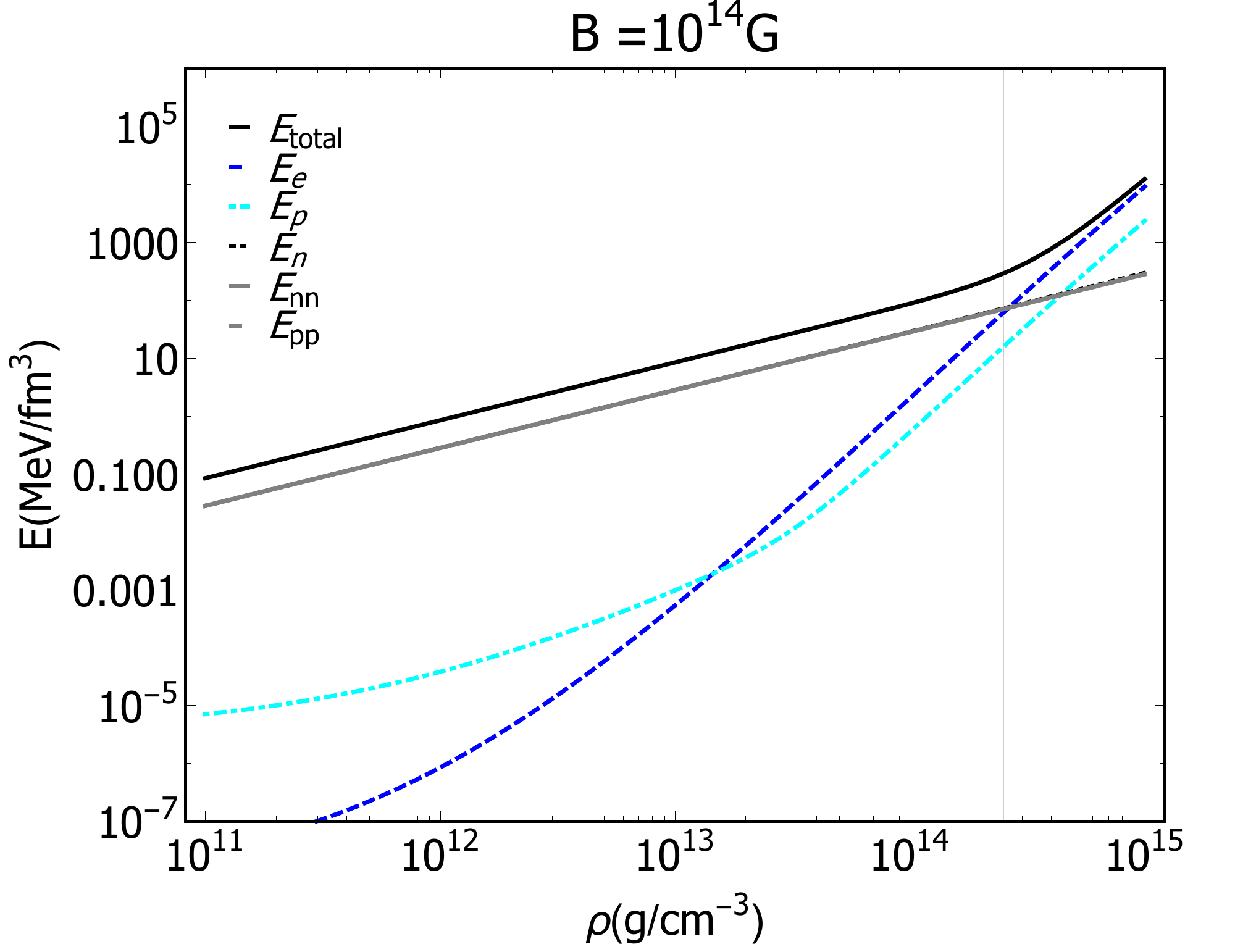}
	\includegraphics[width=0.49\linewidth]{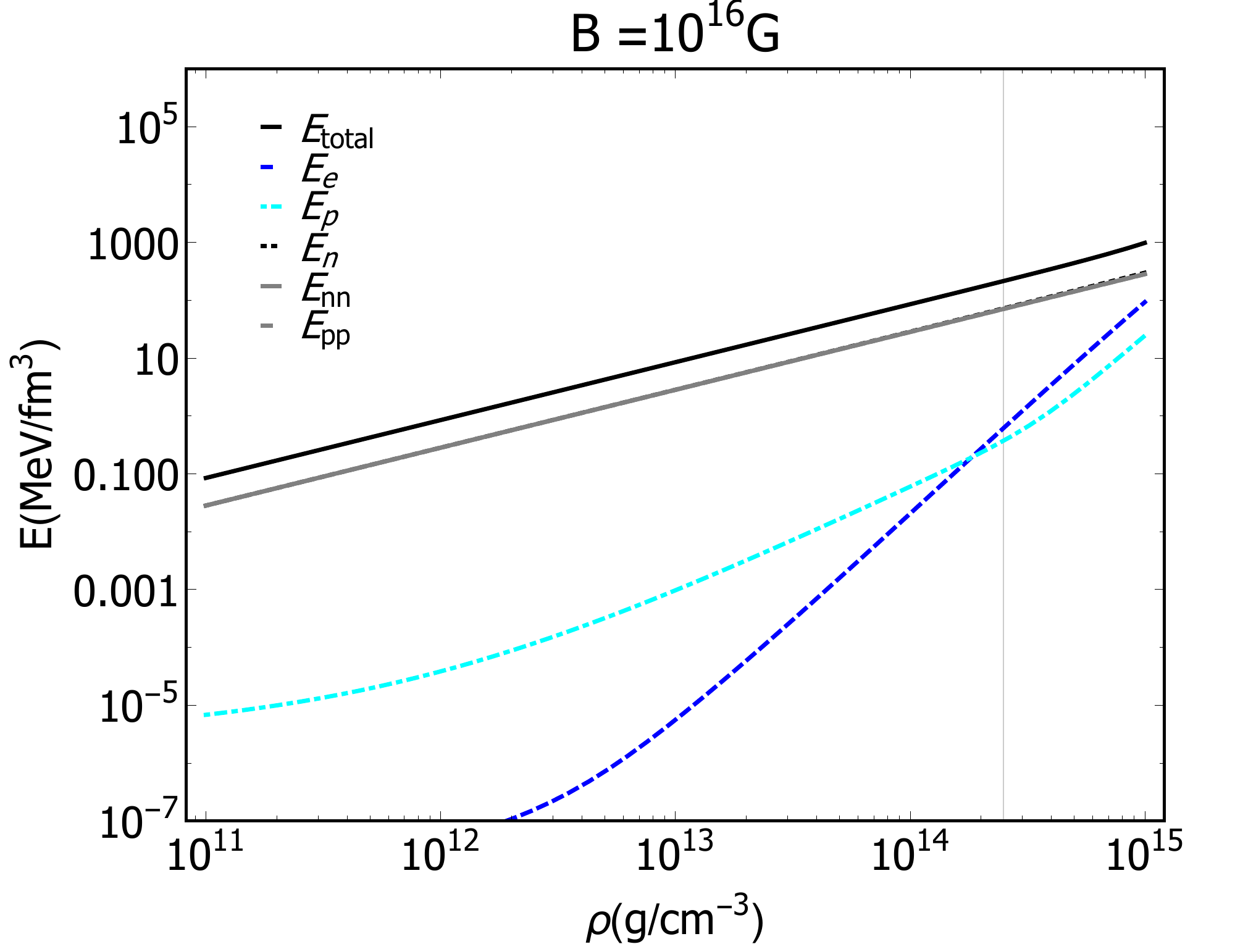}
	\caption{\label{EoSnpe} La densidad de energ\'ia total del \textit{jet} y sus componentes, como funci\'on de la densidad de masa de neutrones para varios valores de $B$.}
\end{figure}

La densidad de energ\'ia interna del \textit{jet} se muestra en la Fig.~\ref{EoSnpe}. Si bien en el caso de las presiones la influencia de los gases bos\'onicos era despreciable, ellos son relevantes a la hora de calcular la densidad de energ\'ia del chorro de materia, y por tanto su presencia ser\'a, al igual que la de los neutrones crucial a la hora de calcular el equilibrio gravitacional del \emph{jet}.

\section{Conclusiones del cap\'itulo}

En este cap\'itulo hemos estudiado los efectos del campo magn\'etico en la estabilidad de la materia que forma el interior de las Estrellas de Neutrones, considerada esta como un gas \textit{npe} parcialmente bosonizado. En las condiciones de campo magn\'etico y densidad de part\'iculas existentes en el interior de las ENs encontramos que:

\begin{itemize}

\item Los gases de electrones, protones, neutrones y neutrones apareados son susceptibles de sufrir un colapso magn\'etico transversal, mientras que el gas de protones apareados es siempre estable.

\item Dadas las condiciones adecuadas de campo magn\'etico y densidad de part\'iculas, los gases de electrones y protones en r\'egimen de colapso magn\'etico podrían superar la gravedad y desencadenar la expulsión de materia hacia el exterior de la estrella.

\item La materia en r\'egimen de colapso  puede automagnetizarse y producir un campo magn\'etico lo suficientemente intenso como para mantener a los gases colimados.

\end{itemize}
	
Los resultados aqu\'i obtenidos validan el mecanismo propuesto para la producci\'on y mantenimiento de los jets astrof\'isicos que se basa en las propiedades de los gases cu\'anticos fuertemente magnetizados. Las EdE de la materia que forma el \textit{jet} fueron estudiadas suponiendo que, dado que el \textit{jet} proviene de una estrella de neutrones, este se compone de un gas \textit{npe} pero con la particularidad de que al menos uno de los gases en la mezcla est\'a en r\'egimen de colpaso. No obstante, el mecanismo que proponemos para su formaci\'on podr\'ia ser v\'alido tambi\'en para otras composiciones con tal de que ellas contengan gases susceptibles de sufrir un colapso magn\'etico. Por tanto, nuestra propuesta podr\'ia ser \'util para explicar la expulsi\'on de materia y la fromaci\'on de \textit{jets} en otros entornos astrof\'isicos.

Para terminar de validar nuestro modelo de \textit{jet} se hace necesario estudiar su estabilidad gravitacional. El estudio de la estabilidad gravitacional del \textit{jet}, es decir, la b\'usqueda de soluciones estables de las ecuaciones de Einstein  con las EdE escritas para el \textit{jet}, requiere o bien ecuaciones de estructura axisim\'etricas, o m\'etodos de integraci\'on num\'erica de las ecuaciones de Einstein que admitan estructuras altamente alejadas de la simetria esf\'erica. En ambos casos, este es a\'un un problema no resuelto.

A lo largo del desarrollo de esta tesis se realizaron dos intentos por encontrar ecuaciones que permitieran la descripci\'on de la estructura del \textit{jet}. En el primero \cite{QuinteroBargueno}, se procedi\'o a la b\'usqueda de soluciones interiores de las ecuaciones de Einstein suponiendo una m\'etrica casi plano-sim\'etrica, ya que este tipo de espacio-tiempo se adapta a la simetr\'ia que el campo magn\'etico impone al tensor de energ\'ia-momento. Pero el uso de los espacio-tiempos antes mencionados impone ciertas condiciones a las presiones y la energ\'ia que solo se cumplen en situaciones muy restrictivas, por ejemplo, en el caso de que las presiones y la energ\'ia del vac\'io predominan sobre las de la materia.

Como resultado del segundo intento, se obtuvieron las ecuaciones de estructura presentadas en \cite{Samantha} para objetos esferoidales que ser\'an discutidas en el Cap\'itulo \ref{cap4}. Sin embargo, estas ecuaciones est\'an restringidas al caso de deformaciones peque\~nas con respecto a la forma esf\'erica y no son apropiadas para estudiar la estabilidad gravitacional del \textit{jet} que como hemos repetido responde a una forma alargada, de m\'axima deformaci\'on con respecto a un esfera.

El estudio de la estabilidad gravitacional de las EdE que estamos proponiendo para un \textit{jet} cuya fuente es una EN, es un problema pendiente en el que pensamos seguir trabajando en el futuro.

%% file: cap3.tex
\chapter{Efectos del campo magn\'etico en las ecuaciones de estado de Estrellas de condensado de Bose-Einstein}
\label{cap3}

En este cap\'itulo se obtienen las ecuaciones de estado para Estrellas de condensado de Bose-Einstein magnetizadas. Para hacerlo, se sigue el procedimiento presentado en \cite{latifah2014bosons}, en el sentido de que las ecuaciones de estado ser\'an obtenidas a partir de un hamiltoniano que es la suma de dos t\'erminos independientes, uno que describe la interacci\'on de las part\'iculas con el campo magn\'etico y otro que describe la interacci\'on de las part\'iculas entre ellas. Las EdE se obtendr\'an para bosones relativistas y no relativistas. En su estudio num\'erico se considerar\'an dos configuraciones del campo m\'agnetico; en una, el campo se toma constante en el interior de la estrella, mientras que en la otra el campo magn\'etico es producido a trav\'es de la automagnetizaci\'on de los bosones y por tanto depende de su densidad. Este cap\'itulo recoge los resultados originales de la autora de la tesis presentados en \cite{Quintero2018BECS} y  \cite{Lismary}.

\section{Modelaci\'on de estrellas de condensado de Bose-Einstein}

Una Estrella de condensado de Bose-Einstein(EBE) es un OC compuesto  por un gas no ideal de bosones vectoriales formados por el apareamiento de dos neutrones con spines paralelos. Al estudiarlas estamos asumiendo que, al menos en alguna etapa de su evoluci\'on, el n\'ucleo de las ENs pudiera estar formado mayoritariamente por materia en esta forma. 

Desde el punto de vista te\'orico, las diferencias entre modelos distintos para EBE vienen dadas, principalmente, por la forma en que se describen las interacciones interpart\'iculas y por el car\'acter relativista o no de estas \cite{Chavanis2012,latifah2014bosons,113}. En \cite{Chavanis2012} una ecuaci\'on de estado politr\'opica es obtenida para estrellas compuestas por gases de bosones escalares a temperatura cero. En este trabajo se analizan dos casos, el de bosones no relativistas (a trav\'es de resolver la ecuaci\'on de Gross-Pitaevskii con interacciones repulsivas de corto alcance tipo Van der Waals) y el de bosones relativistas (a partir de resolver la ecuaci\'on de Klein-Gordon con una interacci\'on del tipo $\lambda \phi^4$). Por otra parte, en \cite{latifah2014bosons}, para obtener las ecuaciones de estado se a\~nade una interacci\'on repulsiva tipo Van der Waals al hamiltoniano de los bosones escalares no relativistas y la temperatura es incluida. Este mismo caso se estudia en \cite{113}, pero a partir de la aproximaci\'on de Hartree-Fock. A pesar de las diferencias, en todos los casos se obtiene que las EBE son objetos relativamente peque\~nos, con radios $R \leq 10$~km, y muy densos, con densidades de masa m\'axima en su centro en el orden de los $10^{16}$g/cm$^3$.

Las EdE de nuestro modelo de EBE magnetizadas ser\'an obtenidas a partir de un hamiltoniano que es la suma de dos t\'erminos independientes, uno que describe la interacci\'on interpart\'iculas y otro que corresponde a la descripci\'on de los bosones como un gas ideal \cite{latifah2014bosons}. La interacci\'on repulsiva entre los bosones ser\'a modelada como una interacci\'on de contacto por pares. Esta forma de describir la interacci\'on, aunque relativamente sencilla, ha demostrado ser muy \'util como punto de partida y de comparaci\'on, pues permiti\'o a los autores de \cite{latifah2014bosons} obtener EdE termodin\'amicamente consistentes cuyas curvas masa-radio son consistentes con otros modelos m\'as elaborados de Estrellas de condensado de Bose-Einstein \cite{Chavanis2012}. 

La novedad de nuestro trabajo con respecto a \cite{latifah2014bosons} consiste en la inclusi\'on del car\'acter vectorial de los bosones as\'i como de su interacci\'on con el campo magn\'etico a trav\'es del t\'ermino de gas ideal del hamiltoniano. En el caso de bosones relativistas, usaremos para ello los resultados obtenidos en el Cap\'itulo \ref{cap2}. Para bosones no relativistas utilizaremos las EdE calculadas por nosotros en \cite{Lismary}. 

\section{Ecuaciones de estado de un Estrella de condensado de Bose-Einstein magnetizada}

El hamiltoniano de un gas de bosones que interactúan entre sí puede escribirse como la suma del hamiltoniano del gas ideal (no interactuante) $\hat{H}$ con el hamiltoniano de interacción $\hat{H}_{int}$:

\begin{equation}\label{hamiltoniantotal}
\hat{H}_{total} = \hat{H} + \hat{H}_{int}.
\end{equation}

En primera aproximaci\'on, supondremos que las part\'iculas que componen el gas no interact\'uan a trav\'es de sus spines y tomaremos la interacci\'on entre los bosones como una interacci\'on de contacto por pares $u_0 \delta(r-\bar{r})$, donde $r$ y $\bar{r}$ denotan ls posiciones de las part\'iculas que interact\'uan \cite{latifah2014bosons}. El par\'ametro $u_0 = 4 \pi a/m$ indica la fortaleza de la interacci\'on; $m$ es la masa del bos\'on y $a$ es la longitud de dispersi\'on ($a=1$fm) \cite{Chavanis2012,latifah2014bosons}. Bajo esta suposic\'on, $\hat{H}_{int}$ queda:

\begin{equation}\label{hamiltonian_int}
\hat{H}_{int} = \frac{1}{2} u_0 \sum_{k,k^{\prime}} \hat{n}_k \hat{n}_{k^{\prime}},
\end{equation}

\noindent siendo $\hat{n}_k$ el operador del n\'umero de ocupaci\'on en el estado $k$. 

Debido a que el n\'umero de part\'iculas en escenarios astrof\'isicos es muy grande mientras que las temperaturas son relativamente bajas, es de esperar que las fluctuaciones no sean muy significativas, de manera que el hamiltoniano de interacci\'on puede aproximarse por su valor de expectaci\'on:

\begin{equation}\label{hamiltonian_int1}
\hat{H}_{int} \approx \langle \hat{H}_{int} \rangle = \frac{1}{2} u_0 N^2,
\end{equation}

\noindent donde, como antes, $N = \sum_{k} \langle \hat{n}_k \rangle$ es la densidad media de part\'iculas. 

Esta forma de modelar la interacci\'on entre part\'iculas es equivalente a conservar solo el primer t\'ermino del desarrollo para bajas energ\'ias de la teor\'ia de dispersi\'on \cite{pethick_smith_2008}. En este caso el choque se decribe exclusivamente en t\'erminos de $a$, que contiene todos los efectos de las interacciones de corto alcance, y las flucutaciones cu\'anticas y t\'ermicas son despreciadas \cite{pethick_smith_2008}. La consecuencia principal de tomar en cuenta el efecto de estas fluctuaciones consiste en un aumento de la presi\'on y la energ\'ia interna del gas \cite{pethick_smith_2008}. Por tanto, para un gas de bosones que interact\'uan entre s\'i, el hamiltoniano Ec.~(\ref{hamiltonian_int1}) genera las EdE con menor presi\'on y energ\'ia de interacci\'on posibles. En este sentido, el modelo que aqu\'i presentamos constituye un caso l\'imite para las EBE, pues cualquier correcci\'on que se haga sobre \'el se traducir\'a en un aumento de las presiones que facilitar\'a el equilibrio gravitacional del OC resultante, aumentando su masa y radios, y atenuando los efectos del campo magn\'etico.

A partir de las Ecs.~(\ref{hamiltoniantotal})-(\ref{hamiltonian_int1}), la funci\'on de partici\'on del sistema puede escribirse como:

\begin{equation}
\Xi_{total} = e^{-\beta V \frac{1}{2} u_0 N^2} \Xi,
\end{equation}

\noindent si llamamos $\Xi$ a la funci\'on de partici\'on del gas ideal de bosones (la calculada a partir de $\hat{H}$). El potencial termodin\'amico por unidad de volumen es:

\begin{equation}\label{termo}
\Omega_{total} =-\frac{1}{\beta V} \ln \Xi_{total} = \frac{1}{2} u_0 N^2 + \Omega,
\end{equation}

\noindent donde $\Omega$ es el potencial termodin\'amico del gas ideal de bosones.

El t\'ermino de interacci\'on entre bosones $1/2 u_0 N^2$ entra en el potencial termodin\'amico Ec.~(\ref{termo}) y por tanto en las ecuaciones de estado, de manera independiente al t\'ermino de gas ideal $\Omega$. En lo que sigue tomaremos ventaja de esto al considerar que las partículas interact\'uan con el campo magnético y entre ellas de manera independiente. En consecuencia, los efectos del campo magn\'etico entrar\'an en las ecuaciones de estado de la estrella solo a trav\'es del t\'ermino de gas ideal $\Omega = \Omega(T,\mu,B)$. Conocida $\Omega$, las EdE del gas magnetizado de bosones que interact\'uan entre s\'i se pueden calcular como \cite{latifah2014bosons,Ferrer}:

\begin{subequations}\label{EoS}
	\begin{align}
	P_{\parallel}&= -\Omega_{total} + N \left(\frac{\partial \Omega_{total}}{\partial N}\right)_{\mu,T,B} = \frac{1}{2}u_0 N^2 -\Omega, \label{ppar} \\
	\nonumber\\
	P_{\perp}& = -\Omega_{total} + N \left(\frac{\partial \Omega_{total}}{\partial N}\right)_{\mu,T,B} + B \left(\frac{\partial \Omega_{total}}{\partial B}\right)_{\mu,T}  = \frac{1}{2}u_0 N^2 -\Omega - B \mathcal M, \label{pper} \\
	\nonumber\\
	E &= \Omega_{total} + \mu N - T \left(\frac{\partial \Omega_{total}}{\partial T}\right)_{\mu,B} = \frac{1}{2}u_0 N^2 + \Omega + \mu N - T \left(\frac{\partial \Omega}{\partial T}\right)_{\mu,B}, \label{energy1}\\
	\nonumber\\
	N &= - \left(\frac{\partial \Omega_{total}}{\partial \mu}\right)_{T,B} = - \left(\frac{\partial \Omega}{\partial \mu}\right)_{T,B}, \\
	\nonumber\\
	\mathcal M &= - \left(\frac{\partial \Omega_{total}}{\partial B}\right)_{\mu,T}= - \left(\frac{\partial \Omega}{\partial B}\right)_{\mu,T}.
	\label{magnetization1}
	\end{align}
\end{subequations}

En las Ecs.~(\ref{EoS}), ambas presiones $P_{\parallel}$ y $P_{\perp}$  tienen tanto la contribuci\'on de la interacci\'on \cite{latifah2014bosons}, como la de los bosones en presencia del campo magn\'etico, incluido el t\'ermino $-\mathcal M B$ que produce la anisotrop\'ia  \cite{Ferrer}. La densidad de energ\'ia interna $E$ del gas ahora contiene adem\'as la densidad de energ\'ia de la interacci\'on.

El potencial termodin\'amico del gas ideal de bosones en interacción con un campo magn\'etico $\Omega$ ser\'a calculado para dos casos: el relativista y su l\'imite no relativista. A primera vista, el caso relativista es m\'as apropiado dadas las altas densidades de part\'iculas y energ\'ia que existen en el interior de los objetos compactos. El caso no relativista, en cambio, es la extensi\'on natural  del modelo de EBE no magnetizada presentado en \cite{latifah2014bosons} y por eso tambi\'en ser\'a estudiado.

Para el caso en el que el gas magnetizado de bosones se considera relativista, el potencial termodin\'amico del gas ideal fue calculado en el Cap\'itulo \ref{cap2} y viene dado por la Ec.~(\ref{Grand-Potential-Tcero}). En el l\'imite no relativista el potencial termodinámico, al que denotaremos $\Omega^{nr}$, se obtiene a partir del espectro correspondiente:

\begin{equation}\label{spectrum1}
	\varepsilon^{nr}(p,B)=m +p^2/2 m - \kappa s B.
\end{equation}

\noindent Una vez hecho el c\'aclulo, el potencial termodin\'amico del gas ideal de bosones no relativistas en interacci\'on con un campo magnético es \cite{Lismary}:

\begin{equation}\label{omeganr}
	\Omega^{nr} = -\frac{m^{3/2}}{(2 \pi)^{3/2} \beta^{5/2}} \{ Li_{5/2}(z_{-})+  Li_{5/2}(z) + Li_{5/2}(z_{+})\},
\end{equation}

\noindent donde $z=e^{\beta \mu}$ es la fugacidad y $z_{\pm} =e^{\beta (\mu \pm \kappa B)}$. 
Con el uso de la Ec.~(\ref{omeganr}), las presiones, la magnetizaci\'on y la densidad de energ\'ia del gas no relativista pueden escribirse como:

\begin{subequations}\label{EoSNBNR}
	\begin{align}
E^{nr} &= - \frac{3}{2} \Omega^{nr} - \kappa B \left (N_{gs} + \frac{Li_{3/2}(z_{+})}{\lambda^3} - \frac{Li_{3/2}(z_{-})}{\lambda^3}\right ),\\
\label{EoSNBNR2}
P^{nr}_{\parallel} &= -\Omega^{nr},\\
\label{EoSNBNR3}
P^{nr}_{\perp} &= -\Omega^{nr}- \mathcal M^{nr} B,\\
\label{EoSNBNR4}
\mathcal M^{nr} &= \kappa\left (N_{gs} + \frac{Li_{3/2}(z_{+})}{\lambda^3} - \frac{Li_{3/2}(z_{-})}{\lambda^3}\right ),\\
\label{EoSNBNR5}
N^{nr} &= N_{gs} + \frac{Li_{3/2}(z_{-}) }{\lambda^3}+ \frac{Li_{3/2}(z)}{\lambda^3}+ \frac{Li_{3/2}(z_{+})}{\lambda^3}.
\end{align}
\end{subequations}

\noindent donde $N_{gs}$ es la fracci\'on de part\'iculas en el condensado y $\lambda = \sqrt{2 \pi/m T}$ es la longitud de onda t\'ermica. 

Aunque los potenciales termodin\'amicos Ecs.~(\ref{Grand-Potential-Tcero}) y (\ref{omeganr}) permiten trabajar a temperatura finita, en este cap\'itulo nos concentraremos en el l\'imite de temperatura cero a fin de estudiar exclusivamente los efectos del campo magn\'etico en las EdE de las Estrellas de condensado de Bose-Einstein. A $T=0$, $\Omega =\Omega_{vac}$ mientras que $\Omega^{nr} =0$. Combinando estos l\'imites con las Ecs.~(\ref{EoS}), las EdE para el gas magnetizado de bosones que interact\'uan entre s\'i son, en el caso relativista:

\begin{subequations}\label{EoSRtotal}
	\begin{align}
	P_{\parallel}^r &= \frac{1}{2}u_0 N^2 -\Omega_{vac},  \label{EoSRtotal1}\\
	P_{\perp}^r &= \frac{1}{2}u_0 N^2-\Omega_{vac}-\mathcal M B,\label{EoSRtotal2}\\
	E^r &=\frac{1}{2}u_0 N^2 +m \sqrt{1-b} N + \Omega_{vac},  \label{EoSRtotal3}\\
	{\mathcal M} &= \frac{\kappa}{\sqrt{1-b}} N, \label{EoSRtotal4}
	\end{align}
\end{subequations}

\noindent y en el no relativista:

\begin{subequations}\label{EoSNRtotal}
	\begin{align}
	P_{\parallel}^{nr} &=  \frac{1}{2}u_0 N^2,  \label{EoSNRtotal1}\\
	P_{\perp}^{nr} &= \frac{1}{2}u_0 N^2 - \mathcal M B ,   \label{EoSNRtotal2} \\
	E^{nr} &=  \frac{1}{2} u_0 N^2 + (m-\kappa B) N,  \label{EoSNRtotal3}
	\\
	{\mathcal M}^{nr} &= \kappa N.  \label{EoSNRtotal4}
	\end{align}
\end{subequations}

\noindent N\'otese que las Ecs.~(\ref{EoSRtotal})-(\ref{EoSNRtotal}) est\'an parametrizadas con respecto a la densidad de part\'iculas $N$ y la intensidad del campo magn\'etico $B$.

Las modificaciones debidas a la presencia del campo magn\'etico con respecto al caso $B=0$ entran en las Ecs.~(\ref{EoSRtotal}) de tres formas. Primero, a trav\'es de la la presi\'on magn\'etica $-\mathcal M B $ que disminuye la presi\'on perpendicular con respecto a la paralela. En segundo lugar, a trav\'es de la presi\'on de vac\'io $-\Omega_{vac}$ que se adiciona a las presiones y se resta a la energ\'ia, y finalmente, a trav\'es del t\'ermino $m \sqrt{1-b} N$ que aparece en lugar de la energ\'ia de reposo $m N$ y expresa el cambio producido en el estado b\'asico por el campo magn\'etico. Por otra parte, las ecuaciones de estado no relativistas Ecs.~(\ref{EoSNRtotal}) tambi\'en son modificadas a trav\'es de la presi\'on magn\'etica y del decrecimiento de la densidad de energ\'ia ahora por la sustitución de la energ\'ia de reposo de los bosones por  $(m-\kappa B)$. 

Con respecto a las ecuaciones de Einstein, las ecuaciones de estado juegan el papel de las fuentes de materia y, por tanto, deben contener las contribuciones energ\'eticas de todas las part\'iculas y campos presentes en el sistema, excepto la correspondiente al campo gravitacional \cite{103}. Por ello, para completar las EdE de la Estrella de condensado de Bose-Einstein magnetizada es necesario a\~nadir a la energ\'ia y las presiones del gas, la energ\'ia y las presiones del campo magn\'etico \cite{Lattimerprognosis}. Desde el punto de vista matem\'atico, para ello basta a\~nadir la llamada contribuci\'on de Maxwell, $B^2/8 \pi$, a $E$ y $P_{\perp}$, y sustraerla de $P_{\parallel}$. En lo adelante, cuando se hable de las EdE del gas magnetizado de bosones vectoriales con interacci\'on, se estar\'a haciendo referencia a las Ecs.~(\ref{EoSRtotal}) o (\ref{EoSNRtotal}) seg\'un sean relativistas o no las part\'iculas, mientras que al hablar de las EdE de la EBE magnetizada se estar\'a haciendo referencia a estas mismas ecuaciones con contribuci\'on de Maxwell.

La relevancia de cada una de las modificaciones que el campo magn\'etico provoca en las ecuaciones de estado y los observables -masas y radios- de la EBE no es obvia y ser\'a analizada de manera num\'erica en las pr\'oximas secciones. Para lograr una comprensi\'on clara del papel que juega cada t\'ermino introducido por el campo magn\'etico, las EdE (as\'i como las curvas masa-radio en el Cap\'itulo \ref{cap5}) ser\'an estudiadas con y sin t\'ermino de Maxwell.

\section{Ecuaciones de estado para las Estrellas de condensado de Bose-Einstein magnetizadas y automagnetizadas}
\label{sec4}

Esta secci\'on est\'a dedicada al estudio num\'erico de las Ecs.~(\ref{EoSRtotal}) y (\ref{EoSNRtotal}), y en particular de las presiones anisotr\'opicas como funci\'on la densidad de masa del gas de bosones en dos configuraciones diferentes del campo magn\'etico. En la primera, la intensidad del campo m\'agnetico ser\'a considerada constante y fijada de manera externa al gas. 
En la segunda, se tomar\'a el campo magn\'etico como producido por la automagnetizaci\'on de los bosones. 

Por lo general, al modelar OCs  sus campos magn\'eticos internos se  suponen dipolares, poloidales o uniformes \cite{Lattimerprognosis}. Su intensidad se toma constante o dependiente de la densidad de part\'iculas o el radio, aunque no siempre estas dependencias obedecen a razones f\'isicas bien fundamentadas \cite{ChatterjeeBprofiles}. Incluso en los casos en que las ecuaciones de Maxwell se acoplan a las de Einstein a fin de garantizar la consistencia f\'isica del campo electromagn\'etico en el interior de la estrella, como fuentes del campo magn\'etico se supone la existencia en la estrella de corrientes el\'ectricas cuya forma matem\'atica garantiza las distribuciones de l\'ineas de fuerza deseadas pero que no guardan relaci\'on alguna con la materia que compone el OC (con las EdE) y cuyos or\'igenes f\'isicos no son nunca justificados \cite{Chatterjee}. En este sentido, al proponer que el campo magn\'etico de la EBE es generado por la automagnetizaci\'on de los bosones se pretende proporcionar un mecanismo para producirlo que est\'a relacionado directamente con la materia que la compone. En tal caso diremos que la EBE est\'a automagnetizada.

Por otra parte, a\'un cuando todo indica que la intensidad del campo magn\'etico  var\'{\i}a en varios \'ordenes de magnitud entre el n\'ucleo de la estrella y su superficie \cite{Lattimerprognosis,Chatterjee,ChatterjeeBprofiles}, suponer el campo constante en el interior de la estrella no deja de ser una buena primera aproximaci\'on debido a que el rango de acci\'on microsc\'opico de la interacci\'on electromagn\'etica es mucho menor que la escala macrosc\'opica de variaci\'on del campo magn\'etico entre el centro y la superficie de la estrella \cite{Lattimerprognosis}. 

\subsection{Ecuaciones de estado con campo magn\'etico constante} \label{sec4A}

\begin{figure}[h]
	\centering
	\includegraphics[width=0.42\linewidth]{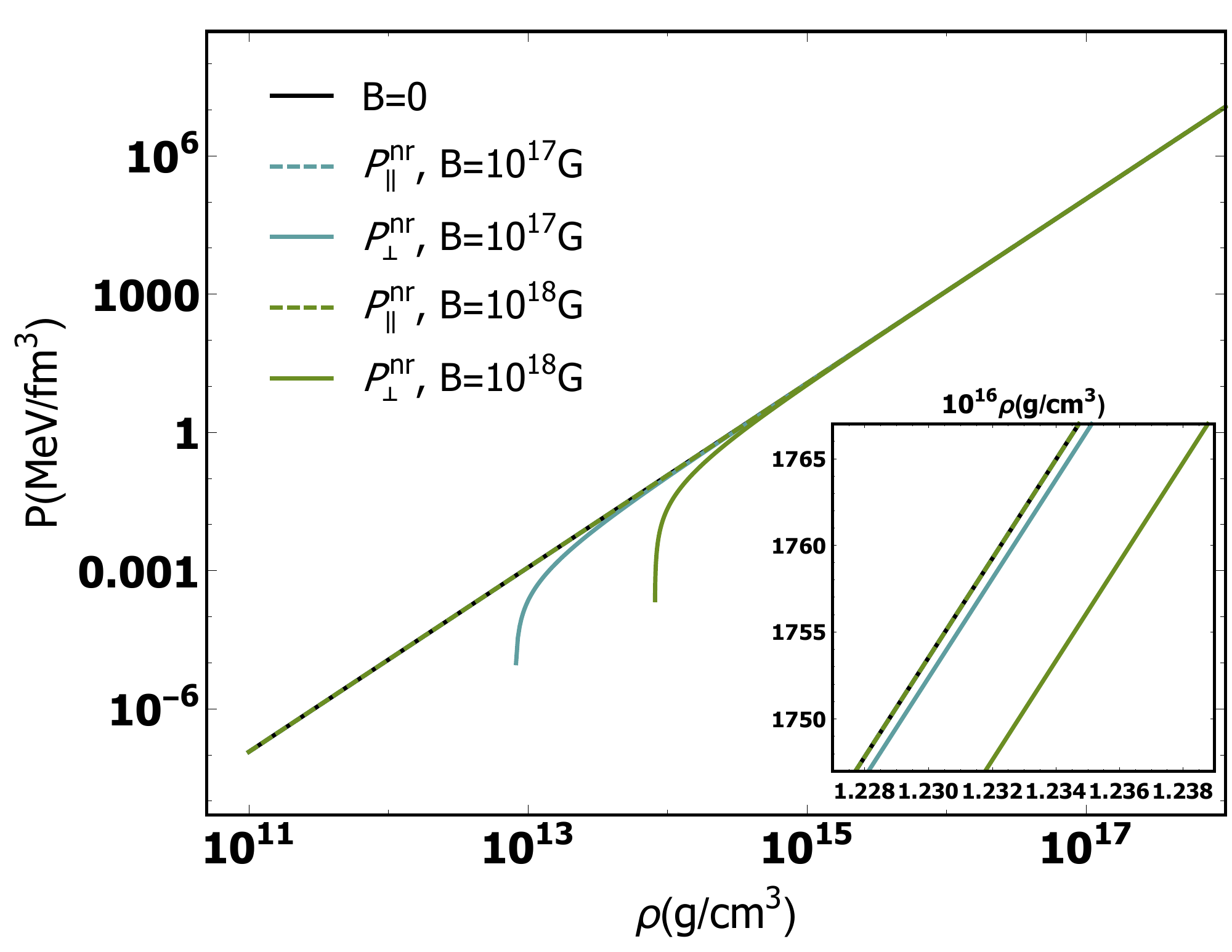}
	\includegraphics[width=0.42\linewidth]{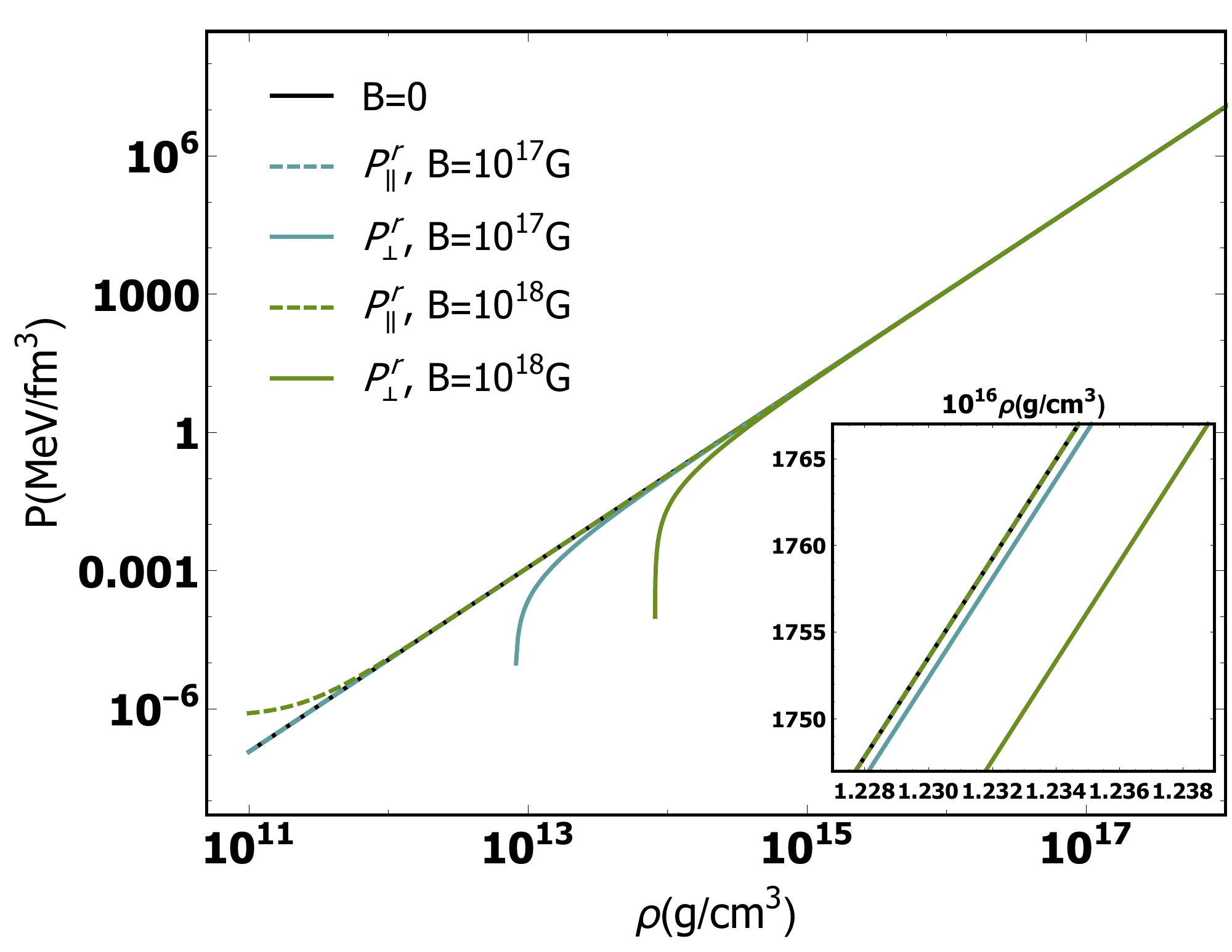}\\
	\includegraphics[width=0.42\linewidth]{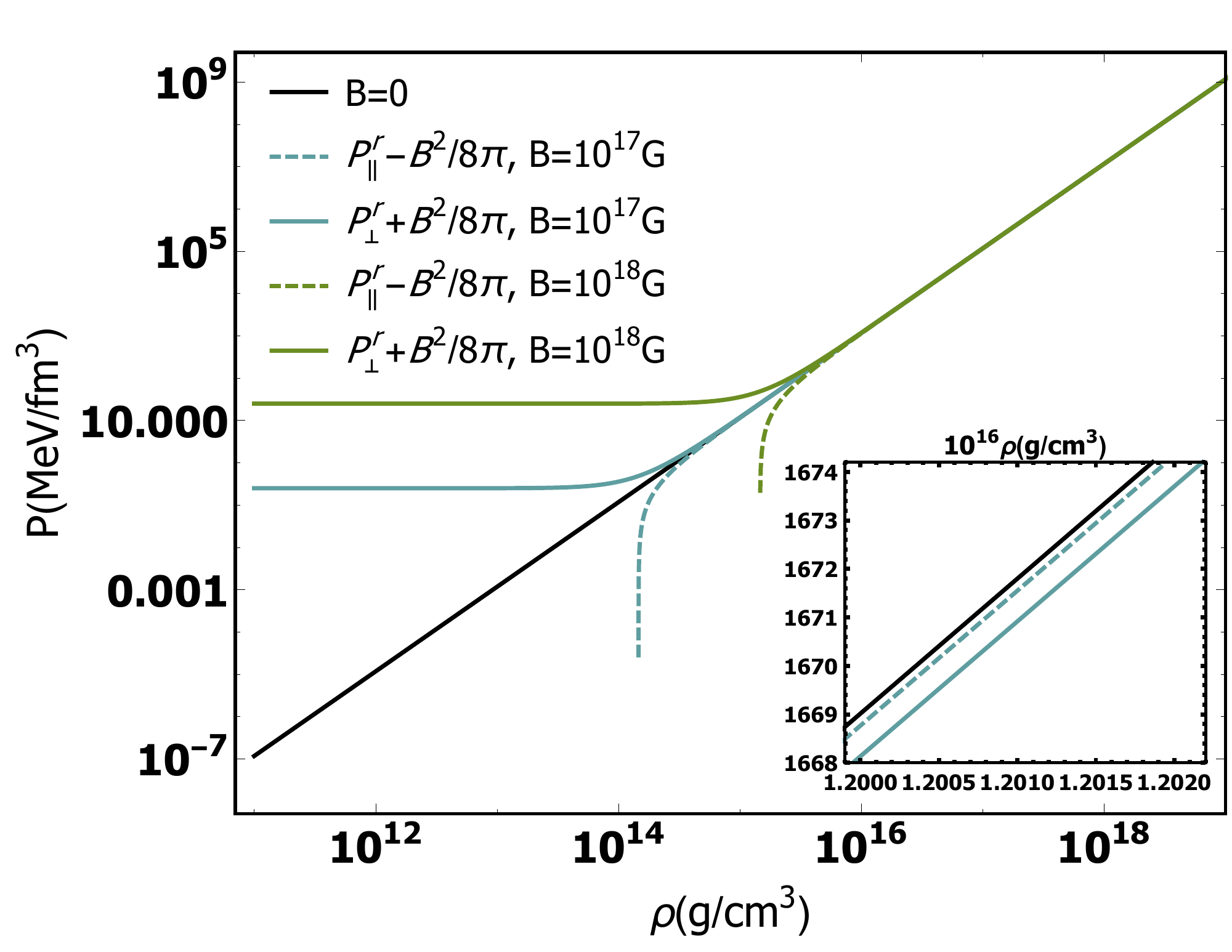}
	\includegraphics[width=0.42\linewidth]{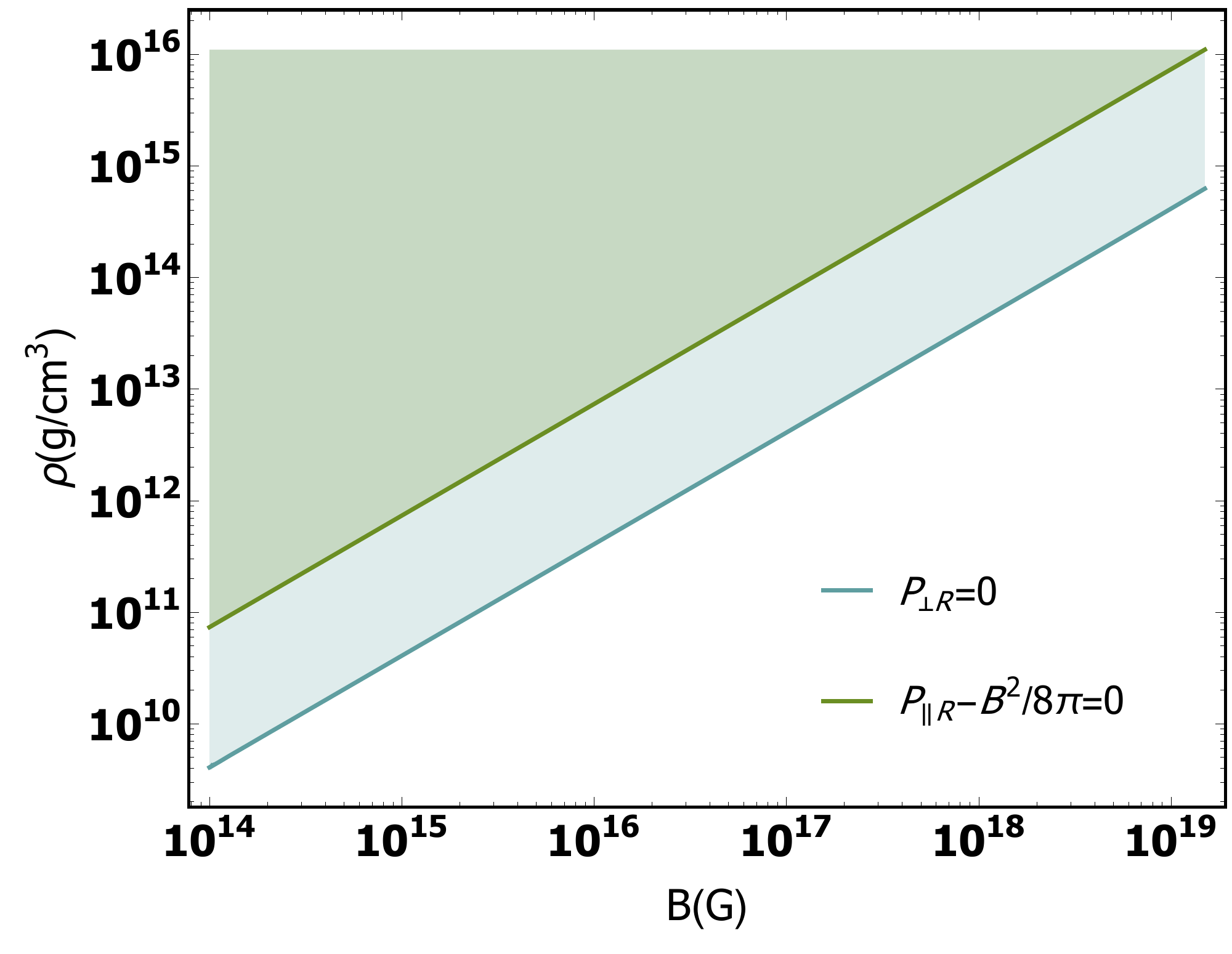}
	\caption{Las presiones como  función de la densidad de masa de los bosones para los casos no-relativista, relativista y relativista con la contribución de Maxwell (paneles superior izquierdo, superior derecho e inferior izquierdo respectivamente). Panel inferior derecho: diagrama de fases para el colapso magn\'etico en el plano $\rho$ vs $B$; en la regiones sombreadas el gas es estable.}\label{EoSplotBcte}
\end{figure}

El efecto de un campo magn\'etico constante en las presiones  de la materia que forma la Estrella de condensado de Bose-Einstein se muestra en la Fig.~\ref{EoSplotBcte} para $B=0$, $B=10^{17}$~G y $B=10^{18}$~G. Los paneles superiores corresponden a las EdE no relativistas y relativistas sin la contribuci\'on de Maxwell, mientras que el panel inferior izquierdo muestra a las EdE relativistas con los t\'erminos de Maxwell. Para las tres EdE graficadas, la regi\'on de altas densidades de masa est\'a dominada por las interacciones entre part\'iculas y las desviaciones con respecto al caso no magnetizado son peque\~nas. En cambio, a medida que la densidad de masa del gas disminuye, la influencia del campo magn\'etico se hace m\'as importante.

Para las EdE no relativistas (panel superior izquierdo de la  Fig.~\ref{EoSplotBcte}), el campo magn\'etico no modifica la presi\'on paralela y por tanto las curvas de $P_{\parallel}^{nr}$ se superponen a la presi\'on del caso no magnetizado. Por el contrario, en la regi\'on de m\'as bajas densidades de masa, la presi\'on perpendicular se separa de la curva de campo magn\'etico cero. La densidad de masa a la cual las curvas de  $P_{\perp}^{nr}$ terminan corresponde al punto en el que $P_{\perp} ^{nr}= 0$ para un campo magn\'etico dado. Para valores menores de $\rho$, $P_{\perp}^{nr} < 0$ y el gas se vuelve inestable. 

El panel superior derecho de la Fig.~\ref{EoSplotBcte} muestra que la presi\'on perpendicular relativista decrece hasta hacerse negativa con la misma dependencia con que lo hace la no relativista. Esto indica que, en lo que concierne a  $P_{\perp}$, la presi\'on magn\'etica  $-{\mathcal M}B$ predomina sobre la de vac\'io $-\Omega_{vac}$. Las similitudes entre $P_{\perp}^{nr}$ y $P_{\perp}^{r}$ indican adem\'as que ${\mathcal M}^{r} \cong{\mathcal M}^{nr}$ para los campos magn\'eticos aqu\'i considerados.

Por otra parte, si la densidad de masa es disminuida por debajo del valor en el cual $P_{\perp}^{r} = 0$, la presión paralela relativista, eventualmente se separa de la curva de $B=0$ y tiende a $-\Omega_{vac}(b)$ que es constante con respecto a la densidad (v\'ease la curva verde punteada). Esta es la \'unica diferencia entre las EdE realativistas y no realtivistas, pero como ella se da para densidades de masa en la regi\'on de presiones inestables, podemos concluir que los efectos relativistas debidos al campo magn\'etico no son relevantes en lo que respecta a las Estrellas magnetizadas de condensado de Bose-Einstein. La similitud entre ambos casos es de esperar porque el campo magn\'etico m\'aximo  que estamos considerando ($10^{18}$G) es dos \'ordenes menor que el campo magn\'etico cr\'itico del gas magnetizado de bosones ($ 10^{20}$)~G. En lo que sigue, utilizaremos solamente las EdE relativistas para los c\'alculos.

Cuando la contribuci\'on de Maxwell se incluye en las EdE (panel inferior izquierdo de la Fig.~\ref{EoSplotBcte}) el papel de las presiones se intercambia. Ahora es la presi\'on paralela la que se hace negativa, mientras que en la regi\'on de bajas densidades de masa la presi\'on perpendicular disminuye hasta que alcanza un valor constante e igual a la contribuci\'on de Maxwell para el valor dado de campo. Ello quiere decir que los t\'erminos de Maxwell dominan sobre las presiones magn\'etica $-M B$ y de vac\'io $-\Omega_{vac}$, cuyo efecto deviene irrelevante. En este caso la inestabilidad en la presi\'on aparece para un valor m\'as alto de la densidad de masa y la diferencia entre las presiones es mayor que en el caso sin contribuci\'on de Maxwell. N\'otese que dado un valor del campo magn\'etico, las densidades de masa para las cuales el gas es inestable est\'an prohibidas en el interior de la estrella, pues una presi\'on negativa no puede balancear la gravedad. En consecuencia, la inestabilidad en la presi\'on impone un l\'imite inferior a las densidades de masa que pueden darse en el interior de una estrella de condensado de Bose Einstein, siendo el valor espec\'ifico de este l\'imite una funci\'on creciente de $B$. Esto \'ultimo puede apreciarse en el panel inferior derecho de la Fig.~\ref{EoSplotBcte} que muestra el diagrama de fases para el colapso magn\'etico en el plano $\rho$ vs $B$ para las EdE relativistas sin y con la contribuci\'on de Maxwell; las diferencias entre las curvas cr\'iticas de ambos casos es de aproximadamente un orden de magnitud. Es de resaltar que para los valores m\'as altos del campo magn\'etico, las densidades de masa l\'imites est\'an en el orden de la densidad nuclear o son mayores que esta. 

\subsection{Ecuaciones de estado con campo magn\'etico autogenerado}
\label{sec4B}

En este ep\'igrafe consideraremos que el campo en el interior de la estrella viene dado por la automagnetizaci\'on de los bosones. Para calcular el campo autogenerado $B_{sg}$ utilizaremos la Ec.~(\ref{selfmag1}), $B_{sg} = 4 \pi {\mathcal M}$,  donde ahora la magnetizaci\'on ${\mathcal M}$ est\'a dada por las Ecs.(\ref{EoSNRtotal}) para el caso no relativista y por las Ecs.~(\ref{EoSRtotal}) para el relativista.

\begin{figure}[h!]
	\centering
	\includegraphics[width=0.42\linewidth]{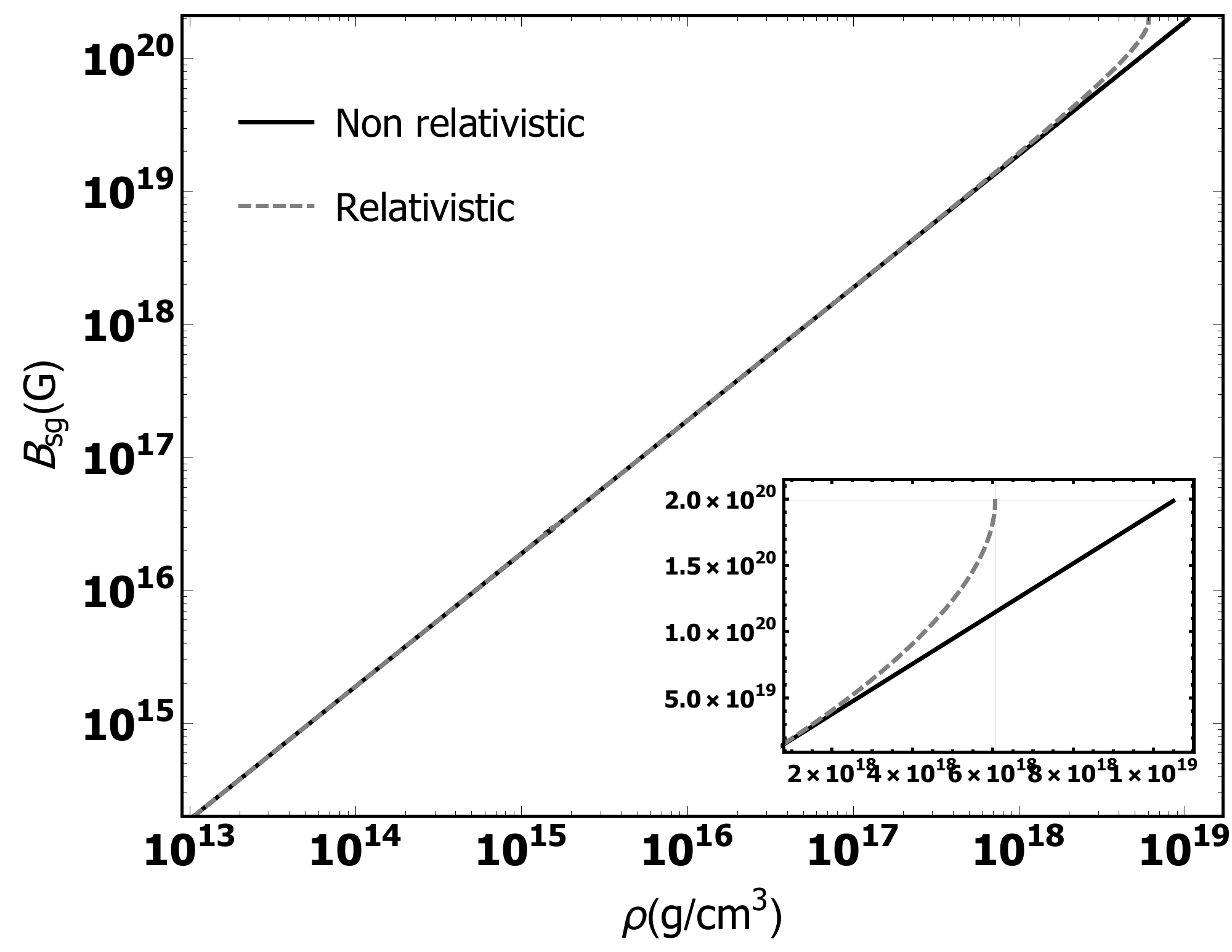}
	\includegraphics[width=0.42\linewidth]{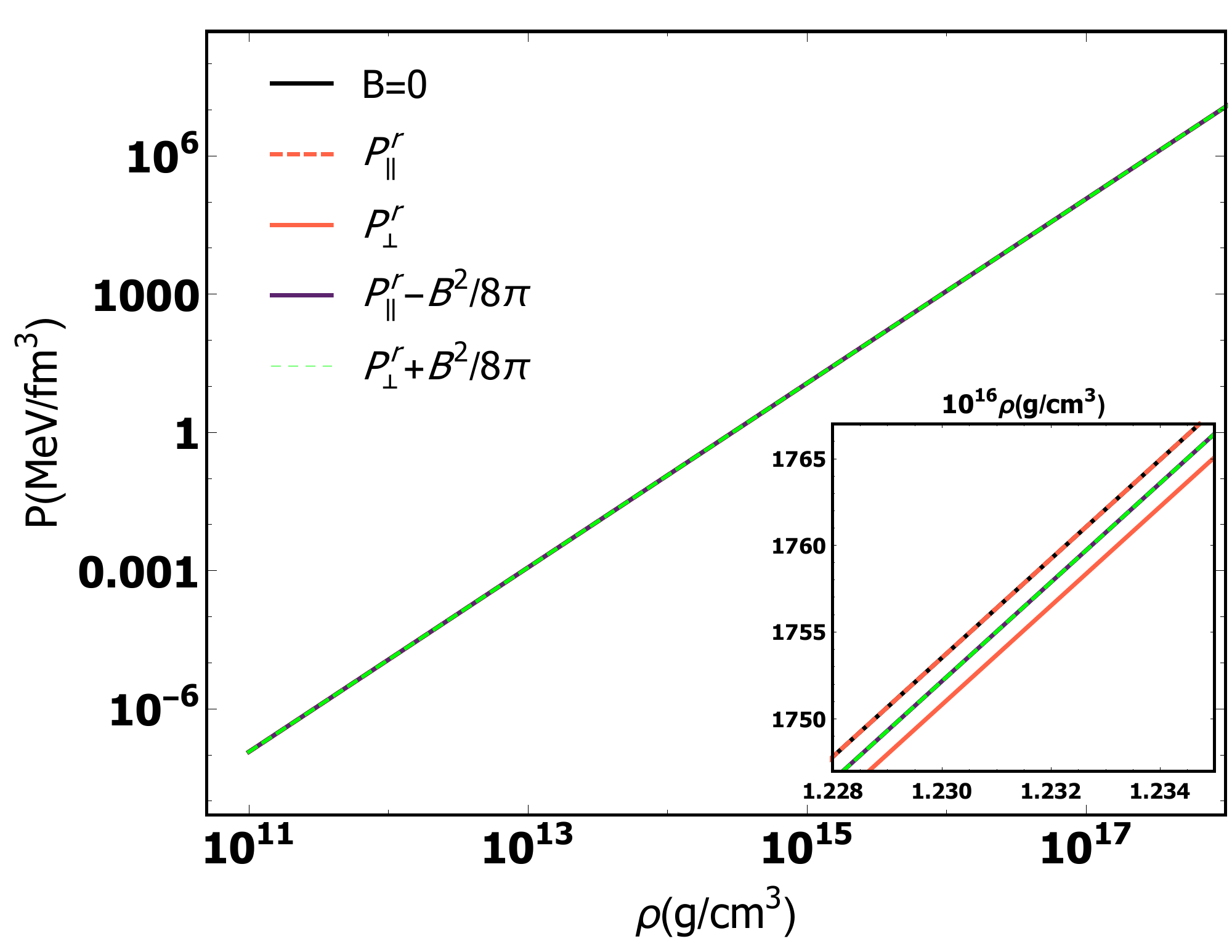}
	\caption{Panel izquierdo: el campo magnético autogenerado como función de la densidad de masa para los casos relativista y no relativista. Panel derecho: las presiones como función de la densidad de masa para la EdE relativista  con y sin contribuci\'on de Maxwell y campo magnético autogenerado.}\label{EoSplotBsg}
\end{figure}

Para la EdE no relativista, la ecuaci\'on para la automagentizaci\'on deviene $B_{sg} = 4 \pi \kappa N$ y $B_{sg}$ aumenta linealmente con la densidad de part\'iculas, como se muestra en el panel izquierdo de la Fig.~\ref{EoSplotBsg}. Evaluando para valores t\'ipicos de la densidad de masa de los bosones, por ejemplo $\rho =m N =10^{14}-10^{16}$g/cm$^{-3}$, puede verse que los valores del campo magn\'etico autogenerado para el caso no relativista, $B=10^{15}-10^{17}$G, est\'an en el orden de los esperados en las ENs.

En el caso relativista, como ya vimos en el Cap\'itulo \ref{cap2}, $B_{sg}$ a\'un aumenta con la densidad de masa (v\'ease la curva punteada en el panel izquierdo de Fig.~\ref{EoSplotBsg}) pero la dependencia ya no es lineal. Adem\'as, en este caso existe una densidad de partículas l\'imite por encima de la cual la condici\'on de automagentizaci\'on deja de cumplirse, y aparece un campo m\'agnetico m\'aximo que puede ser alcanzado por automagnetizaci\'on. Para los bosones aqu\'i considerados, la densidad de masa y el campo magn\'etico m\'aximo son $\rho = 3.61 \times 10^{18}$g/cm$^3$ and $B=2/3 B_c = 1.98 \times 10^{20}$G respectivamente.
Sin embargo, como los valores m\'aximos para el campo magn\'etico en el interior de una estrella de neutrones se estima que est\'an en el orden de los $10^{18}$G, en el caso de automagnetizaci\'on las diferencias entre los casos relativista y no relativista tampoco ser\'an relevantes.

En la pr\'actica, lo que se  hace para estudiar el comportamiento del gas bajo un campo magn\'etico autogenerado, es a\~nadir la Ec.~(\ref{selfmag1}) a las EdE. Las presiones con $B_{sg}$ fueron graficadas en el panel derecho de la Fig.~\ref{EoSplotBsg} como funci\'on de la densidad de masa de los bosones, con y sin contribuci\'on de Maxwell. Una caracter\'istica interesante que muestra este gr\'afico es que la inestabilidad en las presiones desaparece. Esto es consecuencia directa del decrecimiento de $B_{sg}$ con $\rho$. Se puede notar adem\'as en el recuadro, que la contribuci\'on de Maxwell borra la anisotrop\'ia, aunque s\'i mantiene una ligera diferencia con respecto a las EdE no magnetizadas.

\section{Conclusiones del cap\'itulo}

En este cap\'itulo se han obtenido las EdE para un gas magnetizado de bosones vectoriales neutros que interact\'uan entre s\'i y para Estrellas de condensado de Bose-Einstein magnetizadas y automagnetizadas. La descripci\'on termodin\'amica del gas vectorial que compone estas estrellas fue hecha considerando a los  bosones relativistas y no relativistas. En el estudio num\'erico se utilizaron dos configuraciones de campo magn\'etico. En una, el campo magn\'etico se considera constante y fijado externamente, mientras que en la otra es creado por la automagentizaci\'on de los bosones y depende de la densidad de masa del gas. 

Las EdE del gas y de la estrella fueron estudiadas num\'ericamente para valores de campos magn\'eticos y densidades de part\'iculas en los rangos esperados en el interior de las ENs. Del estudio num\'erico se obtuvieron las conclusiones siguientes:

\begin{itemize}
	
\item Aunque la descripci\'on termodin\'amica del gas fue hecha para bosones relativistas y no relativistas la diferencia entre estos reg\'imenes no es apreciable. Esto se debe a que los valores m\'as altos que el campo magn\'etico puede alcanzar en el interior de las ENs ($B\sim10^{18}$~G) est\'an dos \'ordenes por debajo del valor para el cual los efectos relativistas comenzar\'ian a ser importantes en el comportamiento del gas de bosones ($B\sim B_c\sim10^{20}$~G).

\item En el caso en que el campo magn\'etico es constante:

\begin{itemize}
	\item La anisotrop\'ia en las presiones es significativa y aumenta al inlcuir la contribuci\'on de Maxwell.
	
\item En la regi\'on de bajas densidades, la menor de las presiones se hace negativa y el gas deviene inestable. En consecuencia, tomar el campo magn\'etico constante impone un l\'imite inferior a las densidades de masa que pueden sostenerlo en el interior de la estrella.
	
\end{itemize}

\item En el caso en el que el campo magn\'etico es generado por la automagnetizaci\'on de los bosones:

\begin{itemize}
	
	\item La anisotrop\'ia en las presiones es muy peque\~na cuando la  contribuci\'on de Maxwell se ignora.
	
\item En el caso en que la contribuci\'on de Maxwell es tenida en cuenta, la anisotrop\'ia en las presiones desaparece, aunque las diferencias con respecto al caso no magnetizado se mantienen (Las presiones son iguales entre s\'i pero menores que cuando $B=0$).
	
\item Por otra parte, la inestabilidad en las presiones nunca aparece para los valores de campo magn\'etico y densidad de masa aqu\'i considerados. Esto es una consecuencia directa de que la intensidad del campo magn\'etico autogenerado disminuye con la densidad de part\'iculas.
	
	\end{itemize}

\end{itemize}

Es de resaltar que si bien el principal efecto del campo magn\'etico en las EdE es la separaci\'on de la presi\'on en dos componentes, la relevancia de esta separaci\'on depende de c\'omo se modele el campo magn\'etico. 

En cualquier caso, para obtener la estructura de objetos
compactos magnetizados es necesario contar con ecuaciones de equilibrio hidrost\'atico que tengan
en cuenta la anisotropía magnética. Por ello el próximo capítulo estará dedicado a la obtención de
ecuaciones de estructura axisimétricas que tomen en cuenta la anisotropía en las presiones.

%% file: cap4.tex
\chapter{Ecuaciones de estructura anisotr\'opicas}
\label{cap4}

En este cap\'itulo, se discute el problema de determinar la estructura (masa, forma y dimensiones) de objetos compactos anisotr\'opicos. En una primera parte se presenta la manera cl\'asica de atacar este problema en el caso de objetos compactos esf\'ericos a trav\'es de las ecuaciones de Tolman-Oppenheimer-Volkoff. Posteriormente, se describe el proceso para la obtenci\'on y uso de ecuaciones de estructura para objetos esferiodales, partiendo de una m\'etrica axisim\'etrica que es una generalizaci\'on de la m\'etrica de Schwarzschild. Aunque una version previa de las ecuaciones de estructura axisim\'etricas ya hab\'ian sido obtenidas con anterioridad a este trabajo, realizamos dos propuestas novedosas relacionadas con su resoluci\'on. La primera, consiste en calcular la masa total del objeto como la de un esferoide, y la segunda en un \textit{ansatz} que conecta directamente la deformaci\'on del objeto con la anisotrop\'ia en las presiones, que es su origen f\'isico. Este cap\'itulo contiene los resultados originales de la autora recogidos en \cite{Samantha}.

\section{Equilibrio hidrost\'atico estelar en el marco de la Teor\'ia de la Relatividad General}

Debido a las altas masas y densidades que se dan en los objetos compactos y en particular, en las Estrellas de Neutrones (de bosones en nuestro caso), las correcciones de la relatividad general a las ecuaciones de Newton son importantes para la descripci\'on del equilibrio hidrost\'atico de estos objetos \cite{Shapiro,Camezind,Weinberg}. Cuando en el marco de la Teor\'ia de la Relatividad General se habla de modelos de estrellas, a lo que se hace referencia es a una regi\'on interior que es una soluci\'on de las ecuaciones de Eisntein con fuente de materia ($T_{\mu\nu} \neq 0$), y una regi\'on exterior cuya m\'etrica es una soluci\'on asint\'oticamente plana de estas mismas ecuaciones en el vac\'io ($T_{\mu \nu} = 0$) \cite{Camezind}. Estas dos piezas deben ser cuidadosamente empatadas en la superficie de la estrella.

El problema de encontrar soluciones exactas de las ecuaciones de Einstein para las fuentes de materia especificadas por unas EdE no es trivial, incluso en el caso m\'as sencillo de materia en simetr\'ia esf\'erica \cite{Malafarina,MalafarinaHerrera}. Una manera de abordarlo consiste en proponer una m\'etrica interior a partir de ciertas consideraciones generales y luego obtener $T_{\mu \nu}$ a trav\'es de las ecuaciones de Einstein \cite{Malafarina,MalafarinaHerrera}. Sin embargo, las fuentes de materia as\'i obtenidas no tienen por qu\'e corresponderse con ecuaciones de estado conocidas. Por otra parte, esta manera de resolver el problema puede derivar en condiciones sobre las EdE que muchas veces resultan demasiado restrictivas o, incluso, no f\'isicas (v\'ease por ejemplo \cite{QuinteroBargueno}).

Los intentos de dar soluci\'on a este problema de manera inversa, es decir, de determinar la m\'etrica a partir de las EdE, han sido igualmente numerosos \cite{PhysRev55374,PhysRev55364,Malafarina,MalafarinaHerrera,EspositoWitten1975_95,100,Paret2014,Paret2015,ZubairiRomero_2015_89,Zubairi_2015_90,Zubairi2017_88,Zubairi_2017_91,Zubairi_2017_92,Grandclement,Chatterjee}. Si las EdE son isotr\'opicas esta estrategia conduce a las ecuaciones de Tolman-Oppenheimer-Volkoff que permiten la obtenci\'on de observables macrosc\'opicos (masas y radios) \cite{PhysRev55374,PhysRev55364,Camezind}.
En el caso de ecuaciones de estado anisotr\'opicas que conducir\'ian a objetos compactos no esf\'ericos, los estudios se centran m\'as en la obtenci\'on de las propiedades matem\'aticas de las soluciones de las ecuaciones de Einstein dada una simetr\'ia, que en la obtenci\'on de observables macrosc\'opicos derivados de una EdE espec\'ifica \cite{PhysRev55374,PhysRev55364,Malafarina,MalafarinaHerrera,EspositoWitten1975_95,100}.

Entre los objetos compactos no esf\'ericos, los axisim\'etricos llaman la atenci\'on debibo a que este tipo de simetr\'ia emerge de manera natural durante la modelaci\'on de objetos compactos magnetizados o en rotaci\'on.  Intentos por encontrar ecuaciones de estructura axisim\'etricas que puedan ser tratadas de manera (semi-)anal\'itica han sido realizados en \cite{Paret2014,Paret2015}, y en \cite{ZubairiRomero_2015_89,Zubairi_2015_90,Zubairi2017_88,Zubairi_2017_91,Zubairi_2017_92}, pero hasta el momento todos los resultados obtenidos son aproximados. La construcci\'on de m\'etodos totalmente n\'umericos que permitan calcular la estructura de objetos compactos axisim\'etricos tambi\'en ha sido una \'area de investigaci\'on intensa \cite{Bocquet,Konno,Grandclement,Chatterjee}. No obstante, en estos casos la posibilidad de encontrar soluciones estables est\'a limitada por la convergencia de los m\'etodos num\'ericos utilizados y sujeta a una serie de suposiciones y restricciones f\'isico-matem\'aticas que son necesarias para su implementaci\'on. De manera que a pesar de los avances en el campo de la relatividad num\'erica, la obtenci\'on de ecuaciones diferenciales axisim\'etricas para describir la estructura de estrellas deformadas por el campo magn\'etico contin\'ua siendo uno de los problemas m\'as interesantes y dif\'iciles de esta \'area de la f\'isica. 

\section{Objetos compactos esf\'ericos: Ecuaciones de Tolman-Oppenheimer-Volkoff}

Para obtener las ecuaciones de estructura se parte de las ecuaciones de Einstein:

\begin{equation}
G_{\mu\nu}=8\pi G T_{\mu\nu},
\label{TRG}
\end{equation}

\noindent donde $\mu,\nu = 0,1,2,3$, $T_{\mu\nu}$ es el tensor de energía momento de las fuentes de materia, $G =6.711\!\times\!10^{-45}$ MeV$^{-2}$ es la constante de gravitaci\'on universal  y $G_{\mu\nu}$ es el tensor de Einstein, que está relacionado con el tensor métrico $g_{\mu\nu}$ a través de:

\begin{equation}
G_{{\mu \nu}}=R_{{\mu \nu}}-\frac{1}{2}R\, g_{{\mu \nu}},
\end{equation}

\noindent donde

\begin{equation}
R_{\mu\nu}=\Gamma^{\alpha}_{\mu\nu,\alpha}-\Gamma^{\alpha}_{\mu\alpha,\nu}+\Gamma^{\alpha}_{\mu\nu}
\Gamma^{\beta}_{\alpha\beta}-\Gamma^{\beta}_{\mu\alpha}
\Gamma^{\alpha}_{\nu\beta}, \label{TF2}
\end{equation}

\noindent es el tensor de Ricci, $R=R^{\mu}_{\,\,\,\,\mu}$ es el escalar Ricci y

\begin{equation}\label{Christoffel}
\Gamma^{\alpha}_{\mu\nu}=\frac{g^{\alpha\beta}}{2}(g_{\beta\mu ,\nu}+g_{\nu\beta,\mu}-g_{\mu\nu,\beta})
\end{equation}

\noindent son los símbolos de Christoffel.

Como todos los tensores que aparecen en las Ecs.~(\ref{TRG}) son sim\'etricos, ellas  tienen diez componentes independientes que, dada la libertad de elecci\'on de las cuatro coordenadas del espacio-tiempo, se reducen a seis. De manera que las ecuaciones de Einstein son un sistema de seis ecuaciones diferenciales en derivadas parciales no lineales.
 
Las Ecs.~(\ref{TRG}) relacionan la curvatura del espacio-tiempo (a trav\'es de $G_{\mu\nu}$) con las fuentes de masa que la producen ($T_{\mu \nu}$). En rigor, al resolverlas el tensor de energ\'ia-momento deber\'ia ser calculado de manera autoconsistente en el espacio-tiempo curvo creado por la materia que \'el mismo describe. Sin embargo, lo usual cuando las ecuaciones de Einstein se resuelven para estrellas compactas es utilizar un tensor de energ\'ia-momento calculado a priori en el espacio-tiempo plano de Minkowski \cite{Weber:2004kj,1967hea3.conf..259T}. Esta aproximaci\'on puede hacerse porque a pesar de que la estructura y dinámica de los OC se rigen por la interacción entre las fuerzas nucleares, electromagnéticas y gravitacionales, la distancia característica sobre la cual cambia la fuerza gravitatoria ($10^7$cm para $\rho\cong 10^{15}$g/cm$^3$) es veinte órdenes de magnitud mayor que la escala microscópica en la que actúan las fuerzas nucleares y electromagnéticas ($10^{-13}$cm) \footnote{Para que el radio de acci\'on de las fuerzas gravitarotias fuera comparable con el de las fuerzas nucleares y electromagn\'eticas ser\'ian necesarias densidades de alrededor de $10^{49}$g/cm$^3$, m\'as de veinte \'ordenes por encima de las densidades m\'as extremas esperadas en el una EN \cite{1967hea3.conf..259T}.} \cite{Weber:2004kj,1967hea3.conf..259T}. En consecuencia, las propiedades termodin\'amicas de la materia y la radiaci\'on no se ven afectadas por la gravitaci\'on.

En el caso de un objeto compacto est\'atico cuya materia se comporta como un fluido perfecto ($T_{\mu\nu}=diag(-E,P,P,P)$), el alto nivel de simetr\'ia determina que todos los elementos no diagonales de la m\'etrica del espacio-tiempo que ellos generan sean nulos, y que los diagonales dependan \'unicamente de la distancia al origen de coordenadas. En consecuencia, la m\'etrica del espacio-tiempo generado por una estrella est\'atica y esf\'erica puede escribirse como \cite{Camezind}:

\begin{equation}\label{Sch0}
ds^2 = - e^{2 \Phi(r)} dt^2 + e^{2 \Lambda(r)} dr^2
+ r^{2} d\theta^2
+ r^2\sin^2\theta d\phi^2,
\end{equation}

\noindent donde $r,\theta,\phi$ son las coordenadas esf\'ericas usuales. Las funciones $\Phi(r)$ y $\Lambda(r)$ se determinan un\'ivocamente a partir de la distribuci\'on de energ\'ia en el interior de la estrella.

Con el uso de la Ec.~(\ref{Sch0}) los s\'imbolos de Christoffel $\Gamma^{\alpha}_{\mu\nu}$, el tensor y el escalar de Ricci, $R_{\mu\nu}$ y $R=R^{\mu}_{\,\,\,\,\mu}$ respectivamente, y el tensor de Einstein $G_{\mu \nu}$, pueden calcularse como funci\'on de $\Phi(r)$, $\Lambda(r)$ y $r$. Una vez que $G_{\mu\nu}$ es puesto en funci\'on de la m\'etrica, sus componentes se sustituyen en las Ecs.(\ref{TRG}), de conjunto con el tensor energ\'ia-momento de la materia en cuesti\'on, que en este caso es $T_{\mu\nu}=diag(-E,P,P,P)$. Luego de varias transformaciones algebraicas, las ecuaciones de Einstein se reducen a las cuatro ecuaciones diferenciales \cite{Camezind}:

\begin{subequations}\label{TOV}
	\begin{align}
\frac{dm(r)}{dr} &= 4 \pi r^2 E(r),\label{TOV1}\\	
\frac{dP(r)}{dr}&=-G\frac{(E(r)+P(r))\left( 4 \pi r^{3} P(r) + m(r)\right)}{ r^{2}\left(1-\frac{2Gm(r)}{r}\right)}, \label{TOV2}\\
e^{-2\Lambda(r)} &= \left(1-\frac{2Gm(r)}{r}\right), \label{TOV3}\\
\frac{d\Phi(r)}{dr} &=\frac{G}{\left(1-\frac{2Gm(r)}{r} \right)} \left(\frac{m(r)}{r^2} + 4 \pi r P(r)\right) ,\label{TOV$}
\end{align}
\end{subequations}

\noindent que describen la dependencia de la masa $m(r)$, la presi\'on $P(r)$, y las dos funciones m\'etricas $\Phi(r)$ y $\Lambda(r)$ con el radio interno de la estrella $r$. Las cuatro Ecs.~(\ref{TOV}) determinan completamente la estructura del objeto compacto y son conocidas como ecuaciones de Tolman-Oppenheimer-Volkoff (TOV), aunque por lo general, cuando se habla de las TOV se hace referencia solamente a las dos primeras. Esto se debe a que en la pr\'actica, para obtener la masa y el radio de la estrella basta con resolver estas dos ecuaciones.

Para resolver las Ecs.~(\ref{TOV1}) y (\ref{TOV2}) se parte de la condiciones en el centro de la estrella $m(r=0)=0$ y $P(r=0)=P_0$, donde $P_0$ se toma de la ecuaci\'on de estado. Posteriormente se integra en la variable $r$ hasta que la presi\'on se hace igual a cero. La condici\'on $P(R)=0$ define el radio de la estrella, que permite a su vez determinar su masa total $m=m(R)$. Al resolver las Ecs.~(\ref{TOV1}) y (\ref{TOV2}) en el rango de densidades admitido por las EdE, se obtiene una familia \'unica de estrellas, una curva masa-radio, parametrizada por la presi\'on y la densidad de energ\'ia centrales $(E_0,P_0)$. Cada punto $(M(E_0),R(E_0))$ de la familia obtenida respresenta  una estrella de masa $M$ y radio $R$ en equilibrio hidrodin\'amico. En una secuencia estelar, solamente las ramas en las que $dm/dE_0 > 0$ son estables ante perturbaciones radiales \cite{Camezind,Shapiro}.

Aunque las ecuaciones TOV han sido cruciales en el estudio y compresi\'on de la f\'isica de los objetos compactos, el suponer la estrella esf\'erica implica que las ecuaciones (\ref{TOV}) no pueden, en rigor, ser utilizadas para el estudio de objetos compactos magnetizados o en rotaci\'on, pues ellas no pueden tener en cuenta a la vez las presiones paralela y perpendicular. A fin de encontrar ecuaciones de estructura que puedan brindar una descripci\'on m\'as exacta de estrellas no esf\'ericas, en la pr\'oxima secci\'on se intentar\'a su b\'usqueda a partir de una m\'etrica axisim\'etrica.

\section{Ecuaciones de estructura para objetos compactos esferoidales: M\'etrica $\gamma$}

Esta secci\'on est\'a dedicada a la construcci\'on de un modelo general que permita estudiar la estructura de objetos compactos deformados axialmente. Nuestro modelo se basa en las Refs.~\cite{Herrera1999_93,Quevedo2017_99,Zubairi2017_88,ZubairiRomero_2015_89,Zubairi_2015_90,Zubairi_2017_91,Zubairi_2017_92}, donde los autores muestran que un objeto compacto no esf\'erico con simetr\'ia axial puede describirse a trav\'es de la llamada m\'etrica $\gamma$, cuya forma en coordenadas esf\'ericas  $(t,r,\theta,\phi)$ es:

\begin{equation}\label{gammatotal}
ds^2 = - \Delta^\gamma dt^2 + \Delta^{\gamma^2-\gamma-1} \Sigma^{1-\gamma^2} dr^2
+ r^{2} \Delta^{1-\gamma} \Sigma^{1-\gamma^2} d\theta^2
+ r^2\sin^2\theta \Delta^{\gamma^2-\gamma} d\phi^2,
\end{equation}

\noindent con:

\begin{subequations}\label{deltasigma}
	\begin{align}
	\Delta& = \left ( 1 - \frac{2 G m}{r}\right),\\
	\Sigma& = \left ( 1 - \frac{2 G m}{r} + \frac{G^2 m^2}{r^2} \sin^2 \theta \right)
	\end{align}
\end{subequations}

La m\'etrica $\gamma$, tambi\'en conocida como m\'etrica $q$ o m\'etrica de Zipoy-Voorhees, es una familia de soluciones est\'aticas, axisim\'etricas y asint\'oticamente planas de las ecuaciones de Einstein \cite{Herrera1999_93,Malafarina,MalafarinaHerrera,EspositoWitten1975_95,Quevedo2017_99}. Ella depende de dos par\'ametros, $m$ y $\gamma$, y constituye un caso particular de la m\'etrica de Weyl\footnote{La m\'etrica de Weyl es la soluci\'on est\'atica y axisim\'etrica m\'as general de las ecuaciones de Einstein \cite{Herrera1999_93,Malafarina,MalafarinaHerrera,Quevedo2017_99}.}. El sentido f\'isico de los par\'ametros $m$ y $\gamma$ pueden ser investigado a partir de los momentos monopolar $M$ y cuadrupolar $Q$ de la m\'etrica $\gamma$ \cite{Quevedo2017_99}:

\begin{equation}\label{monopolar}
M = \gamma m,
\end{equation}

\noindent que representa la masa gravitacional de la fuente de materia y

\begin{equation}\label{quadrupolar}
Q = \frac{1}{3} m^3 \gamma (1-\gamma^2),
\end{equation}

\noindent que est\'a relacionado con su forma. De las Ecs.~(\ref{monopolar}) y (\ref{quadrupolar}) se sigue que el par\'ametro $m$ est\'a relacionado con la masa total del objeto, mientras que el par\'ametro $\gamma$ tiene que ver con su forma \cite{Herrera1999_93,Malafarina,MalafarinaHerrera}.

Con respecto a la conexi\'on entre $\gamma$ y la forma del objeto \cite{Malafarina,MalafarinaHerrera}, n\'otese que para $\gamma = 0$, la Ec.(\ref{gammatotal}) se reduce al espacio-tiempo plano de Minkowski, con $M=Q=0$. Por otra parte, si $\gamma = 1$, la Ec.(\ref{gammatotal}) se reduce a la m\'etrica de Schwarzschild y la simetr\'ia esf\'erica se recupera. Esto significa que el objeto en cuesti\'on no est\'a deformado pues ahora nuevamente $Q=0$ como corresponde a una fuente de masa esf\'erica.

Desde el punto de vista matem\'atico, la m\'etrica $\gamma$ es la m\'as simple de todas las soluciones exactas conocidas para las ecuaciones de Einstein tales que dichas soluciones son est\'aticas, axisim\'etricas y pueden ser transformadas en la m\'etrica de Schwarzschild variando continuamente uno de sus par\'ametros (en este caso $\gamma$) \cite{Quevedo2017_99}. Por ello se piensa que ella es la mejor candidata para la b\'usqueda de condiciones de equilibrio hidrost\'atico que generalizen las ecuaciones TOV a objetos con momento cuadrupolar \cite{Herrera1999_93,Malafarina,MalafarinaHerrera,Quevedo2017_99}. No obstante, como veremos en los pr\'oximos ep\'igrafes, hasta el momento esto solo ha podido hacerse de manera aproximada.

\subsection{Ecuaciones de estructura anisotr\'opicas}

Como la m\'etrica $\gamma$ puede ser llevada de manera continua hacia la de Schwarzschild, un primer paso razonable hacia la construcci\'on de ecuaciones de estructura para objetos no esf\'ericos, es partir de la Ec.~(\ref{gammatotal}) en el l\'imite $\gamma \cong 1$, lo que es equivalente a considerar objetos poco deformados \cite{Malafarina,MalafarinaHerrera}. Para $\gamma \cong 1$, las Ecs.~(\ref{gammatotal})-(\ref{deltasigma}) se reducen a:

\begin{equation}\label{gammauno}
ds^2 = - \left[1-2G m(r)/r\right]^{\gamma}dt^2 +  \left[1-2G m(r)/r\right]^{-\gamma}dr^2
+ r^2\sin^2\theta d\phi^2 + r^{2}d\theta^2.
\end{equation}

Si siguiendo el procedimiento explicado en la secci\'on anterior, las ecuaciones de Eisntein se resuelven utilizando la m\'etrica Ec.~(\ref{gammauno}) y suponiendo el tensor de energ\'ia-momento isotr\'opico, se obtiene la siguiente ecuaci\'on de estructura para el objeto compacto deformado \cite{ZubairiRomero_2015_89,Zubairi_2015_90}:
\begin{equation}\label{gTOV0}
	\frac{dP}{dr}=- \frac{(E+P)\left[\frac{r}{2}+4\pi r^{3} G  P-\frac{r}{2}\left(1-\frac{2Gm}{r}\right)^{\gamma}\right]}{ r^{2}\left(1-\frac{2Gm}{r}\right)^{\gamma}}.
\end{equation}
En la Ec.~(\ref{gTOV0}), la deformaci\'on entra solo a trav\'es del par\'ametro $\gamma$, y aunque ella se reduce a la Ec.~(\ref{TOV}) cuando $\gamma = 1$, en realidad no aporta ninguna informaci\'on acerca de las dimensiones, m\'as all\'a del radio $R:P(R)=0$. Al obtenerse suponiendo el tensor de energ\'ia-momento isotr\'opico, la Ec.~(\ref{gTOV0}) no permite tampoco, en principio, el uso de presiones anisotr\'opicas.

A fin de obtener un conjunto de ecuaciones de estructura que permita tener en cuenta el car\'acter anisotr\'opico de las ecuaciones de estado, la Ec.~(\ref{gTOV0}) debe ser complementada con la informaci\'on que las EdE aportan sobre el objeto compacto. Siendo el empuje de la presi\'on hacia afuera lo que impide el colapso gravitatorio, que la presi\'on ejercida por el gas sea diferente en las direcciones polar (paralela) y ecualtorial (perpendicular) implica que las dimensiones de la estrella a lo largo de estas direcciones son distintas. La existencia de dos radios distintos indica que los c\'alculos para determinar la estructura del objeto deber\'ian llevarse a cabo, al menos, en dos dimensiones.

En \cite{ZubairiRomero_2015_89}, a fin de incluir las dos presiones en la descripci\'on de la estructura del objeto, se propone en primer lugar, que el objeto compacto es esferoidal. La suposici\'on de un objeto compacto esferoidal es la m\'as simple de las compatibles con la existencia de dos radios en direcciones ortogonales, y la \'unica admisible en nuestro caso si se tiene en cuenta que solo contamos con la informaci\'on relativa a estos dos radios\footnote{Modelos que resultan en objetos compactos axisim\'etricos no esferoidales pueden verse en \cite{Chatterjee}.}. En segundo lugar, los autores de \cite{ZubairiRomero_2015_89} proponen, como un primer paso en la descripci\'on de estos objetos esferoidales, parametrizar la coordenada que describe la distancia en el eje polar $z$ en t\'erminos de $\gamma$ y la coordenada ecuatorial:

\begin{equation}\label{gammazr}
z = \gamma r,
\end{equation}

\noindent aprovechando adem\'as el hecho de que, como se vio en el ep\'igrafe anterior, $\gamma$ est\'a relacionada con la deformaci\'on del objeto. Con ayuda de la Ec.~(\ref{gammazr}), la masa contenida en un esferoide de radio ecuatorial $r$ y radio polar $z=\gamma r$  puede ser calculada como \cite{ZubairiRomero_2015_89,Zubairi_2015_90}:

\begin{equation}\label{masa}
\frac{dm}{dr} = 4 \pi \gamma r^2 E,
\end{equation}

\noindent y la Ec.~(\ref{gTOV0}) puede reescribirse en funci\'on del radio polar $z$ \cite{Zubairi2017_88,Zubairi_2017_91,Zubairi_2017_92}:

\begin{equation}\label{gTOVz}
\frac{dP}{dz}=-\frac{(E+P)\left[\frac{z}{2 \gamma}+4 \pi G  \left(\frac{z}{\gamma}\right)^{3}P-\frac{z}{2 \gamma}\left(1-\frac{2 G m \gamma}{z}\right)^{\gamma}\right]}{ \left(\frac{z}{\gamma}\right)^{2}\left(1-\frac{2 G m\gamma}{z}\right)^{\gamma}}.
\end{equation}

Llegados a este punto, la anisotrop\'ia en las presiones puede introducirse en las ecuaciones de estructura \cite{Zubairi2017_88,Zubairi_2017_91,Zubairi_2017_92}. Para ello se supone que la presi\'on paralela está asociada \'unicamente con la coordenada $z$, mientras que la perpendicular lo estar\'ia solo con $r$. Haciendo esto, se llega al sistema de ecuaciones diferenciales:

\begin{subequations}\label{gTOVpartial}
	\begin{align}
\frac{dm}{dr} &= 4 \pi \gamma r^2 E,\label{gTOVpartial1}  \\
\frac{dP_{\parallel}}{dz}&=-\frac{(E+P_{\parallel})\left[\frac{z}{2 \gamma}+4\pi G \left(\frac{z}{\gamma}\right)^{3}P_{\parallel}-\frac{z}{2 \gamma}\left(1-\frac{2 G m \gamma}{z}\right)^{\gamma}\right]}{ \left(\frac{z}{\gamma}\right)^{2}\left(1-\frac{2 G m\gamma}{z}\right)^{\gamma}}, \label{gTOVpartial2}\\
	\frac{dP_{\perp}}{dr}&=-\frac{(E+P_{\perp})\left[\frac{r}{2}+4 \pi G r^{3}P_{\perp}-\frac{r}{2}\left(1-\frac{2 G m}{r}\right)^{\gamma}\right]}{ r^{2}\left(1-\frac{2 G m}{r}\right)^{\gamma}},\label{gTOVpartial3}	\end{align}
\end{subequations}

\noindent que describen la variaci\'on de la masa y las presiones con las coordenadas espaciales $r$ y $z$ para un objeto compacto esfeoridal. N\'otese que las Ecs.~(\ref{gTOVpartial2}) y (\ref{gTOVpartial3}) est\'an acolpladas entre s\'i a trav\'es de su dependencia con la densidad de energ\'ia $E$, y la masa $m(r)$.

Como las Ecs.~(\ref{gTOVpartial}) fueron obtenidas bajo la suposici\'on de que el objeto que se describe es un esferoide cuya deformaci\'on $\gamma = z/r \cong 1$, es de esperar que ellas funcionen bien siempre y cuando $P_{\parallel} \cong P_{\perp}$. Por otra parte, el sistema de ecuaciones Ecs.~(\ref{gTOVpartial}) difiere del utilizado en \cite{Zubairi2017_88,Zubairi_2017_91,Zubairi_2017_92}, en la forma de calcular la masa total (Ec.~(\ref{gTOVpartial1})) y en algunos detalles de su integraci\'on num\'erica que ser\'an abordados en la pr\'oxima secci\'on.

\subsection{Integraci\'on de las ecuaciones de estructura $\gamma$}

En las Ecs.~(\ref{gTOVpartial}), las coordenadas $z$ y $r$, est\'an ligadas a trav\'es del par\'ametro $\gamma$ que se ha supuesto constante. Una vez que dos de estas cantidades se fijen, la tercera queda determinada autom\'aticamente. Por otra parte, el valor espec\'ifico de $\gamma$ se desconoce.

Para resolver este problema, en \cite{Zubairi2017_88,Zubairi_2017_91,Zubairi_2017_92}, una vez obtenidas las ecuaciones de estructura se hace $\gamma = 1$, y las Ecs.~(\ref{gTOVpartial2}) y (\ref{gTOVpartial3}) se integran en $z,r$ hasta encontrar los radios polar ($Z$) y equatorial ($R$) del esferoide definidos a trav\'es de  las condiciones $P_{\parallel}(Z) = 0$ y $P_{\perp}(R)=0$ respectivamente. El objetivo de hacer $\gamma=1$ es permitir que la anisotrop\'ia del esferoide resultante est\'e dictada \'unicamente por la anisotrop\'ia en la EdE. No obstante, hacer esto incumple la suposici\'on de que $z=\gamma r$ y cancela la dependencia de la masa de la estrella con el radio polar (ver, por ejemplo,\cite{Zubairi2017_88}).

Por otra parte, si $\gamma=1$ las ecuaciones de estructura (\ref{gTOVpartial}) devienen iguales a las TOV y los efectos gravitacionales de tener un objeto deformado desaparecen. Por ello, al enfrentarnos a la resoluci\'on de las ecuaciones de estructura para objetos magnetizados en el presente trabajo decidimos tomar otra estrategia.

Comenzando desde el centro de la estrella con densidad de energ\'ia $E_0 = E(r=0)$, y presiones centrales $P_{\parallel_0} = P_\parallel(r=0)$ y $P_{\perp_0} = P_\perp(r=0)$ tomadas de las ecuaciones de estado, las ecuaciones (\ref{gTOVpartial}) se hacen evolucionar hasta que una de las condiciones $P_\parallel (Z) = 0$ o $P_\perp (R) = 0$ se alcanza. Esto determina el radio correspondiente, ($R$ si $P_{\perp} = 0$ y  $Z$ si $P_{\parallel} = 0 $), a partir del cual el otro radio puede ser calculado a trav\'es de $Z = \gamma R$. Posteriomente, se eval\'ua la masa como $m=m(R)$. Esta manera de proceder permite conservar los efectos de la anisotrop\'ia en la m\'etrica, aunque nos plantea el problema de darle un valor a $\gamma$, que abordaremos en la secci\'on pr\'oxima, pues existen a\'un dos detalles de la integraci\'on de las Ecs.~(\ref{gTOVpartial}) que deben ser aclarados.

El primero es realtivo al significado f\'isico de la condici\'on de contorno utilizada. Que el radio de la estrella venga definido por las condiciones $P_\parallel (Z) = 0$ o $P_\perp (R) = 0$ significa que estamos considerando una EBE desnuda, sin corteza, ni atm\'osfera, ni magnetosfera. Estas, por supuesto, no son las \'unicas condiciones de contorno posibles, y en general su selecci\'on depende del n\'umero de capas que se consideren al construir el modelo de la estrella. La selecci\'on de estas condiciones de frontera espec\'ificas responde a que hasta el momento nuestro inter\'es ha estado dirigido a determinar las propiedades de las Estrellas de condensado de Bose-Einstein magnetizadas desnudas o puras.

El segundo detalle radica en la manera en que se calcula la densidad de energ\'ia $E$ a partir de las EdE durante la integraci\'on de las Ecs.~(\ref{gTOVpartial}). Para aclarar este punto, denotemos por $c_1(N)$ y $c_2(N)$ a las curvas param\'etricas bidimensionales dadas por:
\begin{subequations}
	\begin{eqnarray}
	c_1(N)&=&(E(N),P_{\parallel}(N)) \label{c1}\\
	c_2(N)&=&(E(N),P_{\perp}(N)) \label{c2}
	\end{eqnarray}
\end{subequations}
con $E(N),P_{\parallel}(N)$ y $P_{\perp}(N)$ definidas por las Ecs.~(\ref{EoSRtotal}) con o sin t\'ermino de Maxwell. Dados $\widetilde{P}_{\parallel}$ y $\widetilde{P}_{\perp}$, obtenidos en un paso de la integraci\'on de las Ecs.~(\ref{gTOVpartial}), dos valores param\'etricos, $\widetilde{N}_{\parallel}$ y $\widetilde{N}_{\perp}$ son calculados a partir de interpolar las Ecs.~(\ref{EoSRtotal1}) y (\ref{EoSRtotal2}) respectivamente. Los puntos correspondientes en las curvas (\ref{c1}) y (\ref{c2}) son $c_1(\widetilde{N}_{\parallel})=(\widetilde{E}_{\parallel},\widetilde{P}_{\parallel})$ y $c_2(\widetilde{N}_{\perp})=(\widetilde{E}_{\perp},\widetilde{P}_{\perp})$, donde $\widetilde{E}_{\parallel}=E(\widetilde{N}_{\parallel})$ y $\widetilde{E}_{\perp}=E(\widetilde{N}_{\perp})$. Por tanto, en el pr\'oximo paso de la integraci\'on, el miembro derecho de la Ec.~(\ref{gTOVpartial2}) se actualiza utilizando el punto $c_1(\widetilde{N}_{\parallel})$ con $E=\widetilde{E}_{\parallel}$ y $P_{\parallel}=\widetilde{P}_{\parallel}$, mientras que el miembro derecho de Ec.~(\ref{gTOVpartial3}) se actualiza con $c_2(\widetilde{N}_{\perp})$ a partir de tomar  $E=\widetilde{E}_{\perp}$ y $P_{\perp}=\widetilde{P}_{\perp}$.

La existencia de dos valores de la densidad de energ\'ia en cada paso de integraci\'on, introduce la interrogante de cu\'al seleccionar a la hora de calcular la masa del objeto compacto Ec.~(\ref{gTOVpartial1}). Como estamos trabajando con un objeto anisotr\'opico, la variación de su densidad de masa tambi\'en debe ser diferente en las direcciones paralela y perpendicular al eje magnético. A lo largo de la direcci\'on ecuatorial, la densidad de masa es igual a:
\begin{eqnarray}\label{massdiff1}
dm &=& 4 \pi \gamma r^2 E_{\parallel} dr,
\end{eqnarray}
mientras que en la direcci\'on polar ella es:
\begin{eqnarray}\label{massdiff2}
dm &=& 4 \pi \frac{z^2}{\gamma^2} E_{\perp} dz.
\end{eqnarray}
En las Ecs.~(\ref{massdiff1}) y (\ref{massdiff2}) se han usado las densidades de energ\'ia paralela y perpendicular en dependencia de la direcci\'on en que se realiza la diferenciaci\'on. Si ahora tomamos en cuenta que $z = \gamma r$, la Ec.~(\ref{massdiff2}) puede transformarse en:
\begin{eqnarray}\label{massdiff3}
dM &=& 4 \pi \gamma r^2 E_{\perp} dr.
\end{eqnarray}

Sumando las Ecs.~(\ref{massdiff1}) y (\ref{massdiff3}), se obtiene:
\begin{equation}\label{massdifffinal}
\frac{dM}{dr} = 4 \pi \gamma r^2 \frac{E_{\parallel} + E_{\perp}}{2}.
\end{equation}

La Ec.~(\ref{massdifffinal}) indica que, si no se quiere perder informaci\'on acerca de la anisotrop\'ia en la densidad de masa, el lado derecho de la Ec.~(\ref{gTOVpartial1}) debe ser actualizado con la densidad de energ\'ia media $E=(\widetilde{E}_{\parallel}+\widetilde{E}_{\perp})/2$.

Finalmente, teniendo en cuenta lo que se acaba de discutir, el sistema de ecuaciones de estructura Ecs.~(\ref{gTOVpartial}) puede ser reescrito como:
\begin{subequations}\label{gTOV}
	\begin{eqnarray}
	&& \frac{dm}{dr}=4 \pi r^{2}\frac{(E_{\parallel} +E_{\perp})}{2}\gamma, \label{gTOV1}\\
	&&\frac{dP_{\parallel}}{dz}=\frac{1}{\gamma}\frac{dP_{\parallel}}{dr}=
	-\frac{(E_{\parallel}+P_{\parallel})[\frac{r}{2}+4 \pi  r^{3} G P_{\parallel}-\frac{r}{2}(1-\frac{2 G m}{r})^{\gamma}]}{\gamma r^{2}(1-\frac{2 G m}{r})^{\gamma}}, \label{gTOV3}\\
	&&\frac{dP_{\perp}}{dr}=-\frac{(E_{\perp}+P_{\perp})[\frac{r}{2}+ 4 \pi  r^{3} G P_{\perp}-\frac{r}{2}(1-\frac{2 G m}{r})^{\gamma}]}{ r^{2}(1-\frac{2 G m}{r})^{\gamma}}. \label{gTOV2}
	\end{eqnarray}
\end{subequations}

Las Ecs.~(\ref{gTOV}) contin\'uan acopladas entre s\'i a trav\'es de la masa y su dependencia en las densidades de energ\'ia paralela y perpendicular. N\'otese adem\'as que en el caso en que $\gamma=1$ las ecuaciones TOV son recuperadas.

\subsection{El \textit{ansatz} $\gamma=P_{\parallel_0}/P_{\perp_0}$}


Para estudiar la estructura de estrellas anisotr\'opicas con la ayuda de las Ecs.~(\ref{gTOV}) es necesario fijar de alguna manera el valor del par\'ametro $\gamma$. Desde el punto de vista matem\'atico y geom\'etrico, el par\'ametro $\gamma$ determina la deformaci\'on del objeto compacto, por tanto \'el debe estar relacionado de alguna manera con la anisotrop\'ia en las EdE, que es la causa f\'isica de la deformaci\'on. Como adem\'as $\gamma = z/r$, el problema ahora consiste en establecer una relaci\'on entre los radios polar y ecuatorial con las presiones en las direcciones paralela y perpendicular.

\begin{figure}[h!]
	\centering
	\includegraphics[width=0.49\linewidth]{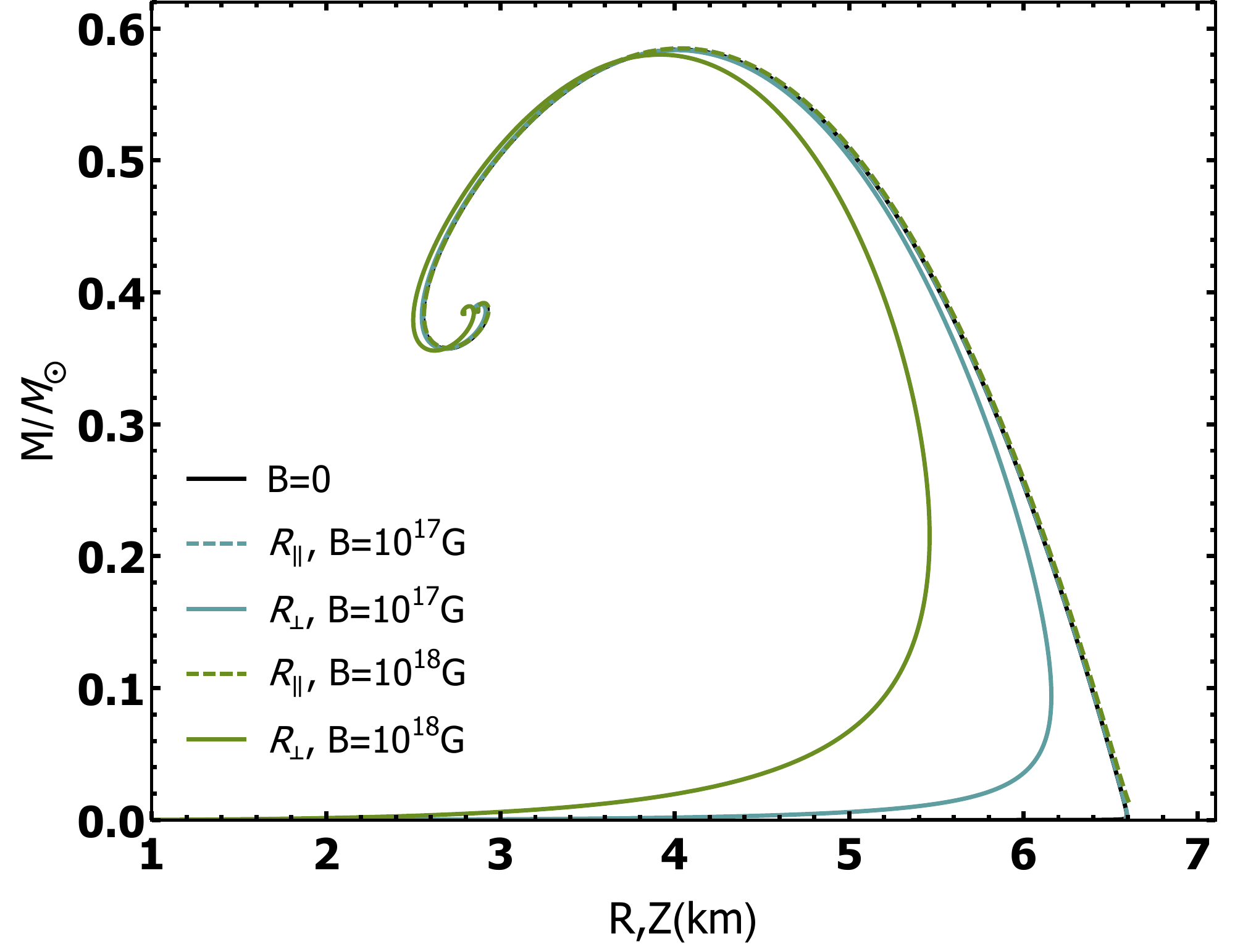}
	\includegraphics[width=0.49\linewidth]{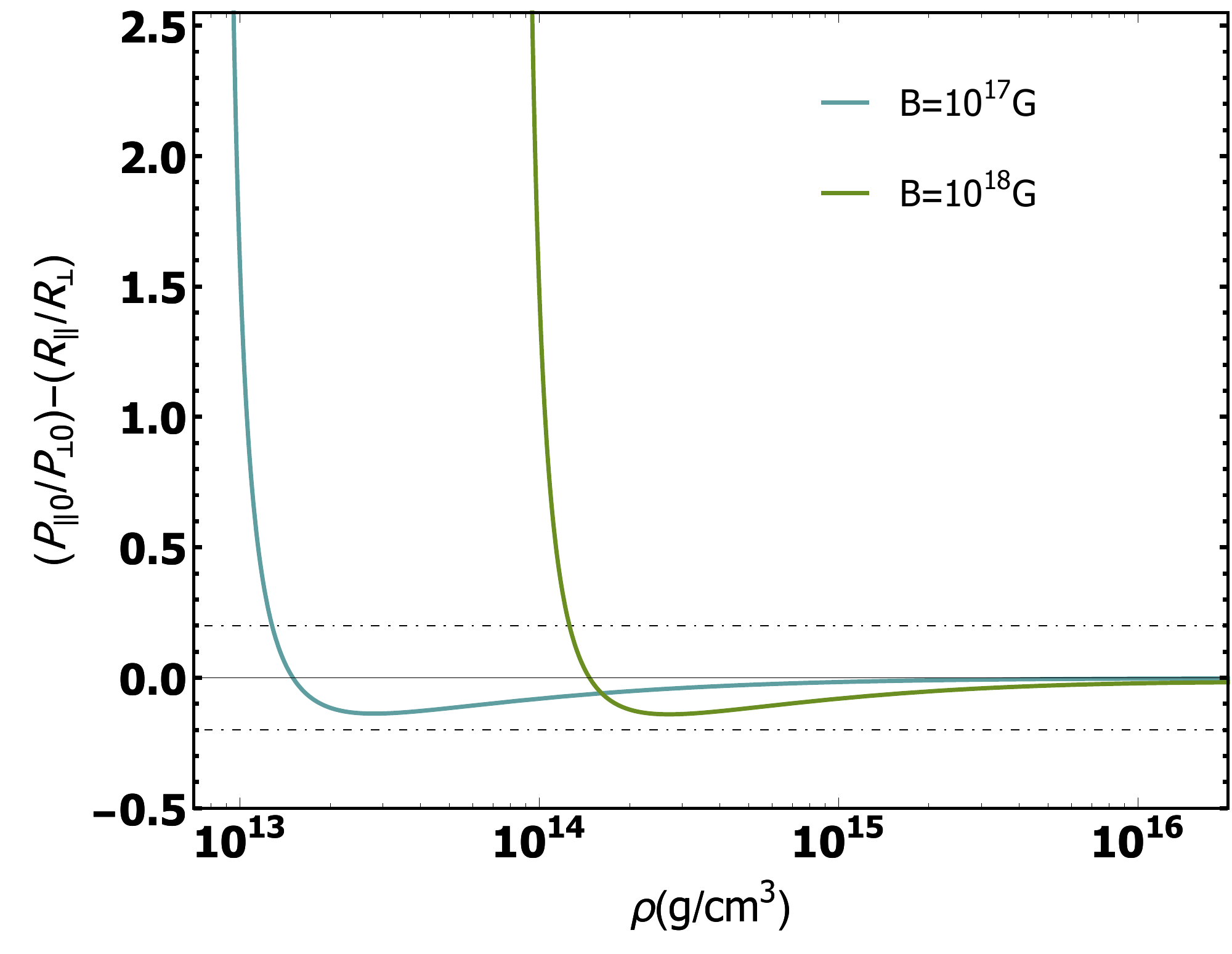}
	\caption{Izquierda: Relaci\'on masa-radio que resulta de resolver de manera independiente para cada una de las presiones las TOV para las EdE de una Estrella de condensado de Bose-Einstein con $B$ constante sin la contribuci\'on de Maxwell. Derecha: La diferencia entre las razones como funci\'on de la densidad de masa de los bosones.}\label{MRGTOV}
\end{figure}

Para ello se apelar\'a a los resultados obtenidos previamente durante el trabajo con objetos compactos con el mismo tipo de anisotrop\'ia magn\'etica \cite{Paret2015,Paret2014}. En dichos trabajos, se utilizan dos aproximaciones distintas para hallar los radios polar y ecuatorial de Enanas Blancas y Estrellas de Quakrs magnetizadas. La primera consiste en utilizar las TOV de manera independiente para cada una de las presiones. La segunda, en el uso de ecuaciones de estructura anisotr\'opicas en simetría cil\'indrica. M\'as all\'a de las diferencias num\'ericas, determinadas tanto por el car\'acter aproximado de ambos procedimientos como por las diferencias entre las ecuaciones de estructura utilizadas, los resultados de ambos m\'etodos coinciden en que, dada una densidad de energ\'ia central $E_0$, el mayor radio corresponde a la mayor presi\'on central. Un comportamiento similar se aprecia si las TOV son resueltas de manera independiente para cada una de las presiones involucradas en las EdE de las Estrellas de condensado de Bose-Einstein magnetizadas, Ecs.~(\ref{EoSRtotal}), como muestra el panel izquierdo de la Fig.~\ref{MRGTOV}. En este caso $B$ se supone constante y el t\'ermino de Maxwell no ha sido tenido en cuenta, por tanto $P_{\parallel 0} > P_{\perp 0} $, y en consecuencia $R_{\parallel} > R_{\perp} $.

Si tenemos en cuenta adem\'as que en el caso isotr\'opico dada una EdE, las masas y radios est\'an un\'ivocamente determinados por los valores de densidad de energ\'ia y presi\'on en el centro de la estrella, es posible establecer la existencia de cierta proporcionalidad entre los radios y las presiones centrales correspondientes. La existencia de una dependencia entre el radio de una estrella y su presi\'on central es un hecho bastante bien establecido en la literatura, aunque su forma espec\'ifica, en general, se desconoce \cite{Lattimer:2000nx}. Esta parece ser muy dependiente del modelo, pues incluso en los casos sencillos que por tener soluci\'on anal\'itica permiten relacionar a $R$ y $P_c$ de forma expl\'icita, dicha relaci\'on cambia notablemente de un modelo a otro.  Por ejemplo, si se toma el l\'imite de compacidad para las ecuaciones de estructura newtonianas en simetr\'ia esf\'erica se obtiene que $R \varpropto \sqrt{P_c} $, mientras que este mismo l\'imite para las TOV arroja $R \varpropto 1/\sqrt{P_c}$, algo que no solo difiere del caso newtoniano, sino que est\'a en contradicci\'on con los resultados mostrados en la Fig.~\ref{MRGTOV}. 

Un primer paso al enfrentar este problema es suponer que la dependencia de los radios de la estrella con las presiones centrales es lineal, de manera que entonces $Z  \varpropto P_{\parallel0}$ y
$ R_{\perp}  \varpropto P_{\perp0}$. Teniendo en cuenta estas relaciones de proporcionalidad de conjunto con la Ec.~(\ref{gammazr}) se llega a:

\begin{equation}
\gamma=\frac{P_{\parallel_0}}{P_{\perp_0}}. \label{gammaa}
\end{equation}

La Ec.~(\ref{gammaa}) conecta la geometr\'ia del OC deformado con la anisotrop\'ia en las presiones del gas magnetizado que lo compone. Ella implica adem\'as que la deformaci\'on del objeto compacto descrito por las Ecs.~(\ref{gTOV})-(\ref{gammaa}) est\'a determinada \'unicamente por la anisitrop\'ia en su centro. No obstante, es bueno aclarar, que los radios no solo dependen de la presi\'on central sino de c\'omo esta evoluciona en el interior de la estrella, y por tanto, de la forma de la EdE. Igualmente es bueno no perder de vista que la Ec.~\ref{gammaa} es un \textit{ansatz} ya que de hecho, la magnitud de la deformaci\'on de la estrella debida al campo magn\'etico deber\'ia poder obtenerse como resultado de integrar las ecuaciones de estructura y no tener que introducirse a priori en ellas. Si esto ha de hacerse as\'i es porque al deducir las Ecs.~(\ref{gTOV}), $\gamma$ se supuso constante. En este sentido, el pr\'oximo paso en la resoluci\'on de este problema ser\'ia intentar la obtenci\'on de nuevas ecuaciones de estructura a partir de la m\'etrica Ec.~(\ref{gammauno}) suponiendo que $\gamma$ depende de $r$.

A fin de tener una idea de cu\'an apropiado es considerar la dependencia lineal de los radios con las presiones centrales, en el panel izquierdo de la Fig.~\ref{MRGTOV} se ha graficado la diferencia $(P_{\parallel0}/ P_{\perp0}) - (R_{\parallel}/ R_{\perp})$ que se obtiene de resolver TOV de manera independiente para las dos presiones dadas por las EdE de las EBE magnetizadas (panel izquierdo de la Fig.~\ref{MRGTOV}). En la regi\'on de altas densidades, en la cual la anistrop\'ia es despreciable, vemos que la diferencia entre ambas razones es tambi\'en despreciable. A medida que la densidad se hace menor, dicha diferencia comienza a aumentar de forma no lineal, pero a\'un se mantiene cercana a cero hasta que las densidades se acercan al valor en el que ocurre el colapso magn\'etico. Es de esperar entonces que sea en las cercan\'ias de las regiones inestables en donde nuestro \textit{ansatz} deje de funcionar bien.

Finalmente, es importante notar que en el caso $B=0$, como $P_\perp = P_\parallel$ el \textit{ansatz} que hemos propuesto sobre el valor de $\gamma$ da autom\'aticamente $\gamma = 1$ y el caso isotr\'opico es recuperado.

La aproximaci\'on propuesta para $\gamma$ da resultados razonables para peque\~nas desviaciones de la configuraci\'on esf\'erica, es decir para $\gamma \cong 1$, como sucede para las densidades de energ\'ia y campos magn\'eticos en las Enanas Blancas \cite{Samantha}. No obstante, una mejora sobre nuestras ecuaciones de estado y sobre el \textit{ansatz} se impone si queremos describir objetos altamente deformados.

\section{Conclusiones del cap\'itulo}

En este cap\'itulo, a partir de la m\'etrica $\gamma$ y de las EdE anisotr\'opicas de un gas magnetizado, se ha obtenido un sistema de ecuaciones de estructura para objetos compactos esferoidales, las Ecs.~(\ref{gTOV}). Dichas ecuaciones describen la variaci\'on de la masa y las presiones internas de la estrella con los radios polar y ecuatorial $z,r$. Ellas son v\'alidas siempre que el OC descrito no se aleje mucho de la forma esf\'erica y sus soluciones dependen del valor del par\'ametro $\gamma = z/r$, que est\'a relacionado con la deformaci\'on del objeto.

Teniendo en cuenta el hecho de que para las TOV, dada una misma densidad de energ\'ia central, mientras menor es la presi\'on, menor es el radio del OC, hemos propuesto interpretar $\gamma$ como la raz\'on entre las presiones centrales perpendicular y paralela de la estrella. Para $B=0$ nuestro modelo da autom\'aticamente $\gamma = 1$ y las Ecs.~(\ref{gTOV}) se reducen a las ecuaciones TOV. Si $B \neq 0$, en cambio, el \textit{ansatz} sobre $\gamma$ permite conectar, al menos en una primera aproximaci\'on, la deformaci\'on del objeto compacto con la f\'isica que la produce.

La combinaci\'on de las Ecs.~(\ref{gTOV}) con el \textit{ansatz} Ec.~(\ref{gammaa}) permite  determinar la masa total y los radios de cualquier objeto compacto magnetizado. Estas ecuaciones tienen adem\'as la ventaja de proporcionar informaci\'on f\'isica relevante sobre el OC a un bajo costo computacional. En el pr\'oximo cap\'itulo, las Ecs.~(\ref{gTOV}) ser\'an utilizadas para estudiar la estructura de Estrellas de condensado de Bose-Einstein magnetizadas.

%% file: cap5.tex
\chapter{Relaci\'on masa-radio y perfiles de campo magn\'etico para Estrellas de condensado de Bose-Einstein magnetizadas y automagnetizadas}
\markright{Capítulo 3: Efectos del MMA en las EdE del gas magnetizado de Fermi.}
\label{cap5}

En este cap\'itulo se estudian los efectos del campo magn\'etico en la estructura, masa, radio y deformaci\'on, de Estrellas de condensado de Bose-Einstein. Para ello se combinan las ecuaciones de estado obtenidas en el Cap\'itulo \ref{cap3} con las ecuaciones de estructura derivadas en el Cap\'itulo \ref{cap4}. Como en el Cap\'itulo \ref{cap3}, el estudio se har\'a suponiendo en un caso que el campo magn\'etico es constante en el interior de la estrella y  en otro que es generado por los bosones a trav\'es de la automagnetizaci\'on, esto \'ultimo con el objetivo contribuir a una modelaci\'on m\'as realista de los objetos compactos. A fin de evaluar si la automagnetizaci\'on es un mecanismo v\'alido para la producci\'on del campo magn\'etico estelar, sus perfiles en el interior de la estrella ser\'an calculados y comparados con los estimados observacionales y te\'oricos. Este cap\'itulo contiene los resultados originales de la autora recogidos en \cite{Quintero2018BECS}.

\section{Relaci\'on masa-radio de Estrellas de condensado de Bose Einstein}

El objetivo \'ultimo de la modelaci\'on de cualquier objeto compacto es la obtenci\'on de sus observables, pues ellos no solo permiten validar el modelo en cuesti\'on a trav\'es de compararlos con datos observacionales, sino que son cruciales para la interpretaci\'on de estos \'ultimos \cite{Chatterjee}. Los observables derivados de un modelo espec\'ifico dependen de las aproximaciones hechas durante la obtenci\'on de las EdE y, muy especialmente, de las ecuaciones de estructura \cite{Weberrotacion}.

Al utilizar las ecuaciones de estructura presentadas en el Cap\'itulo \ref{cap4}, se est\'a suponiendo que las Estrellas de condensado de Bose-Einstein son objetos est\'aticos esferiodales. Esto, en principio, permite determinar la masa, los radios polar y ecuatorial, el momento de inercia, el momento cuadrupolar y el corrimiento al rojo gravitacional \cite{Weberrotacion}. En la tesis nos concentraremos en el estudio de los efectos del campo magn\'etico en la masa y los radios de la EBE. pues estos son los observables m\'as importantes y los dem\'as pueden ser calculados a partir de ellos. En el caso de campo magn\'etico autogenerado, nuestro modelo aporta adem\'as un tercer observable: el campo magn\'etico en la superficie del OC.

\begin{figure}[h!]
	\centering
	\includegraphics[width=1.0\linewidth]{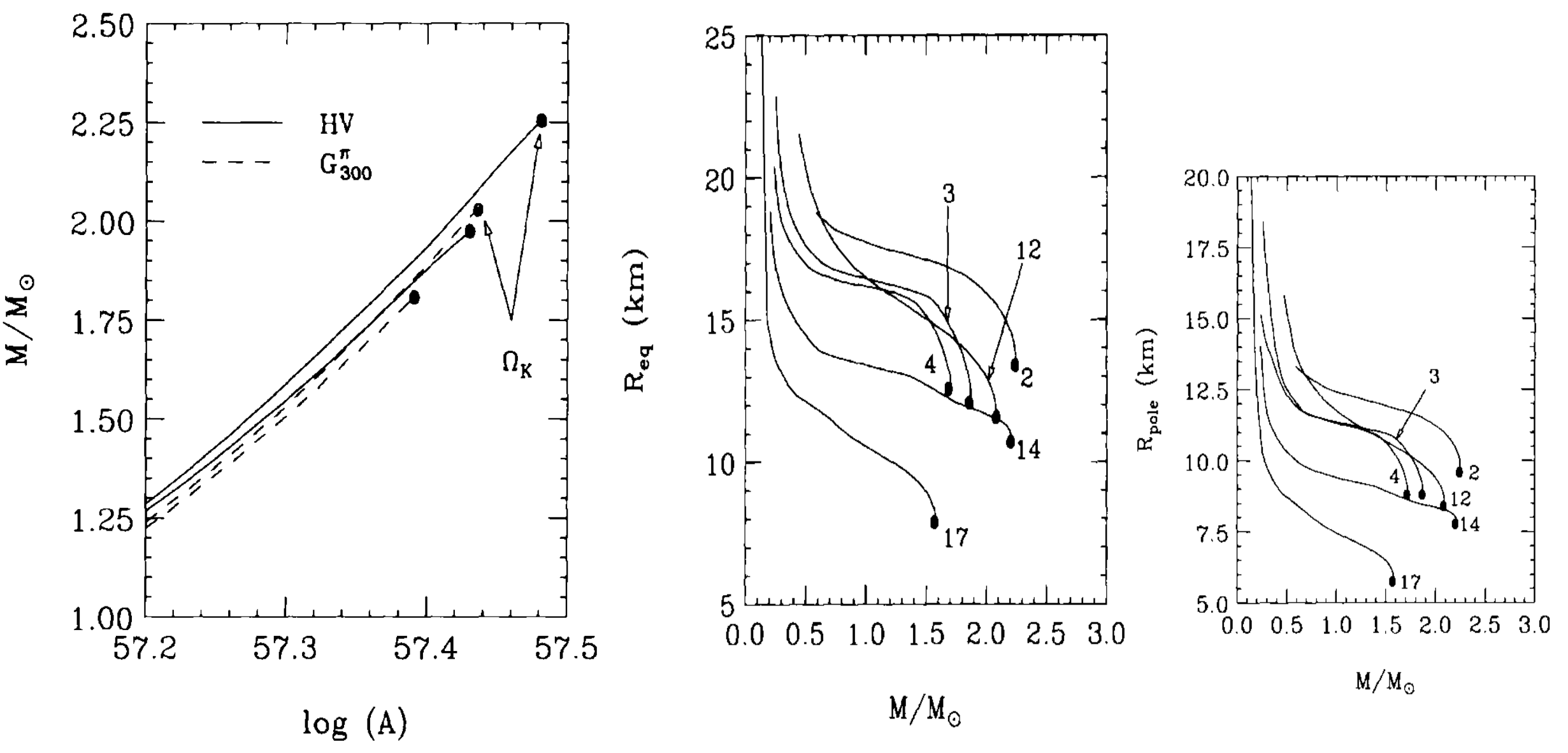}
	\caption{De izquierda a derecha: La masa de la estrella como funci\'on del n\'umero total de bariones en la misma para dos ecuaciones de estado de EN, HV y G$^\pi_{300}$, sin y con rotaci\'on (l\'ineas s\'olidas y punteadas respectivamente). La curvas masa radio para el radio ecuatorial (centro) y polar (derecha) para estas mismas entrellas. N\'otese que la diferencia entre los casos con y sin rotaci\'on es m\'inima. \cite{Weberrotacion}}\label{rotacion}
\end{figure}

Sin embargo, magnitudes igualmente importantes relacionadas con las variaciones del período de rotación de las EBE y la producci\'on de ondas gravitacionales no pueden ser estudiadas con nuestras ecuaciones de estructura porque en ellas est\'an no incluidos los efectos de la rotaci\'on \cite{Weberrotacion}. Desde el punto de vista te\'orico, tener en cuenta la rotaci\'on de la estrella ayuda a estabilizarla y prevenir el colapso gravitacional, al tiempo que permite masas mayores y convierte al OC en un objeto oblato. No obstante, cuantitativamente estos cambios suelen ser muy ligeros con respecto al caso est\'atico, incluso cuando el OC rota a su frecuencia kepleriana\footnote{Se conoce como frecuencia kepleriana a la a m\'axima frecuencia con la que un OC puede rotar sin desintegrarse.} (Fig.~\ref{rotacion}). Por ello, a pesar de que est\'a comprobado que la mayor\'ia de los objetos compactos conocidos rotan, la descripci\'on te\'orica de OC est\'aticos contin\'ua siendo un paso imprescindible en su modelaci\'on. En este sentido es importante resaltar que, dadas unas EdE espec\'ificas, el estudio de las configuraciones estelares estáticas permite saber si la materia que ellas describen autogravita y, por tanto, si es posible la existencia de estrellas compuestas por esta materia; asimismo, a partir de estas soluciones es posible conocer si las magnitudes derivadas de esas EdE son compatibles con las observaciones y, en el caso de estrellas magnetizadas, estudiar los efectos relacionados exclusivamente con el campo magnético.

\section{Estrellas de condensado de Bose-Einstein con campo magn\'etico constante}

En esta secci\'on se discuten los resultados obtenidos al resolver las ecuaciones de estructura, Ecs.~(\ref{gTOV}), para las EdE obtenidas en el Cap\'itulo \ref{cap3} para las Estrellas de condensado de Bose-Eisntein con campo magn\'etico constante. Aunque las EdE de estas estrellas contienen el t\'ermino de Maxwell, en lo que sigue se muestran tambi\'en las soluciones de las ecuaciones de estructura para las EdE sin la contribuci\'on de Maxwell. Esto se ha hecho a fin de tener una mejor valoraci\'on del impacto de la  contribuci\'on de Maxwell en el proceso de construcci\'on de la estrella. Como veremos, las diferencias son apreciables aun en el caso en que se tomen densidades centrales relativamente altas. Esto refuerza el hecho de que la estructura de la estrella depende de toda la EdE y por tanto es siempre sensible a los efectos del campo magn\'etico aunque en las EdE la anisotrp\'ia solo sea apreciable a bajas densidades.

\begin{figure}[h!]
	\centering
	\includegraphics[width=0.42\linewidth]{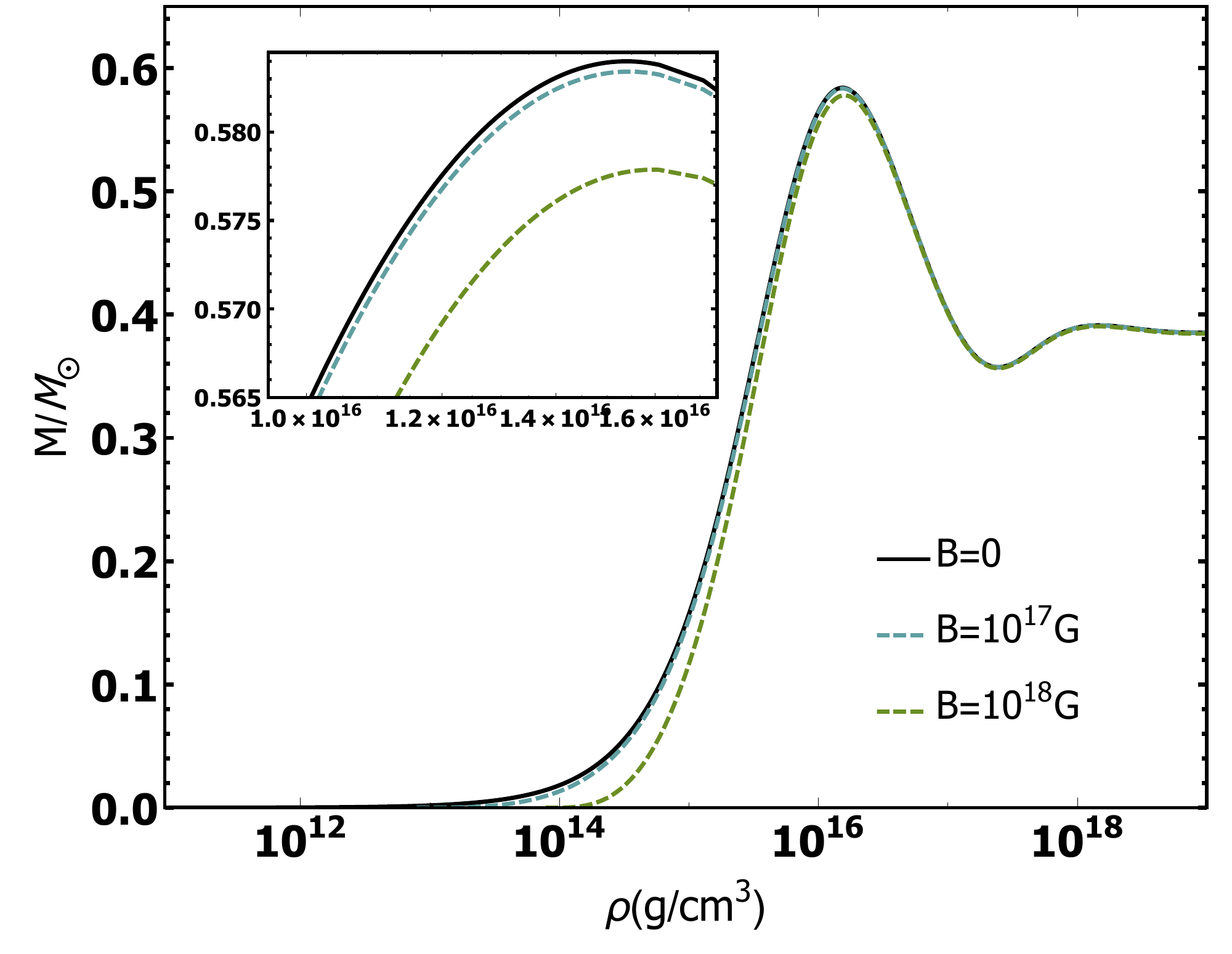}
	\includegraphics[width=0.42\linewidth]{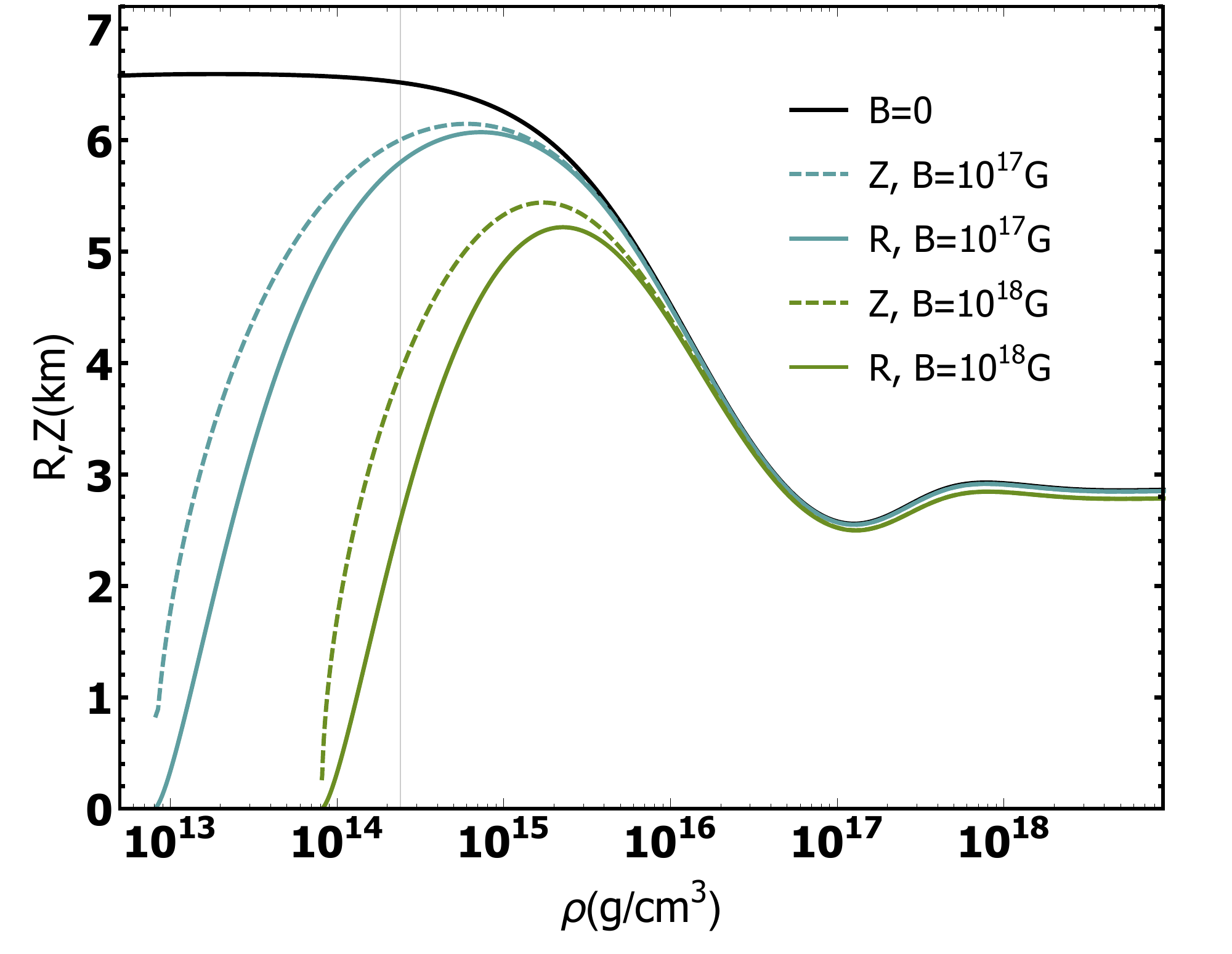}\\
	\includegraphics[width=0.42\linewidth]{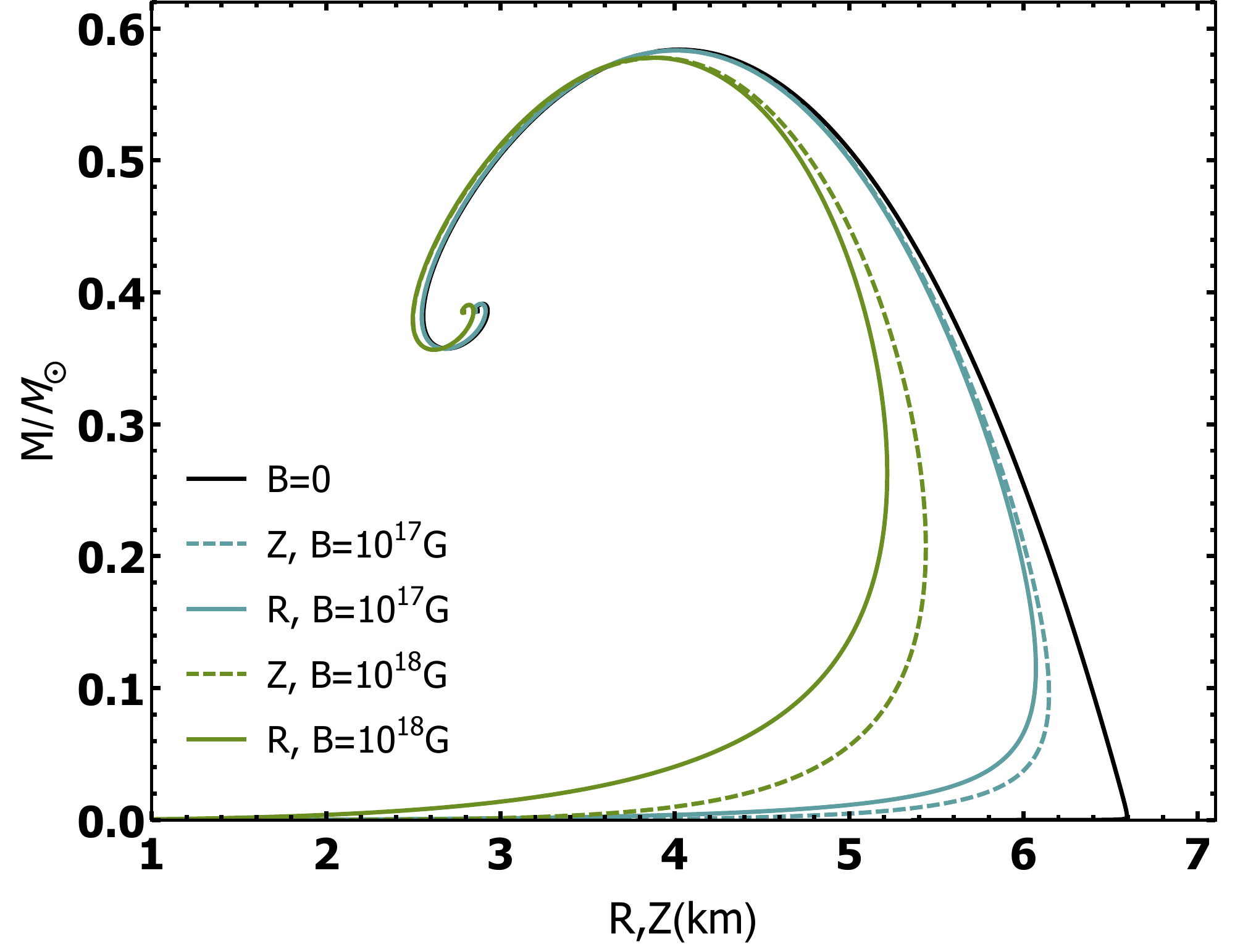}
	\includegraphics[width=0.42\linewidth]{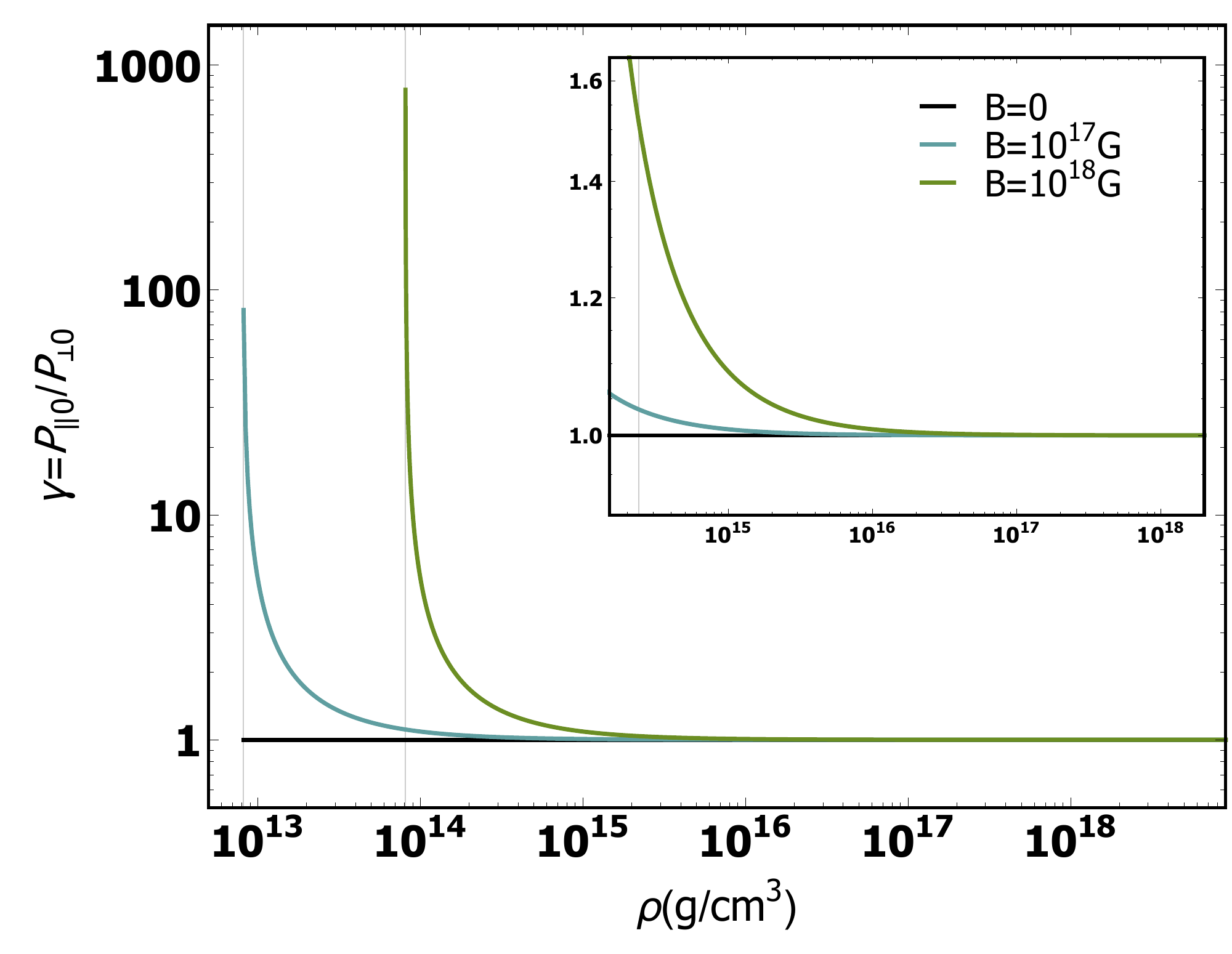}	
	\caption{Los resultados de resolver las ecuaciones de estructura $\gamma$ para la EdE sin la contribución de Maxwell. Paneles superiores: la masa total y los radios ecuatoriales y polares de la estrella en función de la densidad de masa central. Panel inferior izquierdo: relaciones masa-radio con el radio ecuatorial (líneas continuas) y el polar (líneas discontinuas). Panel inferior derecho: el parámetro $\gamma$ como función de la densidad de masa central. Las líneas verticales señalan las densidades en las que $ P_{\parallel} = 0$ y $\gamma\rightarrow\infty$ para un campo magnético dado. La línea vertical en el recuadro se\~nala $\rho_{nuc}$.}\label{mrzro}
\end{figure}

En la Fig.~\ref{mrzro} se muestran los resultados de integrar las ecuaciones de estructura $\gamma$ para $B=0$, $B=10^{17}$G y $B=10^{18}$G para las EdE Ecs.~(\ref{EoSRtotal}) sin la contribuci\'on de Maxwell. Como puede verse en el panel superior izquierdo, en el caso no magnetizado ($B=0$), la curva de $M$ vs $\rho_0$ muestra una regi\'on de configuraciones estelares estables cuya masa m\'axima es $M \approx 0.58 M_{\odot}$, con radio $R \approx 4.2$~km y densidad de masa central $\rho_0 \approx 1.5 \times 10^{16}$~g/cm$^3$. Este resultado est\'a en concordancia con lo obtenido en \cite{latifah2014bosons}. Un campo magn\'etico constante no cambia la forma de la curva $M$ vs $\rho_0$, pero disminuye ligeramente la masa. Dicha disminuci\'on es mayor a bajas densidades aunque a\'un notable para las m\'as altas, como puede ser apreciado en el recuadro del panel izquierdo superior de la Fig.~\ref{mrzro}.

La influencia del campo magn\'etico constante en la la forma y tama\~no de las EBE es m\'as dram\'atica. El campo magn\'etico no solo deforma el objeto, sino adem\'as disminuye su tama\~no. Ambos efectos pueden verse en los paneles superior derecho e inferior izquierdo de la Fig.~\ref{mrzro}. Como $\gamma =z/r=P_{\parallel_0}/P_{\perp_0} $, el radio polar $Z$ es siempre mayor que el ecuatorial $R$ (porque cuando la contribuci\'on de Maxwell es ignorada $P_{\parallel}>P_{\perp}$). Esto significa que la estrella resultante es un objeto prolato. En la regi\'on de bajas densidades hay una enorme desviaci\'on con respecto a la curva de $B=0$ y ambos radios decrecen con la densidad en lugar de tender a un valor constante. Este decrecimiento est\'a relacionado con la regi\'on de inestabilidad que muestran las EdE en forma tal que a medida que $P_{\perp_0}$ se acerca a cero, la estrella deviene menor y menos masiva.

N\'otese que cuando la densidad de masa central para la cual $P_{\perp_0}=0$ es aproximada, $\gamma\rightarrow \infty$. La densidad de masa a la cual $\gamma$ diverge ha sido se\~nalada con las l\'ineas grises verticales en el panel inferior izquierdo de la Fig.~\ref{mrzro}, pero en realidad, como ellas est\'an por debajo de la densidad nuclear no se espera que existan en el n\'ucleo de nuestras estrellas\footnote{Aunque mostramos los resultados de integrar las ecuaciones de estructura para toda la regi\'on de densidades de masa en la que la EdE es estable, no debe olvidarse que el n\'ucleo de una EN est\'a definido por $\rho>\rho_{nuc}$, y por tanto, las soluciones con $\rho_0 < \rho_{nuc}$ deben descartarse.}. Para $\rho_0 = \rho_{nuc}$ los valores de $\gamma$ son bastante moderados, como se muestra en el recuadro del panel inferior derecho de la Fig.~\ref{mrzro}, pero la deformaci\'on es a\'un notable. En las regiones de m\'as alta densidad de masa, $\gamma \rightarrow 1$ y las estrellas son casi esf\'ericas.

\begin{figure}[h!]
	\centering
	\includegraphics[width=0.42\linewidth]{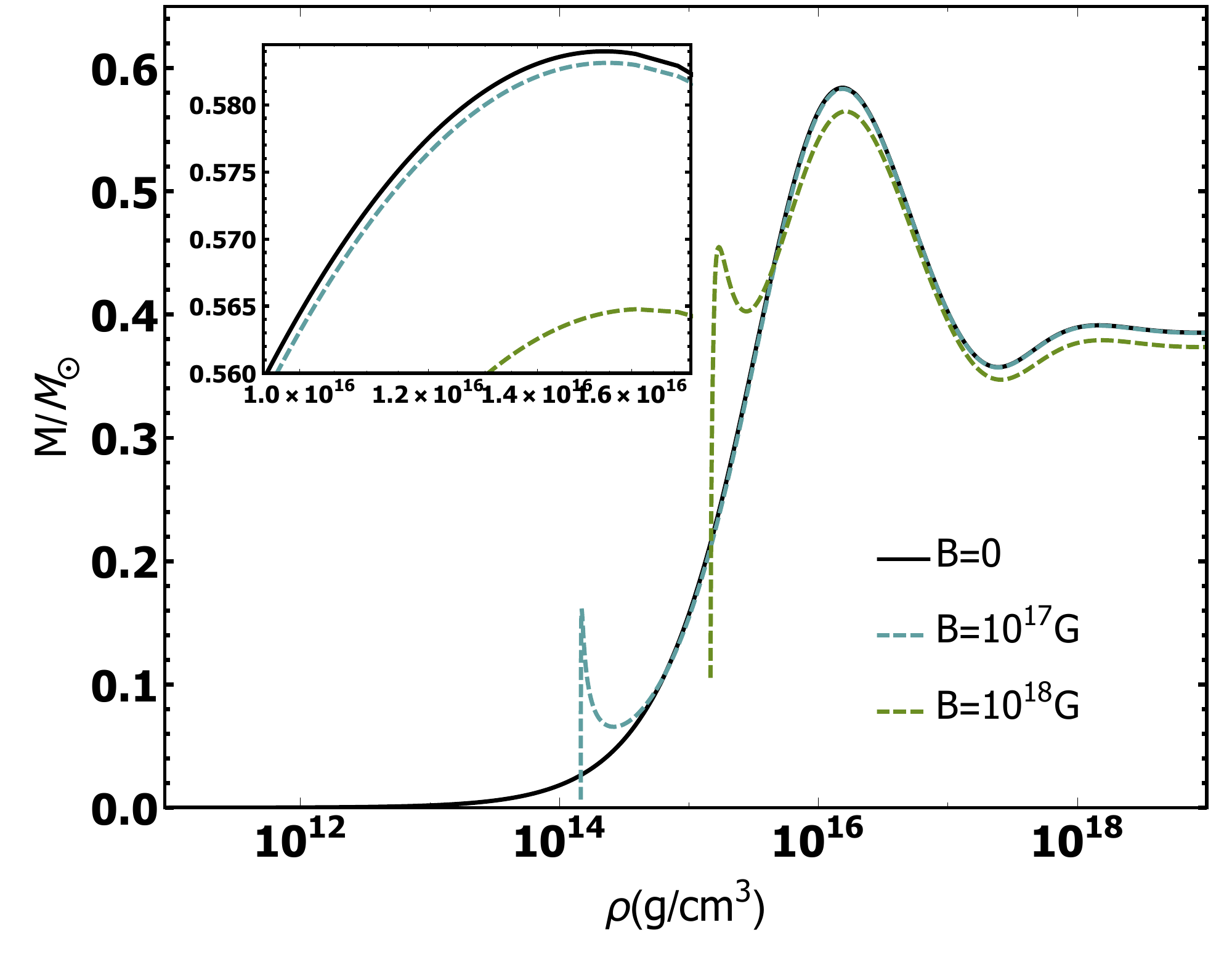}
	\includegraphics[width=0.42\linewidth]{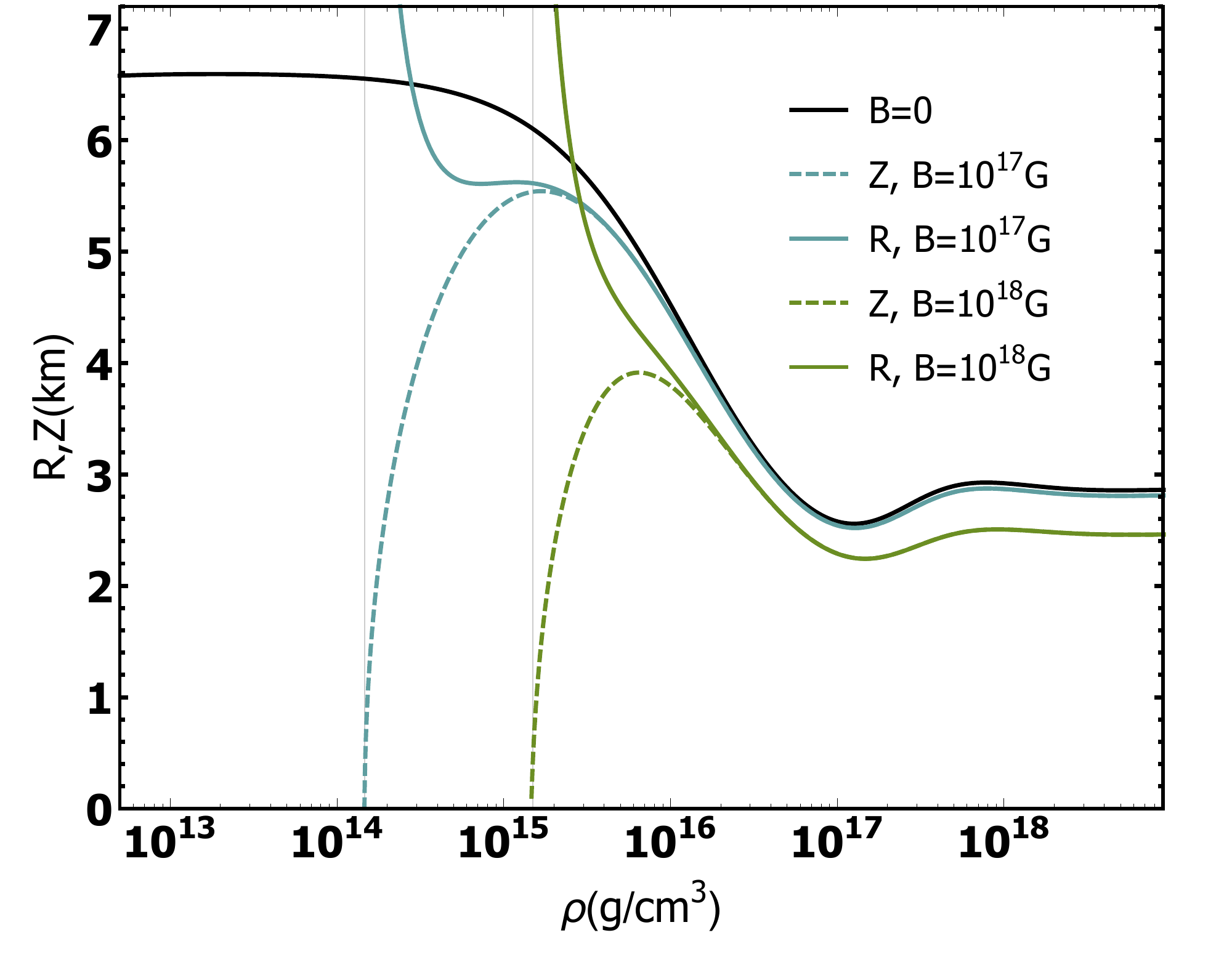}\\
	\includegraphics[width=0.42\linewidth]{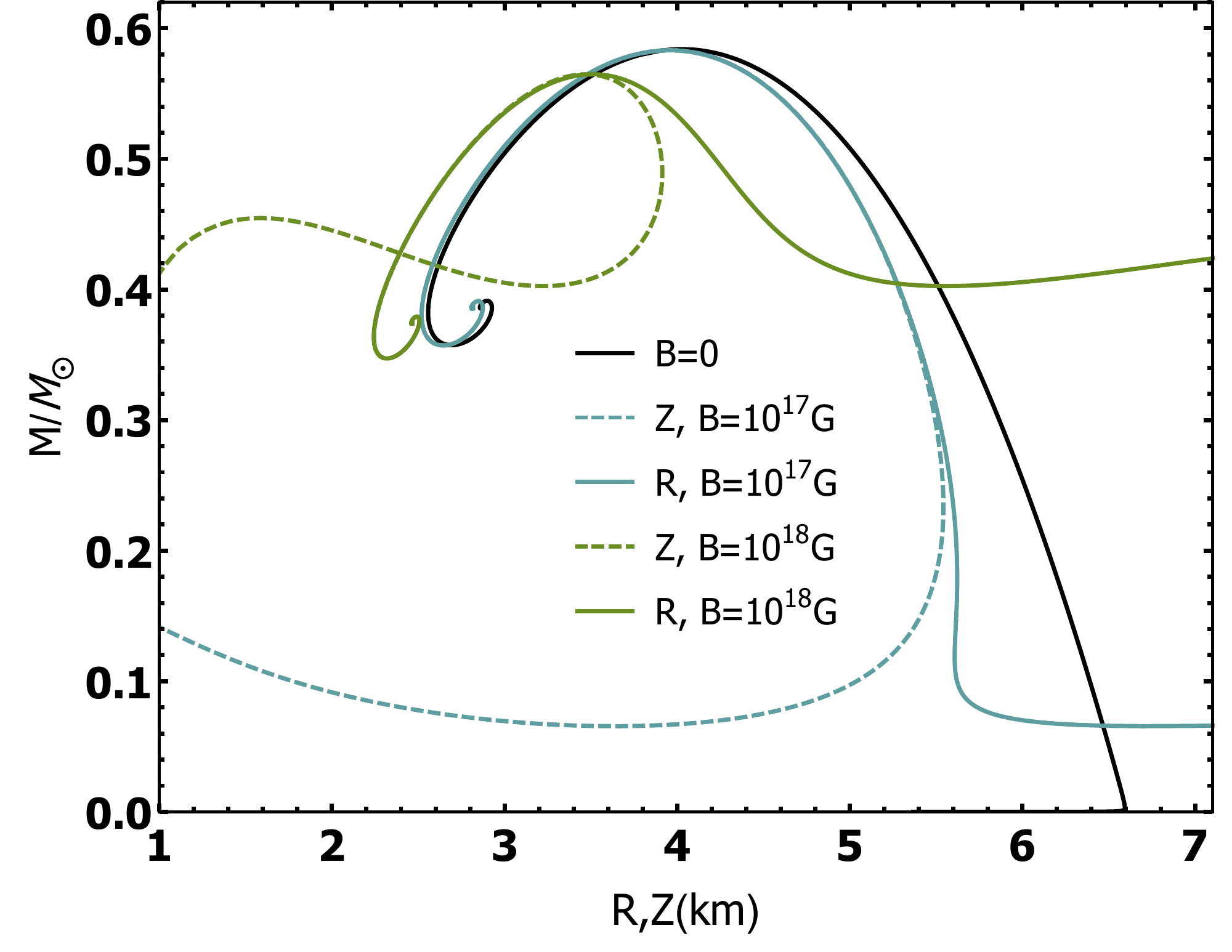}
	\includegraphics[width=0.42\linewidth]{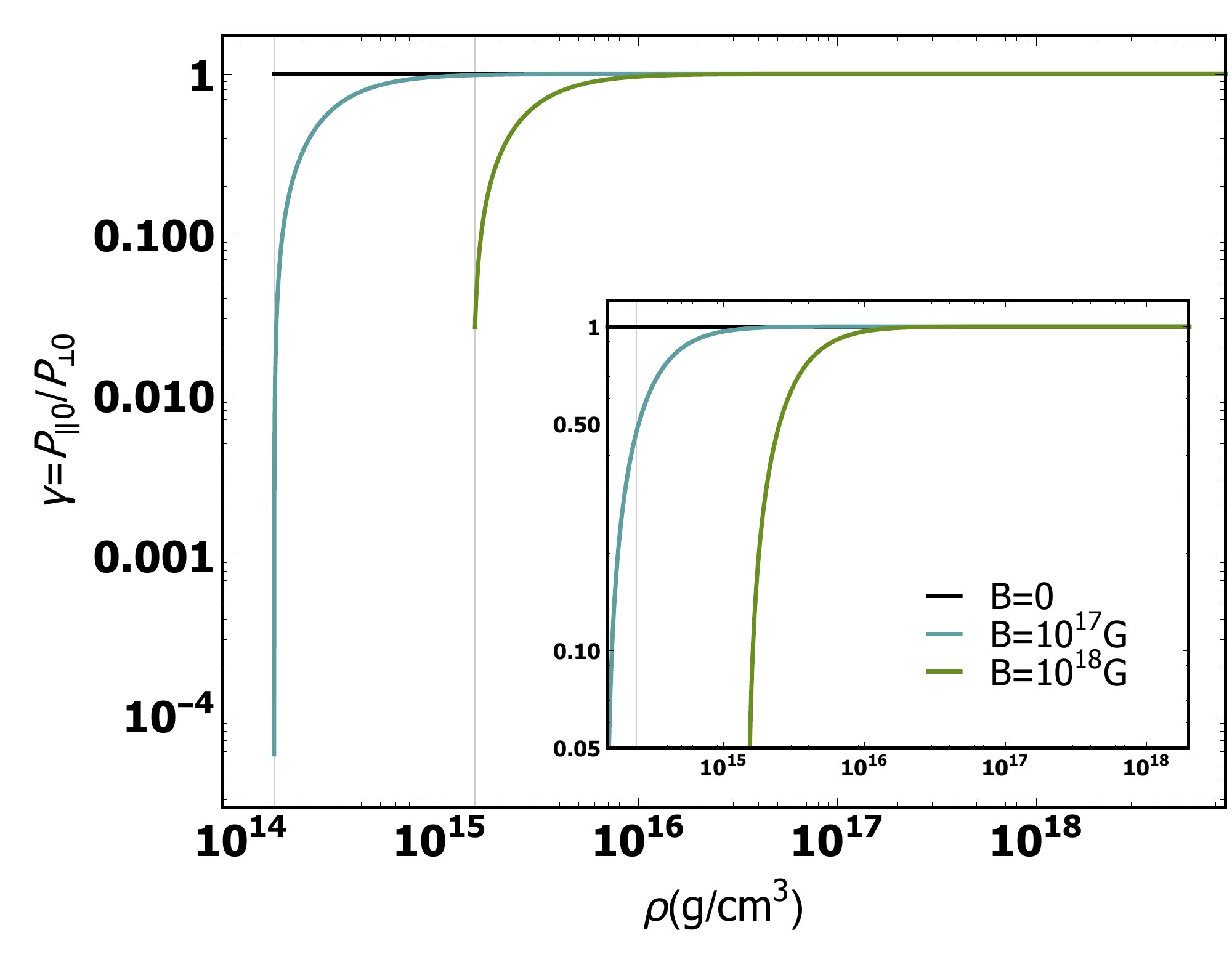}	
	
	\caption{Los resultados de resolver las ecuaciones de estructura $\gamma$ para la EdE con la contribución de Maxwell. Paneles superiores: la masa total y los radios ecuatoriales y polares de la estrella en función de la densidad de masa central. La línea vertical señala $\rho_{nuc} $. Panel inferior izquierdo: relaciones masa-radio con el radio ecuatorial (líneas continuas) y el polar (líneas discontinuas). Panel inferior derecho: el parámetro $ \gamma $ en función de la densidad de masa central. Las líneas verticales señalan las densidades en las que $P _{\parallel} = 0 $ y $\gamma\rightarrow 0 $. La línea vertical en el recuadro señala $\rho_ {nuc}$. }\label{mrzroBcuadrado}
\end{figure}

Las configuraciones estables de masa y radio obtenidas con la contribuci\'on de Maxwell, nuestras EBE magnetizadas, se muestran en la  Fig.~\ref{mrzroBcuadrado}. Como se vio al estudiar las EdE, incluir los t\'erminos de Maxwell refuerza los efectos del campo magn\'etico, de manera que ahora la disminuci\'on de la masa y el tama\~no de las estrellas es mayor y se aprecia fácilmente en todo el rango de densidades considerado. En este caso  $P_{\parallel}<P_{\perp}$, por tanto $\gamma<1$, el radio ecuatorial es menor que el polar y la estrella es oblata. El radio ecuatorial aumenta cuando la densidad de masa disminuye, mientras que el polar disminuye, llegando a lacanzar la diferencia entre ambos hasta cuatro \'ordenes para $B=10^{17}$~G (v\'ease el panel inferior derecho de la Fig.~\ref{mrzroBcuadrado}). Este comportamiento de los radios, al igual que el nuevo pico que aparece en la curva $M$ vs $\rho_o$, est\'a relacionado con el hecho de que cuando $P_{\parallel_0} \rightarrow 0$, $\gamma \rightarrow 0$, como se muestra en el panel inferior derecho de la Fig.~\ref{mrzroBcuadrado}. Hacer $\gamma = 0$ transforma la m\'etrica $\gamma$ en el espacio-tiempo plano de Minkowski \cite{MalafarinaHerrera}, por tanto, las soluciones de las ecuaciones de estructura alrededor de este l\'imite carecen para el presente estudio de inter\'es f\'isico.

\section{Estrellas de condensado de Bose-Einstein automagnetizadas y perfiles de campo magn\'etico}

\begin{figure}[h]
	\centering
	\includegraphics[width=0.42\linewidth]{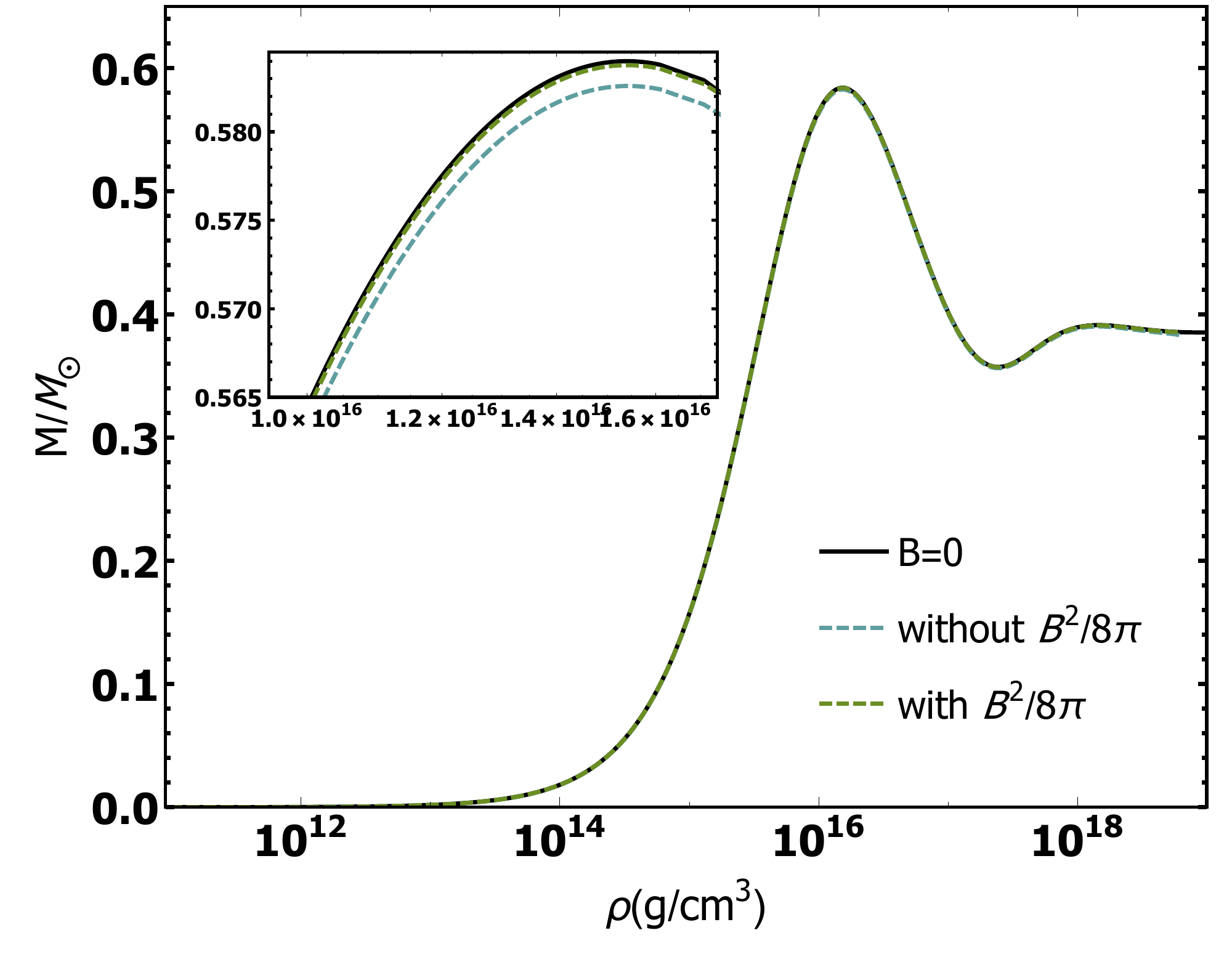}
	\includegraphics[width=0.42\linewidth]{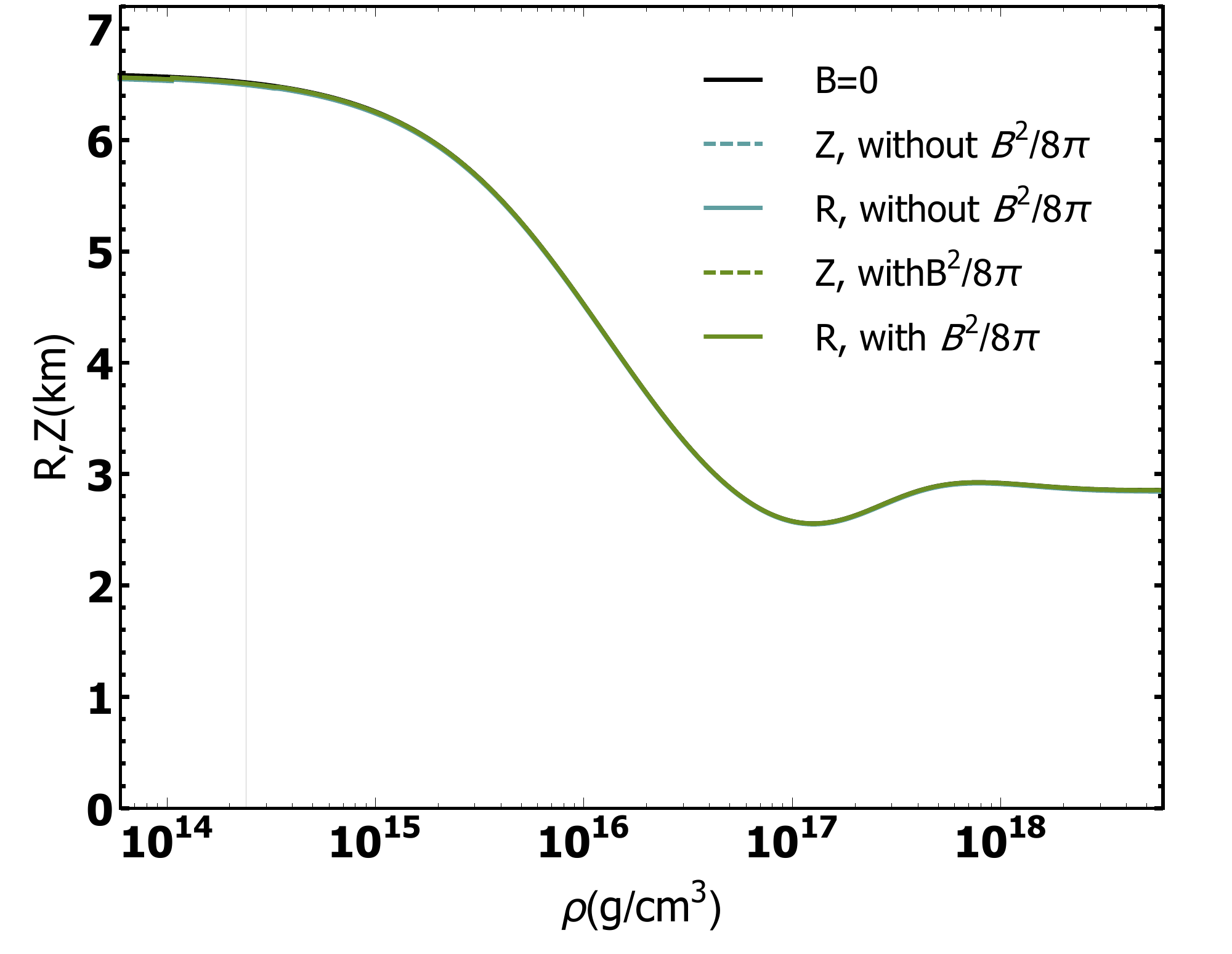}\\
	\includegraphics[width=0.42\linewidth]{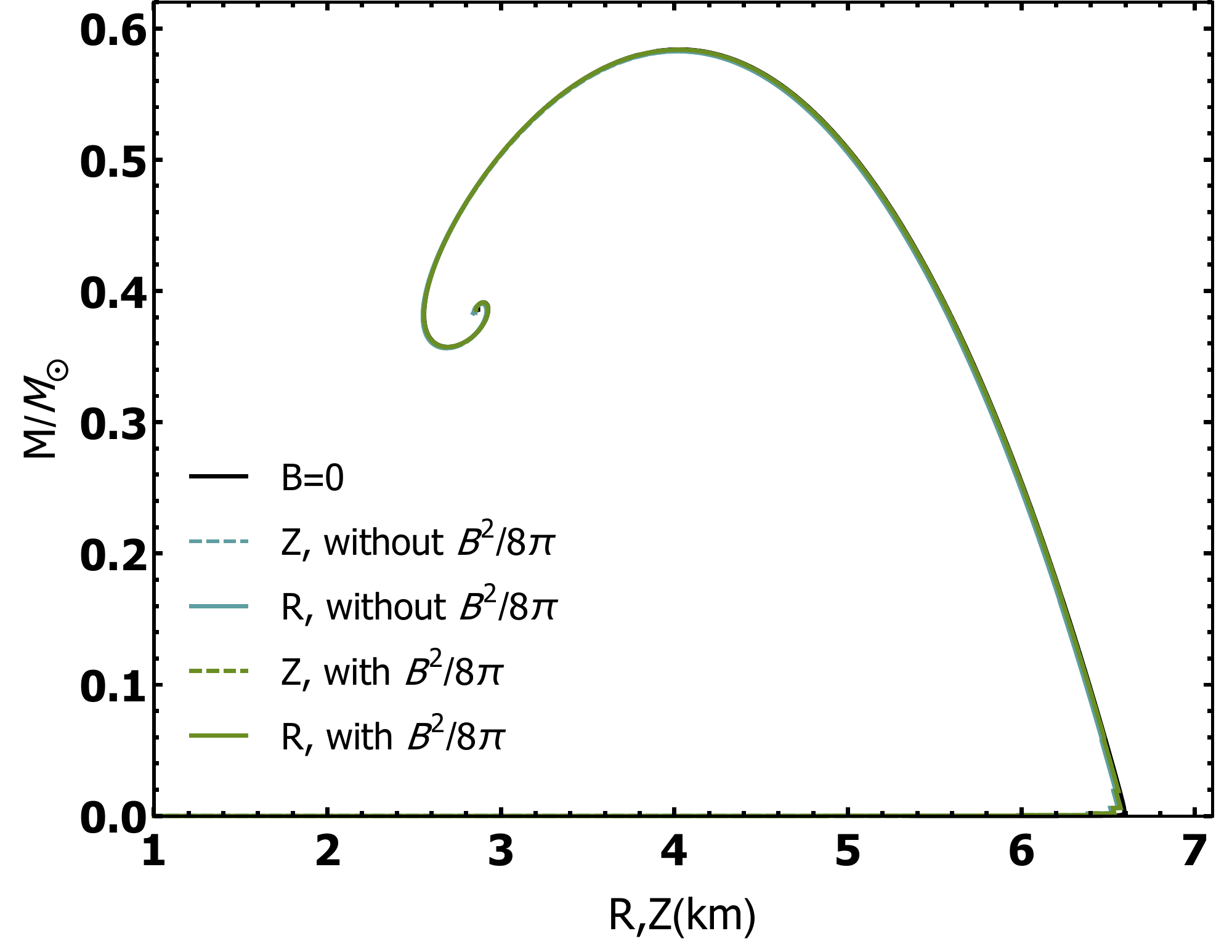}
	\includegraphics[width=0.42\linewidth]{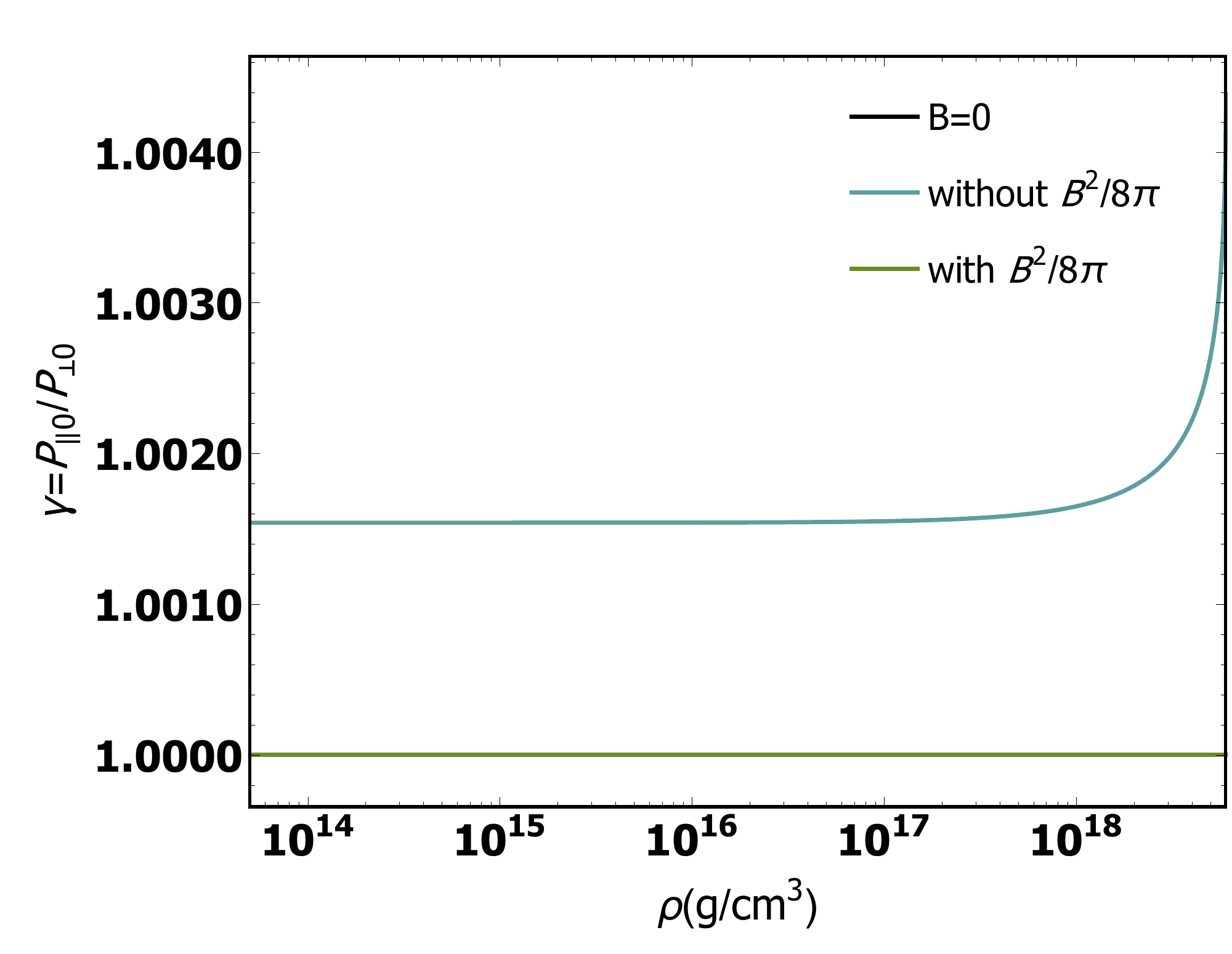}	
	\caption{Los resultados de resolver las ecuaciones de estructura $\gamma$ para la EdE con campo magnético autogenerado. Paneles superiores: la masa total de la estrella y los radios ecuatorial y polar en función de la densidad de masa central. La línea vertical marca $\rho_ {nuc} $. Panel inferior izquierdo: relaciones masa-radio con el radio ecuatorial (líneas continuas) y el polar (líneas discontinuas). Panel inferior derecho: el parámetro $\gamma$ en función de la densidad de masa central.}\label{mrzroBsg}
\end{figure}

En esta secci\'on se resuelven las ecuaciones de estructura Ecs.~(\ref{gTOV}) para las EdE Ecs.~(\ref{EoSRtotal}) con el campo magn\'etico dado como funci\'on de la densidad de masa a trav\'es de Ec.~(\ref{selfmag1}). Los resultados se muestran en la Fig.~\ref{mrzroBsg}.

Las curvas de masa y radios correspondientes a las estrellas de condensado de Bose-Eisntein magnetizadas se superponen casi perfectamente con las de campo cero, y la influencia del campo magn\'etico en la masa, el tama\~no y la forma de la estrella es realmente peque\~na (v\'ease el recuadro en el panel superior derecho de la Fig.~\ref{mrzroBsg}). La raz\'on de esto radica en la disminuci\'on del campo magn\'etico autogenerado con la densidad de part\'iculas, que disminuye la anisotrop\'ia de manera tal que la inestabilidad en las presiones no aparece nunca. Como resultado, las estrellas de condensado de Bose-Eisntein se desv\'ian ligeramente de la forma esf\'erica y sus masa apenas disminuyen con respecto al caso no magnetizado, siendo $\gamma\cong 1$ en todo el rango de densidades centrales considerado.

\begin{figure}[h]
	\centering
	\includegraphics[width=0.42\linewidth]{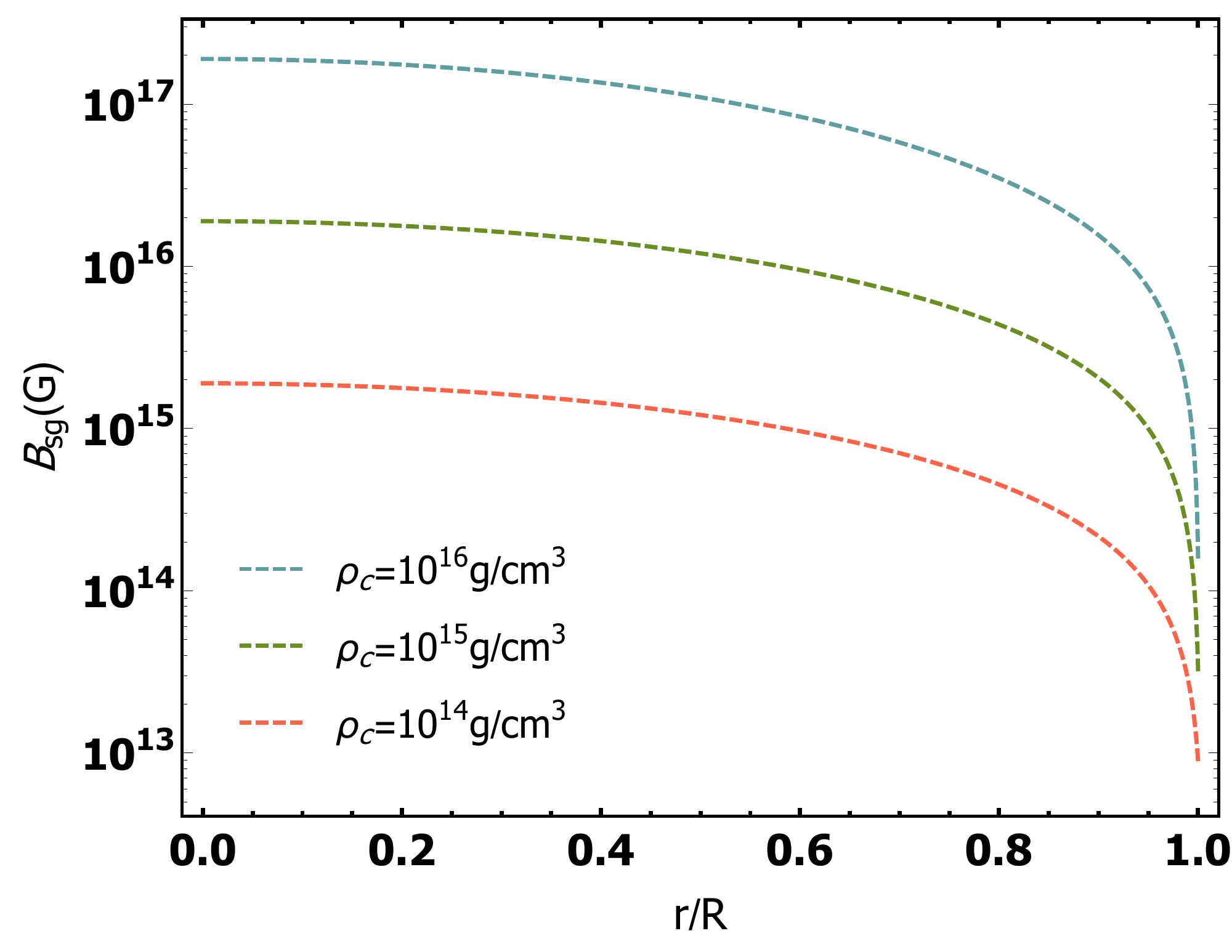}
	\includegraphics[width=0.42\linewidth]{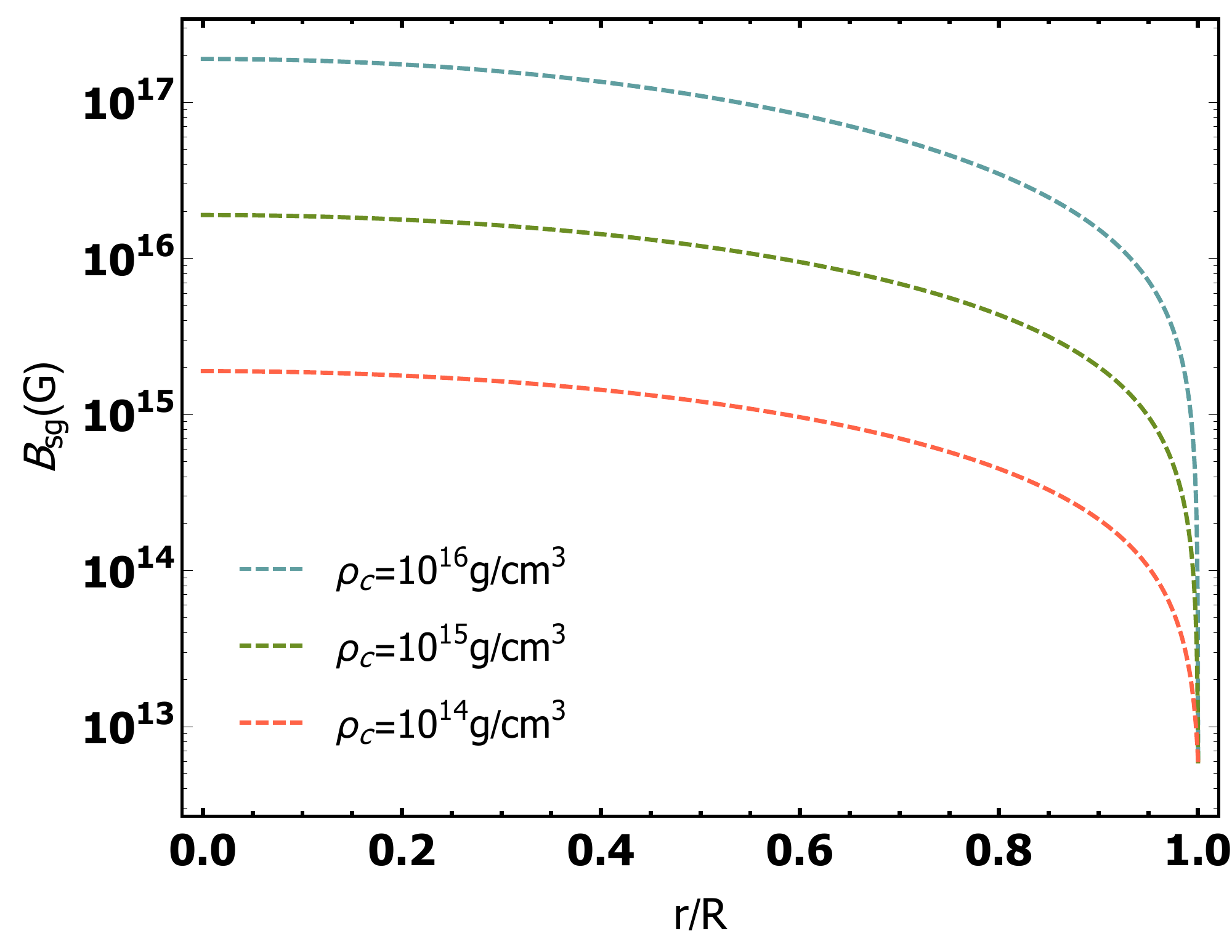}
	\caption{La intensidad del campo magnético en el interior de la Estrella de condensado de Bose-Einstein automagnetizada como función del radio ecuatorial para varios valores de la densidad de masa central. Panel izquierdo: EdE sin la contribución de Maxwell; panel derecho: EdE con la contribución de Maxwell.}\label{Bprofiles}
\end{figure}

El uso de la Ec.~(\ref{selfmag1}) de conjunto con las EdE, permite calcular la intensidad del campo magn\'etico $B$ de manera autoconsistente durante la integraci\'on de las ecuaciones de estructura. Los perfiles de campo magn\'etico interno obtenidos se muestran en la Fig.~\ref{Bprofiles} como funci\'on del radio ecuatorial de la estrella para varios valores de la densidad de masa central. El panel derecho (izquierdo) de la figura muestra las curvas para las EdE sin (con) la contribuci\'on de Maxwell. En el centro de la estrella, dada una misma densidad central, los valores del campo magn\'etico autogenerado son los mismos en ambos casos. Pero el decrecimiento del campo magn\'etico en la superficie de la estrella es mayor cuando se incluye la contribuci\'on de Maxwell, estando la variaci\'on alrededor de  los tres (cuatro) \'ordenes para las EdE con (sin) ella (v\'ease la Tabla~\ref{T1}). En ambos casos los valores de $B$ en el centro y la superficie de la estrella est\'an en el orden de los estimados para ENs \cite{Malheiro:2013loa,1991ApJ...383..745L,Lattimerprognosis}.

Los resultados de esta secci\'on validan la automagnetizaci\'on de los bosones de spin uno como fuente de campo magn\'etico no solo en el caso de Estrellas de condensado de Bose-Einstein magnetizadas, sino tambi\'en para otros modelos de estrellas en los cuales este tipo de part\'iculas est\'e presente. Uno de los m\'eritos de nuestra propuesta es, sin dudas, el hecho de que los bosones generan un campo cuyos valores son consistentes con los estimados observacionales y te\'oricos. A ello s\'umese, que los perfiles de campo magn\'etico se obtienen de manera natural a partir de la soluci\'on de las ecuaciones de estructura. Una vez que se fija su orientaci\'on, el campo magn\'etico se obtiene de primeros principios, libre de suposiciones heur\'isticas (ver \cite{ChatterjeeBprofiles}).

\begin{table}[ht]
	\centering
	\begin{tabular}{|c|c|c|r|}\toprule
		$\rho _0 (\text{g}/\text{cm}^3)$ & $B_0\text{(G)}$&$B_s\text{(G)}$ & Maxwell \\
		\hline
		&                 & $5.99\times10^{12} $ & con    \\
		\cline{3-4} $10^{14}$                       & $1.89\times 10^{15}$ & $8.87\times 10^{12}$    & sin \\
		\hline
		&                 &$5.99\times 10^{12}$ & con\\
		\cline{3-4} $10^{15}$                      & $1.89\times 10^{16}$ & $3.27\times 10^{13}$     & sin \\
		\hline
		&                 &$5.99\times 10^{12}$ & con    \\
		\cline{3-4} $10^{16}$                      & $1.89\times 10^{17}$&  $1.62\times 10^{14}$     & sin \\
		\toprule
	\end{tabular}
	\caption{La intensidad del campo magn\'etico autogenerado en el centro y la superficie de la estrella para distintos valores de la densidad de masa central.}
	\label{T1}
\end{table}

\section{Conclusiones del cap\'itulo}

En este cap\'itulo hemos obtenido las masas y radios de Estrellas de condensado de Bose-Einstein magnetizadas y automagnetizadas. Para ello se resolvieron las ecuaciones de estructura derivadas en el Cap\'itulo \ref{cap4} de conjunto con las EdE obtenidas en el Cap\'itulo \ref{cap3}. 
En el caso de las Estrellas de condensado de Bose-Einstein automagnetizadas, fue posible adem\'as estudiar los perfiles de campo magn\'etico en su interior.  Los resultados obtenidos conducen a las siguientes conclusiones:

\begin{itemize}
	
\item Para campo magn\'etico constante:

\begin{itemize}

\item Las estrellas (no esf\'ericas) de condensado de Bose-Einstein magnetizadas son en general menos masivas y m\'as peque\~nas que sus an\'alogas no magnetizadas (esf\'ericas).

\item El objeto compacto resultante es oblato o prolato en dependencia de en qu\'e direcci\'on se ejerce la mayor de las presiones. En nuestro modelo, debido a que para las EdE de las EBE (con t\'ermino de Maxwell incluido) la presi\'on mayor es la perpendicular al campo, las EBE magnetizadas son oblatas.

\item El comportamiento extremo de las masas, los radios y la forma de la estrella en la proximidad de la zona donde comienza la inestabilidad de la presi\'on est\'a relacionado con la EdE pero tambi\'en con las propiedades de las ecuaciones de estructura $\gamma$ en los l\'imites $\gamma\rightarrow\infty$ y $\gamma\rightarrow 0$. Por ello, un estudio detallado de dichos l\'imites se hace necesario.

\end{itemize}

\item Para campo magn\'etico autogenerado:

\begin{itemize}
	
	\item Los cambios en la masa y la forma de la estrella son peque\~nos en comparaci\'on con el caso no magnetizado.
	
	
	\item Los perfiles de campo magn\'etico obtenidos, as\'i como sus valores extremos, est\'an en el orden de los estimados observacionales y te\'oricos para Estrellas de Neutrones. Esto valida a la automagnetizaci\'on de los bosones vectoriales como mecanismo de generaci\'on de campo magn\'etico en los objetos compactos. Con respecto a esto, nos gustar\'ia resaltar, que la automagnetizaci\'on provee una manera natural de introducir el campo magn\'etico estelar de forma tal que dicho campo es producido directamente por la materia que compone el objeto magnetizado.

\end{itemize}

\end{itemize}

Al igual que en el Cap\'itulo \ref{cap3}, los resultados de este cap\'itulo dependen de la forma en que se modela el campo magn\'etico. En lo que respecta a los observables, esto es una ventaja, pues podr\'ia  ayudar a discriminar entre un modelo y otro a partir de las observaciones.

%% file: conclusiones.tex
\chapter*{Conclusiones}
\addcontentsline{toc}{chapter}{Conclusiones}

Motivados por el vertiginoso desarrollo que las observaciones astrof\'isicas han experimentado en las \'ultimas d\'ecadas y por la necesidad de contar con modelos que permitan su correcta interpretaci\'on, as\'i como por el inter\'es que comprende para la f\'isica dilucidar las leyes que rigen el comportamiento de la materia en las condiciones extremas que se dan en los objetos compactos, al comenzar el trabajo que culmina con la presente tesis, nos propusimos como objetivo general \textbf{estudiar las propiedades termodin\'amicas de un gas ideal magnetizado de bosones vectoriales neutros y aplicarlas al estudio de los efectos del campo magn\'etico en la materia que forma las ENs, considerando como casos particulares los \emph{jets} astrof\'isicos y las Estrellas de condensado de Bose-Einstein.}
Una vez desarrolladas las investigaciones necesarias para la consecuci\'on de este objetivo, consideramos que los principales resultados y conclusiones obtenidos son:

\begin{itemize}
	\item El c\'alculo del espectro energ\'etico y la caracterizaci\'on termodin\'amica del gas magnetizado de bosones vectoriales masivos y neutros en el l\'imite de baja temperatura, y para una y tres dimensiones (Cap\'itulo \ref{cap2} y \cite{Quintero2017IJMP,Quintero2017PRC,Quintero2017AN}).
	
	Este estudio fue realizado en el marco de la teor\'ia de Proca para part\'iculas masivas de spin uno y tiene un c\'aracter general, aunque la motivaci\'on principal para llevarlo a cabo radic\'o en sus posibles aplicaciones astrof\'isicas. En la tesis, sus resultados fueron utilizados para describir la composici\'on y origen de los \textit{jets} astrof\'isicos emitidos por Estrellas de Neutrones, para estudiar los efectos del campo magn\'etico en las ecuaciones de estado y la estructura de Estrellas de condensado de Bose-Einstein, y para explicar la generaci\'on de campos magn\'eticos estelares. 
	
	\item La demostración de que para un gas magnetizado de bosones vectoriales masivos y neutros, tanto la condensaci\'on de Bose-Einstein como la anisotropía en las presiones, el colapso magnético y la automagnetización ocurren para valores de B, N y T en el orden de los típicos para las Estrellas de Neutrones (Cap\'itulo \ref{cap2} y \cite{Quintero2017AN}).  
	
	La relevancia de estos fenómenos para la descripci\'on te\'orica de la materia que compone las Estrellas de Neutrones se comprobó en los dos modelos desarrollados en la tesis (Cap\'itulos \ref{cap2_5}, \ref{cap3} y \ref{cap5}, y \cite{Quintero2019jets,Quintero2018BECS}).
	
	\item La proposici\'on y validación de un  mecanismo para la producción y mantenimiento de los \textit{jets} astrofísicos originados en las ENs a partir de los resultados obtenidos en el estudio del colapso magn\'etico de un gas \textit{npe} parcialmente bosonizado  (Cap\'itulo \ref{cap2_5} y \cite{Quintero2019jets}).
	
	Dicho mecanismo se basa en las propiedades de los gases cu\'anticos fuertemente magnetizados: la expulsi\'on de la materia hacia el exterior de la estrella se produce a ra\'iz del colapso magn\'etico de los gases de electrones y protones, mientras que los fuertes campos m\'agneticos que garantizan la colimaci\'on de la materia una vez que abandona la estrella pueden ser el resultado de la automagnetizaci\'on del gas \textit{npe}.
	
	\item La obtención de las ecuaciones de estado y de la estructura de estrellas de condensado de Bose-Einstein magnetizadas (Cap\'itulos \ref{cap3} y \ref{cap5}, y \cite{Quintero2018BECS}).
	
	Para ello se combinaron en la descripci\'on termodin\'amica los efectos del campo magn\'etico con los de la interacci\'on bos\'on-bos\'on, que fue descrita como una interacci\'on de contacto por pares. Por otra parte, al incluir el campo magn\'etico en las ecuaciones de estado de la estrella  se consideraron dos modelos. En uno el campo magnético es constante y se fija externamente, mientras que en el otro es generado por la automagnetizaci\'on de los bosones.  
	
	\item La construcción de un sistema de ecuaciones de estructura capaz de describir objetos compactos esferoidales con anisotrop\'ia magn\'etica (Cap\'itulo \ref{cap4} y \cite{Samantha}). 
		 
	La obtenci\'on de estas ecuaciones se hizo partiendo de una m\'etrica axisim\'etrica, la m\'etrica $\gamma$, y su principal ventaja es que permiten tomar en cuenta el efecto de las dos presiones simult\'aneamente. Ellas describen la variaci\'on de las presiones paralela y perpendicular, y de la masa de un objeto compacto esferoidal para el cual la raz\'on entre los radios polar y ecuatorial viene dada por el par\'ametro $\gamma$. A fin de conectar la deformaci\'on del objeto compacto con las causas f\'isicas que la producen, el par\'ametro $\gamma$ se supone igual a la raz\'on entre las presiones perpendicular y paralela en el centro de la estrella. Las ecuaciones de estructura obtenidas, a las cuales hemos llamado ecuaciones de estructura $\gamma$, tienen un car\'acter general y sirven para el c\'alculo de la estructura de cualquier objeto compacto esferoidal con tal de que su deformaci\'on no lo aleje mucho de la forma esf\'erica. En la tesis ellas fueron utilizadas para obtener la relación masa-radio de Estrellas de condensado de Bose-Einstein magnetizadas.
	
	\item La comprobaci\'on de que los efectos del campo magn\'etico en las EdE y la estructura de las Estrellas de condensado de Bose-Einstein magnetizadas son muy sensibles a la forma en que el campo magn\'etico es modelado (Cap\'itulos \ref{cap3} y \ref{cap5}, y \cite{Quintero2018BECS}). 
	
	Aunque en general, las EBE magnetizadas (no esf\'ericas) son menos masivas y m\'as peque\~nas que sus an\'alogas sin campo magn\'etico (esf\'ericas), cuantitativamente estos efectos solo se aprecian en el caso de campo magn\'etico constante, pues cuando el campo magnético es producido por los bosones, la anisotrop\'ia en las EdE es despreciable y las curvas masa-radio de las EBE se superponen casi de manera perfecta con las del caso no magnetizado. 
	
	\item La validación de la automagnetización como una fuente de campo magnético factible para los objetos compactos, y la derivaci\'on en nuestro modelo de un nuevo observable que puede ser comparado con los datos observacionales: el campo magn\'etico superficial del objeto compacto. (Cap\'itulos \ref{cap3} y \ref{cap5}, y \cite{Quintero2018BECS}). 
	
	Esto fue posible porque el suponer el campo magn\'etico interno de la estrella generado por los bosones permite calcular sus perfiles internos de manera autoconsistente. Los valores obtenidos están de acuerdo con los que se supone que existen en el interior  y en la superficie de las Estrellas de Neutrones. Esto no solo confirma a la automagnetizaci\'on como una fuente de campo magnético factible para los objetos compactos, sino que tiene además la ventaja de que el campo magnético es producido directamente por la materia que forma la estrella. 
	
\end{itemize} 

\chapter*{Recomendaciones}
\addcontentsline{toc}{chapter}{Recomendaciones}

De acuerdo con los objetivos que nos trazamos al comenzar las investigaciones que culminan con la presentaci\'on de esta tesis y con los resultados obtenidos en ella, formulamos, para la continuaci\'on del trabajo, las siguientes recomendaciones:

\begin{itemize}
	\item Con respecto a la modelaci\'on de los \textit{jets} astrof\'isicos:
	
	\begin{itemize}
		\item Continuar la b\'usqueda de m\'etricas y ecuaciones de estructura est\'aticas que permitan el estudio de la estabilidad gravitacional del \textit{jet}.
		
	\end{itemize}
	
	\item Con respecto a la descripci\'on de la estructura de los objetos compactos no esf\'ericos:
	
	\begin{itemize}
		\item Continuar la b\'usqueda de m\'etricas y ecuaciones de estructura que permitan la descripci\'on de objetos compactos con anisotrop\'ia magn\'etica, sin la restricci\'on de que la deformaci\'on deba ser peque\~na.
		
		\item Extender las ecuaciones de estructura $\gamma$ al caso de objetos compactos anisotr\'opicos en rotaci\'on.
	\end{itemize}
	
	\item Con respecto a la modelaci\'on de las Estrellas de condensado de Bose-Einstein:
	
	\begin{itemize}
		
		\item Incluir los efectos de la temperatura en las ecuaciones de estado de las Estrellas de condensado de Bose-Einstein magnetizadas. 
		
		\item Extender nuestros c\'alculos al caso de estrellas en rotaci\'on a fin de estudiar los efectos de la deformaci\'on magn\'etica en la emisi\'on de ondas gravitacionales.
		
	\end{itemize}
	
\end{itemize}

%% file: appA.tex
\chapter{Unidades y constantes físicas utilizadas}
\label{appA}

En la tesis se ha utilizado el sistema de unidades naturales (UN) para escribir todas la ecuaciones. En este sistema:

\begin{equation}\nonumber
\hbar=c=k_B = 1, 
\end{equation}

\begin{equation}\nonumber
[\text{longitud}]=[\text{tiempo}]= [\text{masa}]^{-1}=[\text{energía}]^{-1} =	[\text{temperatura}]^{-1},
\end{equation}
y
\begin{align} \nonumber
1 \text{ m} & = 5.07\times 10^{13} \text{ MeV}^{-1}, \\ \nonumber
1 \text{ kg} &= 5.61 \times 10^{29} \text{ MeV}, \\ \nonumber
1 \text{ s} &= 1.52 \times 10^{21} \text{ MeV}^{-1}, \\ \nonumber
1 \text{ K} &= 8.617\times10^{-11} \text{ MeV},
\\ \nonumber
1 \text{ J} &= 6.242\times10^{12} \text{ MeV},
\\ \nonumber
1 \text{T} &= 10^4 \text{ G} = 1.954 \times10^{-10} \text{MeV}^2.
\end{align}


En la Tabla \ref{tab:cons} se muestran las constantes físicas utilizadas en la tesis en unidades naturales.

\begin{table}[h]
	\begin{center}
		\begin{tabular}{ll@{}}
			\toprule
			Magnitud Física (Símbolo) 		          		&  UN  \\ \midrule
			Velocidad de la luz ($c$)    			    	& $1$			 \\
	        Constante de Dirac ($\hbar$) 		     		& 1  			\\
            Constante de Boltzmann ($k_B$)					& 1  			\\ \midrule	

            Masa del electrón ($m_e$) 				 	    & $0.511$ MeV  	 \\
            Masa del protón ($m_p$) 				 	    & $938.272$ MeV  	 \\
            Masa del neutrón ($m_n$) 				 	    & $939.565$ MeV  	 \\			
            Carga eléctrica del electrón ($e$)     	 		& $0.085$    	 \\
            Magnetón de Borh ($\mu_B$)                        & $0.083$ MeV$^{-1}$ \\
            Magnetón nuclear ($\mu_N$)                        & $4.528 \times 10^{-5}$ MeV$^{-1}$ \\ \midrule	
		
			Constante de gravitación ($G$)                &  $6.711\!\times\!10^{-45}$ MeV$^{-2}$ \\
			Masa del Sol ($M_{\odot}$)    			     	 	    &  $1.116\!\times\!10^{60}$ MeV  	   \\
			Radio del Sol ($R_{\odot}$)     					    &  $3.528\!\times\!10^{21}$ MeV$^{-1}$
							
			\\ \bottomrule
		\end{tabular}
	\end{center}
	\caption{Constantes físicas utilizadas en la tesis expresadas en unidades naturales.}\label{tab:cons}
\end{table}

%% file: appC.tex
\chapter{Cálculo de $I$ (Ec.(\ref{Grand-Potential-sst2}))}
\label{appC}

Para calcular la integral en el segundo término de la Ec.~ (\ref{Grand-Potential-sst2}):

\begin{equation}\label{I3}
I=\int\limits_{z_0}^{\infty} dz \frac{z^2}{\sqrt{z^2+\alpha^2}} K_1 (y z),
\end{equation}

\noindent se utilizará la función de McDonald $K_1 (yz)$ en la forma:

\begin{equation}\label{k1int}
K_1 (yz) = \frac{1}{y z}\int\limits_{0}^{\infty} dt e^{-t-\frac{y^2 z^2}{4 t}}.
\end{equation}

Si la Ec.~(\ref{k1int}) se sustutuye en la Ec.~(\ref{I3}), la integral sobre $z$ puede realizarse y:

\begin{equation}\label{I31}
I= \frac{\sqrt{\pi}}{y^2} \int\limits_{0}^{\infty} dt \sqrt{t} e^{-t+\frac{y^2 \alpha^2}{4 t}} erfc \left(\frac{y \sqrt{z^2+\alpha^2}}{2 \sqrt{t}}\right).
\end{equation}

A fin de integrar sobre $t$ en la Ec.~(\ref{I31}), la función de error complementaria $erfc(x)$ será reemplazada por su expasión en serie:

\begin{equation}\label{erfc1}
erfc(x) \backsimeq \frac{e^{-x^2}}{\sqrt{\pi} x} \left(1 - \sum_{w=1}^{\infty} \frac{(-1)^w(2 w -1)!!}{(2 x^2)^w}\right).
\end{equation}

Luego de sustituir la Ec.~(\ref{erfc1}) en la Ec.~(\ref{I31}) e integrar por $t$, se llega a la forma final de $I$:

\begin{equation}\label{I32}
I = \frac{z_0^2}{y \sqrt{z_{0}^2 + \alpha^2}} K_2 (y z_0) -
\frac{z_0^2}{y \sqrt{z_{0}^2 + \alpha^2}} \sum_{w=1}^{\infty} \frac{(-1)^w(2 w -1)!!}{(z_0^2+\alpha^2)^w} \left(\frac{z_0}{y}\right)^w K_{-(w+2)} (y z_0).
\end{equation}

%% file: appD.tex
\chapter{Potencial termodinámico del vacío}
\label{appD}

Para obtener la contribución del vacío al potencial termodinámico de un gas de bosones vectoriales neutros, Ec.~(\ref{Grand-Potential-vac}), partimos de su definición:

\begin{equation}\label{Grand-potential-vac-1}
\Omega_{vac}=\sum_{s=-1,0,1}\int\limits_{0}^{\infty}\frac{p_{\perp}dp_{\perp}dp_3}{(2\pi)^2}\varepsilon \nonumber
\end{equation}

\noindent donde $\varepsilon(p_{\perp},p_3, B,s)=\sqrt{p_3^2+p_{\perp}^2+m^2-2\kappa s B\sqrt{p_{\perp}^2+m^2}}$.

Para integrar sobre $p_3$ y $p_{\perp}$ se utilizará la equivalencia:

\begin{equation}
\sqrt{a}= -\frac{1}{2 \sqrt{\pi}} \int\limits_{0}^{\infty} dy y^{-3/2} (e^{- y a}-1)
\end{equation}

\noindent de conjunto con la introducción de la cantidad pequeña $\delta$ como límite inferior de la integral anterior a fin de regularizar la divergencia del término que depende de $a$, y eliminar el término que no depende de $a$:

\begin{equation}
\sqrt{a(\delta)}= -\frac{1}{2 \sqrt{\pi}} \int\limits_{\delta}^{\infty} dy y^{-3/2} e^{-y a}.
\end{equation}

Hagamos ahora $a(\delta) = \varepsilon^2 = p_3^2+p_{\perp}^2+m^2-2\kappa s B\sqrt{p_{\perp}^2+m^2}$. En consecuencia:

\begin{equation}\label{energyintegral}
\varepsilon = -\frac{1}{2 \sqrt{\pi}} \int\limits_{\delta}^{\infty} dy y^{-3/2} e^{- y(p_3^2+p_{\perp}^2+m^2-2\kappa s B\sqrt{p_{\perp}^2+m^2})}.
\end{equation}

Sustituyendo la Ec.~(\ref{energyintegral}) en la Ec.~(\ref{Grand-potential-vac-1}) se obtiene para el potencial termodinámico de vacío la expresión siguiente:

\begin{equation}\label{Grand-potential-vac-2}
\Omega_{vac}=-\frac{1}{8 \pi^{5/2}}\sum_{s=-1,0,1}\int\limits_{\delta}^{\infty} dy y^{-3/2} \int\limits_{0}^{\infty}dp_{\perp} p_{\perp} \int\limits_{-\infty}^{\infty}dp_3 e^{- y(p_3^2+p_{\perp}^2+m^2-2\kappa s B\sqrt{p_{\perp}^2+m^2})}.
\end{equation}

Luego de realizar la integral sobre $p_3$, la Ec.~(\ref{Grand-potential-vac-2}) queda:

\begin{equation}\label{Grand-potential-vac-3}
\Omega_{vac}=-\frac{1}{8 \pi^{2}}\sum_{s=-1,0,1}\int\limits_{\delta}^{\infty} dy y^{-2} \int\limits_{0}^{\infty}dp_{\perp} p_{\perp} e^{- y(p_{\perp}^2+m^2-2\kappa s B\sqrt{p_{\perp}^2+m^2})}.
\end{equation}

Hagamos ahora el cambio de variables $z = \sqrt{m^2+p_{\perp}^2} - s \kappa B$ con el cual la Ec.~(\ref{Grand-potential-vac-3}) puede escribirse como: 

\begin{equation}\label{Grand-potential-vac-4}
\Omega_{vac}=-\frac{1}{8 \pi^{2}}\sum_{s=-1,0,1} \left \{\int\limits_{\delta}^{\infty} dy y^{-3} e^{- y(m^2-2 m s \kappa B)}
+ s \kappa B \int\limits_{\delta}^{\infty} dy y^{-2} \int\limits_{z_1}^{\infty}dz e^{- y(z^2 - s^2 \kappa^2 B^2)}
 \right \},
\end{equation}

\noindent con $z_1 = m - s \kappa B $.

La Ec.~(\ref{Grand-potential-vac-4}) admite todavía una mayor simplificación a partir de hacer un segundo cambio de variables $w=z-z_1$ en su último término, sumar sobre el spin y recordar que $b = B/B_c$ con $B_c = m / 2 \kappa$:

\begin{equation}\label{Grand-potential-vac-5}
\Omega_{vac}=-\frac{1}{8 \pi^{2}}\left \{ \int\limits_{\delta}^{\infty} dy y^{-3} e^{- ym^2} (1+2 \cosh{[m^2 b y]})
+ m b \int\limits_{\delta}^{\infty} dy y^{-2} \int\limits_{0}^{\infty}dw e^{- y(m - w)^2} \sinh[m b (m - w) y] \right \}.
\end{equation}

Para tomar el límite $\delta \rightarrow 0$ sustraeremos de $1+2 \cosh{[m^2 b y]}$ y $ \sinh[m b (m - w) y]$ el primer término de sus expansiones en series alrededor de cero con respecto a las cantidades $m^2 b y$ y $m b (m - w) y$ respectivamente. Con ello se obtiene la siguiente expresión para el potencial termodinámico de vacío:

\begin{eqnarray}\label{Grand-potential-vac-6}
\Omega_{vac}=-\frac{1}{8 \pi^{2}}\int\limits_{0}^{\infty} dy y^{-3} e^{- ym^2} \{2 \cosh{[m^2 b y]} - 2 - m^4 b^2 y^2 \}- &\\
-\frac{m b}{8 \pi^{2}} \int\limits_{0}^{\infty} dy y^{-2} \int\limits_{0}^{\infty}dw e^{- y(m - w)^2}\left \{ \sinh[m b (m - w) y]- m b (m - w) y - \frac{[m b (m - w) y]^3}{6} \right \}. \nonumber
\end{eqnarray}

Integrar sobre $y$ y $w$ en la Ec.~(\ref{Grand-potential-vac-6}) conduce a la Ec.~(\ref{Grand-Potential-vac}), que es la forma final para el potencial termodinámico de vacío.

A través de un procedimiento similar se obtiene la Ec.~ (\ref{Grand-Potential-2Dvacreg}) para la contribución al potencial termodinámico del vacío en el caso del gas de bosones vectoriales en una dimensión.

%% file: appH.tex
\chapter{Potencial termodin\'amico del gas de bosones escalares cargados en presencia de campo magn\'etico}
\label{appH}

En este ap\'endice se muestra el c\'alculo del potencial termodin\'amico $\Omega^{pp}$ del gas de bosones vectoriales cargados  (Ecs.~(\ref{omegasc})) a partir del espectro de dichas part\'iculas \cite{ROJAS1996148}:

\begin{equation}\label{spectrumsc}
	\varepsilon^{pp}(p_3,n,B) =\sqrt{ m_{pp}^2+p_3^2+2qB(n+1/2)}
\end{equation}

\noindent donde $n=0,1,2,...$ denota los niveles de Landau, $m_{pp}$ es la masa y  $q$ es la carga eléctrica de las partículas. 

Al igual que para los bosones vectoriales neutros (Cap\'itulo \ref{cap2}), el potencial termodin\'amico de los bosones escalares cargados puede separarse en sus contribuciones estad\'istica y de vac\'io, $\Omega^{pp}= \Omega^{pp}_{st}+\Omega^{pp}_{vac}$, que vienen dadas por:

\begin{equation}\label{potencialvstsc}
\Omega^{pp}_{st}(B,\mu^{pp},T)= \frac{q B}{4 \pi^2 \beta} \sum_{n=0,1...} \int\limits_{-\infty}^{\infty}dp_3 \ln \left((1-e^{-(\varepsilon^{pp}(p_3,n, B)- \mu^{pp})\beta})(1-e^{-(\varepsilon^{pp}(p_3,n,B)+ \mu^{pp})\beta})\right) ,
\end{equation}

\noindent y: 

\begin{equation}\label{potencialvascc}
\Omega^{pp}_{vac} (B)= \frac{q B}{4 \pi^2}\sum_{n=0,1,...}\int\limits_{-\infty}^{\infty} dp_3  \varepsilon^{pp}(p_3,n,B,s).
\end{equation}

Para calcular $\Omega^{pp}$ tomaremos el límite de baja temperatura $T<<m_{pp}$ y separaremos en dos casos: campo débil (CD) $T >> q B/m_{pp}$ y $B << B_c^{pp} = m^2_{pp}/q$; y campo fuerte (CF) $T << q B/m_{pp}$.

En el l\'imite de baja temperatura $e^{-(\varepsilon^{pp}(p_3,n,B)+ \mu)\beta}<<1$ y la contribuci\'on de las antipart\'iculas puede ser despreciada. En consecuencia $\Omega^{pp}_{st}(B,\mu,T)$ toma la forma:

\begin{equation}\label{potencialvstsc1}
\Omega^{pp}_{st}(B,\mu^{pp},T)= \frac{q B}{4 \pi^2 \beta} \sum_{n=0,1...} \int\limits_{-\infty}^{\infty}dp_3 \ln (1-e^{-(\varepsilon^{pp}(p_3,n, B)- \mu^{pp})\beta}) ,
\end{equation}

\noindent que con el uso del desarrollo en serie del logaritmo puede reescribirse como:

\begin{eqnarray}\label{potencialvstsc2}
\Omega^{pp}_{st}(B,\mu^{pp},T)&=- \frac{q B}{4 \pi^2 \beta}\sum_{k=1,2,...} \sum_{n=0,1...} \int\limits_{-\infty}^{\infty}dp_3 \frac{e^{-k(\varepsilon^{pp}(p_3,n, B)- \mu^{pp})\beta}}{k}\\ \label{potencialvstsc3}
&=-\frac{q B}{4 \pi^2 \beta}\sum_{k=1,2,...} \sum_{n=0,1...} \varepsilon^{pp}(n) \frac{e^{k \beta \mu^{pp}} K_{1}(k \beta \varepsilon^{pp}(n))}{k},
\end{eqnarray}

\noindent con $K_{l}(x)$ la funci\'on de McDonald de orden $l$ y $\varepsilon^{pp}(n) = \sqrt{m^2_{pp}+2 q B(n+1/2)}$. 

En el caso de campo d\'ebil, como $b^{pp} =B/B_c^{pp} <<1 $, la suma sobre los niveles de Landau puede aproximarse por una integral y la Ec.~(\ref{potencialvstsc3}) se transforma en \cite{Khalilov1997}:

\begin{eqnarray}\label{potencialvstsc4}
\Omega^{pp}_{st}(B,\mu^{pp},T) &=- \frac{3 (\varepsilon^{pp})^3}{2 \pi^2 \beta}\sum_{k=1,2,...} \frac{e^{k \beta \mu^{pp}}}{k}  \int\limits_{1}^{\infty}dz z^2 K_{1}(z k \beta \varepsilon^{pp})\\\label{potencialvstsc5}
&= - \frac{3 (\varepsilon^{pp})^2}{2 \pi^2 \beta^2}\sum_{k=1,2,...} \frac{e^{k \beta \mu^{pp}}}{k}  K_{2}(k \beta \varepsilon^{pp}),
\end{eqnarray}

\noindent donde $\varepsilon^{pp} = m_{pp} \sqrt{1+b^{pp}}$. Para pasar de la Ec.~(\ref{potencialvstsc3}) a la Ec.~(\ref{potencialvstsc4}) se hizo adem\'as el cambio de variables $z=\sqrt{1+2nb^{pp}/(1+b^{pp})}$. 

Teniendo en cuenta nuevamente que $T<<m_{pp}$, la Ec.~(\ref{potencialvstsc5}) puede aproximarse por:

\begin{eqnarray}\label{potencialvstsc6}
\Omega^{pp}_{st}(B,\mu^{pp},T) \cong - \frac{3 (\varepsilon^{pp})^{3/2}}{(2 \pi)^{3/2} \beta^{5/2}} \sum_{k=1,2,...} \frac{e^{k \beta (\mu^{pp}-\varepsilon^{pp})}}{k^{5/2}}, 
\end{eqnarray}

\noindent con lo cual la parte estad\'istica del potencial termodin\'amico de los bosones escalares cargados en el l\'imite de campo d\'ebil es:

\begin{equation}\label{potencialvstsc7}
\Omega^{pp}_{st}(B,\mu^{pp},T) = -\frac{3 (\varepsilon^{pp})^{3/2} Li_{5/2}(e^{\beta(\mu^{pp}-\varepsilon^{pp})})}{(2 \pi)^{3/2} \beta^{5/2}},
\end{equation}

\noindent siendo $Li_{n}(x)$ la funci\'on polilogar\'itmica de orden $n$.

Para el caso de campo fuerte, como $T<<qB/m_{pp}$, todas las part\'iculas se encuentran en el primer nivel de Landau, $n=0$ y por tanto:

\begin{eqnarray}\nonumber
\Omega^{pp}_{st}(B,\mu^{pp},T)&=- \frac{q B}{4 \pi^2 \beta}\sum_{k=1,2,...} \sum_{n=0,1...} \int\limits_{-\infty}^{\infty}dp_3 \frac{e^{-k(\varepsilon^{pp}(p_3,n, B)- \mu^{pp})\beta}}{k}\\\nonumber &=- \frac{q B}{4 \pi^2 \beta}\sum_{k=1,2,...}\int\limits_{-\infty}^{\infty}dp_3 \frac{e^{-k(\varepsilon^{pp}(p_3,0, B)- \mu)\beta}}{k}\\\label{potencialvstsc8}
&=-\frac{q B  \varepsilon^{pp}}{2 \pi^2 \beta}\sum_{k=1,2,...} \frac{e^{k \beta \mu^{pp}} K_{1}(k \beta (\varepsilon^{pp})}{k^2}.
\end{eqnarray}

Si adem\'as recordamos que $T<<m_{pp}$, en el caso de campo fuerte $\Omega_{st}^{pp}(B,\mu^{pp},T)$ es:

\begin{eqnarray}\nonumber
\Omega^{pp}_{st}(B,\mu^{pp},T)& \cong -\frac{q B (\varepsilon^{pp})^{1/2}}{(2 \pi^2)^{3/2} \beta^{3/2}}\sum_{k=1,2,...} \frac{e^{k \beta (\mu^{pp}-\varepsilon^{pp})}}{k^{3/2}}\\\label{potencialvstsc9}
&=-\frac{m^2_{pp} b (\varepsilon^{pp})^{1/2}}{(2 \pi)^{3/2} \beta^{3/2}}Li_{3/2}(e^{k \beta (\mu^{pp}-\varepsilon^{pp})}).
\end{eqnarray}

Por otra parte, siguiendo el procedimeinto presentado en el Ap\'endice \ref{appD} es posible comprobar que, tanto para campo fuerte como para campo d\'ebil, la contribuci\'on de vac\'io $\Omega_{vac}^{pp}(B)$ tiende a cero al ser regularizada. De forma que el potencial termodin\'amico total del gas magnetizado de bosones escalares cargados queda:

\begin{equation}\nonumber
\centering
\Omega^{pp} = \left \{
\begin{array}{ll}
-\frac{3 (\varepsilon^{pp})^{3/2} Li_{5/2}(e^{\beta(\mu^{pp}-\varepsilon^{pp})})}{(2 \pi)^{3/2} \beta^{5/2}} , & \, {\bf CD} \\[10pt]
-\frac{3 m_{pp}^2 b (\varepsilon^{pp})^{1/2} Li_{3/2}(e^{\beta(\mu^{pp}-\varepsilon^{pp})})}{(2 \pi)^{3/2} \beta^{3/2}}, & \, {\bf CF}
\end{array}
\right.,
\end{equation}

\noindent que es precisamente la Ec.~(\ref{omegasc}). A partir de ella, las expresiones dadas por las Ecs.~(\ref{EoSCSB1}) para las EdE del gas de bosones escalares cargados en presencia de un campo magn\'etico se obtienen de manera an\'aloga a las del gas vectorial neutro (ver Cap\'itulo~\ref{cap2}).